\documentclass[10pt,b5paper,twoside,openright]{scrreprt} 

\usepackage{float}            % more control over floating objects
\usepackage{alltt}            % verbatim environment 

\usepackage{amssymb}          % AMS-Symbols, e.g. \mathbb{N} 
\usepackage{amsmath}          % 
\usepackage{amstext}          % 
\usepackage{graphicx}         % 
\usepackage{graphics}         %
\usepackage[dvips]{color}     %
\usepackage{mathbbol}

\newcommand{\Op}[1]{\hat{#1}}
\newcommand{\M}[1]{{\bf #1}}

\newfont{\tensy}{cmsy10}

\newcommand{\bra}[1]{\big\langle#1\big|}
\newcommand{\ket}[1]{\big|#1\big\rangle}
\newcommand{\bracket}[2]{\big\langle#1 \bigm| #2\big\rangle}
\newcommand{\sbra}[1]{\langle#1|}
\newcommand{\sket}[1]{|#1\rangle}

\newcommand{\abs}[1]{\left|#1\right|}
\newcommand{\su}{\uparrow}
\newcommand{\sd}{\downarrow}
\newcommand{\Tr}{{\rm Tr}}
\newcommand{\dprime}{{\prime\prime}}

\begin{document}

%%%%%%%%%%%%%%%%%%%%%%%%%%
%% Title page with date %%
%%%%%%%%%%%%%%%%%%%%%%%%%%
\title{Spin transport in nanocontacts and nanowires}

\author{
  {\bf Tesis doctoral} \\
  \\
  por \\
  \\
  David Jacob \\ 
  \\
  Director de tesis: Juan Jos\'e Palacios Burgos \\
  Departamento de F\'isica Aplicada \\
  Universidad de Alicante
}

\date{Alicante en febrero de 2007}

\maketitle

%%\cleardoublepage

%%%%%%%%%%%%%%
%% Abstract %%
%%%%%%%%%%%%%%
\begin{abstract} 
In this thesis we study electron transport through magnetic nanocontacts and 
nanowires with ab initio quantum transport calculations. The aim is to gain a 
thorough understanding of the interplay between electrical conduction and 
magnetism in atomic-size conductors and how it is affected by different aspects 
as e.g. the atomic structure and the chemical composition of the conductor. To 
this end our ab initio quantum transport program ALACANT which combines the 
non-equilibrium Green's function formalism (NEGF) with density functional theory 
(DFT) calculations has been extended to describe spin-polarized systems. We 
present calculations on nanocontacts made of Ni as a prototypical magnetic material. 
We find that atomic disorder in the contact region strongly reduces the a priori high 
spin-polarization of the conductance leading to rather moderate values of the so-called 
ballistic magnetoresistance (BMR). On the other hand, we show that the adsorption of 
oxygen in the contact region could strongly enhance the spin-polarization of the 
conduction electrons and thus BMR by eliminating the spin-unpolarized s-channel. Finally, 
we show that short atomic Pt chains suspended between the tips of a nanocontact are magnetic 
in contrast to bulk Pt. However, this emergent nanoscale magnetism barely affects the overall 
conductance of the nanocontact making it thus difficult to demonstrate by simple conductance 
measurements. In conclusion, we find that spin-transport through atomic-scale conductors is 
quite sensitive to the actual atomic structure as well as to the chemical composition of the 
conductor. This presents both, opportunities and challenges for the realization of future 
nanoscale spintronics devices.
\end{abstract}

%%%%%%%%%%%%%%%%%%%%%%%%
%% Resumen en espanol %%
%%%%%%%%%%%%%%%%%%%%%%%%
%%\include{resumen}
%%%%%%%%%%%%%%%%%%%%%%%%

%%\cleardoublepage

%%%%%%%%%%%%%%%%%%%%%%
%% Acknowledgements %%
%%%%%%%%%%%%%%%%%%%%%%
%%\newpage{\pagestyle{empty}\cleardoublepage}

\chapter*{Acknowledgments}

I am especially grateful to Prof. Juanjo Palacios and Dr. Joaquin 
Fern\'andez-Rossier for directing this work. The discussions on 
physical problems were always both, enlightening and enjoyable. 
Their enthusiasm for the physics was really inspiring and motivating. 
I also would like to thank Dr. Carlos Untiedt, Dr. Mar\'ia Jos\'e Caturla,
Prof. Enrique Louis, Prof. Jos\'e Antonio Verg\'es and Dr. Guillermo 
Chiappe for fruitful discussions. Special thanks to James McDonald
who built the Beowulf cluster facility here in the Applied Physics 
Department without which this work would not have been possible.
I thank the MECD for financial support under grant No. UAC-2004-0052. 
I feel grateful to all my colleagues, Cristophe, Deborah, Eladio, Federico, 
Fernando, Giovanni, Igor, Lo\"ic, Martin, Natamar, Pedro, Reyes and 
Richard which have made my life here in Alicante really enjoyable by sharing 
a lot of beers, barbecues, paellas, tapas, and a lot more with them. 
I would like to thank my family for their constant support. 
Finally but most importantly, I would like to thank Mar\'ia for her 
love and support, and especially for her patience during the last months.

%%%%%%%%%%%%%%%%%%%%%%%%%%%%%%%%%%%%%%%%%%%%%
%% Table of contents                       %%
%%%%%%%%%%%%%%%%%%%%%%%%%%%%%%%%%%%%%%%%%%%%%
\tableofcontents
%%%%%%%%%%%%%%%%%%%%%%%%%%%%%%%%%%%%%%%%%%%%%

%%%%%%%%%%%%%%%%%%%%%%%%%%%%%%%%%%%%%%%%%%%%%
%% Chapter 1 - Introduction                %%
%%%%%%%%%%%%%%%%%%%%%%%%%%%%%%%%%%%%%%%%%%%%%

%%%%%%%%%%%%%%%%%%%%%%
\chapter{Introduction}
%%%%%%%%%%%%%%%%%%%%%%

%% Microelectronics
The invention of the integrated circuit (IC) in 1959 by J. Kilby 
(Nobel price in 2000 together with H. Kroemer and Z. I. Alferov)
triggered the stunning and still ongoing development of computer
technology which has lead to ever faster, cheaper, and smaller 
computers. An IC is an electronic circuit where all electronic 
components (transistors, capacitors, interconnects) are integrated 
on a single silicon chip. The successive improvement of the 
fabrication techniques and the introduction of new materials 
allowed to constantly decrease the sizes of the IC components
so that an increasing number of them could be integrated on a 
single IC. This development has lead to increasingly powerful 
and faster computer chips. 
%% Mass storage devices
The other basic ingredient of modern computer systems are 
non-volatile mass storage devices presented in today's computers by 
hard drives which store data in form of magnetic bits 
on magnetic disks (hard disks). Here the improvement of fabrication 
techniques, introduction of new materials and new concepts for the
the read and write mechanism of the hard drives allowed to constantly 
decrease the minimum area needed to record a magnetic bit making them faster 
and increasing their data capacities by several orders of magnitude since
the invention in 1979.

%% Atomic limit
Thus miniaturization has become the leading paradigm of today's 
computer industry and a lot of effort in industrial research and 
development is dedicated to further decreasing the minimum feature 
size of semiconductor chips or the areas of the magnetic bits on
hard disks. However, this miniaturization trend 
cannot go on forever since the atomic scale presents an ultimate 
limit which can never be surpassed and possibly not even reached. 
This is not a kind of hypothetical scenario but rather relevant for 
the development of microprocessor chips in the near future as is 
impressively demonstrated by today's most advanced microprocessors 
and memory chips whose smallest features (the gate length) have 
already reached a size in the order of 30 nm, and IC transistors 
with a gate length of only 10nm are currently under investigation
\cite{Doyle:itj:02}. The miniaturization of IC components has in fact 
reached a level where atomic structure effects like \emph{electro-migration} 
processes \cite{Black:radc:68} become an issue because they can seriously 
limit the usability and lifetime of ICs. Another serious problem are the so-called
subthreshold leakage currents between the electrodes of the transistor which increase 
considerably as the size of the transistors shrinks and are responsible for 
a grand part of the power consumption loss in microchips \cite{Kao:02}. 
Similar problems arise when reducing the dimensions of magnetic bits.
Moreover, when finally reaching the nanoscale, inevitably quantum 
effects will come into play and seriously challenge current silicon
based IC technology possibly demanding radically different approaches. 
{\it Molecular electronics} and {\it spintronics} are two promising 
examples of radically new strategies for information processing.

%% Molecular electronics
Molecular electronics aims at employing single molecules as the ultimate 
electronic components for realizing nanoscale electronic circuits.
Indeed the use of single molecules as rectifiers for electrical current
has been proposed as early as 1974 by Aviram and Ratner \cite{Aviram:cpl:74}.
But not until the advent of the scanning probe microscopes was it possible 
to study and manipulate material properties at the molecular or atomic level
let alone electrically contact individual molecules.
The invention of the scanning tunneling microscope (STM) by G. Binning and 
H. Rohrer in 1981 (Nobel price in 1986 together with E. Ruska) 
\cite{Binning:prl:82} made it for the first time possible to study 
(metallic) surfaces and molecules adsorbed on them with atomic resolution. 
By contacting an individual molecule adsorbed on a metal surface with the STM 
tip it is possible to measure the conductance of this molecular conductor. 
This technique was first employed to measure the conductance of individual 
C$_{60}$ molecules \cite{Joachim:prl:95,Joachim:cpl:97}. Since then more 
conductance measurements of individual molecules have been reported in the 
literature \cite{Reed:science:97,Kergueris:prb:99,cui:science:01,
  Smit:nature:02,Champagne:nl:05}. However, establishing electrical 
contacts with individual molecules and measuring their conductance remains a 
formidable task and care must be taken to assure that a molecule has indeed
been contacted \cite{Champagne:nl:05}. 

%% Nanocontacts
Using an STM it is also possible to fabricate atomic-size nanocontacts where two 
sections of a metal wire are connected via a constriction of just a few atoms in 
diameter which thus represent the ultimate electrical conductors with respect to 
size \cite{Agrait:physrep:03}. Nanocontacts are formed when an STM tip is pressed 
into the substrate and then slowly retracted until an atomic-size neck is formed 
\cite{Gimzewski:prb:87}. Another technique to fabricate very stable and 
reproducible nanocontacts in an efficient and cheap way is the mechanically 
controllable break junction technique (MCBJ)\cite{Muller:prl:92} where a notched 
thin wire mounted on a bending beam is broken in controlled way. The bending can 
be fine-controlled by a piezo element allowing a very precise control over the 
separation of the two sections of the wire. Just a third method to fabricate 
nanocontacts is by electrodeposition \cite{Morpurgo:apl:99,Li:nanotech:99} where 
the nanocontacts are electro-chemically deposited between two macroscopic 
electrodes. Interesting phenomena for atomic-size nanocontacts have been observed 
in experiments like the formation of monatomic chains suspended between Gold or 
Platinum contacts, see e.g. \cite{Untiedt:prb:02}.

%% Spintronics
Another approach that promises to revolutionize conventional electronics is the field
of spintronics \cite{Wolf:science:01} which aims to combine the traditionally separated 
fields of magnetic information storage and semiconductor electronics in order to build 
more powerful electronic devices that exploit the electron spin in addition to the 
electron charge. The example that best illustrates the spintronics philosophy is the 
so-called Magnetic Random Access Memory (MRAM) \cite{Parkin:jap:99} which combines the 
advantages of conventional magnetic data storage (hard drives) and conventional electronic 
random excess memory (RAM) into a single device in order to achieve at the same time 
non-volatile memory cells that are as fast as conventional RAM cells.

%% Spin-valves and GMR effect
An essential ingredient for spintronics is the generation of spin-polarized electron 
currents which is typically accomplished by passing the electrical current through
a ferromagnetic metal. The other basic ingredient are spin-valves which are devices 
that change their resistivity depending on the polarization of the spin-current
and thus allow to detect the spin-polarization of an electrical current. 
An example for an actual spin-valve device are the giant magneto-resistance (GMR) devices
consisting of alternating magnetic and non-magnetic metal multilayers that display a 
strong sensitivity of the electrical current to the relative orientation of the 
magnetizations of the magnetic layers \cite{Baibich:prl:88}. Soon after its discovery
in 1988, the GMR effect was exploited to improve the sensitivity of read heads in hard drives 
which before had been based on simple magnetic induction or the much smaller anisotropic 
magneto-resistance effect displayed by bulk metals. This in turn allowed to decrease the 
size of the magnetic bits and thus to increase the data storage density of the hard disks
dramatically. Consequently, GMR read heads can now be found in the hard drives of every 
modern computer. 

%% Magnetic tunnel junctions and TMR effect
Magnetic tunnel junctions (MTJs) are similar to GMR devices but feature an insulating layer 
instead of the non-magnetic metal layer separating the ferromagnetic metal layers which presents 
a tunnel barrier for the electrons flowing between the ferromagnetic layers. MTJs display a MR 
known as tunneling magneto-resistance (TMR) which was first demonstrated by Julliere 
\cite{Julliere:pl:75}. The TMR spin-valve is a crucial ingredient for an efficient realization 
of the above described MRAM device \cite{Gallagher:ibmjrd:06}. In the earlier experiments TMR 
was found to be much smaller than GMR, but very recently new material combinations (Fe-MgO-Fe) 
for the MTJs motivated by theoretical studies \cite{Butler:prb:01,Mathon:prb:01} have lead to 
a dramatic increase of the TMR which now actually exceeds GMR values 
\cite{Djayaprawira:apl:05,Gallagher:ibmjrd:06}. 

%% Nanoscale spintronics
GMR and TMR spin-valves are nanoscale devices in the sense that the thicknesses of the layers 
making up the spin-valve structures are on the nanometer scale, and can be made as thin as 
5\r{A} with modern growth methods, so that electron transport is governed by quantum effects.
In fact, these devices actually only work because of quantum effects, i.e. the coherent 
transmission of a spin-polarized current through the non-magnetic layer. Moreover, when finally 
shrinking the other two dimensions of a spin-valve to the nanoscale, quantum and atomic structure 
effects will become even more important. Thus it is of fundamental importance to gain a 
thorough understanding of the interplay between magnetism and electrical conduction at the 
atomic scale. Nanocontacts and nanowires made from ferromagnetic metals allow to study this 
interplay between magnetism and electrical conduction or in more fundamental terms the 
interplay between electron spin and charge flow in the smallest possible magnetic conductors. 

An important question is whether GMR or TMR effects survive when the other two device 
dimensions are scaled down to the nanoscale, or whether other magneto-resistance (MR) effects 
emerge at the nanoscale which could be exploited for the realization of nanoscale spintronics 
devices. Therefore, measuring the MR of ferromagnetic nanocontacts has recently attracted a lot 
of interest \cite{Oshima:apl:98,Garcia:prl:99,Ono:apl:99,Chung:prl:02,Viret:prb:02,Chopra:prb:02,
Hua:prb:03,Untiedt:prb:04,Sullivan:prb:05,Gabureac:prb:04,Bolotin:nl:06,Keane:apl:06}. 
Indeed, some groups have found a huge MR (exceeding even the GMR effect) for Ni nanocontacts 
which was coined ballistic magneto-resistance (BMR) for its supposed origin in the ballistic
scattering of spin polarized electrons on a sharp domain wall (DW) which should form at the 
atomic neck of the nanocontact\cite{Bruno:prl:99,Tatara:prl:99,Imamura:prl:00}. However, the 
possibility of huge BMR in ferromagnetic nanocontacts has been a controversial topic since its 
discovery, and is one of the principal topics of this thesis (See Ch. \ref{ch:Ni-nanocontacts}
and Ch. \ref{ch:NiO-chains}).

Integrating the fields of molecular electronics with that of spintronics is another promising 
approach for realizing nanoscale spintronics devices because of the expected very long 
spin-decoherence times in organic molecules as compared to bulk metals and semiconductors. 
Strong indications on the possibility of integrating the two fields come from a few recent 
experiments indicating e.g. a long spin-flip scattering length for carbon nanotubes 
\cite{Tsukagoshi:nature:99}, spin injection from strongly spin polarized materials into carbon 
nanotubes \cite{Hueso:nature:07}, and spin transport through organic molecules 
\cite{Ouyang:Science:03}. Since ferromagnetic nanocontacts are a basic ingredient for molecular 
spintronics for injecting spin-polarized currents into the molecules, it is also in this
context important to gain a solid understanding of electrical conduction and magnetism in 
nanocontacts.

In this thesis electron transport through magnetic nanocontacts and nanowires is investigated 
theoretically. The starting point for the theoretical investigation of electron transport through
nanocontacts and nanowires is the Landauer formalism, which is introduced in Ch. \ref{ch:transport}.
The Landauer approach assumes that electron transport through nanostructures is phase coherent,
i.e. decoherence by phase-breaking scattering processes is neglected. This turns out to be a
reasonable assumption at low temperatures and for small bias voltages. Indeed, the Landauer approach 
has been successfully applied for studying electrical transport through metallic nanocontacts, so 
that now we have a good understanding of atomic scale conductors \cite{Brandbyge:prb:97,Cuevas:prl:98:81}.

In order to predict the transport properties of atomic scale nanocontacts and nanowires it is important 
to have a realistic description of the electronic and magnetic structure of the nanocontact taking into 
account its actual atomic structure. This can be achieved most conveniently by ab initio 
electronic structure methods based on localized atomic orbitals like e.g. GAUSSIAN \cite{Gaussian:03} 
or SIESTA \cite{Ordejon:prb:96}. Ch. \ref{ch:ab-initio} shows how ab initio quantum transport 
calculations based on density functional theory (DFT) \cite{Hohenberg:pr:64} are implemented in the 
ALACANT package \cite{Palacios:prb:01,ALACANT:06} which as a part of this thesis was extended in order 
to describe spin transport in magnetic systems and to incorporate one-dimensional leads calculated 
from first principles in addition to the semi-empirical Bethe lattice electrodes \cite{Palacios:ctcc:05}. 
Other DFT based quantum transport methods are developed by various groups around the world 
\cite{Brandbyge:prb:02,Rocha:prb:06}. 
In Ch. \ref{ch:spin-transport}, the Landauer formalism is generalized to describe transport of
spin-polarized electrons (or in short ``spin transport'') in magnetic nanostructures. In order 
to understand basic aspects of spin transport in nanoscopic conductors like the scattering of 
spin-polarized electrons on domain walls as well as the formation of domain walls in nanoscopic
conductors, the spin-resolved transport formalism is applied to simplified models of magnetic 
materials.

In Ch. \ref{ch:Ni-nanocontacts} spin transport through Ni nanocontacts is investigated theoretically
with ab initio quantum transport calculations using the afore mentioned ALACANT package. In order to
asses the above discussed possibility of huge BMR in Ni nanocontacts, we calculate the magneto resistance 
due to the formation of a DW at the atomic neck of the nanocontact for different contact geometries. 
We find that BMR of {\it pure} Ni nanocontacts is rather moderate \cite{Jacob:prb:05}, i.e. much smaller 
than the famous GMR effect. This is contrary to the claims of huge BMR found in the first experiments on 
Ni nanocontacts but in agreement with some recent experiments measuring very clean samples under very 
controlled conditions \cite{Bolotin:nl:06,Keane:apl:06}. 

In Ch. \ref{ch:NiO-chains}, we study the electronic and magnetic structure, and transport properties 
of atomic NiO chains, both ideal infinite ones and short ones suspended between the tip atoms of Ni 
nanocontacts. It is found that the presence of a single oxygen atom between the tip atoms of a Ni 
nanocontact increases the spin-polarization of the conduction electrons dramatically, converting
the nanocontact into an almost perfect half-metallic conductor, and leading to huge values of the 
BMR \cite{Jacob:prb:06}. It is also discussed to what extent these results could explain the huge 
MR of Ni nanocontacts obtained in the experiments mentioned above.

In Ch. \ref{ch:Pt-nanowires}, the electronic structure and transport properties of {\it magnetic} Pt 
nanowires is studied. It had been shown before that atomic Pt nanowires can actually become magnetic in 
contrast to bulk Pt due to the lower coordination of the Pt atoms in the atomic chain compared to the 
bulk \cite{Delin:ss:04,Delin:jpcm:04}. We reproduce this result for infinite chains, and find that also
short Pt chains suspended between the tips nanocontacts become magnetic. However, the overall conductance
of the Pt nanocontact is barely affected by the magnetism of the chain, so that simple conductance measurements
of Pt nanocontacts cannot probe the magnetism \cite{Fernandez-Rossier:prb:05}.

Finally, Ch. \ref{ch:conclusions} concludes this thesis with a general discussion of the obtained results,
and their significance in the broader context of spin transport and spintronics in nanoscopic systems. 
Moreover, we will point out open questions and give an outlook on possible future lines of work.

%%% Outline 
%The thesis is organized as follows: In Ch. \ref{ch:transport} the 
%\begin{itemize}
%\item Chapter 2: Quantum theory of electron transport
%\item Chapter 3: Simple Models
%\item Chapter 4: Ab initio transport calculations
%\item Chapter 5: Ni nanocontacts
%\item Chapter 6: Effect of oxygen
%\item Chapter 7: Spin-orbit coupling
%\item Chapter 8: Conclusions and Outlook
%\end{itemize}

%%% Local Variables: 
%%% mode: latex
%%% TeX-master: "thesis"
%%% End: 

%%%%%%%%%%%%%%%%%%%%%%%%%%%%%%%%%%%%%%%%%%%%%

%%%%%%%%%%%%%%%%%%%%%%%%%%%%%%%%%%%%%%%%%%%%%
%% Chapter 2 - Theory of quantum transport %%
%%%%%%%%%%%%%%%%%%%%%%%%%%%%%%%%%%%%%%%%%%%%%

\chapter{Quantum theory of electron transport}
\label{ch:transport}

\begin{figure}[b]
   \begin{center}
     \includegraphics[width=0.95\linewidth]{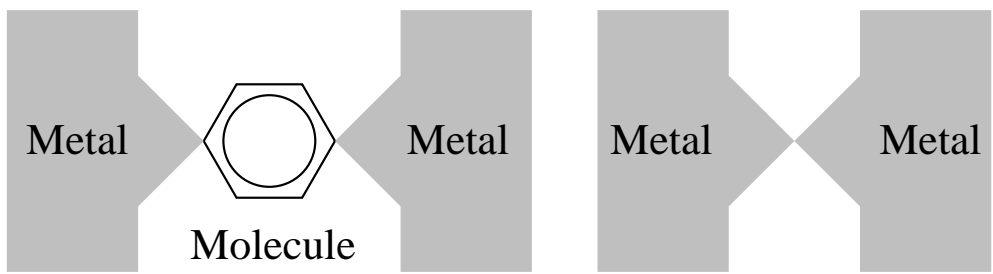}
   \end{center}
   \setcapindent{0cm}   
   \caption{Sketch of typical nanoscopic conductors. Left: A molecule connecting the tips
     of two bulk metal electrodes. Right: A nanoscopic constriction of a metal wire connecting 
     two bulk metal electrodes.
   }
   \label{fig:nanoconductor}
 \end{figure}

The description of electrical conduction through a nanoscopic conductor like a molecule bridging 
the tips of two metal electrodes or a nanoscopic constriction in a metal wire as shown in Fig. 
\ref{fig:nanoconductor} is a challenging problem. In systems of such small size the dimension of 
the conductor becomes comparable to the Fermi wavelength of the conduction electrons, so that the 
transport properties of the conductor are governed by quantization effects demanding a full 
quantum treatment of the transport process. Moreover, for molecular- and atomic-size conductors 
the actual atomic structure of the conductor has a strong effect on the electronic structure and 
transport properties.
%%The nanoscopic conductor presents the 
%%``bottleneck'' for the electrons coming from the electrodes and thus predominantly determines 
%%the resistance of the system. 

%   

Our starting point for a quantum description of electrical conduction is the Landauer formalism
\cite{Landauer:ibmjrd:57,Landauer:philmag:70, Buettiker:prl:86,Buettiker:ibmjrd:88,Datta:book:95} 
which is introduced in Sec. \ref{sec:Landauer}. In Sec. \ref{sec:NEGF} we describe the Landauer 
formalism within the framework of non-equilibrium Green's functions (NEGF) formalism. In Sec. 
\ref{sec:models} we illustrate the derived formalism by applying it to simple model systems.
Finally, in Sec. \ref{sec:discussion} we discuss the validity of the Landauer approach, pointing 
out its problems and limitations, and indicating ways of improving it as well as alternative 
approaches.

%To take into account the effect of the 
%atomic structure on the electronic and transport properties of the conductor it is convenient to 
%combine the NEGF approach with standard {\it ab initio} electronic structure methods like the 
%Hartree-Fock approximation (HFA)\cite{Szabo:book:89} or density functional theory (DFT) \cite{Hohenberg:pr:64} 
%which will be explained in Sec. \ref{sec:ab-initio-trans}. 
%Since the aim of this work is to describe spin-resolved transport through magnetic nanostructures
%a spin-unrestricted formulation of the transport theory is described in Sec. \ref{sec:spintrans}. 
%Finally, in Sec. \ref{sec:discussion} we discuss the validity, limitations and problems of the DFT based 
%transport theory developed in this chapter, and point out some ways for improving the DFT based transport
%theory in a systematic way.

\section{Landauer formalism}
\label{sec:Landauer}

In the Landauer formalism electron transport is considered as a scattering process where the nanoscopic conductor 
acts as a quantum mechanical scatterer for the electrons coming in from the leads. It is further assumed that the 
electrons scatter only elastically 
on the nanoscopic sample, i.e. inelastic scattering e.g. by phonons or by other electrons is neglected so that 
transport becomes phase coherent. Thus electron transport through a nanoscopic conductor is described in terms of 
\emph{non-interacting} quasi particles coming in from the leads and being scattered elastically on the 
nanoscopic device. 

 \begin{figure}[h]
   \begin{center}
     \includegraphics[width=0.4\linewidth]{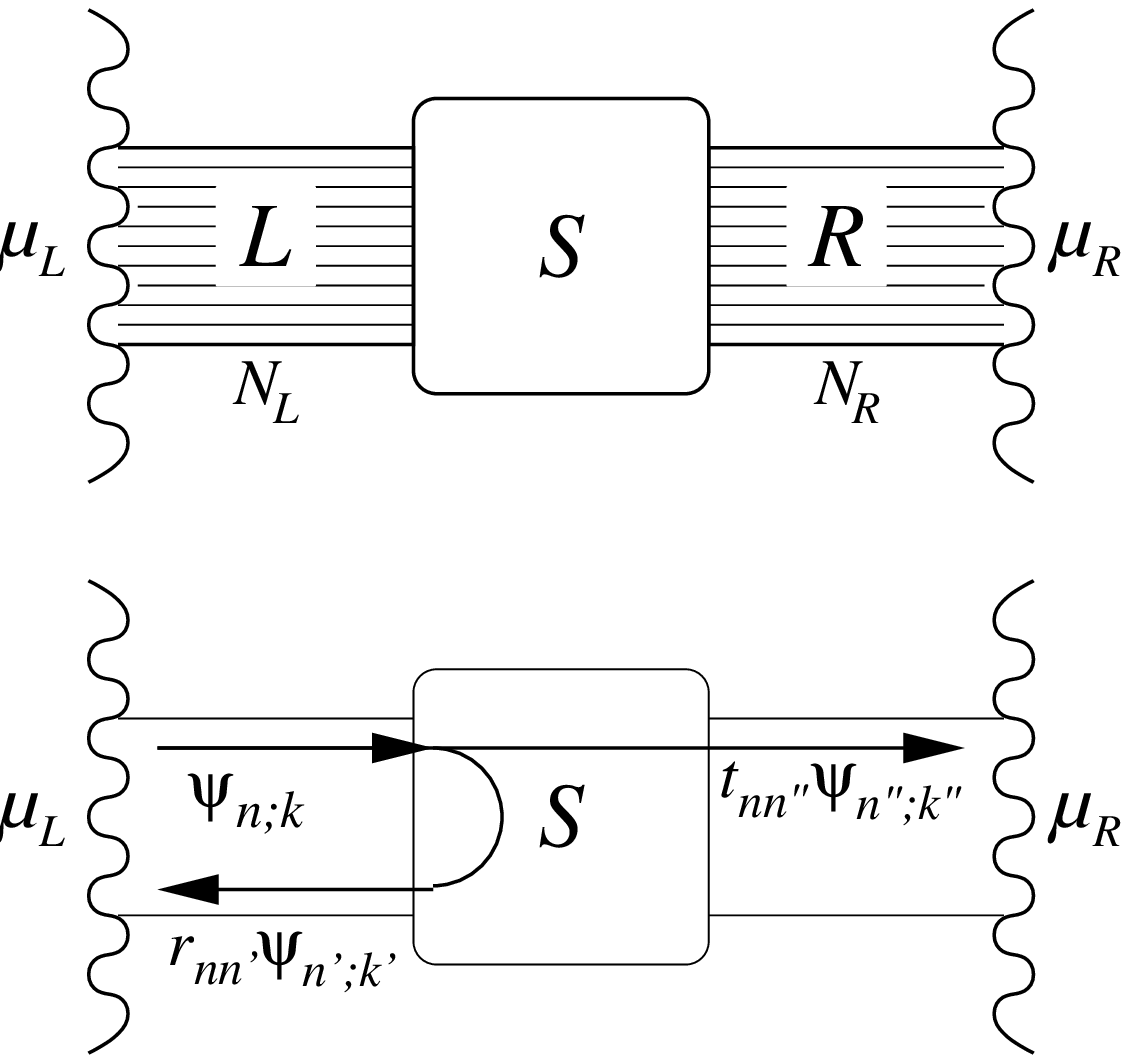}
     \includegraphics[width=0.5\linewidth]{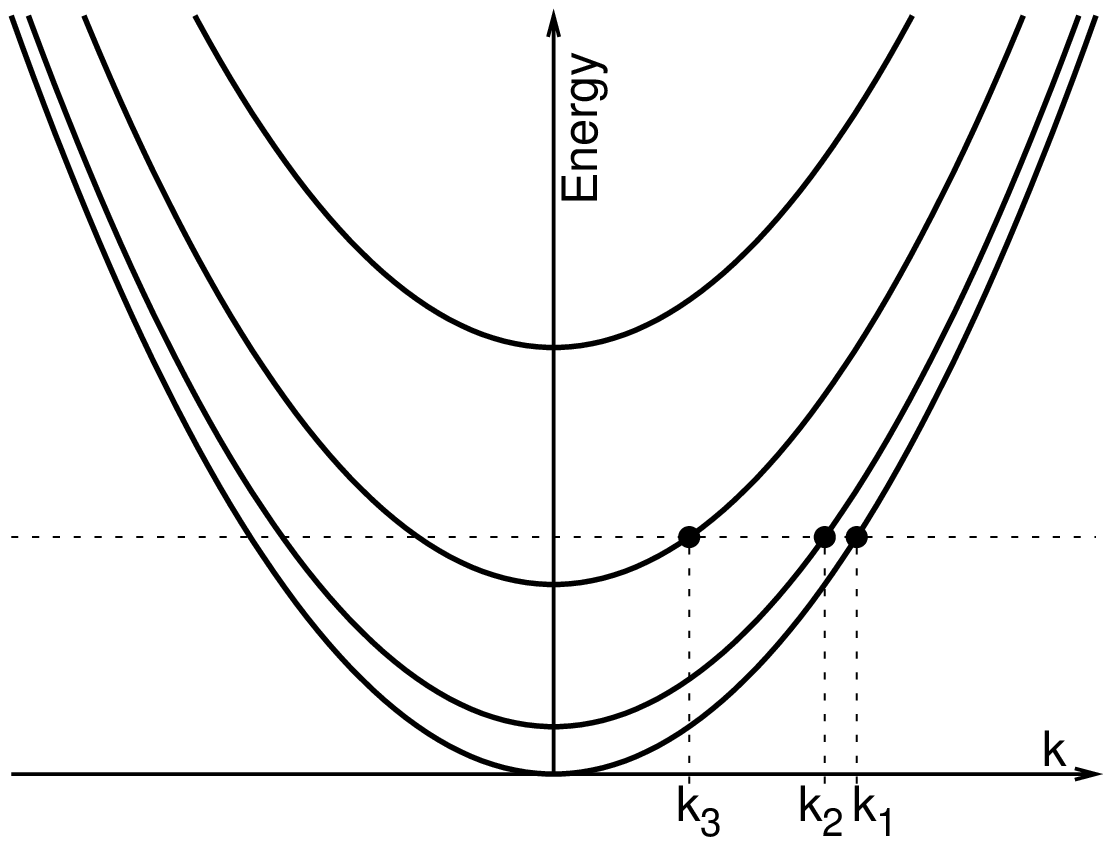}
   \end{center}
   \setcapindent{0cm}   
   \caption{
     Left: Schematic illustration of Landauer formalism (See text for explanation). 
     Right: Energy dispersion of propagating modes for electrons moving freely in one direction
     and confined in the other directions.
   }
   \label{fig:Landauer}
 \end{figure}

Fig. \ref{fig:Landauer} shows how the transport problem is modeled in the Landauer formalism: The central scattering 
region ($S$) containing the nanoscopic conductor is connected via two ideal semi-infinite leads of some finite width 
to two electron reservoirs that are each in thermal equilibrium but at different chemical potentials 
$\mu_L$ and $\mu_R$. The reservoirs are assumed to be reflection-less, i.e. incoming electrons are not reflected 
back to the leads, so that the two reservoirs are independent of each other. 
Due to the finite width of the leads the motion of the electrons perpendicular to the direction of the leads is 
quantized giving rise to a finite number of propagating modes or bands $\{\psi_{n,k}\}$ for a given energy $E$. 
This is illustrated on the right hand side of Fig. \ref{fig:Landauer} for the case of a two-dimensional electron 
gas confined in the vertical direction but moving freely in the horizontal direction. In that case a propagating
wave is given by a plane wave in the $z$-direction $e^{ikz}$ modulated by a transverse wavefunction $\phi_{n}(x,y)$: 
\[
\psi_{n;k}(x,y) = \phi_{n}(x,y) e^{ik z}.
\]
In the more general case of a periodic potential, $V(x,y,z+a)=V(x,y,z)$, generated e.g. by the atomic nuclei in the 
direction of the lead the propagating modes are Bloch waves
\[
\psi_{n;k}(x,y,z) = \sum_{j,\alpha} c_{n;\alpha}(k) \, \phi_{j;\alpha}(x,y,z) \, e^{i k a j},
\]
where the sum goes over all unit cells $j$ of the lead and $\phi_{j\alpha}(x,y,z)$ is a localized wavefunction centered 
in unit cell $j$. From the fact that $\phi_{j\alpha}(x,y,z+a)=\phi_{j-1\alpha}(x,y,z)$, it follows that the Bloch functions 
also have the periodicity of the potential, apart from a trivial phase factor:
\begin{eqnarray*}
  \lefteqn{\psi_{n;k}(x,y,z+a) = \sum_{j,\alpha} c_{\alpha;n}(k) \, \phi_{j;\alpha}(x,y,z+a) \, e^{i k a j} } \\
  && = e^{i k a} \sum_{j,\alpha} c_{\alpha;n}(k) \, \phi_{j-1;\alpha}(x,y,z) \, e^{i k a (j-1)}
  = e^{i k a} \psi_{n;k}(x,y,z).
\end{eqnarray*}

We define $k_n(E)$ as the wave vector corresponding to the band $n$ at the energy $E$ that gives rise to a current 
in the \emph{positive} $z$-direction. The current associated with the propagating mode $n$ is given by its group velocity 
which in turn is given by the derivative of the dispersion relation $\frac{1}{\hbar}\frac{dE_n}{dk}$ for that band:
\begin{equation}
  \label{eq:mode-current}
  j_n(k) := \frac{e}{\lambda} \, v_n(k) = \frac{e}{\lambda\,\hbar} \, \frac{dE_n}{dk}(k),
\end{equation}
where $\lambda$ is the length of the conductor. This is trivial to proof for the case of free electrons where the 
group velocity is directly proportional to the wave vector, $v_n(k)=\hbar k /m_e$. For Bloch waves the situation is 
more complicated, and the group velocity can even have the opposite sign of the wave vector. A proof can be found
e.g. in the book by Ashcroft and Mermin \cite{Ashcroft:book:76}.

Elastic scattering means that an electron with some energy $E$ coming from one of the reservoirs will be scattered 
to some out-moving state with the \emph{same} energy $E$ of one of the leads, so that the process is phase coherent.
Thus an incoming wave at some energy $E$ on the left lead will give rise to a coherent superposition with outgoing 
states of the {\it same} energy $E$ on both leads:
\begin{equation}
  \psi_{n;k_n(E)}^L + \sum_{n^\prime \in N_L} r_{n n^\prime}(E) \psi_{n^\prime;-k_{n^\prime}(E)}^L + 
  \sum_{n^\dprime \in N_R} t_{n n^\dprime}(E) \psi_{n^\dprime;k_{n^\dprime}(E)}^R,
\end{equation}
where $r_{n n^\prime}(E)$ is the probability amplitude for an incoming electron on mode $n$ at energy $E$ to be reflected into the 
outgoing mode $n^\prime$ of the left lead, and $t_{n n^\dprime}(E)$ is the amplitude for the electron to be transmitted into the 
mode $n^\dprime$ of the right lead. Thus an incoming electron on mode $n$ of the left lead will be transmitted with a probability of 
$\sum_{n^\dprime} \|t_{n n^\dprime}(E)\|^2$ to the right lead giving rise to a current density in that lead of magnitude
\begin{equation}
  j_{n}^t(E) := \sum_{n^\dprime \in N_R} \|t_{n n^\dprime}(E)\|^2 j_{n^\dprime}(k_{n^\dprime}(E)).
\end{equation}
%%

%Analogically, an incoming wave on the right lead gives rise to the following superposition:
%%%
%\begin{equation}
%  \psi_{n_i,k(E)}^r + \sum_{n_r} r_{n_i,n_r}^r(E) \psi_{n_r,k^\prime(E)}^r + \sum_{n_r} t_{n_i,n_l}^l(E) \psi_{n_l,k^\prime(E)}^l.
%\end{equation}
%%%
%The superposition of an incoming wave $n_r$ with the backscattered modes in the right lead gives rise to a current density
%in the right lead:
%%%
%\begin{equation}
%   j_{n_r}^r(k) = j_{n_r^\prime}(k) + \sum_{n_r^\prime} \|r_{n_r,n_r^\prime}^r(E)\|^2 j_{n_r^\prime}(k).
%\end{equation}

The left electron reservoir injects electrons into the \emph{right}-moving modes of the left lead up to the chemical
potential $\mu_L$. Thus the transmission of electrons through the nanoscopic conductor gives rise to a current to the 
right in the right electrode. 
%%On the other hand, the injection of electrons from the right electron reservoir up to the chemical
%%potential $\mu_R$ gives rise to a current to the left \emph{and} also to the right due to the backscattering of electrons 
%%on the nanoscopic conductor. Hence the total current in the right lead is given by:
%%
\[
  I_t = \sum_{n \in N_L} \int_{E_{n}(k)<\mu_L} dk \, j_{n}^t(E_n(k)) 
  = \sum_{n \in N_L, n^\prime \in N_R} \int_{-\infty}^{\mu_L} dE \, {\mathcal D}^R_{n^\prime}(E) 
  \|t_{n n^\prime}(E)\|^2 j_{n^\prime}(k_{n^\prime}(E)) 
  %%+ \sum_{n_r} \int_{E_{n_r}(k)<\mu_R} dk \, j^r_{n_r}(k)
  %%- \sum_{n_r} \int_{E_{n_r}(k)<\mu_R} dk \, j_{n_r}(k) \nonumber \\
\]
In the second step we have converted the integral over the wave vectors into an integral over the energy
by making use of the density of states (DOS) ${\mathcal D}^R_n(E)$ projected onto band $n$ of the right lead. For 
one-dimensional systems the DOS is given by the inverse derivative of the dispersion relation of the band, 
${\mathcal D}_n(E) = \frac{1}{2\pi}\frac{dk_n}{dE}$, so that it cancels exactly with the group velocity of the band:
\begin{equation}
  I_t =  \sum_{n \in N_L,n^\prime \in N_R} \frac{e}{h} \int_{-\infty}^{\mu_L} dE \, \|t_{n n^\prime}(E)\|^2
  =  \frac{e}{h} \sum_{n \in N_L} \int_{-\infty}^{\mu_L} dE \, T_n(E),
\end{equation}
where we have defined the {\it transmission per conduction channel} $T_n(E)$ as 
\begin{equation}
  \label{eq:chan-transm}
  T_n(E)= \sum_{n^\prime \in N_R} \|t_{n n^\prime}(E)\|^2.  
\end{equation}

On the other hand the right electron reservoir injects electrons into the \emph{left}-moving 
modes of the right lead up to the chemical potential $\mu_R$ and the transmission of electrons 
through the device region gives rise to a left-directed current in the left lead:
\begin{equation}
  I_t^\prime =  \frac{e}{h} \sum_{n \in N_R} \int_{-\infty}^{\mu_R} dE \, T^\prime_n(E),
\end{equation}
where $T^\prime_{n^\prime}(E)$ now is the transmission probability of channel $n^\prime$ of 
the right lead:
\begin{equation}
  T^\prime_{n^\prime}(E)= \sum_{n \in N_L} \|t^\prime_{n^\prime n}(E)\|^2,
\end{equation}
and $t^\prime_{n^\prime n}(E)$ is the transmission amplitude for a mode $n^\prime$ of the right
lead to be transmitted into mode $n$ of the left lead. Because of time inversion symmetry the
amplitude $t^\prime_{n^\prime n}(E)$ is the same as the the amplitude $t_{n n^\prime}(E)$ for 
the transmission from the left to the right electrode apart from a trivial phase factor. Hence 
the total transmission probability from the left to the right lead $T(E):=\sum_n T_n(E)$ is equal to 
the total transmission probability from the right to the left lead  
$T^\prime(E):=\sum_{n^\prime} T^\prime_{n^\prime}(E)$:
\begin{equation}
  \sum_{n \in N_L} T_n(E) 
  = \sum_{n \in N_L, n^\prime \in N_R} \|t_{n n^\prime}(E)\|^2 
  = \sum_{n \in N_L, n^\prime \in N_R} \|t^\prime_{n^\prime n}(E)\|^2  
  = \sum_{n^\prime \in N_R} T^\prime_{n^\prime}(E).
\end{equation}
Furthermore the summed reflection probability $R^\prime(E)$ for all electrons injected from the 
right reservoir at some energy $E$ is $R^\prime(E) = N_R - T^\prime(E) = N_R - T(E)$. Therefore 
the current composed of \emph{backscattered} electrons (originating from the right reservoir) 
and \emph{transmitted} electrons (originating from the left reservoir) cancels exactly the 
current of the \emph{incoming} electrons coming in from the right electron reservoir. Analogously
the same holds true for the left lead. Thus the net current at some energy $E$ is zero when the current
there is current injection from \emph{both} electrodes at that energy.

Thus assuming a positive bias voltage $V$, so that $\mu_L = \mu_R+eV > \mu_R$, only electrons above 
$\mu_R$ give a net contribution to the total current. Since for energies above $\mu_R$ the only 
contribution to the net current through the right lead is the transmission current $I_t$ of electrons 
coming from the left, the total current for a given bias voltage $V$ is given by the famous Landauer 
formula:
\begin{equation}
  I(V)   =  \frac{e}{h} \sum_{n \in N_L} \int_{\mu_R}^{\mu_L} dE \, T_n(E).
\end{equation}
Taking the derivative with respect to the bias voltage one obtains the corresponding conductance:
\begin{equation}
  G(V) = \frac{\partial I}{\partial V} = \frac{e^2}{h} \sum_{n \in N_L}  T_n(eV),
\end{equation}
where $\frac{e^2}{h}$ is half the fundamental conductance quantum $G_0$. The spin-degree of freedom of the 
electrons is contained in the index $n$ for the channels. Assuming a spin-degenerate system the transmissions 
for up- and down channels are equal, so that a factor of two appears when summing transmissions over the 
spin-degree of freedom, and one obtains the usual Landauer formula with $G_0$ as the proportionality constant. 
Here however, we are interested in magnetic systems, so the transmissions are spin-dependent. 

The transmission amplitudes $t_{nm}(E)$ define an in general non-quadratic matrix $\M t(E) := (t_{nm}(E))$. The 
square of this matrix defines a (quadratic) hermitian matrix called the transmission matrix:
\begin{equation}
  \label{eq:TMat}
  \M T(E) := \M t^\dagger(E) \M t(E) \mbox{ or } T_{nm}(E) = \sum_m t^\ast_{m^\prime n}(E) t_{n^\prime m}(E)
%%  \Rightarrow \M T^\dagger(E) = \M t^\dagger(E) (\M t^\dagger(E))^\dagger = \M T(E).
\end{equation}
The channel transmissions $T_n(E)= \sum_{n^\prime \in N_R} \|t_{n n^\prime}^r(E)\|^2$  are now just the diagonal elements 
of this transmission matrix, and summing up over all channel transmissions in the Landauer formula now corresponds to taking
the trace of the transmission matrix:
\begin{equation}
  G(V) = \frac{e^2}{h} \times \Tr[\M{T}(eV)] \, \mbox{ and }
  \, I(V) =  \frac{e}{h} \sum_{n \in N_L} \int_{\mu_R}^{\mu_L} dE \, \Tr[\M{T}(E)].
\end{equation}

The transmission matrix is the central quantity in the Landauer formalism since it allows to calculate the 
electrical conductance and current-voltage characteristics of a nanoscopic conductor. Depending on the actual 
system a variety of methods exists for obtaining this quantity. For our purpose of describing transport 
through atomic- and molecular-size conductors the NEGF described in the next section is the most appropriate 
approach since it can be combined in a straight-forward manner with {\it ab initio} electronic structure 
methods like density functional theory (DFT) or the Hartree-Fock approximation (HFA) as implemented in 
standard quantum chemistry codes employing atomic orbital basis sets.

\section{Non-equilibrium Green's function formalism}
\label{sec:NEGF}

In this section we will restate the Landauer approach to quantum transport in the language of one-body Green's 
functions \cite{Economou:book:83,Datta:book:95}. To this end we will first introduce the Hamiltonian and overlap 
matrices for the transport problem in a basis set of localized atomic orbitals. We will then derive a Green's 
function for the finite scattering region connected on both sides to semi-infinite leads. From the Green's function (GF)
one can then obtain the (reduced) density matrix and the transmission matrix. Here we will mainly follow the arguments 
presented by Paulsson in his introductory paper on the NEGF \cite{Paulsson:02} but generalize them to non-orthogonal 
basis sets (NOBS) as commonly employed in quantum chemistry packages. A few other derivations of the Landauer formalism 
within the NEGF framework taking into account non-orthogonality of basis sets can be found in the recent literature 
\cite{Viljas:prb:05,Thygesen:prb:06}.

%%We will take into account overlap 
%%between non-orthogonal atomic orbitals explicitly since we would like to combine the NEGF with {\it ab initio} 
%%electronic structure calculations later (Ch. \ref{ch:ab-initio}) which are usually implemented with non-orthogonal 
%%atomic basis sets. The derivations presented here become much more simpler when neglecting the overlap, and many 
%%derivations of the Landauer approach from the NEGF neglecting the overlap can be found in the literature
%%\cite{}. The opposite seems to be true for derivations taking into account overlap. That is the reason why we take 

\subsection{Hamiltonian and overlap}

\begin{figure}[h]
   \begin{minipage}[b][][b]{0.5\linewidth}
     \begin{flushright}
       \includegraphics[width=\linewidth]{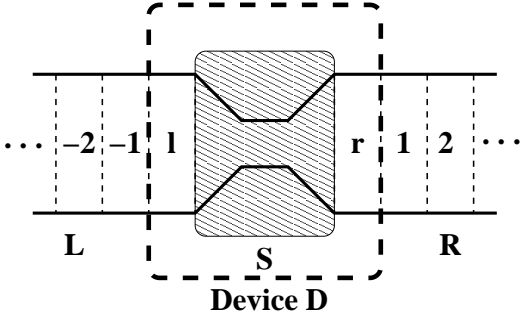}
     \end{flushright}
   \end{minipage}
   \hfill
   \begin{minipage}[b][][b]{0.45\linewidth}
     \setcapindent{0cm}
     \caption{Sketch of the transport problem.
       L: Left (bulk) lead. D: Device. R: right 
       (bulk) Lead. S: Central scattering region
       containing nanoconstriction or contacted
       molecule.See text for further explanations.
     }
     \label{fig:LDR}
   \end{minipage}
 \end{figure}

We divide the system into 3 parts as shown in Fig. \ref{fig:LDR}: The left lead (L), 
the right lead (R), and the intermediate region called device (D) containing the
central scattering region (S). This scattering region can be given by e.g. a molecule 
coupled to metallic contacts, or simply a nanoscopic constriction as indicated
in the figure.

We assume that the leads are only coupled to the scattering region but not to each other.
Thus the device region must be chosen sufficiently large for that to be true.
The Hamiltonian $\hat H$ describing the system is then given by the matrix
\begin{equation}
  \label{eq:HLDR}
  {\bf H} = \left( 
    \begin{array}{ccc}
      {\bf H}_{L}  & {\bf H}_{LD} & {\bf 0}_{LR} \\
      {\bf H}_{DL} & {\bf H}_{D}  & {\bf H}_{DR} \\
      {\bf 0}_{RL} & {\bf H}_{RD} & {\bf H}_{R} 
    \end{array}
  \right).
\end{equation}
Furthermore we assume a non-orthogonal localized basis set. Assuming
again no overlap between atomic orbitals in different leads the overlap 
of the atomic-orbitals of the system is given by the following overlap 
matrix:
\begin{equation}
  \label{eq:SLDR}
  {\bf S} = \left( 
    \begin{array}{ccc}
      {\bf S}_{L}  & {\bf S}_{LD} & {\bf 0}_{LR} \\
      {\bf S}_{DL} & {\bf S}_{D}  & {\bf S}_{DR} \\
      {\bf 0}_{RL} & {\bf S}_{RD} & {\bf S}_{R} 
    \end{array}
  \right).
\end{equation}
As indicated in Fig. \ref{fig:LDR} we subdivide the leads into unit cells (UCs)
which must be chosen sufficiently large so that the coupling between 
non-neighboring unit cells can be neglected.
Thus in general a UC consists of several primitive unit cells (PUCs).
The Hamiltonian matrix ${\bf H}_L$ of the left lead can be subdivided into
sub-matrices in the following manner:
\begin{eqnarray}
  \label{eq:H_L}
  \M{H}_L 
  &=& 
  \left( 
    \begin{array}{cccc}
      & \vdots       & \vdots       & \\
      \cdots & \M H_{-2,-2} & \M H_{-2,-1} & \\
      \cdots & \M H_{-1,-2} & \M H_{-1,-1} &
    \end{array} \right) 
  = \begin{pmatrix}
    \ddots & \ddots         & \ddots         &                & \M 0   \\
    \,     & \M H_1^\dagger & \M H_0         & \M H_1         &        \\
    \,     &                & \M H_1^\dagger & \M H_0         & \M H_1 \\
    \M 0   &                &                & \M H_1^\dagger & \M H_0
  \end{pmatrix}
\end{eqnarray}
Analogously the Hamiltonian of the right lead is given by the following
matrix:
\begin{equation}
  \label{eq:H_R}
  \M{H}_R 
   = \left( 
     \begin{array}{cccc}
       \M H_{1,1} & \M H_{1,2} & \cdots \\
       \M H_{2,1} & \M H_{2,2} & \cdots \\
       \vdots     & \vdots     &        
     \end{array} \right)
  = \begin{pmatrix}
    \M H_0         & \M H_1         &        &        & \M 0   \\
    \M H_1^\dagger & \M H_0         & \M H_1 &        &        \\
    \,             & \M H_1^\dagger & \M H_0 & \M H_1 &        \\
    \M 0           &                & \ddots & \ddots & \ddots
    \end{pmatrix}.
\end{equation}
In a similar way, the overlap inside the leads is given by the matrices
\begin{eqnarray}
  \label{eq:S_L}
  \M{S}_L 
  &=& \left( 
    \begin{array}{cccc}
      & \vdots       & \vdots       & \\
      \cdots & \M S_{-2,-2} & \M S_{-2,-1} & \\
      \cdots & \M S_{-1,-2} & \M S_{-1,-1} &
    \end{array} \right) 
  = \begin{pmatrix}
    \ddots & \ddots         & \ddots         &                & \M 0   \\
    \,     & \M S_1^\dagger & \M S_0         & \M S_1         &        \\
    \,     &                & \M S_1^\dagger & \M S_0         & \M S_1 \\
    \M 0   &                &                & \M S_1^\dagger & \M S_0  
  \end{pmatrix}
\end{eqnarray}
and
\begin{equation}
  \label{eq:S_R}
  \M{S}_R 
   = \left( 
     \begin{array}{cccc}
       \M S_{1,1} & \M S_{1,2} & \cdots \\
       \M S_{2,1} & \M S_{2,2} & \cdots \\
       \vdots     & \vdots     &        
     \end{array} \right)
  = \begin{pmatrix}
    \M S_0         & \M S_1         &        &        & \M 0   \\
    \M S_1^\dagger & \M S_0         & \M S_1 &        &        \\
    \,             & \M S_1^\dagger & \M S_0 & \M S_1 &        \\
    \M 0           &                & \ddots & \ddots & \ddots
    \end{pmatrix}
\end{equation}

Furthermore the unit cell of each lead that is immediately connected to the scattering region 
(unit cell ``$l$'' for the left and unit cell ``$r$'' for the right lead)  is included into 
the device part of the system.:
\begin{equation}
  \label{eq:Device}
  \M{H}_D 
   = \left( 
     \begin{array}{ccc}
      \M H_l     & \M H_{l,S} & \M 0_{l,r} \\
      \M H_{S,l} & \M H_S     & \M H_{S,r} \\
      \M 0_{r,l} & \M H_{r,S} & \M H_r   
    \end{array} \right)
  \mbox{ and }
  \M{S}_D 
   = \left( 
     \begin{array}{ccc}
      \M S_l     & \M S_{l,S} & \M 0_{l,r} \\
      \M S_{S,l} & \M S_S     & \M S_{S,r} \\
      \M 0_{r,l} & \M S_{r,S} & \M S_r  
    \end{array} \right).
\end{equation}

In the next subsection we will show how to calculate the GF for the device
part of the system as defined by the above Hamilton and overlap matrices.

\subsection{Green's function for the open system}

The one-body GF operator $\hat G(E)$ of a system is defined as the solution to the 
generalized inhomogeneous Schr\"odinger equation (see e.g the book by E. N. Economou, Ref. \cite{Economou:book:83}):
\begin{equation}
  \label{eq:GF-def}
  (z-\hat H) \hat G(z) = \hat 1,
\end{equation}
where $\hat H$ is an (effective) one-body Hamiltonian, and $z$ is a complex number. When $z$ does not coincide with the 
eigenvalues $\epsilon_k$ of the Hamiltonian $\hat{H}$, the GF operator has the following formal solution:
\begin{equation}
  \label{eq:GF-op}
  \hat G(z) =  (z-\hat H)^{-1} \mbox{ for } z \ne \epsilon_k.
\end{equation}
Obviously, for $z=\epsilon_k$ the GF operator has a pole and is thus not well defined. 
In this case one can define two GFs which are both solutions to eq. (\ref{eq:GF-def})
by a limiting process. 1) The {\it retarded} GF is defined as:
\begin{eqnarray}
  \label{eq:RetGF}
  \hat G(E) &:=& \lim_{\eta \to 0} (E+i\eta - \hat H)^{-1},
\end{eqnarray}
and 2) the {\it advanced} GF is defined as the hermitian conjugate of the {\it retarded} GF:
\begin{eqnarray}
  \label{eq:AdvGF}
  \hat G^\dagger(E) &:=& \lim_{\eta \to 0} (E-i\eta - \hat H)^{-1}.
\end{eqnarray}

$E$ is a real number which can be any of the energy eigenvalues $\epsilon_k$ of the Hamiltonian $\hat H$.
When $E$ does not coincide with an eigenvalue of $\hat H$ the two GFs reduce to the GF operator defined in eq.
(\ref{eq:GF-op}). In the basis of eigenstates of the Hamiltonian the GF operator becomes diagonal:
\begin{equation}
  \label{eq:GF-op-eigenstates}
  \hat{G}(z) = \sum_k \frac{\sket{k}\sbra{k}}{z-\epsilon_k} \, \mbox{ with } \, \hat{H}\sket{k} = \epsilon_k \, \sket{k}.
\end{equation}
The GF yields the complete information of a one-body system. 
For example, eq. (\ref{eq:GF-op-eigenstates}) makes clear that 
the poles of the GF along the real axis represent the eigenvalues 
of the Hamiltonian $\hat H$. Thus by plotting $\hat G(E)$ 
along the real axis one can find all the eigenvalues in a certain 
energy range. 

A very important quantity is the density of states (DOS) 
$\mathcal{D}(E)$. The DOS can be calculated from the trace of the 
imaginary part of the GF along the real axis:
\begin{eqnarray}
  \label{eq:DOS}
  \lefteqn{
    {\rm Im}\;\Tr[\Op{G}(E)] = \sum_k \lim_{\eta\to 0} {\rm Im}\frac{1}{E + i\eta-\epsilon_k}= 
  }
  \nonumber\\
  & &
  = \sum_k \lim_{\eta\to 0} \frac{ -\eta }{(E-\epsilon_k)^2+\eta^2} 
  = -\pi \sum_k \delta(E-\epsilon_k) = -\pi \, \mathcal{D}(E).
\end{eqnarray}
Another important quantity of the GF formalism is the 
{\it spectral density} $\hat{A}(z)$ which is defined as
the difference between the {\it retarded} and the 
{\it advanced} GF:
\begin{eqnarray}
  \hat{A}(E) := i \, (\hat{G}(E) - \hat{G}^\dagger(E)).
\end{eqnarray}
It is straight forward to show that the eigenstate representation 
of the spectral function on the real axis is given by:
\begin{eqnarray}
  \hat{A}(E) := \sum_k \delta( E - \epsilon_k) \ket{k}\bra{k},
\end{eqnarray}
so that the trace of the spectral function directly gives the DOS.
The spectral function can thus be seen as a generalized DOS.

The grand advantage of the GF formalism is that it allows one to 
calculate all properties of a one-body system without having to 
calculate the eigenstates of the Hamiltonian explicitly. Instead 
the GF can be calculated in any basis set by a matrix inversion for 
any value $z$, eq. (\ref{eq:GF-op}). It turns out that in many 
situations this is more convenient than to solve the whole 
eigenvalue problem. This is the case for the transport problem 
defined by the Hamiltonian and overlap matrices given in the 
previous subsection, eqs. (\ref{eq:HLDR}) and (\ref{eq:SLDR}).

The matrix $\M{G}_{\alpha\beta}(E):=\bra{\alpha} \Op{G}(E) \ket{\beta}$ 
corresponding to the GF operator in a {\it non-orthogonal} basis set 
$\{\ket{\alpha}\}$ is given by:
\begin{equation}
  \label{eq:GF-mat-eq}
  (E \, {\bf S}- {\bf H}) {\bf S}^{-1} {\bf G}(E) = {\bf S},
\end{equation}
where $\M{S}$ is the overlap matrix of the NOBS: $S_{\alpha\beta}=\bracket{\alpha}{\beta}$.
This, however, is a somewhat inconvenient definition for the calculation of the GF
since it involves the inversion of the $\bf S$ matrix.
Instead, we define a new GF matrix ${\bf \widetilde{G}}(E)$ by 
\begin{equation}
  \label{eq:GTilde-def}
  {\bf \widetilde{G}}(E) := {\bf S}^{-1} {\bf G}(E) {\bf S}^{-1}.
\end{equation}
This GF can be calculated more conveniently from the much simpler equation
\begin{equation}
  \label{eq:GTilde-eq}
  (E \, {\bf S} - {\bf H}) {\bf \widetilde{G}}(E) = {\bf 1}.
\end{equation}

For convenience we introduce the following energy-dependent matrix:
\begin{eqnarray}
  \M{W}(E) := \M{H} - E \, \M{S}
  = \left(
      \begin{array}{ccc}
        {\bf H}_{L} -E\,{\bf S}_{L}   & {\bf H}_{LD} -E\,{\bf S}_{LD} & {\bf 0}                      \\
        {\bf H}_{DL} -E\,{\bf S}_{DL} & {\bf H}_{D} -E\,{\bf S}_{D}   & {\bf H}_{DR}-E\,{\bf S}_{DR} \\
        {\bf 0}                       & {\bf H}_{RD} -E\,{\bf S}_{RD} & {\bf H}_{R} -E\,{\bf H}_{R}
      \end{array}
    \right).
%%  \left(
%%    \begin{array}{ccc}
%%      {\bf W}_{L}(E)  & {\bf W}_{LD}(E) & {\bf W}_{LR}(E)\\
%%      {\bf W}_{DL}(E) & {\bf W}_{D}(E)  & {\bf W}_{DR}(E)\\
%%      {\bf W}_{RL}(E) & {\bf W}_{RD}(E) & {\bf W}_{R}(E)
%%    \end{array}
%%  \right).
\end{eqnarray}

Thus we obtain for the GF for the transport problem the following matrix equation
which defines a system of equations for each sub-matrix of the total GF.
\begin{eqnarray}
%%  \lefteqn{
%%    \left(
%%      \begin{array}{ccc}
%%        E\,{\bf S}_{L} -{\bf H}_{L}  & E\,{\bf S}_{LD}-{\bf H}_{LD} & {\bf 0}                      \\
%%        E\,{\bf S}_{DL}-{\bf H}_{DL} & E\,{\bf S}_{D} -{\bf H}_{D}  & E\,{\bf S}_{DR}-{\bf H}_{DR} \\
%%        {\bf 0}                      & E\,{\bf S}_{RD}-{\bf H}_{RD} & E\,{\bf H}_{R} -{\bf H}_{R}     
%%      \end{array}
%%    \right) \times }
%%  \nonumber \\
%%  \nonumber \\
%%  & & \hspace{2.2cm} \times 
  \lefteqn{
    -\left(
      \begin{array}{ccc}
        \M{W}_{L}(E)  & \M{W}_{LD}(E) & \M{W}_{LR}(E) \\
        \M{W}_{DL}(E) & \M{W}_{D}(E)  & \M{W}_{DR}(E) \\
        \M{W}_{RL}(E) & \M{W}_{RD}(E) & \M{W}_{R}(E)       
      \end{array}
    \right) \times
  }
  \nonumber\\
  & & \times \left(
    \begin{array}{ccc}
      \widetilde{\M G}_{L}(E)  & \widetilde{\M G}_{LD}(E) & \widetilde{\M G}_{LR}(E) \\
      \widetilde{\M G}_{DL}(E) & \widetilde{\M G}_{D}(E)  & \widetilde{\M G}_{DR}(E) \\
      \widetilde{\M G}_{RL}(E) & \widetilde{\M G}_{RD}(E) & \widetilde{\M G}_{R}(E)       
    \end{array}
  \right) 
  = \left(
  \begin{array}{ccc}
    \bf 1 & \bf 0 & \bf 0 \\
    \bf 0 & \bf 1 & \bf 0 \\
    \bf 0 & \bf 0 & \bf 1
  \end{array}
\right),
\end{eqnarray}
This matrix equation can be solved for each of the matrix elements (see App. \ref{app:partitioning}). 
For the device part of the GF $\widetilde{\M G}_D$ we obtain:
\begin{equation}
  \label{eq:GD}
  \widetilde{\M G}_{D}(E) = (E {\bf S}_{D} - {\bf H}_{D} 
  - {\bf \widetilde{\Sigma}}_L(E) - {\bf \widetilde{\Sigma}}_R(E))^{-1},
\end{equation}
where we have introduced the \emph{self-energies} of the leads 
${\bf \widetilde{\Sigma}}_L$ and ${\bf \widetilde{\Sigma}}_R$ 
which describe the influence of the leads on the electronic 
structure of the device. They can be calculated from the GFs 
of the \emph{isolated} left and right semi-infinite lead, 
$\M g_L(E) = (E\,{\bf S}_{L} -{\bf H}_{L})^{-1}$
and $\M g_R(E) = (E\,{\bf S}_{R} -{\bf H}_{R})^{-1}$, 
respectively:
\begin{eqnarray}
  \label{eq:SigmaL}
  \widetilde{\M\Sigma}_L(E) &=& \M{W}_{DL}(E) \, \widetilde{\M g}_L(E) \, \M{W}_{LD}(E) \\
  \label{eq:SigmaR}
  \widetilde{\M\Sigma}_R(E) &=& \M{W}_{DR}(E) \, \widetilde{\M g}_R(E) \, \M{W}_{RD}(E) 
\end{eqnarray}
For later use, we also define the so-called coupling matrices:
\begin{eqnarray}
  \widetilde{\M\Gamma}_L(E) 
  &:=& i \, \left(\widetilde{\M\Sigma}_L(E) - \widetilde{\M\Sigma}_L^\dagger(E)\right)
  \nonumber\\
  &=& i \, \M{W}_{DL}(E) \, (\widetilde{\M g}_L(E)-\widetilde{\M g}_L^\dagger(E)) \, \M{W}_{LD}(E), 
  \\
  \widetilde{\M\Gamma}_R(E) 
  &:=& i \, \left( \widetilde{\M\Sigma}_L(E) - \widetilde{\M\Sigma}_R^\dagger(E) \right)
  \nonumber\\
  &=& i \, \M{W}_{DR}(E) \, (\widetilde{\M g}_R(E)-\widetilde{\M g}_R^\dagger(E)) \, \M{W}_{RD}(E). 
\end{eqnarray}

Thus we have expressed the GF of the device region in terms of the device Hamiltonian and the Green's
functions of the \emph{isolated} leads. The term ${\bf H}_{D} + {\bf \widetilde{\Sigma}}_L(E) + {\bf \widetilde{\Sigma}}_R(E)$
can be interpreted as an effective Hamiltonian for the device region, and its energy dependence stems from the fact, that the 
lifetime of an electron in the device region is now finite due to the coupling to the leads.
The technique of calculating the GF in parts by dividing the underlying Hilbert space into subspaces
is called {\it partitioning technique} and is shown in more detail in App. \ref{app:partitioning} where 
we also list the other matrix elements of the GF.

Since the coupling $\M{H}_{DL}$ between the left lead and the device is only due to coupling between the last unit cell ($-1$) 
of left lead and the unit cell $l$ included in the device region, only the surface GF of the left lead, i.e. the
GF projected into the unit cell $-1$, is necessary for calculating the self-energy $\widetilde{\M\Sigma}_L$. 
The same holds true for the right lead in an analogous manner. Furthermore, $\widetilde{\M\Sigma}_L$ is different from zero 
only in the $l$-region of the device, and analogously $\widetilde{\M\Sigma}_R$ in the $r$-region:
\begin{equation}
  \label{eq:Sigma}
  \widetilde{\M\Sigma} := \widetilde{\M\Sigma}_L + \widetilde{\M\Sigma}_R 
   = \left( 
     \begin{array}{ccc}
      \widetilde{\M\Sigma}_l & \M 0_{l,S} & \M 0_{l,r} \\
      \M 0_{S,l}             & \M 0_S    & \M 0_{S,r} \\
      \M 0_{r,l}             & \M 0_{r,S} & \widetilde{\M\Sigma}_r
    \end{array} \right), 
\end{equation}
where the non-zero matrix elements $\widetilde{\M\Sigma}_l$ and $\widetilde{\M\Sigma}_r$ of $\widetilde{\M\Sigma}$
can be expressed in terms of the surface GFs of the left and right lead, $\widetilde{\M g}_{-1,-1}$ and $\widetilde{\M g}_{1,1}$:
\begin{eqnarray}
  \label{eq:sigma_l}
  \widetilde{\M\Sigma}_l(E) &=& (\M{H}_1^\dagger - E\M{S}_1^\dagger) \, \widetilde{\M g}_{-1,-1}(E) \, (\M{H}_1 - E\M{S}_1),
  \\
  \label{eq:sigma_r}  
  \widetilde{\M\Sigma}_r(E) &=& (\M{H}_1 - E\M{S}_1) \, \widetilde{\M g}_{+1,+1}(E) \, (\M{H}_1^\dagger - E\M{S}_1^\dagger).
\end{eqnarray}

The self-energies can be calculated iteratively by Dyson equations, as shown in App. \ref{app:self-energy-1D}:
\begin{eqnarray}
  \label{eq:DysonL}
  \M{\widetilde{\Sigma}}_{l}(E) &=& 
  (\M{H}_1^\dagger - E \, \M{S}_1^\dagger) \, 
  (E\,\M{S}_0-\M{H}_0-\M{\widetilde{\Sigma}}_{l}(E))^{-1} \, 
  (\M{H}_1 - E\,\M{S}_1),
  \\ \nonumber \\
  \label{eq:DysonR}
  \M{\widetilde{\Sigma}}_{r}(E) &=& 
  (\M{H}_1 - E\,\M{S}_1) \,
  (E\,\M{S}_0 - \M{H}_0 - \M{\widetilde{\Sigma}}_{r}(E))^{-1} \, 
  (\M{H}_1^\dagger - E \, \M{S}_1^\dagger).
\end{eqnarray}

Now we have achieved a description of the electronic structure of the device region in terms of one-body GFs
which take into account the effect of the coupling of the leads to the device region. In the next section we will show how
to obtain the (reduced) density matrix and subsequently the electron density of the device region from this GF.
Thereafter we demonstrate, how to calculate the transmission matrix from the GF of the device region and the 
self-energies of the leads.

%%The selfenergies are given by the surface GFs of the \emph{isolated} semi-infinite
%%leads $\M{\widetilde g}_L^{(0)} := (\M{\widetilde g}_L)_{(-1,-1)} $ and  $\M{\widetilde g}_R^{(0)}:=(\M{\widetilde g}_R)_{(1,1)}$, 
%%respectively:
%
% \begin{eqnarray}
%   \label{eq:SigmaL}
%   \M{\widetilde{\Sigma}}_{L}(E) &=& 
%   (\M{H}_{L}^{(1)}-E\,\M{S}_{L}^{(1)}) \, \M{\widetilde{g}}_L^{(0)}(E)\, (\M{\widetilde H}_{L}^{(1)}-E\,\M{\widetilde S}_{L}^{(1)}),
%   \nonumber \\ \\
%   \label{eq:SigmaR}
%   \M{\widetilde{\Sigma}}_{R}(E) &=& 
%   (\M{H}_{R}^{(1)}-E\,\M{S}_{R}^{(1)}) \, \M{\widetilde{g}}_R^{(0)}(E)\, (\M{\widetilde H}_{R}^{(1)}-E\,\M{\widetilde S}_{R}^{(1)}).
%   \nonumber \\
% \end{eqnarray}
%%

\subsection{Calculation of density matrix and electron number at equilibrium}

The reduced density matrix of first order (in quantum chemistry often called charge density matrix) is obtained 
by tracing out all but one of the one-particle subspaces from the many-body density matrix $\hat{\rho}$:
\begin{equation}
  \label{eq:red-densmat1}
  \hat{P} = N \times \sum_{n_2,\ldots,n_N} \bra{n,n_2,\ldots,n_N}\hat{\rho}\ket{n^\prime,n_2,\ldots,n_N} \ket{n}\bra{n^\prime},
\end{equation}
where $\{\ket{n}\}$ is a set of one-body states.

%In general, a system in contact with a reservoir at some finite temperature $T$ and chemical potential $\mu$
%resides in an incoherent mixture of states described by the density matrix of the system:
%%%
%\begin{equation}
%  \label{eq:densmat}
%  \hat{P} = \sum_m p_m(T,\mu) \ket{\Psi_m}\bra{\Psi_m},
%\end{equation}
%%%
%where $p_m(T,\mu)$ are the occupation probabilities for a pure state $\ket{\Psi_m}$ in the grand-canonical 
%ensemble which depend on the temperature and chemical potential. The pure states are in general many-body
%states, i.e. superpositions of slater determinants. Thus the density matrix is a many-body operator acting
%on the many-body Hilbert space, and which contains all the information on the system. 

At zero temperature the system will be in its (many-body) ground state $\Psi$, so that the density matrix becomes
\begin{equation}
  \label{eq:densmat}
  \hat{\rho}_\Psi = \ket{\Psi}\bra{\Psi}.
\end{equation}
It is straight forward to show that in this case the reduced density matrix can be expressed as 
\begin{equation}
  \label{eq:red-densmat2}
  \hat{P} = \sum_{k,l} \ket{k} \, \bra{\Psi}\hat{c}_k^\dagger\hat{c}_l \ket{\Psi} \, \bra{l}.
\end{equation}
One can obtain a lot of information from the reduced density matrix about a many-body system in spite of the fact 
that it is actually a one-body operator. For example, one can calculate the expectation value of any one-body observable
from the trace of the product of the reduced density matrix and the observable:
\begin{equation}
  \langle \hat A \rangle = \Tr[ \hat{P} \hat{A} ].
\end{equation}
The electron density is given by the diagonal elements of the reduced density matrix in the real space representation:
\begin{equation}
  n(\vec r) = \bra{\vec{r}} \hat{P} \ket{\vec{r}}.
\end{equation}
Finally, the number of electrons is the trace of the density matrix. 

In the Landauer approach we consider the electrons as a system of non-interacting quasi-particles, i.e. the Coulomb interaction
between the electrons is only taken into account on the mean-field level. In the case of a system of non-interacting particles 
the many-body ground state of the system $\ket{\Psi}$ is given by a single Slater determinant:
\begin{equation}
  \ket{\Psi} = \prod_{k,\epsilon_k \le \mu} \hat c_k^\dagger \ket{0},
\end{equation}
where $\ket{k}$ are the eigenstates of the one-body Hamiltonian, $\ket{0}$ is the vacuum ground state, and $\mu$ the chemical 
potential, so that the Slater determinant consists of all states with energies less than or equal to the chemical potential.
In this case the reduced density matrix is diagonal in the one-body eigenstates $\ket{k}$:
\begin{equation}
  \hat{P} = \sum_k \,  n_k \, \ket{k} \bra{k},
\end{equation}
where the occupation number $n_k$ of the eigenstate $\ket{k}$ is given by the Fermi distribution function,
$n_k=f(\epsilon_k-\mu)$, and thus is either one for states below the chemical potential or zero for states 
above the chemical potential at zero temperature. Since we will make use of the {\it reduced} density matrix
only but not of the full many-body density matrix we will refer to the {\it reduced} density matrix for the
sake of simplicity from here on simply as the density matrix, unless otherwise stated. 
 
Using the eigenstate representation of the GF, eq. (\ref{eq:GF-op-eigenstates}), it is straight forward to show
that the density matrix can actually be obtained from the GF of the system by integrating the imaginary part of the 
GF up to the chemical potential $\mu$ of the system:
\begin{eqnarray}
   \lefteqn{
     -\frac{1}{\pi} {\rm Im} \int_{-\infty}^{\mu} dE \, \M{\hat G}(E) 
     = \lim_{\eta\to 0} -\frac{1}{\pi} \int_{-\infty}^{\mu} dE \, {\rm Im} 
     \sum_k \frac{\ket{k}\bra{k}}{E-\epsilon_k+i\eta} 
   }
   \nonumber\\
   & & 
   = \sum_k \int_{-\infty}^{\mu} dE \, \lim_{\eta\to 0}\frac{1}{\pi} \frac{\eta}{(E-\epsilon_k)^2+\eta^2}\ket{k}\bra{k}
   = \sum_k \int_{-\infty}^{\mu} dE \, \delta(E-\epsilon_k) \ket{k}\bra{k}
   \nonumber\\
   & &
   = \sum_k f(\epsilon_k-\mu)\ket{k}\bra{k} \equiv \hat{P}.
\end{eqnarray}

Subsequently, the (standard) density matrix in a NOBS $\{\ket{\alpha}\}$, 
is obtained by integration of the standard non-orthogonal GF matrix,
$G_{\alpha\beta}(E) := \bra{\alpha}\hat{G}(E)\ket{\beta}$:
\begin{equation}
  {P}_{\alpha\beta} := \bra{\alpha}\hat{P}\ket{\beta} 
  = -\frac{1}{\pi} {\rm Im} \int_{-\infty}^{\mu} dE \, G_{\alpha\beta}(E).
\end{equation}
Analogously to the GF matrix $\widetilde{\M G}$ we can define a new density matrix by
\begin{equation}
  \label{eq:rho-tilde}
  \widetilde{\M {P}} = \M{S}^{-1} \M{{P}} \M{S}^{-1},
\end{equation}
which is thus obtained by integration of the non-standard GF $\widetilde{\M G}(E)$:
\begin{equation}
  \M{\widetilde {P}} = -\frac{1}{\pi} {\rm Im} \int_{-\infty}^{\mu} dE \, \M{\widetilde G}(E).
\end{equation}

One can obtain the number of electrons from the trace of the density matrix. However, the trace of
an operator has to be taken in an orthogonal basis set, but $\M{\widetilde {P}}$ has been defined in
terms of the density matrix in a NOBS $\{\ket{\alpha}\}$. The standard procedure
for orthogonalizing atomic orbitals is the so-called \emph{symmetric orthogonalization} 
\cite{Szabo:book:89}. A density matrix $\M{{P}}$ in a NOBS will be transformed
to the density matrix $\M{{P}}^\perp$ in the orthogonalized basis according to:
\begin{equation}
  \label{eq:orthogonalization}
  \M{{P}}^\perp = \M{S}^{-1/2} \, \M{{P}} \, \M{S}^{-1/2} = \M{S}^{+1/2} \, \M{\widetilde{{P}}} \, \M{S}^{+1/2}.
\end{equation}
In the last step we have used the definition of the non-standard density matrix $\M{\widetilde{{P}}}$ to obtain
the transformation to the orthogonalized density matrix.

Applying the above transformation for the density matrix $\M{\widetilde{{P}}}$ we find for the number of electrons 
the following expression:
\begin{eqnarray}
  N_e &=& \Tr[ \M S^{1/2} \, \M{\widetilde {P}} \, S^{1/2} ] = \Tr [ \M{\widetilde {P}} \, \M S ]
  =   \Tr [ \M{\widetilde {P}}_D \, \M S_D ] + \Tr [ \M{\widetilde {P}}_{DL} \, \M S_{LD} ] 
  + \Tr [ \M{\widetilde {P}}_{DR} \, \M S_{RD} ] 
  \nonumber\\
  &+& \Tr [ \M{\widetilde {P}}_L \, \M S_L ] + \Tr [ \M{\widetilde {P}}_{LD} \, \M S_{DL} ] 
  +   \Tr [ \M{\widetilde {P}}_R \, \M S_R ] + \Tr [ \M{\widetilde {P}}_{RD} \, \M S_{DR} ] 
\end{eqnarray}
where we have first exploited the invariance of the trace under commutation of matrices and then applied the above explained
division into sub-matrices of the density matrix and the expression (\ref{eq:SLDR}) for the overlap matrix. In analogy to the
Mulliken analysis we can identify the number of electrons in the device, $N_D$, in the left lead, $N_L$, and in the right lead
$N_R$ as follows:
\begin{eqnarray}
  N_D &=& \Tr [ \M{\widetilde {P}}_D \, \M S_D ] + \Tr [ \M{\widetilde {P}}_{DL} \, \M S_{LD} ] + \Tr [ \M{\widetilde {P}}_{DR} \, \M S_{RD} ] 
  \nonumber\\
  N_L &=& \Tr [ \M{\widetilde {P}}_L \, \M S_L ] + \Tr [ \M{\widetilde {P}}_{LD} \, \M S_{DL} ] 
  \nonumber\\
  N_R &=& \Tr [ \M{\widetilde {P}}_R \, \M S_R ] + \Tr [ \M{\widetilde {P}}_{RD} \, \M S_{DR} ] 
\end{eqnarray}
Thus due to the overlap of the device orbitals with the lead orbitals one also has to calculate the off-diagonal elements
$DL$ and $DR$ of the density matrix $\widetilde{\M {P}}$ in order to calculate the number of electrons in the device part.
However, computing the off-diagonal elements of the density matrix would require calculation of the corresponding 
off-diagonal elements of the GF matrix $\widetilde{\M G}(E)$ in terms of the GF matrix of the device and the self-energies
(see App. \ref{app:partitioning}) which are quite tedious expressions. Instead we will require charge neutrality only for the scattering
region and not for the entire device region. That is the reason why we have included one unit cell of each lead 
into the device region in addition to the scattering region. Since the scattering region does not have an overlap with
the two leads it is sufficient to calculate the density matrix of the device region to obtain the number of electrons
inside the scattering region:
\begin{equation}
  N_S = \Tr_S [ \M{\widetilde {P}}_D \, \M S_D ] = \sum_{\alpha\in S,\beta\in D}  \widetilde{{P}}_{\alpha\beta} \, S_{\beta\alpha} 
\end{equation}

Finally, we also have to define an appropriate DOS $\mathcal D$ and projected DOS (PDOS) $\mathcal D_i$ for 
the case of non-orthogonal orbitals. In order to be coherent with the above definition of the electron numbers 
in the respective subspaces of the leads and the device, it is best to adapt a similar Mulliken-like definition 
for the PDOS:
\begin{equation}
  \label{eq:NOB-PDOS}
  \mathcal{D}_\alpha(E) = -\frac{1}{\pi} {\rm Im}(\M{\widetilde G}(E)\,\M{S})_{\alpha\alpha}
\end{equation}
where $\alpha$ denotes some of the atomic orbitals of the system. If we are interested in the DOS projected
onto some orbital $\alpha$ of the scattering region, it suffices to calculate the GF and overlap matrix for the 
device region instead for the entire system, 
$\mathcal{D}_\alpha(E) = -\frac{1}{\pi} {\rm Im}(\M{\widetilde G}_D(E)\,\M{S}_D)_{\alpha\alpha}$.
And the DOS projected onto the entire scattering region is just 
$\mathcal{D}_S(E) = -\frac{1}{\pi} \sum_{\alpha \in S} {\rm Im}(\M{\widetilde G}_D(E)\,\M{S}_D)_{\alpha\alpha}$.

\subsection{Non-equilibrium density matrix}

In the previous section we have assumed that the system is in thermal equilibrium, i.e. the chemical
potentials of the two electron reservoirs are equal: $\mu_L=\mu_R=\mu$. Although in many cases one can
estimate the conductance at small bias from the zero-bias conductance, i.e. in equilibrium, it would be
very interesting to study the interplay between an electron current and the electronic structure of the 
nanoscopic conductor. Thus we have to derive an expression for the density matrix when the system is out 
of equilibrium, i.e. $\mu_L\ne\mu_R$. 

%We start from the important observation that the (retarded) GF gives the response of the system subject
%to an external perturbation given by some state $\ket{v}$:
%%
%\begin{equation}
%  \label{eq:GF-response}
%  (E - \hat H) \ket{\psi} = \ket{v} \, \Rightarrow \, \ket{\psi} = \hat{G}^{(+)}(E) \ket{v}.
%\end{equation}
%%
%One can now think of the electrons coming from the reservoirs incident on the sample as perturbations of the electronic
%structure of the sample which can be calculated from the retarded GF of the device. 

First, we calculate the response $\ket{\psi^+}$ of the entire system to an incoming wave on the left lead $\ket{\psi^i}$
which we assume is a solution of the \emph{isolated} left lead, i.e. $\hat{H}_L \ket{\psi^i} = E \ket{\psi^i}$.
\begin{equation}
  \hat H ( \ket{\psi^+} + \ket{\psi^i} )  = E ( \ket{\psi^+} + \ket{\psi^i} )
\end{equation}
Because of the non-orthogonality of the lead orbitals with the device orbitals we now switch to a matrix presentation 
of the problem as done in the previous section. The incoming wave as a solution of the left isolated lead is given as 
an expansion over the lead orbitals:
\begin{equation}
  \ket{\psi^i} = \sum_{\alpha\in L} \psi^i_\alpha \ket{\alpha} 
  \Leftrightarrow \vec{\psi}^i = 
  \left(
    \begin{array}{c}
      \vec\psi^i_L \\
      \vec{0}_D \\
      \vec{0}_R
    \end{array}
  \right)
\end{equation}
The response wave function $\ket{\psi^+}$ on the other hand expands over the whole system:
\begin{equation}
  \vec{\psi}^+ = 
  \left(
    \begin{array}{c}
      \vec\psi^+_L \\
      \vec\psi^+_D \\
      \vec\psi^+_R
    \end{array}
  \right)  
\end{equation}
Thus in matrix representation the above Schr\"odinger equation for the response to an incoming wave reads:
\begin{eqnarray}
  \M{H} ( \vec\psi^+ + \vec\psi^i ) &=& E \M{S} ( \vec\psi^+ + \vec\psi^i ) \nonumber \\
  \Rightarrow (\M{H} - E \M{S}) \vec\psi^+ &=& (E \M{S} - \M{H} )\vec\psi^i 
  = \left(
    \begin{array}{c}
       (E \M{S}_L - \M{H}_L ) \vec\psi^i_L \\
       (E \M{S}_{DL} - \M{H}_{DL} ) \vec\psi^i_L \\
      \vec{0}_R
    \end{array}
  \right)  \nonumber\\
  \Rightarrow 
  \left(
    \begin{array}{c}
      \vec\psi^+_L \\
      \vec\psi^+_D \\
      \vec\psi^+_R
    \end{array}
  \right) &=& -\widetilde{\M{G}}(E) 
  \left(
    \begin{array}{c}
       (E \M{S}_L - \M{H}_L ) \vec\psi^i_L \\
       (E \M{S}_{DL} - \M{H}_{DL} ) \vec\psi^i_L \\
      \vec{0}_R
    \end{array}    
  \right).
\end{eqnarray}
Since $\ket{\psi^i}$ is a solution for the isolated left lead,
\begin{equation}
  \M{H}_L \vec\psi^i_L = E \M{S}_L \vec\psi^i_L
  \Rightarrow
  (\M{H}_L - E \M{S}_L ) \vec\psi^i_L = 0
\end{equation}
we obtain finally:
\begin{equation}
  \label{eq:response}
  \left(
    \begin{array}{c}
      \vec\psi^+_L \\
      \vec\psi^+_D \\
      \vec\psi^+_R
    \end{array}
  \right) = -\widetilde{\M{G}}(E) 
  \left(
    \begin{array}{c}
       \vec{0}_L \\
       (E \M{S}_{DL} - \M{H}_{DL} ) \vec\psi^i_L \\
      \vec{0}_R
    \end{array}    
  \right) 
  = \left(
    \begin{array}{l}
      \widetilde{\M{G}}_{LD}(E) \\
      \widetilde{\M{G}}_D(E)    \\
      \widetilde{\M{G}}_{RD}(E)
    \end{array}    
  \right) 
  \M{W}_{DL}(E) \, \vec\psi^i_L 
\end{equation}

The left electron reservoir fills the incoming electron waves $\ket{\psi_{L,n}(k)}$ on the left lead up to the 
chemical potential $\mu_L$ of that reservoir which result in wavefunctions expanded over the whole system
$\ket{\Psi_n(k)}=\ket{\psi^+_n(k)}+\ket{\psi_{L,n}(k)}$. Thus the density matrix due to injection of electrons
from the left reservoir is given by:
\begin{eqnarray}
  \hat{P}^{(L)} 
  &=& \sum_{n\in N_L,k} f(\epsilon_{L,n}(k)-\mu_L) \ket{\Psi_n(k)}\bra{\Psi_n(k)} \nonumber \\
  &=& \sum_{n\in N_L,k} \int dE \, \delta(E-\epsilon_{L,n}(k)) f(E-\mu_L) \ket{\Psi_n(k))}\bra{\Psi_n(k)} \nonumber \\
  &=& \int dE \, f(E-\mu_L) \sum_{n\in N_L,k} \ket{\Psi_n(k))} \delta(E-\epsilon_{L,n}(k)) \bra{\Psi_n(k)}
\end{eqnarray}
%%
%%where in the last step we have cast the summation of the k-vectors into an integral over the energy
%%by help of the band-projected DOS $D_{L,n}(E) = \sum_{k} \delta(E-\epsilon_{L,n}(k))$, and 
%%where we have defined $\ket{\Psi_n(E))}:=ket{\Psi_n(k_n(E)))}$.
In matrix representation this density matrix is given as
\begin{equation}
  {P}^{(L)}_{\alpha\beta} = \bra{\alpha}\hat{P}^{(L)}\ket{\beta} = 
  \sum_{n\in N_L,k} f(\epsilon_{L,n}(k)-\mu_L) \, 
  S_{\alpha\alpha^\prime} \Psi_{n\alpha^\prime}(k) 
  {\Psi_{n\beta^\prime}(k)}^\ast S_{\beta^\prime\beta}
\end{equation}
Thus the non-standard matrix representation is
\begin{equation}
  \label{eq:PL}
  \widetilde{\M{{P}}}^{(L)} = \M{S}^{-1} \M{{P}} \M{S}^{-1}  
  = \int dE \, f(E-\mu_L) \sum_{n\in N_L,k} \vec{\Psi}_{n}(k) \, \delta(E-\epsilon_{L,n}(k)) \, [\vec{\Psi}_{n}(k)]^\dagger.
\end{equation}
%%
%%where we have made the identification: $\vec{\Psi}_{n}(E) := \vec{\Psi}_{n}(k_n(E))$.
From eq. (\ref{eq:response}) we see that the device part is given by 
\begin{eqnarray}
  \widetilde{\M{{P}}}^{(L)}_D
  &=& \int dE \, f(E-\mu_L) \sum_{n\in N_L,k} \vec{\Psi}_{D,n}(k) \, \delta(E-\epsilon_{L,n}(k)) \, [\vec{\Psi}_{D,n}(k)]^\dagger
  \nonumber \\
  &=& \int dE \, f(E-\mu_L) \widetilde{\M{G}}_D(E) \, \M{W}_{DL}(E)
  \nonumber \\
  && \underbrace{\sum_{n\in N_L,k} \vec\psi_{L,n}(k) \, \delta(E-\epsilon_{L,n}(k)) \,
    [\vec\psi_{L,n}(k)]^\dagger}_{ \widetilde{\M{a}}_L(E)/2\pi = i(\widetilde{\M{g}}_L(E)-\widetilde{\M{g}}_L^\dagger(E))/2\pi}
  \M{W}_{LD}(E) \, \widetilde{\M{G}}_D^\dagger(E) 
  \nonumber \\
  &=& \frac{i}{2\pi}\int dE \, f(E-\mu_L) \widetilde{\M{G}}_D(E) 
  \underbrace{(\widetilde{\M{\Sigma}}_L(E)-\widetilde{\M{\Sigma}}_L^\dagger(E))}_{\widetilde{\M\Gamma}_L(E)/i} 
  \widetilde{\M{G}}_D^\dagger(E),
\end{eqnarray}
where in the 2nd step we have made use of the spectral density of the isolated left lead, 
$\widetilde{\M a}_L(E)= i(\widetilde{\M{g}}_L(E)-\widetilde{\M{g}}_L^\dagger(E))$.

For the injection of electrons from the right we obtain an analogous expression. 
Summing up both contributions we obtain the density matrix out of equilibrium:
\begin{equation}
  \label{eq:NEQ-densmat}
   \widetilde{\M{{P}}}_D^{neq} = \widetilde{\M{{P}}}^{(L)}_D + \widetilde{\M{{P}}}^{(R)}_D
   = \frac{1}{2\pi} \int dE \, \widetilde{\M{G}}_D(E) [ f(E-\mu_L) \widetilde{\M\Gamma}_L(E) + f(E-\mu_R) \widetilde{\M\Gamma}_R(E)] \widetilde{\M{G}}_D^\dagger(E) 
\end{equation}

Finally, we define the non-equilibrium GF matrix $\widetilde{\M{G}}^<(E)$ of the device as:
\begin{equation}
  \label{eq:Gless}
  \widetilde{\M{G}}^<(E) := i \widetilde{\M{G}}_D(E) [ f(E-\mu_L) \widetilde{\M\Gamma}_L(E) + 
  f(E-\mu_R) \widetilde{\M\Gamma}_R(E)] \widetilde{\M{G}}_D^\dagger(E)
\end{equation}
Then the non-equilibrium density matrix can be written as:
\begin{equation}
  \widetilde{\M{{P}}}_D^{neq} = - \frac{i}{2\pi} \int dE \, \widetilde{\M{G}}^<(E)
\end{equation}
In equilibrium, i.e. when $\mu_L=\mu_R=\mu$, the GF matrix $\widetilde{\M{G}}^<(E)$ reduces to the imaginary part
of the retarded GF matrix:
\begin{eqnarray}
  \widetilde{\M{G}}^<(E) &=& f(E-\mu) \, i \widetilde{\M{G}}_D(E) 
  [ \widetilde{\M\Gamma}_L(E) + \widetilde{\M\Gamma}_R(E)]
  \widetilde{\M{G}}_D^\dagger(E) 
  \nonumber \\
  &=& f(E-\mu) \, \widetilde{\M{G}}_D(E) 
  [ (\widetilde{\M{G}}_D(E))^{-1} - (\widetilde{\M{G}}^\dagger_D(E))^{-1} ]
  \widetilde{\M{G}}_D^\dagger(E) 
  \nonumber \\
  &=& f(E-\mu) [ \widetilde{\M{G}}^\dagger_D(E) - \widetilde{\M{G}}_D(E)]
  = -2i  \, f(E-\mu) \, {\rm Im} [  \widetilde{\M{G}}_D(E) ],
\end{eqnarray}
and one arrives again at the expression for the equilibrium density matrix derived before:
\begin{equation}
  \label{eq:EQ-densmat}
  \widetilde{\M{{P}}} = -\frac{1}{\pi} \int dE \, f(E-\mu) \, {\rm Im} [  \widetilde{\M{G}}_D(E) ].
\end{equation}

The same reasoning shows that for an out of equilibrium situation the density matrix can be calculated 
using the much simpler expression (\ref{eq:EQ-densmat}) for the energy integration up to the lowest of 
the two chemical potentials, e.g. for $\mu_L > \mu_R$ we have:
\begin{equation}
  \widetilde{\M{{P}}}_D^{neq} = -\frac{1}{\pi} \int_{-\infty}^{\mu_R} dE \, {\rm Im} [ \widetilde{\M{G}}_D(E) ] 
  - \frac{i}{2\pi} \int_{\mu_R}^{\mu_L} dE \, \widetilde{\M{G}}^<(E).
\end{equation}

\subsection{Current and transmission}

Now we derive an expression for the electrical current through the device region
in terms of the GFs by calculating the time derivative of the electron
charge in the device region. The temporal change of the number of electrons inside
the device region is equal to the sum of the currents from the left and from the 
right lead:
\begin{eqnarray}
  \label{eq:dN_D/dt}
  \frac{\partial N_D}{\partial t} &=& 
  \Tr_D\left[ \frac{\partial\widetilde{\M P}}{\partial t} \, \M S \right] = I_L + I_R,
\end{eqnarray}
where we have taken a current as positive when it increases the number of electrons 
inside the device region, i.e. when the current flows from the leads to the device. 
Typically we are interested in steady-state situations, so that the electron charge 
of the device will be conserved, meaning that $I_L = -I_R$. Then the net current 
$I$ {\it through} the device is given by $I \equiv I_L = -I_R$. 

We will now derive an expression for the total current into the device 
$\frac{\partial N_D}{\partial t}=I_L+I_R$ by taking the time derivative of the density 
matrix $\widetilde{\M P}$, and divide the obtained expression for the total current
$I_L+I_R$ into the device into its individual contributions $I_L$ and $I_R$. 
%%Since $\widetilde{\M P}=\widetilde{\M P}^{(L)}+\widetilde{\M P}^{(R)}$ 
%%the net current $I$ can be divided into two contributions, one ($I^{(L)}$) due to electron 
%%injection from the left reservoir and one ($I^{(R)}$) due to electron injection from the 
%%right reservoir. Thus we have in steady state: $I=I^{(L)}+I^{(R)}$.
We start from the expression (\ref{eq:PL}) in order to calculate the net current 
through the device due to the injection of electrons from the left electrode:
\begin{eqnarray}
  \label{eq:dN_D^L/dt}
    \frac{\partial\widetilde{\M N}_D^{(L)}}{\partial t} := \Tr_D\left[ \frac{\partial\widetilde{\M P}^{(L)}}{\partial t} \, \M S \right]
    &=& \int dE \, f(E-\mu_L) \sum_{n\in N_L,k} \delta(E-\epsilon_{L,n}(k)) \times
  \\
  & & \times 
  \Tr_D\left[ 
    \M S \, \frac{\partial\vec{\Psi}_{n}(k)}{\partial t} [\vec{\Psi}_{n}(k)]^\dagger
    + \vec{\Psi}_{n}(k) \frac{\partial[\vec{\Psi}_{n}(k)]^\dagger}{\partial t} \, \M S 
  \right].
  \nonumber
\end{eqnarray}
For the first term of the trace we find:
\begin{eqnarray}
  \label{eq:trace1}
  \lefteqn{
%%    \Tr_D\left[\frac{\partial\vec{\Psi}_{n}(k)}{\partial t} [\vec{\Psi}_{n}(k)]^\dagger \, \M S \right]
%%    = 
    \hspace{0.5cm} \Tr_D\left[\M S \, \frac{\partial\vec{\Psi}_{n}(k)}{\partial t} [\vec{\Psi}_{n}(k)]^\dagger \right] 
    = \frac{1}{i\hbar}\Tr_D\left[\M H \, \vec{\Psi}_{n}(k) [\vec{\Psi}_{n}(k)]^\dagger \right]
  }
  \nonumber\\
%%  & & 
%%  = \frac{1}{i\hbar}\Tr\left[ [\M H \, \vec{\Psi}_{n}(E)]_D [\vec{\Psi}_{n}(E)]^\dagger_D \right]
%%  \\
  & &
  = \frac{1}{i\hbar}\Tr\left[ 
    \M H_{DL} [\vec{\Psi}_{n}(k)]_L [\vec{\Psi}_{n}(k)]^\dagger_D
    + \M H_D  [\vec{\Psi}_{n}(k)]_D [\vec{\Psi}_{n}(k)]^\dagger_D 
%%  \right.
%%  \nonumber\\ 
%%  & & \textcolor{white}{
%%    = \frac{1}{i\hbar}\Tr\left[
%%      \M H_{DL} [\vec{\Psi}_{n}(E)]_L [\vec{\Psi}_{n}(E)]^\dagger_D 
%%    \right.
%%  }
%%  + \left.  
    + \M H_{DR} [\vec{\Psi}_{n}(k)]_R [\vec{\Psi}_{n}(k)]^\dagger_D 
  \right] 
%% \hspace{3cm}
 \nonumber \\
\end{eqnarray}
where we have made use of the invariance of the trace under cyclic permutations and the fact that 
$\vec{\Psi}_{n}(k)$ is an eigenstate of the system, so that 
$i\hbar \, \M S \, \frac{\partial\vec{\Psi}_{n}(k)}{\partial t} = \M H \, \vec{\Psi}_{n}(k)$. 

Analogously, we find for the second term of the trace (\ref{eq:dN_D^L/dt}):
\begin{eqnarray}
  \label{eq:trace2}
 \lefteqn{
   \hspace{0.5cm}\Tr_D\left[ \vec{\Psi}_{n}(k) \frac{\partial[\vec{\Psi}_{n}(k)]^\dagger}{\partial t} \, \M S \right]
   = -\frac{1}{i\hbar}\Tr_D\left[\vec{\Psi}_{n}(k) [\vec{\Psi}_{n}(k)]^\dagger \, \M H \right]
 }
 \nonumber\\
 & & = -\frac{1}{i\hbar}\Tr\left[ 
   [\vec{\Psi}_{n}(k)]_D [\vec{\Psi}_{n}(k)]^\dagger_L \M H_{LD}
%% \right]
%% -\frac{1}{i}\Tr\left[ 
   + [\vec{\Psi}_{n}(k)]_D [\vec{\Psi}_{n}(k)]^\dagger_D \M H_{D} 
%%   \right.
%% \right]
%% \nonumber\\
%% & & \textcolor{white}{=}
%% -\frac{1}{i}\Tr\left[ 
%% \nonumber\\
%%  & & \textcolor{white}{ 
%%    = -\frac{1}{i}\Tr\left[ 
%%      [\vec{\Psi}_{n}(E)]_D [\vec{\Psi}_{n}(E)]^\dagger_L \M H_{LD}
%%    \right.
%%    }
%%  + \left. 
    + [\vec{\Psi}_{n}(k)]_D [\vec{\Psi}_{n}(k)]^\dagger_R \M H_{RD}
 \right]
 \nonumber\\
\end{eqnarray}

When adding up the two contributions to the trace in (\ref{eq:dN_D^L/dt}), 
the term involving $\M H_D$ in (\ref{eq:trace1}) cancels with the term involving
$\M H_D$ in (\ref{eq:trace2}). For later convenience we add the following
term which is equal to zero to the trace: 
\begin{eqnarray}
  0 &=& 
  \frac{1}{i\hbar}\Tr_D\left[E_n(k) \M S \, \vec{\Psi}_{n}(k) [\vec{\Psi}_{n}(k)]^\dagger \right] 
  -\frac{1}{i\hbar}\Tr_D\left[\vec{\Psi}_{n}(k) [\vec{\Psi}_{n}(k)]^\dagger \, E_n(k) \M S \right]
  \nonumber\\
  &=& \frac{1}{i\hbar}\Tr\left[ 
    E_n(k) \M S_{DL} \, [\vec{\Psi}_{n}(k)]_L [\vec{\Psi}_{n}(k)]^\dagger_D
  \right]
  + \frac{1}{i\hbar}\Tr\left[
    E_n(k) \M S_{DR} \, [\vec{\Psi}_{n}(k)]_R [\vec{\Psi}_{n}(k)]^\dagger_D 
  \right]
  \nonumber \\
  &-& \frac{1}{i\hbar}\Tr\left[ 
    [\vec{\Psi}_{n}(k)]_D [\vec{\Psi}_{n}(k)]^\dagger_L \, E_n(k) \M S_{LD}
  \right]
  -\frac{1}{i\hbar}\Tr\left[ 
    [\vec{\Psi}_{n}(k)]_D [\vec{\Psi}_{n}(k)]^\dagger_R \, E_n(k) \M S_{RD}
  \right],
  \nonumber
\end{eqnarray}
where the two terms involving $S_D$ have canceled out each other.
Summing up all contributions to the trace in (\ref{eq:dN_D^L/dt}) 
and grouping together on the one hand terms involving hopping between 
the left electrode and the device and on the other hand terms involving 
hopping between the right electrode and the device, we get in total:
\begin{equation}
  \Tr_D\left[ 
    \M S \, \frac{\partial\vec{\Psi}_{n}(k)}{\partial t} [\vec{\Psi}_{n}(k)]^\dagger
    + \vec{\Psi}_{n}(k) \frac{\partial[\vec{\Psi}_{n}(k)]^\dagger}{\partial t} \, \M S 
  \right] = j_{L,n}(k) + j_{R,n}(k)
\end{equation}
where
\begin{equation*}
  j_{L,n}(k) := \frac{1}{i\hbar} \Tr\left[     
    [\vec{\Psi}_{n}(k)]^\dagger_D \M W_{DL}(E_n(k)) \, [\vec{\Psi}_{n}(k)]_L 
    -[\vec{\Psi}_{n}(k)]^\dagger_L \, \M W_{LD}(E_n(k)) [\vec{\Psi}_{n}(k)]_D \,
  \right],
\end{equation*}
and
\begin{equation*}
  j_{R,n}(k) := \frac{1}{i\hbar} \Tr\left[
    [\vec{\Psi}_{n}(k)]^\dagger_D \M W_{DR}(E_n(k)) \, [\vec{\Psi}_{n}(k)]_R 
    - [\vec{\Psi}_{n}(k)]^\dagger_R \, \M W_{RD}(E_n(k)) [\vec{\Psi}_{n}(k)]_D \,
  \right].
\end{equation*}
The term $j_{L,n}(k)$ gives the current coming from the left lead, while $j_{R,n}(k)$
gives the current coming from the right electrode. We proceed by expressing the states 
$\ket{\Psi_n(k)}$ in terms of the incoming waves $\ket{\Psi_n^i(k)}$ by the help of 
expression (\ref{eq:response}) for the response to an incoming wave:
\begin{eqnarray}
  {[\vec{\Psi}_n(k)]}_D &=& \widetilde{\M{G}}_D(E) \, \M{W}_{DL}(E) \, [\vec{\Psi}_n^i(k)] 
  \\
  \Rightarrow {[\vec{\Psi}_n(k)]}_D^\dagger &=& [\vec{\Psi}_n^i(k)]^\dagger \, \M{W}_{LD}(E) \, \widetilde{\M{G}}^\dagger_D(E) 
  \\
  {[\vec{\Psi}_n(k)]}_R &=& \widetilde{\M{G}}_{RD}(E) \, \M{W}_{DL}(E) \, [\vec{\Psi}_n^i(k)]
  \nonumber\\
  &\stackrel{(\ref{eq:gf:GTilde_RD})}{=}& 
  \widetilde{\M g}_{R}(E) \, \M W_{RD}(E) \, \widetilde{\M G}_{D}(E) \M{W}_{DL}(E) [\vec{\Psi}_n^i(k)]
  \\
  \Rightarrow {[\vec{\Psi}_n(k)]}_R^\dagger 
  &=& [\vec{\Psi}_n^i(k)]^\dagger \M{W}_{LD}(E) \widetilde{\M G}^\dagger_{D}(E) \, 
  \M W_{DR}(E) \, \widetilde{\M g}^\dagger_{R}(E) 
\end{eqnarray}
where the energy argument of the GFs $E$ is the band energy of the right lead: $E\equiv E^R_n(k)$.
Inserting these expressions into the expression for $j_{R,n}(k)$ defined above we get:
\begin{eqnarray}
  \label{eq:j_Rn(k)}
  \lefteqn{
    j_{R,n}(k) = \frac{1}{i\hbar} 
    \Tr\left[ [\vec{\Psi}_n^i(k)]^\dagger \, \M{W}_{LD}(E) \, \widetilde{\M{G}}^\dagger_D(E) \times
    \right.
  }
  \nonumber\\
  & &
  \times \underbrace{
    \M W_{DR}(E) \, 
    \left\{
      \widetilde{\M g}_{R}(E) - \widetilde{\M g}^\dagger_{R}(E) 
    \right\}
    \, \M W_{RD}(E)
  }_{-i\widetilde{\M\Gamma}_R(E)} \,
  \widetilde{\M{G}}_D(E)) \, \M{W}_{DL}(E) 
  \vec{\Psi}_n^i(k)
  \Big]  
  \nonumber\\
%  & & = \frac{1}{i\hbar} \Tr\left[ [\vec{\Psi}_n^i(k)]^\dagger \times \right.
%  \nonumber\\
%  & &
%  \left.
%    \M{W}_{LD}(E_n(k)) \,
%    \widetilde{\M{G}}^\dagger_D(E_n(k)) \, 
%    \left\{ \widetilde{\M \Sigma}_{R}(E_n(k)) - \widetilde{\M \Sigma}^\dagger_{R}(E_n(k)) \right\} \,
%    \widetilde{\M G}_{D}(E_n(k)) \ \M{W}_{DL}(E_n(k)) \,
%    \vec{\Psi}_n^i(k)
%  \right]
%  \nonumber\\
  & &
  = -\frac{1}{\hbar} 
  \Tr\left[
    \widetilde{\M{G}}^\dagger_D(E) \, 
    \widetilde{\M \Gamma}_{R}(E) \,
    \widetilde{\M G}_{D}(E) \, \M{W}_{DL}(E) \,
    \vec{\Psi}_n^i(k)  [\vec{\Psi}_n^i(k)]^\dagger \,
    \M{W}_{LD}(E)
  \right].
\end{eqnarray}
Thus summing over all states $n$
resulting from electron injection from the left electrode and integrating over energy
we get the contribution $I_R^{(L)}$ to the current $I_R=I_R^{(L)}+I_R^{(R)}$ resulting 
from electron injection from the left electrode.
\begin{eqnarray}
  I^{(L)}_R 
  &=& \int dE \, f(E-\mu_L) \sum_{n\in N_L,k}  \delta(E-E_n(k)) \, j_{R,n}(k)
  \nonumber\\
  &=& -\frac{1}{\hbar} \int dE \, f(E-\mu_L) \,
  \Tr\bigg[
    \widetilde{\M{G}}^\dagger_D(E) \, 
    \widetilde{\M \Gamma}_{R}(E) \,
    \widetilde{\M G}_{D}(E)
  \nonumber\\
  & & 
    \M{W}_{DL}(E) \,
    \sum_{n\in N_L,k}  \delta(E-E_n^R(k)) \vec{\Psi}_n^i(k)  [\vec{\Psi}_n^i(k)]^\dagger \,
    \M{W}_{LD}(E)
  \bigg] 
  \nonumber\\
  &=& -\frac{1}{h} \int dE \, f(E-\mu_L)
  \Tr\left[
    \widetilde{\M{G}}^\dagger_D(E) \, 
    \widetilde{\M \Gamma}_{R}(E) \,
    \widetilde{\M G}_{D}(E) \, 
    \widetilde{\M \Gamma}_{L}(E)
  \right].
\end{eqnarray}
In a similar way we get for the current $I^{(R)}_L$ through the left lead resulting from 
electron injection from the right reservoir: 
\begin{eqnarray}
  I^{(R)}_L
  &=& -\frac{1}{h} \int dE \, f(E-\mu_R)
  \Tr\left[
    \widetilde{\M{G}}^\dagger_D(E) \, 
    \widetilde{\M \Gamma}_{R}(E) \,
    \widetilde{\M G}_{D}(E) \, 
    \widetilde{\M \Gamma}_{L}(E)
  \right].
\end{eqnarray}

In steady-state the contributions to the current due to electron injection
from the left reservoir on the one hand and from the right reservoir on the
other hand are individually conserved, i.e. $I^{(L)}_L=-I^{(L)}_R\equiv I^{(L)}$ 
and $I^{(R)}_L=-I^{(R)}_R \equiv I^{(R)}$. Thus in total the net current 
{\it through} the device is $I = I^{(L)} + I^{(R)} =  -I^{(L)}_R+I^{(R)}_L$,
and we finally obtain the famous Landauer formula for the current expressed
with GFs.:
\begin{eqnarray}
  \label{eq:Landauer-GF}
  I &=& \frac{1}{h} \int dE \, ( f_L(E)-f_R(E) ) \,
  \Tr\left[
    \widetilde{\M{G}}^\dagger_D(E) \, 
    \widetilde{\M \Gamma}_{R}(E) \,
    \widetilde{\M G}_{D}(E) \, 
    \widetilde{\M \Gamma}_{L}(E)
  \right].
\end{eqnarray}

The trace in (\ref{eq:Landauer-GF}) corresponds to the total transmission function
$T(E)$ which is the sum over all channel transmissions $T_n(E)$ defined earlier:
\begin{equation}
  \label{eq:Caroli}
  T(E) := \sum_n T_n(E) \equiv
  \Tr\left[
    \widetilde{\M{G}}^\dagger_D(E) \, 
    \widetilde{\M \Gamma}_{R}(E) \,
    \widetilde{\M G}_{D}(E) \, 
    \widetilde{\M \Gamma}_{L}(E)
  \right].
\end{equation}
This is the so-called Caroli expression for the transmission function $T(E)$ named
after C. Caroli who first derived it for the simple case of a one-dimensional tight-binding 
chain \cite{Caroli:jphysc:71}.

Obviously, the (non-hermitian) expression inside the trace of (\ref{eq:Caroli}) must 
have to do with the (hermitian) transmission matrix $\M T(E)$ defined earlier in eq. 
(\ref{eq:TMat}). Exploiting the invariance of the trace under cyclic permutations
we find an equivalent expression for the transmission function now involving a hermitian
expression that we can identify with the transmission matrix:
\begin{equation}
  \label{eq:NEGF-TMat}
  \M T(E) \equiv  
  \widetilde{\M \Gamma}^{1/2}_{L}(E) \,
  \widetilde{\M{G}}^\dagger_D(E) \, 
  \widetilde{\M \Gamma}_{R}(E) \,
  \widetilde{\M G}_{D}(E) \, 
  \widetilde{\M \Gamma}^{1/2}_{L}(E).
\end{equation}
Now we can also identify the transmission amplitude 
$\M t(E)=(t_{mn}(E))$ with:
\begin{equation}
  \M t(E) \equiv   
  \widetilde{\M \Gamma}_{R}^{1/2}(E) \,
  \widetilde{\M G}_{D}(E) \, 
  \widetilde{\M \Gamma}^{1/2}_{L}(E).
\end{equation}
The generalized power of a hermitian matrix $\M A$ is defined as $\M A^q = \M U \, {\rm diag}(a_i^q) \, \M U^\dagger$
where the $a_i$ are the eigenvalues of the matrix $\M A$ and $\M U$ is the unitary transformation that diagonalizes 
$\M A$.

We note that the dimension of the transmission matrix (\ref{eq:TMat}) obtained in Sec. \ref{sec:Landauer} 
is equal to either the number of modes of either the left or the right lead, which in turn is equal to the 
dimension of the unit cell of either lead. But the transmission matrix derived here has the dimension of the
device subspace which is bigger than the dimensions of the unit cells of either lead. So the transmission
matrix (\ref{eq:NEGF-TMat}) derived from the NEGF is strictly speaking not equal to the original one 
(\ref{eq:TMat}) of the Landauer formalism. However, both formulations are of course equivalent. In fact,
all non-zero matrix elements of the transmission matrix in the NEGF correspond to the original Landauer 
transmission matrix. All other matrix elements of the NEGF transmission matrix are zero.

%%Remarks:
%%\begin{itemize}
%%\item Equivalent definition: 
%%  \[
%%  \M T(E) \equiv  
%%  \widetilde{\M \Gamma}^{1/2}_{R}(E) \,
%%  \widetilde{\M{G}}^\dagger_D(E) \, 
%%  \widetilde{\M \Gamma}_{L}(E) \,
%%  \widetilde{\M G}_{D}(E) \, 
%%  \widetilde{\M \Gamma}^{1/2}_{R}(E)
%%  \]
%%  Non-zero only in $r$-subspace of device.
%%\end{itemize}

\section{Application to simple models}
\label{sec:models}

In order to illustrate some aspects of the NEGF formalism derived above, it is applied here to some 
simple model systems. We start with the well-known example of the simple tight-binding chain with one 
orbital per atom. While the standard tight-binding model neglects the overlap between neighboring atomic
orbitals, the overlap is included here as an additional parameter $s$ in order to illustrate its effect on
the electronic structure. Thus we have for the matrix elements of the Hamiltonian:
\begin{eqnarray}
  \label{eq:tb-model}
  \bra{m} \hat{H} \ket{n} = H_{mn} =
  \left\{  
    \begin{array}{l}
      \epsilon_0 \mbox{ for } m=n, \\
      t \mbox{ for } \abs{m-n} = 1, \\
      0 \mbox{ otherwise } 
    \end{array}
  \right. 
\end{eqnarray}
and for the overlap matrix elements:
\begin{eqnarray}
  \bracket{m}{n} = S_{mn} = 
  \left\{  
    \begin{array}{l}
      1\mbox{ for } m=n, \\
      s \mbox{ for } \abs{m-n} = 1, \\
      0 \mbox{ otherwise } 
    \end{array}
  \right. 
\end{eqnarray}
Note, that the representation of the Hamiltonian in operator form in the NOBS 
is {\it not} straight forward, since it involves the inversion of the overlap matrix 
$\M{S}=(S_{mn})$ as pointed out in App. \ref{app:NOBS}, eq. (\ref{eq:NOB-Op}):
\begin{eqnarray}
    \Op H = \sum_{mn} \ket{m} (\M{S}^{-1} \M{H} \M{S}^{-1} )_{mn} \bra{n}.
\end{eqnarray}
Thus one introduces hopping terms beyond the first neighbor into the Hamiltonian operator 
although the corresponding matrix elements are zero. Orthogonalizing the basis set for example
by the L\"owdin scheme one would also obtain hopping terms beyond the the nearest neighbor
hopping. But we will not pursue this any further. Instead we will apply the above developed 
NEGF formalism for a NOBS to the model and study the effect of the overlap
on the DOS.

\begin{figure}
  \begin{center}
    \includegraphics[width=0.9\linewidth]{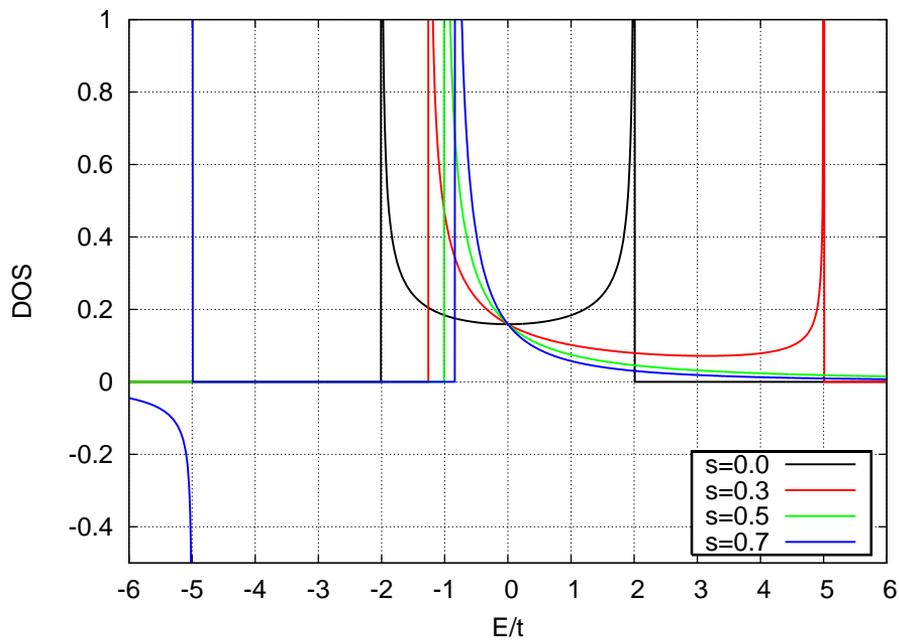}
  \end{center}
  \setcapindent{0cm}   
  \caption{
    PDOS of tight-binding chain with overlap for different values of the overlap parameter $s$
    but for fixed on-site energy parameter $\epsilon_0$ and hopping parameter $t$.
  }
  \label{fig:tb-chain}
\end{figure}

The PDOS for a NOBS is given by eq. (\ref{eq:NOB-PDOS}). The PDOS of the
tight-binding chain is plotted in Fig. \ref{fig:tb-chain} for various parameters of the nearest
neighbor overlap $s$, while the on-site energy and the hopping are kept fixed: $\epsilon_0:=0$
and $t:=-1$. We observe, that increasing the overlap leads to a displacement towards higher
energies and a simultaneous broadening of the energy band until at $s=1/2$ the band width becomes
infinite. This is due to a singularity of the energy dispersion at the Brillouin zone boundary
$ka = \pm\pi$:
\begin{equation}
  \label{eq:NOB-tb-dispersion}
  E(k) = \frac{\epsilon_0 + 2t \cos( ka )}{1+2s \cos( ka)} \longrightarrow \infty \mbox{ for } 
  ka \rightarrow \pm\pi \mbox{ and } s=\frac{1}{2}.
\end{equation}
Increasing the overlap $s$ beyond this value, the DOS becomes even negative for some values of
the energy. This pathological behavior can be traced back to the appearance of singularities inside
the first Brillouin zone, and relates to the fact that the tight-binding model is actually 
pathological for overlaps $s>1/2$. In fact, it is physically impossible to have a nearest neighbor 
overlap of $s>1/2$ but no overlap between second nearest neighbors. The pathological behavior can 
be repaired quite easily by including overlap beyond the nearest neighbor approximation. This can 
be seen directly from the energy dispersion of the generalized tight-binding model including hopping 
and overlaps beyond the first nearest neighbor:
\begin{equation}
  \label{eq:eq:NOB-gen-tb-dispersion}
  E(k) = \frac{\epsilon_0 + \sum_{n} 2t_n \cos( k n a )}{1+ \sum_{n} 2 s_n \cos( k n a )},
\end{equation}
where $s_n$ and $t_n$ denote overlap and hopping, respectively, between $n$-th nearest neighbors.
Thus in order to avoid unphysical results it is important to include overlap between atomic orbitals 
up to a sufficiently high degree of neighborhood. This will become very important in the context of
combining {\it ab initio} electronic structure calculations with the NEGF formalism as the 
the basis sets employed in common quantum chemistry packages like GAUSSIAN \cite{Gaussian:03}
are typically highly non-orthogonal.

\begin{figure}
  \begin{center}
    \includegraphics[width=0.5\linewidth]{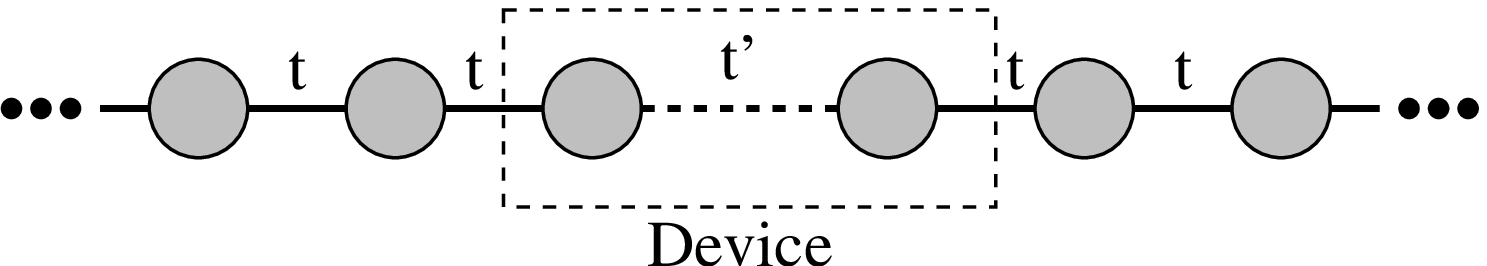}
    \includegraphics[width=0.75\linewidth]{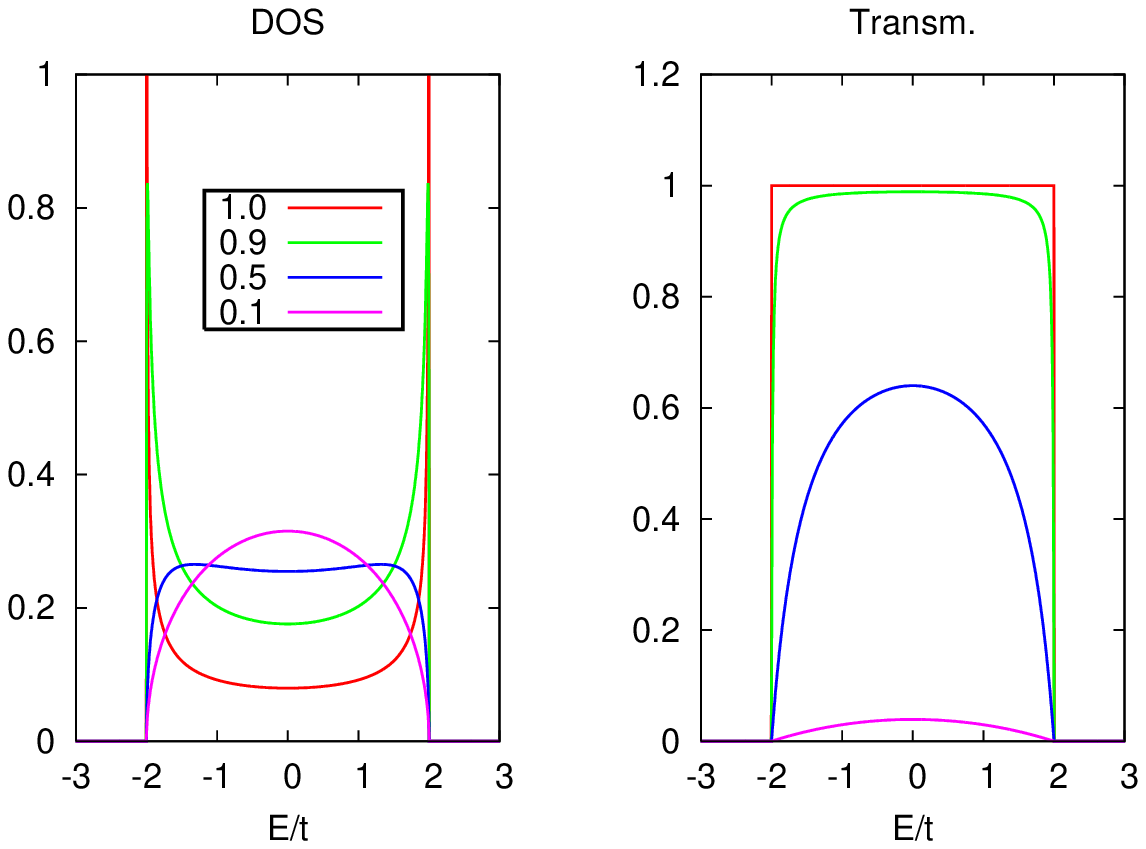}
  \end{center}
  \setcapindent{0cm}   
  \caption{
    Tunneling between two semi-infinite chains. The graphs show the PDOS projected onto a
    tip atom (left) and the transmission (right) for different values of the hopping $t^\prime$
    between the chains in units of the hopping parameter $t$ of the tight-binding chains. 
    Overlap between atomic orbitals is completely neglected.
  }
  \label{fig:tunneling}
\end{figure}

Next, we will have a look at the electronic structure and the transport properties of two tunnel-coupled
semi-infinite tight-binding chains. In Fig. \ref{fig:tunneling} the PDOS projected onto one of the tip atoms
(bottom left) and the transmission function (bottom right) is shown for the model depicted at the top of
the figure for various values of the tunnel coupling $t^\prime$ between the two chains. We observe, how
the PDOS is slowly transformed from the DOS of the 1D chain with the characteristic van Hoft singularities
at the band edges to the semicircular ``surface'' DOS of the semi-infinite chain when the tunneling coupling
$t^\prime$ is decreased. Simultaneously, the perfect transmission of the 1D chain within the energy band
is reduced considerably when decreasing the tunnel coupling as is expected. For $t^\prime=0.1 t$ we enter the 
tunneling regime where the transmission is described by the Bardeen formula for tunneling 
$T \approx t^\prime/t \mathcal{D}(\epsilon_F)$ \cite{Bardeen:prl:61}.

\begin{figure}
  \begin{center}
    \includegraphics[width=0.6\linewidth]{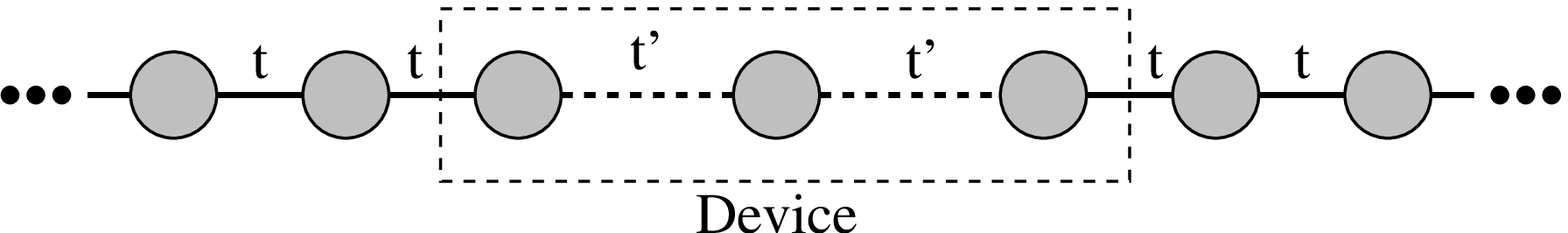}
    \includegraphics[width=0.75\linewidth]{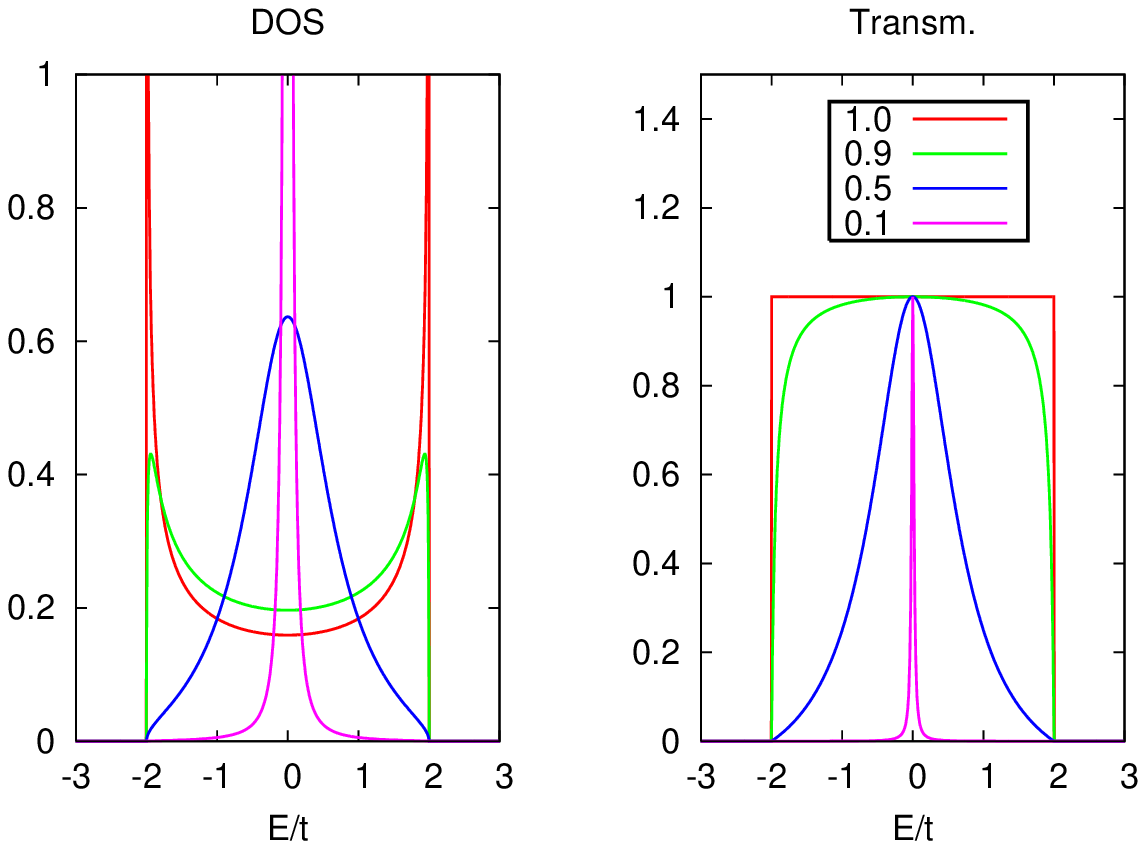}
  \end{center}
  \setcapindent{0cm}   
  \caption{
    Tunneling through a dot connected to two semi-infinite leads. The graphs show the 
    PDOS projected onto the central dot (left) and the transmission (right) for different
    values of the coupling $t^\prime$ of the dot to the two leads  in units of the hopping 
    parameter $t$ of the tight-binding chains. 
  }
  \label{fig:dot}
\end{figure}

Finally, we have a look at a simplified model of a typical situation in mesoscopic physics
and molecular electronics: An island (which could be a quantum dot in a mesoscopic system or 
a molecule in molecular electronics) coupled to two semi-infinite leads. The island in the 
simple model is a single site coupled to two semi-infinite tight-binding chains (see the 
illustration at the top of Fig. \ref{fig:dot}). We start again with strong coupling to the 
leads, $t^\prime \equiv t$, and decrease $t^\prime$. We observe that the PDOS projected onto 
the island is tranformed from the DOS of the perfect 1D chain for ($t^\prime=t$) to a 
Lorentz peak for weak coupling and also the transmission is now given by a Lorentz peak
in that limit. We further note that the transmission remains perfect exactly at the on-site
energy of the island although the coupling has decreased by one order of magnitude. This 
phenomenon is called resonant transmission through a molecular (or atomic) level.

\section{Discussion of the Landauer approach}
\label{sec:discussion}

The Landauer formalism introduced above, assumes that electron transport in nanoscopic
conductors is phase coherent. Inelastic scattering of electrons e.g. in electron-electron
or electron-phonon scattering processes is neglected. This turns out to be a rather good
approximation for low temperature and small bias voltages. The reason for this is that 
for small bias voltages the transport properties are principally determined by the 
conduction electrons at the Fermi level. At low temperatures the conduction electrons
at the Fermi level cannot loose energy because the Fermi sea is completely occupied.
On the other hand they cannot gain energy because there are no phonons available at low
temperatures. Therefore electron transport is elastic, and thus phase coherent at low
temperature and for small bias.

A strict generalization of the Landauer formula to interacting electron systems
(i.e. including electron-electron interaction and/or electron-phonon interaction)
is given by the Meir-Wingreen formula \cite{Meir:prl:92} which assumes an interacting
device coupled to non-interacting leads. The Meir-Wingreen formula actually reduces to 
the Landauer formula in the limit of zero temperature and zero bias, thus demonstrating
the above considerations strictly. Thus in principle the Landauer formula can be considered
a strict result (for low temperature and small bias) under the condition that the GF of the
device is the true interacting GF (see e.g. the book by Mahan \cite{Mahan:book:00} for an introduction
to many-body GFs). However, the GF employed here is the one-body GF describing non-interacting
electrons. Thus the electron-electron interaction can only be taken into account on a mean-field
level. The task is now to find a one-body GF that approximates the true interacting many-body GF
sufficiently well. In Ch. \ref{ch:ab-initio} we show how to combine the NEGF formalism with 
{\it ab-initio} electronic structure calculations on the level of DFT,
i.e. the one-body GF is calculated from the Kohn-Sham Hamiltonian. Although DFT has been quite successful 
in describing many materials it is not clear {\it a priori}, whether it can give a reasonable 
description of the transport properties of nanoscopic conductors. This issue is discussed in Ch. 
\ref{ch:ab-initio}. 

The effect of inelastic scattering by electron-phonon interaction on the electrical transport through Au 
nanocontacts \cite{Viljas:prb:05} and atomic Au wires \cite{Frederiksen:prl:04} has been studied recently
with microscopic models on the basis of the Meir-Wingreen formula with different approximations for the 
electron-phonon coupling. Viljas {\em et al.} treated the electron-phonon interaction in lowest order perturbation 
theory in combination with a tight-binding model of the electronic structure, while Frederikson {\em et al.} 
treated the interaction within the self-consistent Born approximation (SCBA) \cite{Mahan:book:00} in 
combination with an {\it ab initio} description of the electronic structure on the DFT level. 

In principle, it is also possible to introduce decoherence by inelastic scattering into the Landauer formalism 
phenomenologically by introducing ``conceptual voltage probes'' which act as phase-breaking scatterers as was 
first realized by B\"uttiker\cite{Buettiker:ibmjrd:88}. However, this can only be done in a meaningful way for 
very simple models systems \cite{DAmato:prb:90}. To obtain meaningful results for realistic models of nanoscopic conductors 
it is necessary to develop a microscopic theory for the effect of inelastic scattering on the transport
like the ones mentioned above.

\chapter{{\it Ab initio} quantum transport}
\label{ch:ab-initio}

Density functional theory (DFT) has become a standard method for electronic structure calculations
in condensed matter physics and quantum chemistry \cite{Koch:book:01}. This chapter explains how the 
transport theory described in Ch. \ref{ch:transport} can be combined with DFT to describe transport 
through nanostructures from first principles. Therefore, I will first briefly introduce the basic 
principles of DFT in Sec. \ref{sec:DFT}. An extensive review of DFT can be found e.g. in Ref.
\cite{Jones:rmp:89}. Finally, in Sec. \ref{sec:KS-NEGF}, I will describe how the DFT based quantum 
transport approach is implemented in the ALACANT package.

\section{Density functional theory}
\label{sec:DFT}

The electrons in a solid or a molecule are described by a many-body wavefunction $\Psi(\{\vec r_i\})$ 
of the electron coordinates $\{\vec r_i\}$, which is a solution of the many-body Schr\"odinger equation:
\begin{equation}
  \label{eq:MB-Schroedinger}
  \hat{\mathcal H} \, \ket{\Psi} = E \ket{\Psi}.
\end{equation}
$\hat{\mathcal H}$ is the Hamiltonian of the interacting electron system, containing the kinetic energy 
$\hat{\mathcal T}$, the Coulomb attraction of the atomic nuclei (ion cores) $\hat{\mathcal V}_{e-i}$
and the electron-electron interaction $\hat{\mathcal V}_{e-e}$:
\begin{equation}
  \label{eq:MB-Hamiltonian}
  \hat{\mathcal H} = \hat{\mathcal T} + \hat{\mathcal V}_{e-i} + \hat{\mathcal V}_{e-e}.
\end{equation}
The many-body Schr\"odinger equation can in principle be solved by expanding the wavefunction
in a basis of Slater determinants, thus converting it to a problem of diagonalizing matrices.
However, in practice this can only be done for systems with very few electrons as the dimension 
of the Slater basis grows with the number $N$ of electrons as $N!$. Instead, before the advent of
DFT the standard approach in solid state physics and in quantum chemistry has been to approximate
the many-body wave function by a single Slater determinant that minimizes the total energy. This
is the Hartree-Fock approximation (HFA) which is the basis for more refined approaches in quantum
chemistry, like perturbation theory, or configuration interaction \cite{Szabo:book:89}. However, 
the HFA gives a rather crude description of the electronic structure of solids and molecules. The 
main reason is that the single-determinant approach neglects electron correlations, but instead
describes the electrons as independent of each other interacting via an effective mean field. 
In particular, the description of metallic solids within the HFA is very bad, describing them
often as insulators or semiconductors. The above mentioned systematic improvement of HFA by 
perturbation theory or configuration interaction on the other hand becomes computationally too 
expensive for larger number of electrons just as the exact diagonalization approach described 
above, limiting these methods to systems with a relatively small number of electrons.

An alternative approach is density functional theory (DFT) where the basic variable is the 
electron density instead of the full many-body electron wave function $\Psi(\{\vec r_i\})$:
\begin{equation}
  n(\vec r) = \bra{\vec{r}} \hat{P} \ket{\vec{r}} = N \int 
  \abs{ \Psi( \vec r, \vec r_2,\ldots\vec r_N) }^2 d\vec r_2 \ldots d\vec r_N,
\end{equation}
which is a much simpler quantity  to handle than the full many-body wavefunction.

\subsection{Hohenberg-Kohn theorem}
\label{sub:HK-theorem}

The basis for DFT is the Hohenberg-Kohn (HK) theorem \cite{Hohenberg:pr:64} which reduces
the fully interacting $N$-electron problem to determining the ground state electron
density $n(\vec r)$: 

\begin{enumerate}
\item The non-degenerate ground state energy of an $N$ electron system, $E_N$, is
  a unique, universal functional of the electron density $n(\vec r)$. 
\item The electron density that minimizes the energy functional is the exact
  ground state electron density
\end{enumerate}

So if we know the energy functional of the electron density $E[n]$ then we can determine
the electron density of the ground state by simply minimizing the energy functional, which
would be much simpler than resolving the many-body Schr\"odinger equation. It is important to 
note, that no approximations has been made so far. Thus unlike the HFA, DFT is in principle 
an exact theory which allows us to calculate the exact ground state density of an interacting 
electron system. However, the drawback is that the exact energy functional is not known. So 
one has to come up with approximations for the energy functional. Different approximations to 
the energy functional are discussed after the next subsection. 

\subsection{Kohn-Sham equations}
\label{sub:Kohn-Sham}

The HK theorem actually does not give a practical recipe for calculating the energy functional
of the electron density, but only states the existence of that functional. A practical
way for calculating the energy functional is given by the Kohn-Sham (KS) equations \cite{Kohn:pr:65}
which map the interacting electron system with some electron density $n(\vec r)$ onto an
auxiliary non-interacting system with the same electron density as the interacting system.
The KS equations are thus the starting point for any practical implementation of DFT.

We start by decomposing the energy functional into different contributions:
\begin{equation}
  E[n] = T[n] + E_H[n] + E_{xc}[n] + E_{ext}[n].
\end{equation}
$T[n]$ is the kinetic energy for that electron density
\begin{equation}
  T[n] = \bra{\Psi} \hat{\mathcal T} \ket{\Psi},
\end{equation}
where $\ket{\Psi}$ is the many-electron wavefunction corresponding to the electron
density $n$ for that system. Note, that there is no explicit formula for computing the 
kinetic energy from the electron density $n$. This is one of the reasons why there is 
no explicit formula for the energy functional.
$E_H[n]$ is the Hartree term describing the classical Coulomb repulsion of the electron cloud.
For this term an explicit formula can be written for its functional dependence on the
electron density:
\begin{equation}
  E_H[n] = \frac{1}{2} \int \frac{n(\vec r)n(\vec r^\prime)}{\abs{\vec r - \vec r^\prime}} d\vec{r} d\vec{r}^\prime
\end{equation}
$E_{xc}[n]$ is the so-called exchange-correlation (XC) term and no explicit exact formula
exists for this term either. It is thus the other term besides the kinetic energy 
term responsible for the impossibility of finding an explicit expression for the total energy 
functional. The XC term contains pure quantum effects like the exchange interaction due
to the Pauli principle and electron correlation effects. This term is usually expressed
in terms of an integral over the electron density and an unknown XC energy density 
$\epsilon_{xc}[n](\vec r)$:
\begin{equation}
  E_{xc}[n] = \int d\vec{r} \, n(\vec{r}) \epsilon_{xc}[n](\vec r).
\end{equation}
Finally, the last term $E_{ext}[n]$ is the interaction of the electron cloud with the external 
potential due to the atomic nuclei. This term can again be written as an explicit functional
of the electron density:
\begin{equation}
  E_{ext}[n] = \int d\vec{r} \, n(\vec{r}) V_{e-i}(\vec{r})
\end{equation}

The basic idea of the KS method is to introduce an auxiliary set of one-electron 
wavefunctions $\{\phi_i(\vec{r})\}$ that give rise to the same electron density $n(\vec r)$ 
as the full many-body wavefunction $\Psi(\vec r_1,\ldots,\vec r_N)$:
\begin{equation}
  n(\vec{r}) = \sum_{i=1}^{N} \phi_i^\ast(\vec{r}) \phi_i(\vec{r}) \, \mbox{ with } 
  \bracket{\phi_i}{\phi_j} = \delta_{ij}
\end{equation}

We can then define a KS energy functional $T_{KS}[n]$ that can easily be calculated 
from the auxiliary KS wavefunctions $\phi_i$:
\begin{equation}
  T_{KS}[n] = \sum_{i=1}^N \bra{\phi_i} \hat{\mathcal{T}} \ket{\phi_i}.
\end{equation}
However, the KS kinetic energy term is not identical with the kinetic energy term
of the interacting electron system $T[n]$. The difference between the KS kinetic energy 
and the true kinetic energy is again unknown, and is absorbed into the also unknown
XC functional which is redefined in the KS method as:
\begin{equation}
  E_{XC}[n] \rightarrow E_{XC}^\prime[n] = E_{XC}[n] + T[n] -T_{KS}[n].
\end{equation}
%%
%%In analogy to the Hartree term and the term for the energy due to the external potential, 
%%one often writes the XC functional in terms of an integral over the XC potential and the 
%%electron density:
%%
%%\begin{equation}
%%  E_{XC}[n] = \frac{1}{2} \int n(\vec{r}) \epsilon_{XC}[n](\vec{r}) d\vec{r},
%%\end{equation}
%%
%%where the XC potential is defined as the functional derivative of the XC energy functional:
%%
%%\begin{equation}
%%  \epsilon_{XC}[n] := \frac{\delta E_{XC}[n]}{\delta n}(\vec{r}).
%%\end{equation}

Thus all many-body effects have now been shifted to the XC functional,
which is the only contribution to the total energy functional that needs to be 
approximated. All other terms can be calculated exactly within the KS method.
The main advantage of the KS method is that we get a set of effective 
one-body Schr\"odinger equations --the KS equations-- for the auxiliary KS 
wavefunctions $\phi_i$ which in turn give the ground state electron density of the true 
interacting electron system. From the variational principle it follows that in the ground 
state the total energy must be stationary with respect to variations of the KS wavefunctions:
\begin{equation}
  \frac{\delta E[n]}{\delta\phi_i} - \epsilon_i \phi_i = 0
\end{equation}
where the $\epsilon_i$ are Lagrange multipliers which ensures the orthogonality of 
the KS wavefunctions $\phi_i$ and will give the effective KS eigenenergies. The KS 
equations follow directly:
\begin{equation}
  \left( 
    -\frac{\hbar^2}{2m} \nabla^2 + V_H[n](\vec{r}) + V_{ext}(\vec{r}) + V_{XC}[n](\vec{r})
  \right) \phi_i(\vec{r}) = \epsilon_i \phi_i(\vec{r})
\end{equation}
where 
\begin{equation}
  \label{eq:VH}
  V_H[n](\vec{r}) := \int d\vec{r}^\prime \frac{n(\vec{r}^\prime)}{\abs{\vec{r}-\vec{r}^\prime}}
\end{equation}
is the Hartree potential due to the direct electron-electron interaction and 
\begin{equation}
  V_{XC}[n](\vec{r}) := \frac{\delta E_{XC}[n]}{\delta n}(\vec{r}).
\end{equation}
is the effective exchange-correlation potential.

The interacting problem has now been reduced to a set of non-interacting Schr\"odinger 
equations. One should emphasize that up until now no approximations have been made. The
KS equations are exact if one knows the exact XC functional. The KS Hamiltonian
\begin{equation}
  \hat{H}_{KS} = -\frac{\hbar^2}{2m} \nabla^2 + \hat{V}_{ext} + \hat{V}_H[n] + \hat{V}_{XC}[n]
\end{equation}
depends on the electron density $n(\vec r)$ which is the quantity to be determined by
the KS wavefunctions. Thus the KS equations are a non-linear eigenvalue problem which 
has to be solved self-consistently. The central task of the KS formulation of DFT is 
to find suitable approximations for the unknown XC functional.

\subsection{Energy functionals}
\label{sub:functionals}

\textbf{Local density approximation (LDA)}. The simplest approximation is to start from the homogeneous electron 
gas, and assume that the electron density is only slightly modulated by the potential of the ion cores. In this case 
the XC energy density $\epsilon_{XC}[n](\vec{r})$ at point $\vec{r}$ can be assumed to be that of a homogenous electron 
gas with the same density as the local density of the non-homogeneous electron gas, i.e. 
$\epsilon_{XC}^{LDA}[n](\vec{r}):=\epsilon_{XC}^{HEG}(n(\vec{r}))$. 
\begin{equation}
  \label{eq:LDA}
  E_{XC}^{LDA}[n] = \frac{1}{2}\int n(\vec{r})\epsilon_{XC}^{HEG}(n(\vec{r})) \, d\vec{r}.
\end{equation}
Thus in LDA the XC energy density $\epsilon_{XC}^{LDA}$ is simply a function of the local electron density $n(\vec{r})$ 
and not a functional. In spite of its apparent simplicity LDA has been quite successful in describing metallic
systems. One of its major shortcomings is that it cannot give a reasonable description of strongly localized 
electrons like e.g. the $d$- and $f$-states of transition metals. In this case the electron density is strongly 
modulated and the approximation of a slowly varying density is obviously a bad one.

\textbf{Generalized Gradient Approximation (GGA)}. A logical step for improving LDA is to make the XC
energy density also a function of the gradient of the electron density in addition to the local electron 
density:
\begin{equation}
  \label{eq:GGA}
  E_{XC}^{GGA}[n] = \frac{1}{2}\int n(\vec{r})\epsilon_{XC}(n(\vec{r}),\vec\nabla n(\vec{r})) \, d\vec{r}  
\end{equation}
While LDA works well for simple metals, GGA improves the description of transition metals considerably. However,
in general even GGA does not give a good description of semiconductors and insulators. And in particular, it fails to
give a good description of the so-called strongly correlated materials like the transition metal oxides. This is due 
to the insufficient cancellation of the self-interaction error by the approximate XC functionals of both LDA and
GGA (although the self-interaction error is smaller for GGA). This self-interaction error raises artificially the
 energy of occupied localized states like the $d$- and $f$-states which are of outmost importance in the transition 
metal oxides \cite{Leung:prb:91}.

\textbf{Hybrid functionals}. As said before the main problem of the LDA and GGA functionals
is the spurious self-interaction error for localized electron states like e.g. the $d$- and $f$-states of transition metals,
and the orbitals of molecules. On the other hand in the HFA the exchange interaction term cancels exactly the self-interaction
present in the Hartree-term. Thus a possible way of correcting the spurious self-interaction in the GGA functional is to 
reintroduce Hartree-Fock exchange (HFX) into the KS theory by mixing some HFX with the GGA exchange functional, an idea that was first 
put forward by Becke\cite{Becke:jcp:93}. Although a deeper theoretical justification was given for this ad hoc correction, the 
exact amount of HFX that needs to be reintroduced in order to give a sufficient correction of the self-interaction error 
in the GGA functional is unfortunately unknown and thus needs to be fitted to experiments\cite{Becke:jcp:93}. Thus strictly speaking 
DFT using hybrid functionals does not really represent an ab-initio method. However, it turns out that the parameterization of the
B3LYP hybrid functional which was determined by fitting to experimental results of a certain class of molecules also works 
surprisingly well for many molecules which had not been included into the fitting data set. Moreover, it also gives a reasonable 
description of the electronic structure and of the magnitude of the band gap of insulating solids (where GGA normally fails) 
and even of some strongly correlated materials like e.g. NiO \cite{Moreira:prb:02}. 

\textbf{LDA+U method}. The idea behind the LDA+U method \cite{Anisimov:prb:91} is similar to the hybrid functional method: 
For the strongly localized atomic orbitals of a material e.g. the $d$-orbitals of the transition metals and the $f$-orbitals 
of the rare earth metals, a Hubbard U parameter presenting the screened electron-electron interaction in these orbitals is 
added to the LDA KS Hamiltonian and treated in the HFA thus correcting the self-interaction error of these orbitals. 
In principle it is possible to extract the U parameter from a previous LDA calculation. However, in practice this is seldom 
done since it does not reproduce the experiments sufficiently well. Instead the U parameter is normally fitted in order to
reproduce certain material properties just as the amount of HFX in the hybrid functionals has been fitted to reproduce certain 
properties of molecules. As the hybrid functionals, LDA+U gives a reasonable description of the electronic and magnetic 
structure of strongly correlated materials like NiO, and improves considerably the magnitude of the band gap of these materials 
in comparison with LDA \cite{Anisimov:prb:93}. 

\textbf{Exact Exchange functionals}. Another approach to correct the self-interaction error of standard DFT methods is to 
derive a local KS exchange potential from the non-local exact exchange energy of the HFA. The Exact Exchange (EXX) scheme 
is self-interaction free since the exact exchange energy of the HFA cancels exactly the self-interaction of the Hartree-term. 
On the other hand, the EXX potential is local as required from the KS theory in contrast to the HFX term added by hand
in the Hybrid functional approach which is inherently non-local. Therefore the EXX together with a good approximation of
the correlation energy (as is the case in LDA and GGA) comes very close to an exact implementation of DFT within the KS scheme,
and thus gives an excellent description of the electron density. Moreover, no empirical parameters are needed so that EXX is a 
real {\it ab initio} method. Unfortunately, the computational cost for the calculation of the EXX functional is very high
compared to other methods 
%%which makes this method unsuitable for big and complex materials at the moment
\cite{Goerling:prb:95,Staedele:prl:97,Staedele:prb:99}.

\section{Kohn-Sham based NEGF formalism}
\label{sec:KS-NEGF}

We now extend the Landauer formalism presented in Ch. \ref{ch:transport} to the case 
of interacting electrons. The system is described in an effective single-particle 
picture by treating the electron-electron interaction by a mean-field method like 
the KS scheme or the HFA. The problem treated here, is to apply a mean-field method 
(established only for finite or periodic systems) to the open (i.e. infinite and 
non-periodic) system of the transport problem consisting of the leads and the device.
The arguments are presented for the KS formalism, but are easily applied to the HFA.

Far away from the scattering region the electronic structure of the leads has relaxed 
to that of the bulk (perfect) lead. So if the device region is sufficiently big (i.e. 
a sufficiently big part of the electrodes is included into the device region) the lead 
electrons can be described as  non-interacting quasi-particles moving in the effective
potential landscape (generated by the ion-core potentials and the effective mean-field 
potential of the other electrons of the bulk leads. Thus we describe the leads by the
fixed effective one-body Hamiltonians of the bulk leads, and the electron-electron
interaction is taken into account explicitly only the device region. The (many-body) 
device Hamiltonian $\hat{\mathcal{H}}_{D}$ comprises a single-body term describing the 
kinetic energy $\hat{\mathcal{T}}_D$ and the external potential $\hat{\mathcal{V}}_{e-i}$ 
inside the device, and the electron-electron interaction $\hat{\mathcal{V}}_{e-e}$ between 
the electrons. Applying the KS scheme the resulting KS Hamiltonian for the \emph{entire} 
system is given by
\begin{equation}
  \label{eq:KS-H_LDR}
  \M{H}_{KS}[n] = \left(
  \begin{array}{ccc}
    \M{H}_{L}  & \M{H}_{LD}   & \M{0}      \\
    \M{H}_{DL} & \M{H}_{D}[n] & \M{H}_{DR} \\
    \M{0}      & \M{H}_{RD}   & \M{H}_{R}
  \end{array}
  \right),
\end{equation}
and depend on the electron density $n$ of the entire system. In order to keep things simple
we will not treat overlap between basis functions explicitly. The arguments presented here
are easily generalized to non-orthogonal basis sets by using the formulas presented in Ch.
\ref{ch:transport}.

We define a Green's function (GF) matrix corresponding to the KS Hamiltonian which thus also 
becomes a functional of the electron density $n(\vec{r})$
\begin{equation}
  \label{eq:KS-G}
  \M{G}_{KS}[n](E) = ( E - \M{H}_{KS}[n] + i \eta^{0+} )^{-1}.
\end{equation}
On the other hand, we can obtain the electron density by integrating the GF up to 
the Fermi energy:
\begin{equation}
  n(\vec{r}) = -\frac{1}{\pi} {\rm Im} \int_{-\infty}^{\epsilon_F} {\rm d}E \, 
  \bra{\vec{r}}\hat{G}_{KS}[n](E)\ket{\vec{r}}.
\end{equation}
Thus, by self-consistently calculating the density matrix from the Green's
function, we solve the mean-field problem and thus minimize the energy of 
the entire system. 

Up until now we have only restated the KS formalism in terms of GFs instead of 
wavefunctions. Now we have to apply it to the situation of the open system described
by the Hamiltonian (\ref{eq:KS-H_LDR}). The difficulty lies in determining the Fermi 
energy $\epsilon_F$ for the open system. However, as we will see now, this is only a 
conceptual problem, that can solved in a relatively easy manner.

The Fermi energy is that of the entire system in equilibrium and in general will be 
different from the Fermi energy of any of the two (isolated) bulk leads due to the 
electric field caused by charges present in the device region. However, this electric 
field will only cause a shift in the electrostatic potential of the leads if the device 
region is chosen sufficiently big, since only the direct Coulomb interaction (i.e. 
classical electrostatic interaction) between charges is long range while the pure 
quantum mechanical part of the Coulomb interaction (i.e. exchange and correlation 
contributions) is relatively short range. On the other hand the Fermi level must be 
the same throughout the entire system. Thus we can shift the Hamiltonian of the two 
leads to a common arbitrary Fermi level which we choose to be zero for convenience. 
The electrostatic shift of the two leads is equivalent to a shift $\Delta_D$ of 
the Hamiltonian of the device region:
\begin{equation}
  \label{eq:KS-G_D}
  \M{G}_{D}[n](E) = ( E - \M{H}_{D}[n]-\Delta_D -\M{\Sigma}_L(E)-\M{\Sigma}_R(E) )^{-1}.
\end{equation}
The negative of that shift corresponds to the real Fermi energy of the entire system: $\epsilon_F=-\Delta_D$.
Next, in order to find the real Fermi level of the system we impose charge neutrality on the device region 
since we have assumed that the device contains a sufficient part of the electrodes so that it has relaxed to 
its bulk electronic structure. The number of electrons in the device region corresponding to a certain energy
shift is found by integrating the Greens function of the device part up to the adjusted Fermi energy of the
leads, i.e. zero:
\begin{equation}
  N_D(\Delta_D) = -\frac{1}{\pi} {\rm Im} \int_{-\infty}^{0} {\rm d}E \, {\rm Tr}[\M{G}_{D}[n](E)].
\end{equation}
Thus by imposing charge neutrality on the device region we obtain the energy shift 
$\Delta_D$ of the device Hamiltonian and thus the Fermi energy of the entire system.

The KS Hamiltonian of the device $\M{H}_{D}$ is given by the single-body term 
$\M{H}_{D}^0 = \M{T}_{KS}+\M{V}_{\rm ext}$ and the effective single-particle 
potential of the electron-electron interaction comprising the Hartree- and
the XC term $\M{V}_C[n] = \M{V}_{H}[n]+\M{V}[n]$ which are both functionals
of the electron density $n(\vec{r})$.
But since we have assumed that the electrons do not interact with each other outside 
the device region the Hamiltonian and consequently the GF only depend on the electron 
density $n_D(\vec{r})$ inside the device region. In turn $\rho_{\rm D}$ can be 
calculated by integrating the device part of the GF:
\begin{equation}
  n_D(\vec{r}) = -\frac{1}{\pi} {\rm Im} \int_{-\infty}^{0} {\rm d}E \, 
  \bra{\vec{r}}\M{G}_{D}[n](E)\ket{\vec{r}}.
\end{equation}
The self-consistent calculation of the electron density from the GF
to minimize the energy of the whole system can thus be restricted to the device's 
subspace. 

The energy of the system can be calculated by summing up the KS energies $\epsilon_i$
corrected for the double-counting:
\begin{equation}
  \label{eq:KS-Energy}
  E[n] = \sum_i \epsilon_i - J[n_D] + E_{xc}[n_D] - \int d\vec{r} \, V_{xc}[n_D](\vec{r}) \, n_D(\vec{r}).
\end{equation}
where we have made use of the fact that the electrons are assumed to be interacting 
only inside the device region so that only there we need to account for the 
double-counting error made in summing up the KS energies. In order to separate the 
energy contributions of the leads from the energy contribution of the device in the 
sum we rewrite the sum over the KS energies in terms of the KS Hamiltonian and the
corresponding density matrix $\hat{P}$ defined by the KS eigenstates:
\begin{equation}
  \label{eq:KS-energy-sum}
  E^\prime[\M{P}_{KS}] := \sum_i \epsilon_i = {\rm Tr}[ \M{P}_{KS} \, \M{H}_{KS} ].
\end{equation}

By dividing the KS Hamiltonian and the density matrix into sub-matrices 
as before we can separate the different contributions to the energy, and
write the total energy as
\begin{eqnarray}
  \label{eq:E_tot}
  E^\prime[\M{P}_{KS}] =  E_{L} + E_{LD} + E_{D} + E_{RD} + E_{R},
\end{eqnarray}
where we have defined the energy of the device,
\begin{equation}
  \label{eq:E_D}
  E_{D} = {\rm Tr}[\M{P}_{D} \, \M{H}_{D}] 
\end{equation}
the energies of the left lead and right lead,
\begin{eqnarray}
  \label{eq:E_L,E_R}
  E_{L} = {\rm Tr}[\M{P}_{L} \, \M{H}_{L}] &\mbox{ and }& E_{R}[\M{P}_{R}] = {\rm Tr}[\M{P}_{R} \M{H}_{R}]
\end{eqnarray}
and the coupling energies between left lead and device,
\begin{eqnarray}
  \label{eq:E_LD}
  E_{LD} = {\rm Tr}[\M{P}_{LD} \, \M{H}_{DL}] + {\rm Tr}[\M{P}_{DL} \, \M{H}_{LD}] = 2 {\rm Re}\left({\rm Tr}[\M{P}_{LD} \, \M{H}_{DL}]\right),
\end{eqnarray}
and right lead and device
\begin{eqnarray}
  \label{eq:E_RD}
  E_{RD} = {\rm Tr}[\M{P}_{RD} \, \M{H}_{DR}] + {\rm Tr}[\M{P}_{DR} \, \M{H}_{RD}] = 2 {Re}\left({\rm Tr}[\M{P}_{RD} \, \M{H}_{DR}]\right).
\end{eqnarray}

Since we have assumed that the device region is sufficiently big so that
the electronic structure in the leads has relaxed to that of the bulk, 
$\M{P}_L$ and $\M{P}_R$ remain constant during the self-consistent procedure, 
and thus also the energy contributions of the leads, $E_{\rm L}$ and $E_{\rm R}$.
On the other hand the coupling energies actually do change because the GFs 
$\M{G}_{\rm LD}$ and $\M{G}_{\rm RD}$ directly depend on $\M{G}_{\rm D}$ (see eqs. 
(\ref{eq:gf:GTilde_LD}) and (\ref{eq:gf:GTilde_RD})). So the two energy terms $E_{\rm LD}$ 
and $E_{\rm RD}$ always have to be included into the energy calculation.

\begin{figure}
  \begin{center}
    \includegraphics[width=0.98\linewidth]{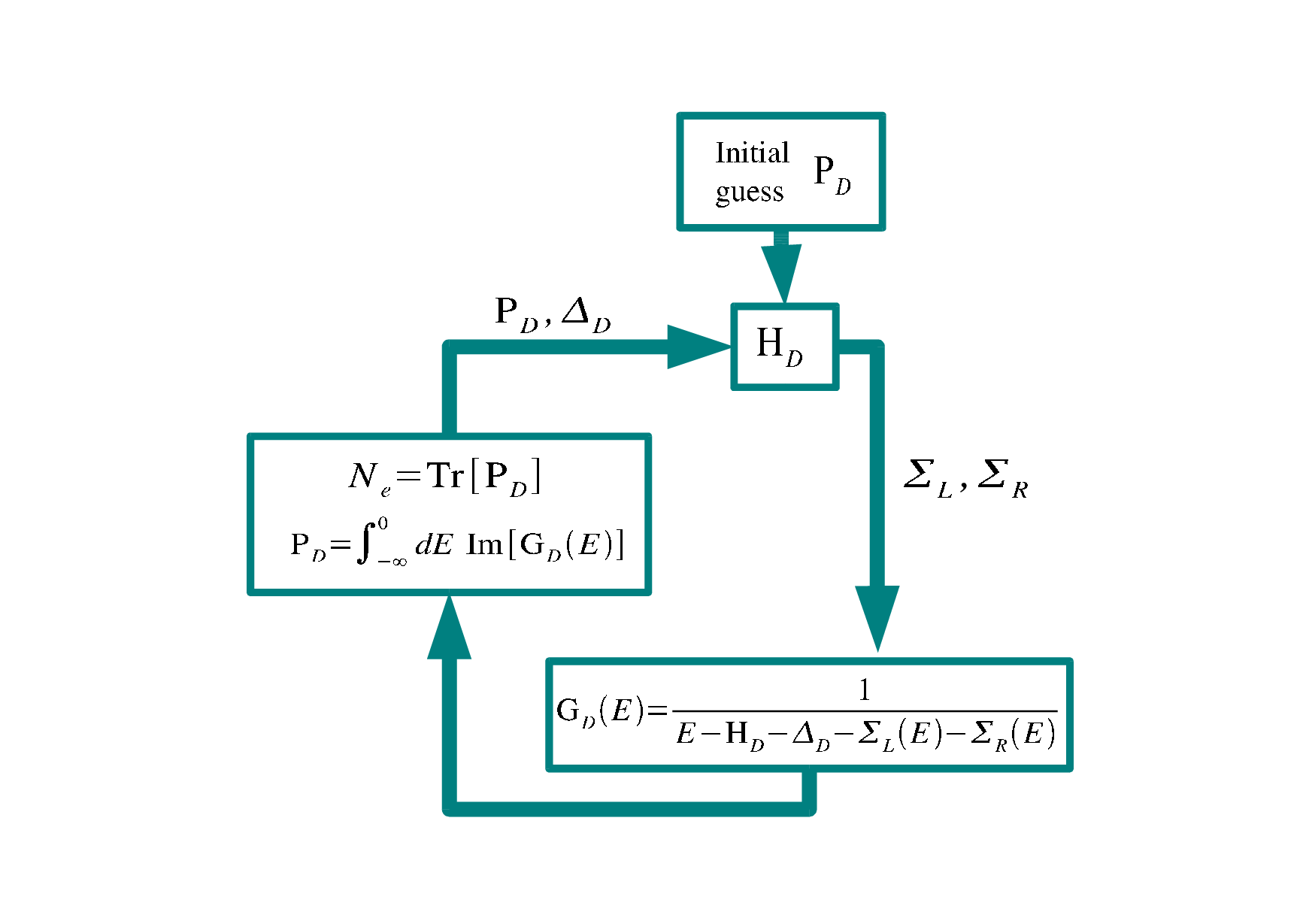}
  \end{center}
  \setcapindent{0cm}
  \caption{
    Diagram illustrating the ALACANT self-consistent procedure for KS based
    NEGF formalism as explained in the text.
  }
  \label{fig:ALACANT-SCF}
\end{figure}

The arguments presented in this section are easily adapted to the HFA. Instead of the KS Hamiltonian 
$\M{H}_{KS}$ the Fock matrix $\M{F}$ is used to calculate the GF of the device which depends on the 
density matrix $\M{P}$ which in turn can be calculated from the Hartree-Fock GF of the device thus 
defining again an iterative procedure for the self-consistent solution of the HF equation. In the HFA 
the expression for calculating the total energy from summing up the HF eigenenergies corrected for 
the double-counting is much simpler than the corresponding expression in the KS scheme.
\begin{equation}
  E_{\rm tot} = \sum_i \epsilon_i - \frac{1}{2} {\rm Tr}[\M{P}_D \M{V}_{HF}]
  = {\rm Tr}[\M{P} \M{F}] - \frac{1}{2} {\rm Tr}[\M{P}_D \M{V}_{HF}],
\end{equation}
where $\M{V}_{HF}$ is the effective Coulomb interaction in the HFA.

Finally, Fig. \ref{fig:ALACANT-SCF} shows a schematic picture of the implementation of 
the self-consistent transport method presented in this section. This approach is 
implemented in the ALACANT package which interfaces the quantum chemistry program 
GAUSSIAN\cite{Gaussian:03} to implement the NEGF formalism.

\subsection{Modeling of the bulk electrodes}
\label{sec:electrodes}

As explained in Ch. \ref{ch:transport}, part of the electrodes has to be included 
in the device region, and in the case of nanocontacts the nanoscopic device actually 
consists only of the atomically sharp tips of the two metal electrodes. In the KS based
NEGF described before and implemented in the ALACANT package the electronic structure of
the device region and thus the part of the electrodes included in the device is calculated 
self-consistently within the KS scheme while the electronic structure of the rest of the
bulk electrodes is assumed to be fixed. Their effect on the electronic structure of
the device region is taken into account via self-energies in the KS GF of the device region,
eq. (\ref{eq:KS-G_D}).

Since the exact atomic structure of the macroscopic electrodes in a real experiment is 
unknown the question arises how to model the part of the left and right semi-infinite 
electrodes that has not been included in the device region, and how to calculate the
corresponding self-energies. The ignorance of the exact atomic structure of the electrodes
in an experiment is a not negligible source of uncertainty in the transport calculations.
Therefore an appropriate model for the bulk electrodes should have the property that the 
calculated conductance does not depend too much on the actual atomic structure of the electrode.
A first problem that one encounters is that the self-energy of the bulk electrodes can only 
be calculate for idealized situations. 

%% Nanowire electrodes
A possible choice is to model the electrodes as perfect semi-infinite crystalline leads of some 
finite thickness, i.e. nanowires \cite{Taylor:prb:01:63:2,Brandbyge:prb:02,Damle:prb:01}. In this 
case the self-energies can be calculated as shown in Ch. \ref{ch:transport} and App. 
\ref{app:self-energy-1D}. However, it has been shown that the results of a transport calculation 
using nanowires as electrode models depend strongly on the actual thickness of the nanowire 
and the actual atomic structure \cite{Ke:jcp:05,Thygesen:prb:05}. Moreover, the electrodes
in real experiments are very far from being perfect nanowires but are actually substantial 
polycrystalline electrodes. Nevertheless, using perfect nanowires as the electrodes is interesting for 
performing model calculations, e.g. to study the effect of a single impurity or vacancy on the transport 
properties of an otherwise perfect nanotube or the effect of a constriction in a perfect graphene
nanoribbon \cite{Munoz-Rojas:prb:06}.

%% Bulk crystalline electrodes
In order to avoid the that the resulting conductance reflects the finite thickness of the nanowire
one can go to the crystalline limit by making the wire infinitely thick. In practice this can be done
by making the system periodic in the directions perpendicular to the transport direction so that the 
self-energy matrices become dependent on the wave vector perpendicular to the transport direction
\cite{Thygesen:prb:05,Rocha:prb:06}. However, this approach is computationally expensive and the 
resulting conductance still reflects the crystal direction of the electrodes.
%% Jellium electrodes
Alternatively, jellium models have been used in methods based on a description in terms of scattering 
states\cite{Lang:prb:95,Hirose:prb:95}. In this case the electrodes are completely 
featureless and thus do not reflect any of the chemical properties of the material. Thus this choice 
is also not free from controversy \cite{Fujimoto:prb:03}. 

\begin{figure}
  \begin{center}
    \includegraphics[width=0.75\linewidth]{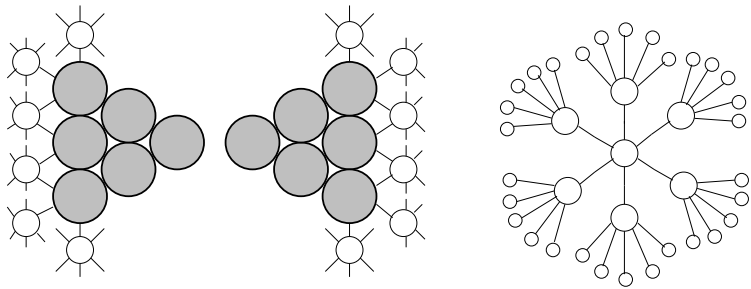} 
  \end{center}
  \setcapindent{0cm}   
  \caption{
    Left: Schematic 2D illustration of a nanocontact (big grey circles) with the 
    first atoms of the BL (small white circles) attached to the 
    outer planes of the nanocontact. Right: Finite section of Bethe lattice (BL) with 
    coordination 6. All atoms of the BL have the same coordination as in the 
    corresponding crystalline structure giving rise to short range order. But 
    there is no long range order in the BL due to the absence of closed loops.
  }
  \label{fig:nc-bethe-lattice}
\end{figure}

For these reasons the choice in the original implementation of the ALACANT package (then called GECM for
Gaussian Embedded Cluster Method) \cite{Palacios:ctcc:05} was to describe the bulk electrodes with a Bethe 
lattice parameterized tight-binding model with the coordination and parameters appropriate for the chosen 
electrode material.  The advantage of choosing a Bethe lattice (BL) resides in that, although it does not 
have long-range order, the short-range order is captured and it reproduces fairly well the bulk density 
of states of most commonly used metallic electrodes. The right hand side of Fig. \ref{fig:nc-bethe-lattice} 
depicts schematically a BL of coordination 6. The left hand side of that same figure illustrates
schematically how the device (depicted here as a nanocontact) is connected to the BL electrodes: 
For each atom in the outer planes of the device, a branch of the BL is added in the direction of 
any missing bulk atom (including those missing in the same plane). The directions in which tree branches are 
added are indicated by white small circles which represent the first atoms of the branch in that direction. 
Assuming that the most important structural details of the electrode are included in the central cluster, 
the BLs should have no other relevance than that of introducing the most generic bulk electrode 
for a given metal. In App. \ref{app:bethe-lattices} it is explained how to calculate BL self-energies. 

\begin{figure}
  \begin{tabular}{cc}    
    \includegraphics[width=0.38\linewidth]{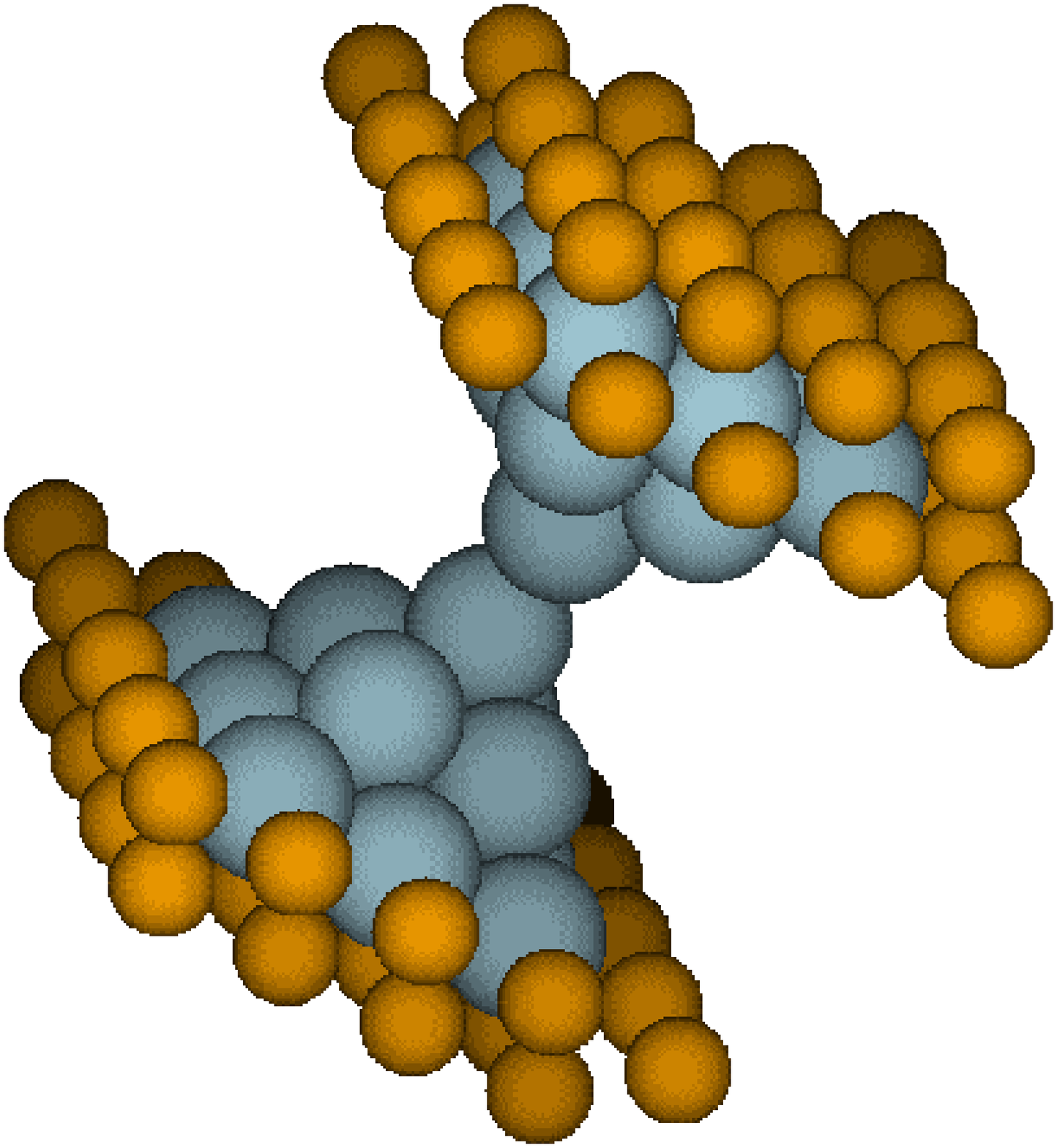} &
    \includegraphics[width=0.6\linewidth]{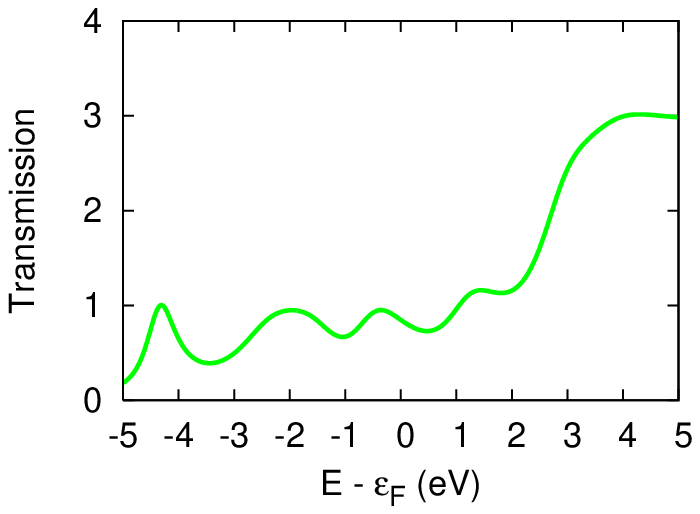}
  \end{tabular}
  \setcapindent{0cm}   
  \caption{
    Left: Atomic structure of an Al nanocontact consisting of two perfect pyramids
    along the (001)-direction made up of 14 Al atoms (shown in grey) each. The 
    surface atoms of the BL connected to the outermost planes of the 
    nanocontact are also shown in a gold-like color. Right: Transmission of the 
    Al-nanocontact shown on the left. The electronic structure of the device region
    was calculated on the LDA level of DFT employing the CRENBS minimal basis set
    \cite{Pacios:jcp:85}. The tight-binding parameters of the BL are fitted 
    to reproduce LDA electronic structure calculation of crystalline bulk Al and were 
    taken from the handbook by Papaconstantopoulos \cite{Papaconstantopoulos:book:86}.
  }
  \label{fig:al28-nc-transm}
\end{figure}

Fig. \ref{fig:al28-nc-transm} shows the result of an actual transport calculation for the Al 
nanocontact as shown on the left hand side of that same figure. The left hand side of Fig. 
\ref{fig:al28-nc-transm} also shows how the surface atoms of the BL electrodes are
connected to the outermost atomic planes of the nanocontact. The electronic structure of the 
device region has been calculated on the LDA level of DFT and the CRENBS minimal basis set and 
effective core pseudo potential (ECP) by Christiansen and coworkers \cite{Pacios:jcp:85} has 
been employed. 

Apart from giving a reasonable generic description of the bulk electrodes the computation of 
the self-energies for the BL model is also computationally very efficient compared 
to computing the self-energies for perfect crystalline systems due to the absence of loops. 
In this respect the BL model for the description of electrodes is advantageous 
as compared to the perfect crystalline models of the electrodes. However, if one is interested
in studying transport in an idealized model system in order to understand fundamental aspects
it might be important to avoid any kind of scattering that occurs at the electrode-device 
interface, e.g. to study the intrinsic transport properties of nanotubes, graphene nanoribbons 
or metallic nanowires neglecting the otherwise inevitable scattering at the interface between 
the bulk electrode and the nanotube or nanowire. Therefore a part of this thesis has been dedicated
to the implementation of a module for calculating self-energies of perfect one-dimensional electrodes
into the ALACANT package.

\begin{figure}
  \begin{flushright}
    \includegraphics[width=0.98\linewidth]{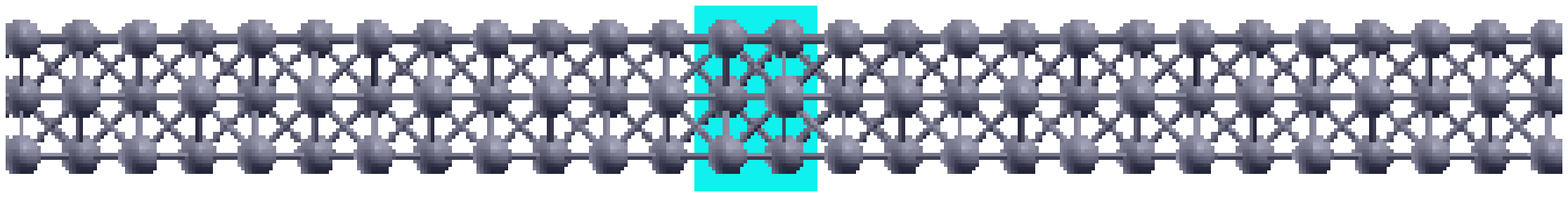}
  \end{flushright}
  \includegraphics[width=\linewidth]{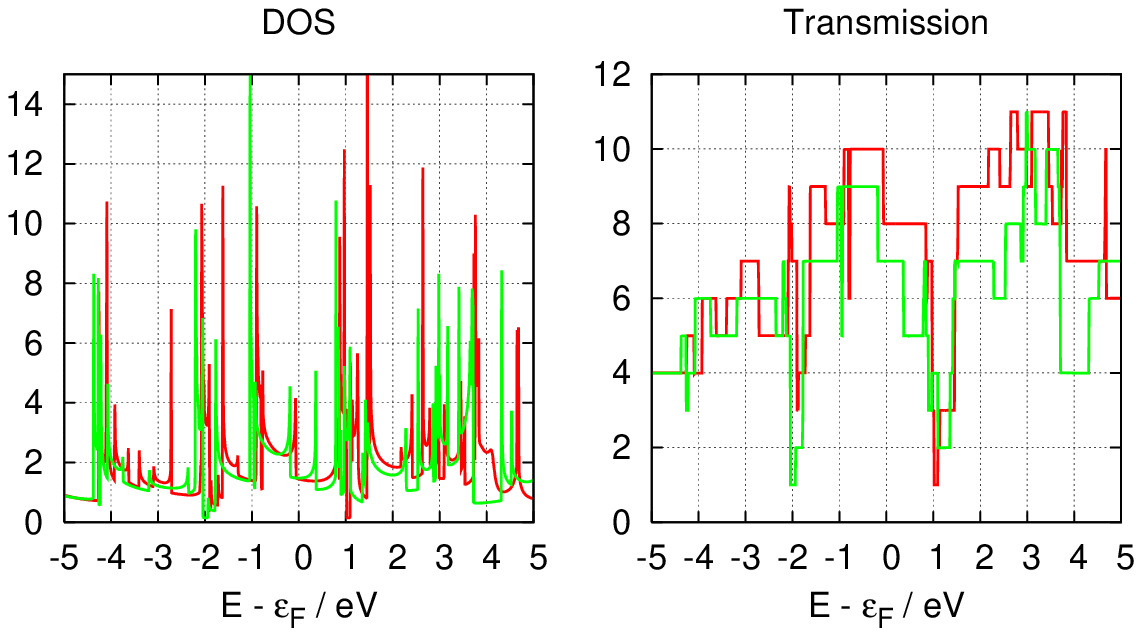} 
  \setcapindent{0cm}   
  \caption{
    DOS and transmission for the infinite Al-nanowire shown above. Red curves in the DOS and 
    transmission have been calculated taking into account interactions between nearest neighbor 
    and next-nearest neighbor planes only while green curves have been calculated taking into 
    account inter-plane interactions up to the fourth order. 
  }
  \label{fig:al5-al4-wire-transm-dos}
\end{figure}
   
In Ch. \ref{ch:transport} and App. \ref{app:self-energy-1D} we have explained how to calculate 
the self-energies in the case of ideal one-dimensional leads of some finite width taking into 
account the overlap between the atomic orbitals of the atoms making up the one-dimensional lead. 
We now apply the developed theory to calculate the transport properties of an Al nanowire (Fig. 
\ref{fig:al5-al4-wire-transm-dos}) based on previous DFT electronic structure calculation of the 
nanowire on the LDA level. We use the CRYSTAL03 {\it ab initio} program for periodic systems 
\cite{Crystal:03} and employ the same CRENBS minimal basis set with ECP as before \cite{Pacios:jcp:85}. 
The nanowire is along the 001 direction of the bulk crystal and the atoms have approximately bulk 
distances (nearest neighbor distance $\approx 2.8$\r{A}). The primitive unit cell of the nanowire 
(colored section in top panel of Fig. \ref{fig:al5-al4-wire-transm-dos}) consists of two planes 
containing 5 and 4 atoms, respectively. By defining a unit cell for the transport calculations
consisting just of one primitive unit cell we can take into account interactions between nearest
and next-nearest neighbor planes. With a unit cell consisting of two two primitive unit cells
we can take into account inter-plane interactions up to fourth order.

Fig. \ref{fig:al5-al4-wire-transm-dos} shows the DOS and transmission of the nanowire in the 
case that a) only interactions between up to next-nearest neighbor layers (red curves) and b) 
interactions between up to fourth nearest-neighbor layers (green curves) are taken into account.
and we observe that DOS and transmission change when taking into account interactions of higher
order. However, beyond fourth order the interactions become negligible, so that DOS and transmission
do not change further when taking into account interactions beyond fourth order.

\begin{figure}
  \begin{center}
    \includegraphics[width=0.8\linewidth]{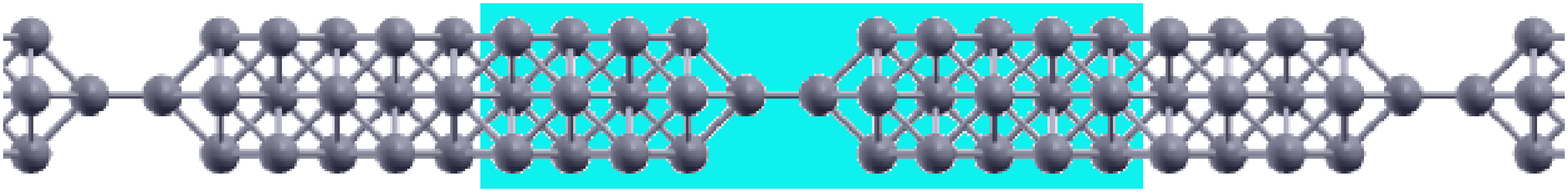}
  \end{center}
  %%  \vspace{0.5cm}
  \begin{center}
    \includegraphics[width=0.8\linewidth]{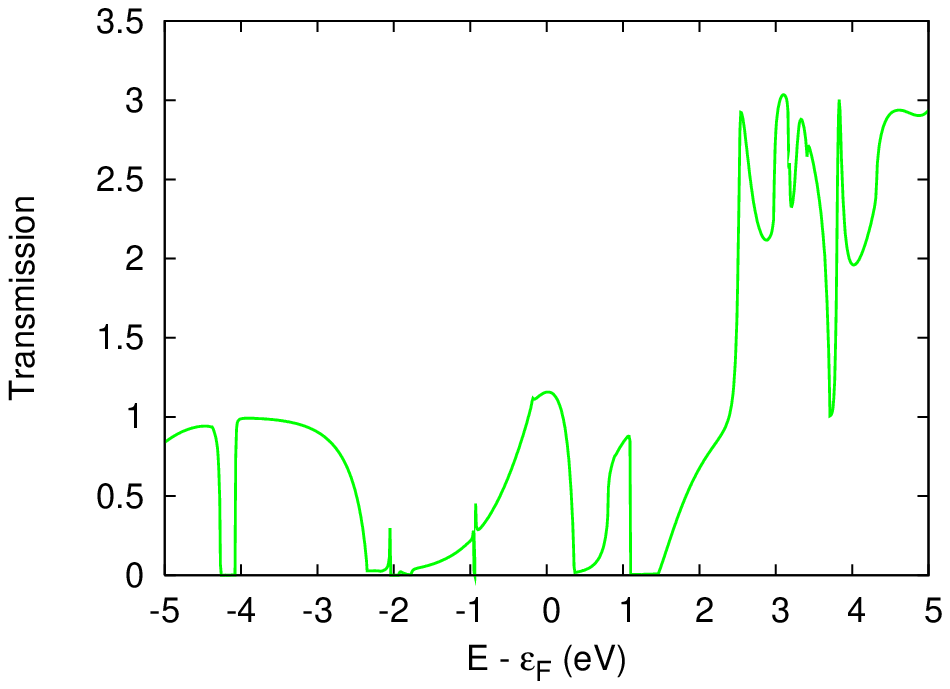}
  \end{center}
  \setcapindent{0cm}   
  \caption{
    Transmission of the Al nanocontact shown above. The electronic structure of the
    nanocontact was calculated using a super-cell which is periodically repeated in 
    one direction using the CRYSTAL03 package for periodic systems. 
    %%The device 
    %%Hamiltonian for the transport calculation was extracted from the KS-Hamiltonian 
    %%of a single super-cell, while for the calculation of the self-energies of the leads 
    %%the Hamiltonian of the unit cell and the coupling between neighboring unit cells 
    %%were extracted from the electronic structure calculation of the perfect infinite 
    %%nanowire shown in Fig. \ref{fig:al5-al4-wire-transm-dos}.
  }
  \label{fig:al42-nc-transm}
\end{figure}

In order to compare with transport calculations describing the electrodes with the
BL model, we calculate now the transport properties of a similar nanocontact
as above but connected to two semi-infinite Al nanowires instead of the BL bulk
electrodes. As in the case of the perfect nanowire we calculate the electronic structure
with the CRYSTAL03 program for periodic systems by defining a super-cell containing the 
nanocontact as shown at the top of Fig. \ref{fig:al42-nc-transm}. The super-cell has been
chosen large enough so that the electronic structure of the outermost atoms of the super-cell 
has relaxed to its bulk value (i.e. of the perfect nanowire). The device Hamiltonian is then
taken as the (on-site) super-cell Hamiltonian of the converged KS calculation. For the 
calculation of the self-energies of the semi-infinite leads the Hamiltonian of the unit cell 
and the coupling between neighboring unit cells are extracted from the electronic structure 
calculation of the perfect infinite nanowire.

From a comparison of Fig. \ref{fig:al28-nc-transm} and Fig. \ref{fig:al42-nc-transm} we see 
that although the two transmission curves differ from each other in the detail, some overall
features in the two curves are actually very similar. For example, for energies of 2.5 eV
above the Fermi level we observe a plateau of about 3 in the transmission and around the 
Fermi level the transmission is roughly one. On the other hand, in the case of the nanowire 
electrodes we see that the transmission function is much more irregular than in the case
of the BL electrodes. The strong oscillations of the transmission function in
the case of the nanowire electrodes are interference effects due to the finite width of the
electrodes which also dependent very strongly on the actual width of the nanowire, so that
the transmission curve changes strongly in dependence of the nanowire width. Only in the limit
of very thick electrodes do these interference effects disappear \cite{Ke:jcp:05}. On the other
hand the transmission function in the case of the BL electrodes and is essentially 
independent of the details of the details of the electrode-device interface as long as the 
interface is not too close to the nanoscopic constriction\cite{Jacob:07}.

\section{Beyond density functional theory}
\label{sec:beyond-DFT}

The methodology presented in this chapter for calculating the transport properties of nanoscale 
conductors by combining DFT based electronic structure calculations and the NEGF works quite well 
for simple nanoscale conductors like metallic nanocontacts and nanowires, but has difficulties to 
cope with more complicated nanoscopic systems, like molecular conductors. One of the problems clearly 
is the spurious electronic self-interaction inherent in the standard approximations to the density 
functional like the LDA and GGA as discussed in Subsec. \ref{sub:functionals}. As explained there, 
this problem can actually be cured by semi-empirical methods like using hybrid functionals 
\cite{Becke:jcp:93} or the LDA+U method \cite{Anisimov:prb:91} that correct the spurious 
self-interaction, but have the disadvantage of introducing empirical parameters which have to be 
fitted to the material properties and thus there predictive power is limited. 

A further problem of the DFT based quantum transport approach is that DFT is a ground state theory for 
the electron density while transport properties also involve the excited states of a system. Thus even 
if we knew the exact density functional (which we don't) it is not clear that the DFT based transport 
approach could actually give a good description of the transport properties. For the same reason there 
is no guarantee that an exact DFT would yield the magnitude of the band gap of insulators or the HOMO-LUMO
gap of molecules. Indeed, it seems to be the case that EXX functionals which are already very close to 
an exact implementation of DFT improve only slightly on band gap in comparison with GGA. The hybrid 
functionals and LDA+U yield reasonable values of the HOMO-LUMO gap and of the band gaps of insulating 
solids because they go beyond DFT. 

Finally, while DFT is in principle a many-body theory in the sense that DFT 
yields the exact ground state electron density of the correlated many-body ground state if we knew the exact 
density functional, our KS based transport theory is ultimately an effective one-body theory as it makes 
use of the effective one-body KS orbitals and energies. Therefore the KS based transport theory presented 
here does not capture the true many-body effects due to electron correlations in the transport properties
of nanoscopic conductors like the Kondo effect \cite{Kondo:ptp:64,Anderson:pr:61}.

%% TDFT
There are several possibilities to solve these problems. The first is to use the time-dependent density 
functional theory (TDDFT) \cite{Runge:prl:84} which is a time-dependent extension of DFT. TDDFT allows to compute the 
time-dependent electron density of an interacting electron system in a time-dependent external potential. 
Since a time-dependent external field gives rise to electron excitations one can obtain the electron 
densities and energies of the excited many-body states from a TDDFT calculation. Thus TDDFT in 
combination with a good approximation for the density functional (i.e. self-interaction free) like e.g. 
EXX should give a very good description of the energy spectrum of metals, semiconductors and insulators 
as well as of molecules. Accordingly, combining the NEGF transport formalism with the KS formulation of 
TDDFT \cite{Koentopp:07,Stefanucci:06} should improve systematically the description of the transport 
properties of molecular conductors where 
the transport approach based on ground-state DFT with the standard approximations to the functionals 
normally fails. However, TDDFT is computationally considerably more demanding than ground state DFT, and 
so is the EXX method in comparison with the standard approximations to DFT as explained before. 
Consequently, up until now TDDFT transport calculations have only been demonstrated for relatively 
simple model systems.

%% KS Green's function
Alternatively, one can try to systematically improve the KS GF by many-body techniques. As explained at 
the end of Ch. \ref{ch:transport}, the Landauer formula becomes exact in the limit of small bias voltages 
and low temperatures even in the presence of strong electron correlations if the exact GF calculated from 
the many-body ground state is known. On the other hand, given a good approximation to the energy functional, 
DFT only guarantees a good approximation for the electron density but not for the GF of the many body 
ground-state. Thus it is not clear whether the KS GF calculated from the effective KS Hamiltonian actually 
gives a good approximation to the exact GF of the many-body ground state. But the KS GF is a good starting
point for applying many-body techniques.

%% GW
A systematic improvement of LDA is the GW perturbation theory where the electron self-energy is 
approximated by the product of the GF $\hat G$ and the screened Coulomb interaction $\hat W$. 
GW gives an excellent description of many metallic systems and improves considerably the description 
of semiconductors and insulators improving e.g. the magnitude of the band gap. Therefore, various 
groups have recently made an effort to implement the GW method for the description of transport in 
nanoscopic conductors \cite{Thygesen:06,Darancet:06}. 

%% DMFT
However, when dealing with strongly correlated materials like e.g. transition metal oxides a perturbative 
approach like the GW method is no longer appropriate and more sophisticated methods are needed like e.g. the
Dynamical Mean Field Theory (DMFT) \cite{Kotliar:rmp:05}. The DMFT is based on the observation that electron 
correlation effects are most important for the strongly localized electrons of the $d$- and $f$-shells while
they can normally be neglected for the delocalized $s$- and $p$-electrons. Therefore it is a good approximation
to treat the electrons interactions only locally exact while the interaction with the rest of the system can
be described on a mean-field level and thus enters only as an effective bath via a self-energy. Thus the DMFT 
maps the interacting electron problem of an infinite system onto the Anderson impurity model with a single
interacting site connected to an infinite systems \cite{Anderson:pr:61}. In this way the DMFT approach calculates 
a locally exact GF that neglects the long-range correlations of the electrons which are normally
unimportant. By extending the interacting region of the DMFT over atoms of the complete infinite system (Cluster-DMFT) 
one can in principle make DMFT as exact as necessary. However, the cost in computation increases exponentially
with the size of the cluster.

%%% Local Variables: 
%%% mode: latex
%%% TeX-master: "~/suficiencia/report/transport"
%%% End: 

%%%%%%%%%%%%%%%%%%%%%%%%%%%%%%%%%%%%%%%%%%%%%

%%%%%%%%%%%%%%%%%%%%%%%%%%%%%%%%%%%%%%%%%%%%%
%% Chapter 3 - Spin transport              %%
%%%%%%%%%%%%%%%%%%%%%%%%%%%%%%%%%%%%%%%%%%%%%

\chapter{Spin transport}
\label{ch:spin-transport}

We are primarily interested in electron transport through magnetic nanocontacts. 
Therefore we need to generalize the above formalism to systems without 
spin-degeneracy. Introducing the spin-degree of freedom the Hamiltonian and the 
Green's function (GF) of the device are now further subdivided into the following
spin-dependent sub-matrices:
\begin{equation}
  \label{eq:tt:spin-H_D/G_D}
  \M{H}_{\rm D} = \left( 
    \begin{array}{cc}
      \M{H}_D^{\su\su} & \M{H}_D^{\su\sd} \\
      \M{H}_D^{\sd\su} & \M{H}_D^{\sd\sd}
    \end{array}
  \right)
  \hspace{1cm}
  \M{G}_D(E) = \left( 
    \begin{array}{cc}
      \M{G}_D^{\su\su}(E) & \M{G}_D^{\su\sd}(E) \\
      \M{G}_D^{\sd\su}(E) & \M{G}_D^{\sd\sd}(E)
    \end{array}
  \right)
\end{equation}

The magnetization of the electrodes L and R is assumed to be along the $z$-direction only. 
Therfore the spin-mixing parts $\su\sd$ and $\sd\su$ of the self-energy matrices of the 
electrodes and hence the coupling matrices become zero:
\begin{eqnarray}
  \label{eq:spin-Sigma/Gamma}
  \M{\Sigma}_{L,R}(E) = \left( 
    \begin{array}{cc}
      \M{\Sigma}_{L,R}^\su(E) & 0 \\
      0 & \M{\Sigma}_{L,R}^\sd(E)
    \end{array}
  \right) \, \Rightarrow \,
  \M\Gamma_{L,R}(E) = \left( 
    \begin{array}{cc}
      \M\Gamma_{L,R}^\su(E) & 0 \\
      0 & \M\Gamma_{L,R}^\sd(E)
    \end{array}
  \right).
\end{eqnarray}

\begin{figure}
  \begin{minipage}[b][][b]{0.5\linewidth}
     \begin{flushright}
       \includegraphics[width=\linewidth]{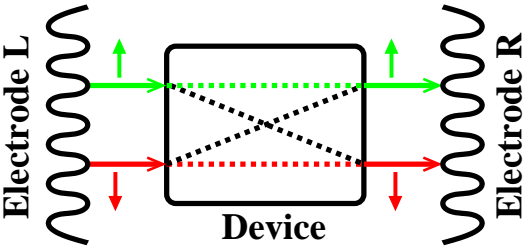}
     \end{flushright}
   \end{minipage}
   \hfill
   \begin{minipage}[b][][b]{0.45\linewidth}
     \setcapindent{0cm}
     \caption{Scattering of spin channels. An electron with a certain spin-state 
       $\sigma$ coming in from the left can be scattered to another spin-state $\sigma^\prime$ 
       of the right lead or conserve its spin when passing the device.}
     \label{fig:spin-channels}
   \end{minipage}
\end{figure}

Thus the transmission can be resolved into the different spin-channels:
\begin{eqnarray}
  \label{eq:spin-transm}
  T(E) &=& {\rm Tr}[\M\Gamma_L(E)\,\M{G}_D^\dagger(E)\,\M\Gamma_R(E)\,\M{G}_D(E)]
  \nonumber\\
  &=& \sum_{\sigma_1} {\rm Tr}[(\M\Gamma_L(E)\,\M{G}_D^\dagger(E)\,\M\Gamma_R(E)\,\M{G}_D(E))_{\sigma_1\sigma_1}]
  \nonumber\\
  &=& \sum_{\sigma_1,\sigma_2} 
  {\rm Tr}[\M\Gamma^{\sigma_1}_L(E)\,(\M{G}_D^{\sigma_1\sigma_2}(E))^\dagger\,\M\Gamma_R^{\sigma_2}(E)\,\M{G}_D^{\sigma_2\sigma_1}(E)]
  \nonumber\\
  &=& T^{\su\su}(E) +  T^{\su\sd}(E) +  T^{\sd\su}(E) +  T^{\sd\sd}(E).
\end{eqnarray}
where we have defined the spin-resolved transmission probabilities
\begin{equation}
  \label{eq:spin-caroli}
  T^{\sigma_1\sigma_2}(E) := 
  {\rm Tr}[\M\Gamma^{\sigma_1}_L(E) \, (\M{G}_D^{\sigma_1\sigma_2}(E))^\dagger \, \M\Gamma_R^{\sigma_2}(E) \, \M{G}_D^{\sigma_2\sigma_1}(E)].
\end{equation}
$T^{\sigma_1\sigma_2}$ is the probability that an electron enters the device from the 
left electrode with spin $\sigma_1$ and is emitted to the right with spin $\sigma_2$.
The meaning of the spin-channels is illustrated in Fig. \ref{fig:spin-channels}. 

Obviously, when there are no spin-mixing terms in the device Hamiltonian 
($\M{H}_D^{\su\sd}=\M{H}_D^{\sd\su}=0$) then the cross terms $T^{\su\sd}$ and $T^{\sd\su}$ 
must be zero, in real nanocontacts, leaving only the most important interactions to capture the essential physics 
of magnetic materials.

\section{Domain wall scattering in the classical $sd$-model} 
\label{sec:t-J-model}

\begin{figure}
  \begin{center}
    \includegraphics[angle=270,width=0.99\linewidth]{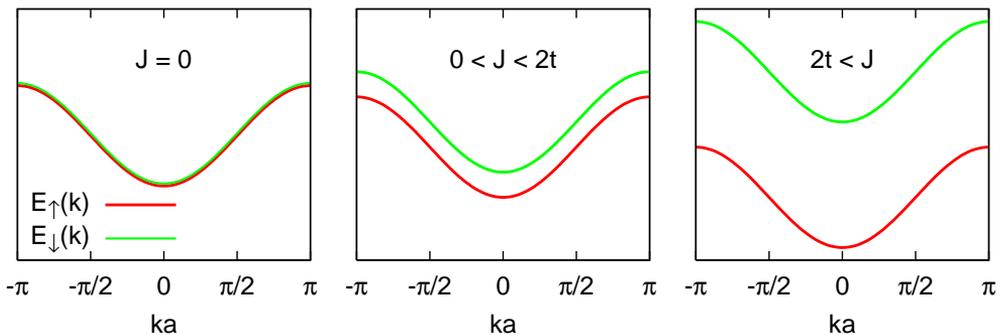}
  \end{center}
  \setcapindent{0cm}
  \caption{
    Energy dispersion for classical $sd$-model for homogeneous magnetization and for different 
    limits of     the exchange splitting $J$. For $J=0$ (left) the chain is paramagnetic and 
    for $J>0$ it     is magnetic. When $J$ becomes larger than half of the band width $2t$ a 
    gap opens between the two spin bands.}
  \label{fig:t-J-model}
\end{figure}

What is the effect of a domain wall i.e. a magnetization reversal on a spin-polarized current? And 
how does this effect depend on the concrete magnetization profile? To answer this question we consider 
a tight binding chain with an exchange splitting $J$ for the spin, i.e. the so-called classical $sd$-model:
\begin{equation}
  \label{eq:t-J-model}
  \hat{\mathcal{H}} = -t \sum_{i,\sigma=\su,\sd} \left( 
    \hat{c}_{i\,\sigma}^\dagger\hat{c}_{i+1\,\sigma} + 
    \hat{c}_{i+1\,\sigma}^\dagger\hat{c}_{i\,\sigma} 
  \right)
  + \sum_i \vec{M}_i \cdot \vec{S}_i,
\end{equation}
where $\vec{M}_i$ is the magnetic moment of site $i$ which interacts with the spin 
$\vec{S}_i=(\hat\sigma_x,\hat\sigma_y,\hat\sigma_z) \otimes \ket{i}\bra{i}$. This 
gives rise to an exchange splitting for the spin at that site:
\begin{equation}
  \vec{M}_i \cdot \vec{S}_i = -J ( \sin\theta_i \, \hat\sigma_x + \cos\theta_i \, \hat\sigma_z) \otimes \ket{i}\bra{i}.
\end{equation}
To keep things simple we have assumed that the magnetization lies in the $xz$-plane, i.e. the
$y$-component of the of the magnetization is zero, and that the exchange splitting $J$ is the
same for all sites. $\theta_i$ is the angle of the local magnetization with the $z$-axis. For
a homogeneous magnetization (e.g. $\theta_i = 0$) we get two spin-bands that are split
by an amount $2J$: $E_{\su/\sd}(k) = -2t \cos(ka) \mp J$. This situation is depicted in Fig.
\ref{fig:t-J-model} for different values of the parameter $J$. For $J=0$ we have of course a 
the situation of a paramagnetic metal, while for $J>0$ we have a ferromagnet. For the latter 
we can further distinguish two situations. For $J<2t$ the two spin-bands overlap. This is the 
typical situation in ferromagnetic metals. On the other hand for $J>2t$ a gap opens between
the two spin-bands. And the model either describes a magnetic insulator (at half filling) or
a half-metallic conductor where the conduction electrons are 100\% spin-polarized.

\subsection{Abrupt domain wall}
\label{sub:abrupt-dw}

\begin{figure}
  \begin{center}
    \includegraphics[width=0.5\linewidth]{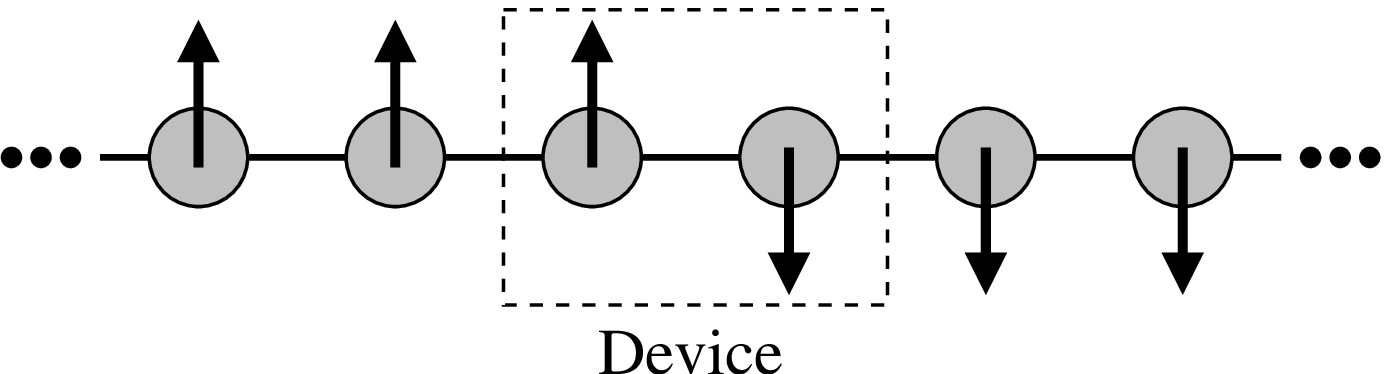}
    \includegraphics[width=0.75\linewidth]{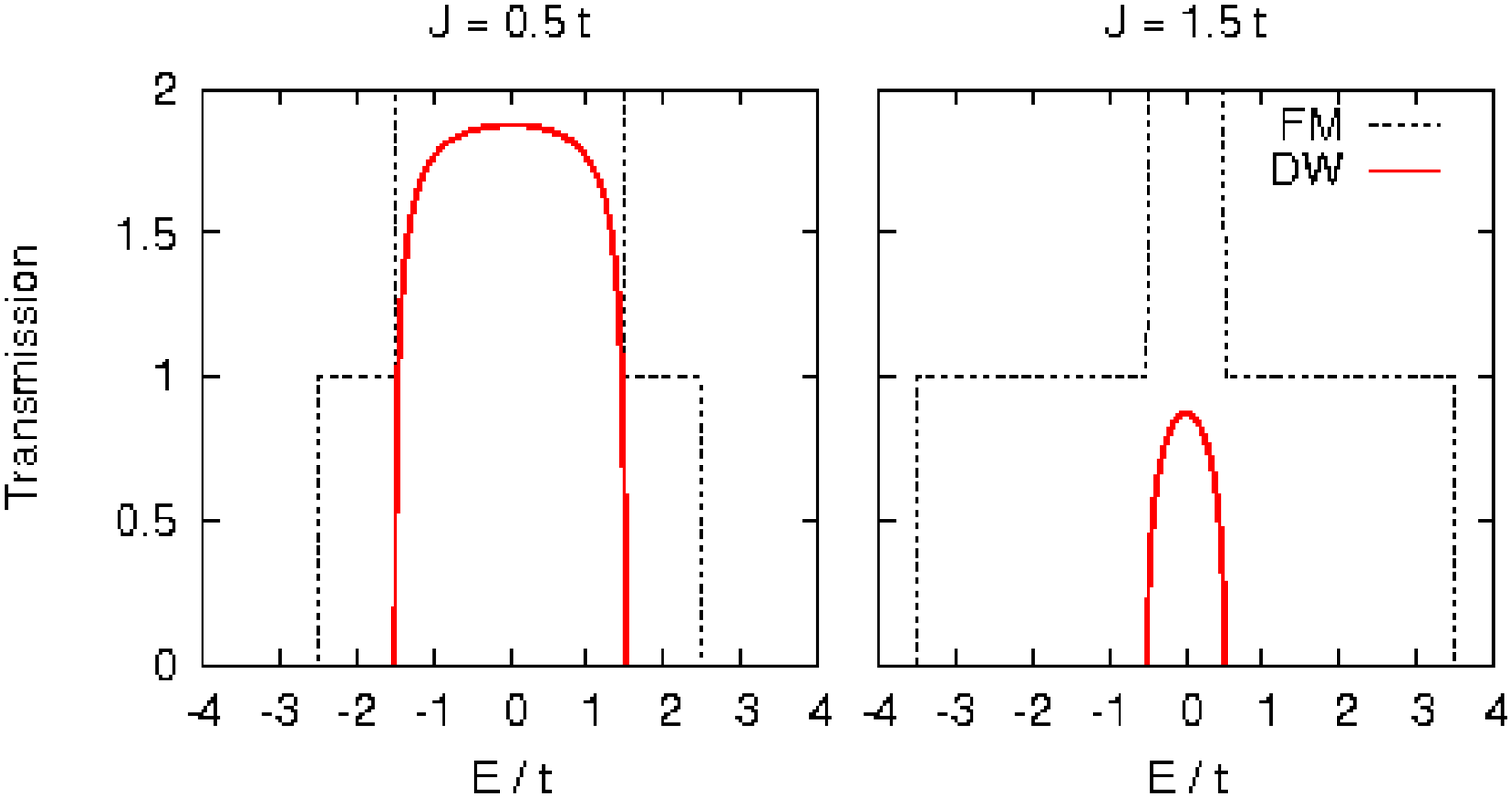}
  \end{center}
  \setcapindent{0cm}
  \caption{Abrupt domain wall in classical $sd$-model. Top graph: Sketch of model. 
    Bottom left: Domain wall transmission (solid red line) for $J=0.5 t$ compared to transmission
    of homogeneous ferromagnetic chain for that parameter. Bottom right: The same as left graph
    but for $J=1.5 t$.
    }
  \label{fig:dw_abrupt}
\end{figure}

We now discuss the effect of an abrupt domain wall on the transport properties for the ferromagnetic
phase of the classical $sd$-model, i.e when $J>0$. As depicted in Fig. \ref{fig:dw_abrupt} the magnetization
of each of the leads is homogenous, but aligned opposite to each other so that there is a sharp 
magnetization reversal in the device region. However, the magnetization axis is the same for all atoms
(we choose $z$, i.e. $\theta_i\in\{0,\pi\}$), so that the spin-mixing term $\propto\hat\sigma_x$ in the 
Hamiltonian (\ref{eq:t-J-model}) is zero. 

When $J>2t$ there is no overlap between the two spin bands so that the transmission is simply zero for 
all energies since an electron coming in on one of the leads cannot be transmitted through the device to 
the other lead conserving also its spin. But since there is no spin-mixing term in the Hamiltonian the  
transmission probabilities involving a spin flip are zero, i.e. the spin is conserved. Thus in the case
of the abrupt domain wall there can only be a nonzero transmission when the two spin bands do overlap.
Note, that even in the case where there is an overlap between the two spin-bands the transmission becomes
zero for energies where the two bands do not overlap as can be seen from the two lower panels of Fig.
\ref{fig:dw_abrupt}. Thus for a half-metallic conductor, (i.e. spin-polarization is 100\% near the Fermi
level), the effect of a domain wall is maximal, giving theoretically an infinite resistance for an abrupt
domain wall. This is what makes half-metals the ideal materials for spintronics applications. 

\subsection{Smooth domain walls}
\label{sub:smooth-dw}

\begin{figure}
  \begin{tabular}{lll}
    \includegraphics[scale=0.21]{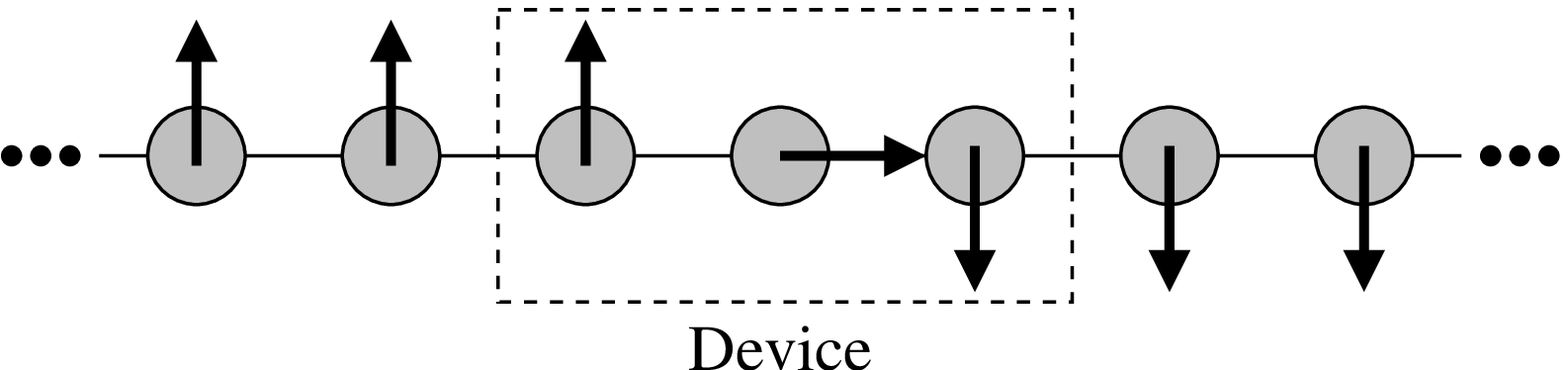} &
    \includegraphics[scale=0.21]{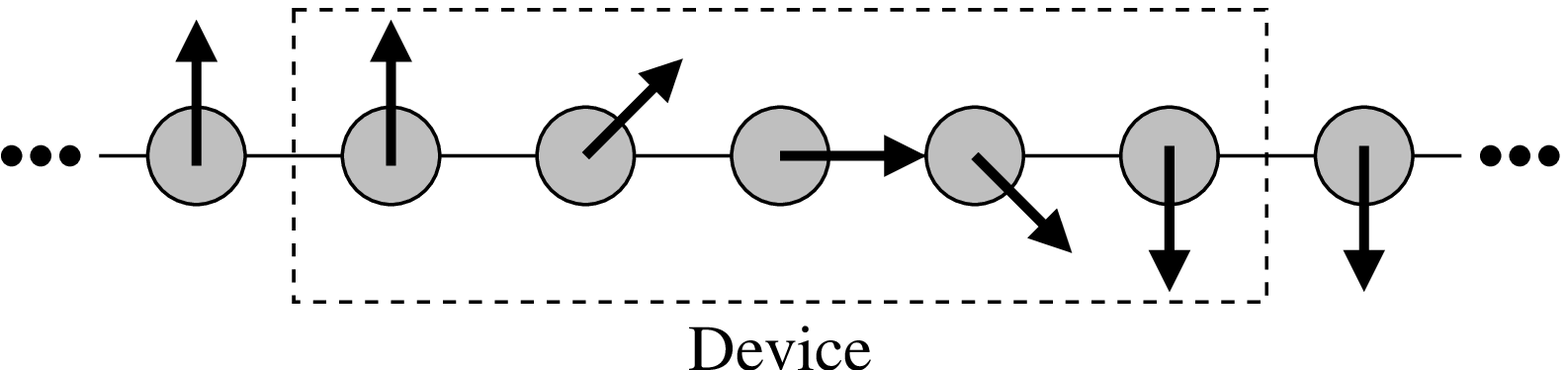} &
    \includegraphics[scale=0.21]{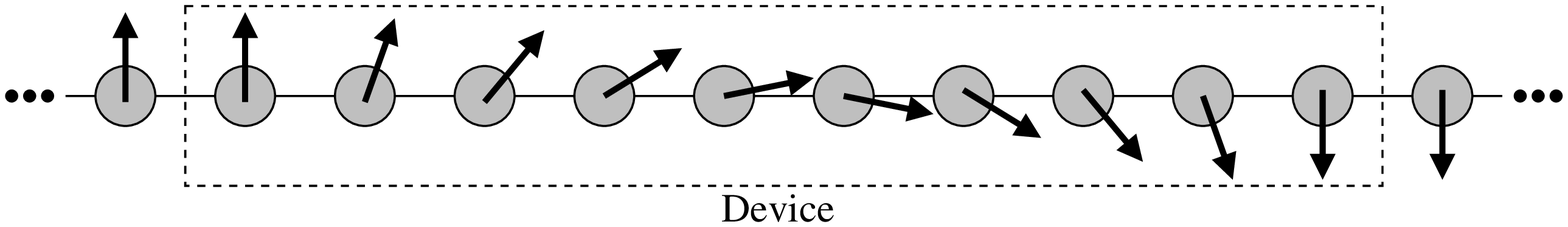}
  \end{tabular}
  \begin{center}
    \includegraphics[width=\linewidth]{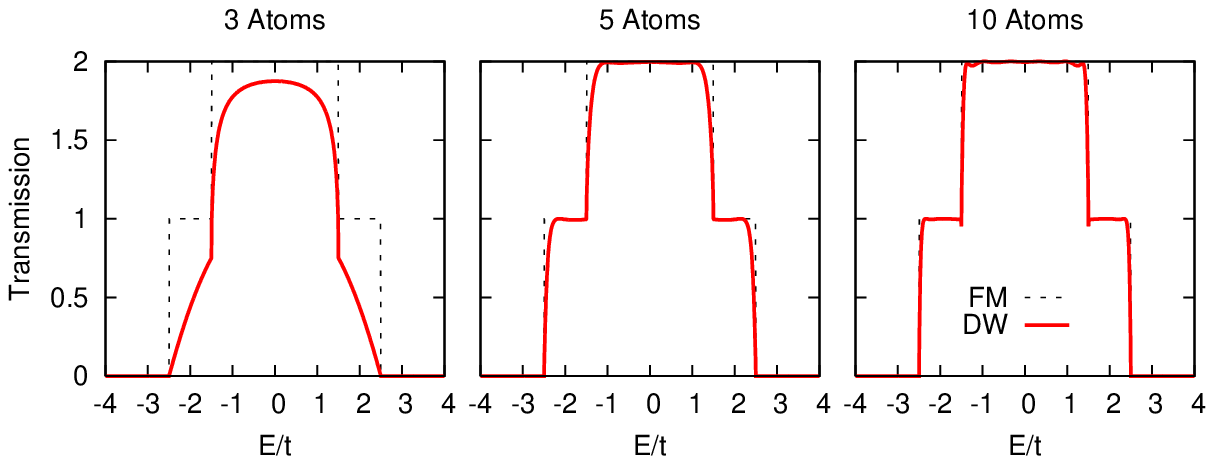}
  \end{center}
  \setcapindent{0cm}
  \caption{Smooth domain walls. The upper panels show 3 domain walls of different lengths.
    Upper left: Domain wall with 3 atoms. Upper center: Domain wall with 5 atoms. Upper right: 
    Domain wall with 10 atoms. The magnetization angle $\theta_i$ changes linearly with the atom 
    position $i$ inside the domain. The lower panels show the corresponding transmissions of
    the domain walls (solid red lines) in comparison with the homogeneous ferromagnetic chain
    (dashed black lines) with same parameter $J$. $J=0.5 t$ in all 3 cases.
  }
  \label{fig:smooth-dw}
\end{figure}

Instead of the abrupt magnetization reversal the magnetization could change smoothly, as illustrated 
in the top panels of Fig. \ref{fig:smooth-dw}. What would be the effect of this on the transmission of the 
domain wall? Due to the smooth change of magnetization in the domain wall, the spin-mixing term 
$\propto\hat\sigma_x$ will not be zero anymore and therefore the spin-flipping channels will open, i.e. 
$T^{\su\sd}$ and $T^\sd\su$ will become finite. Thus one expects that the domain wall resistance
decreases. This is exactly what happens, as can be seen from the bottom panels of Fig. \ref{fig:smooth-dw} 
which show the transmissions for different lengths of the domain walls when the magnetization angle $\theta_i$
changes linearly with the position.

We observe that in general the transmission increases with increasing smoothness (i.e. increasing domain length) 
of the domain. In particular, it becomes different from zero at energies where there is no overlap between the 
two spin-bands. This is due to the spin-mixing by the non-collinear magnetization profile of the domain wall
which allows electrons to spin-flip when crossing the domain wall. The spin-flip process becomes more and more
adiabatic when increasing the smoothness of the domain wall thus resulting in fewer backscattering for incoming
electrons. Therefore the perfect transmission of the chain with homogenous magnetization is finally recovered 
for very smooth domain walls as we can see in Fig. \ref{fig:smooth-dw} for the case of the domain wall with a 
length of 10 atoms.

\section{One-dimensional Hubbard model}
\label{sec:Hubbard}

The classical $sd$-model discussed in the previous section is a rather simplistic model for a magnetic
material. There the spin-polarization of the conduction electrons is due to the interaction of
the electron spin with a classical magnetic moment localized on an atom that gives rise to an
exchange splitting $J$ of the electrons. However, in reality the spin-polarization of the electrons
in a ferromagnetic metal does not arise from the interaction with a local magnetic moment but
is a direct consequence of the Coulomb interaction between the electrons and the Pauli exclusion
principle. The Coulomb interaction gives rise to the so called exchange term when the two spins
of the two electrons are both in the same direction. This term is usually negative, i.e. associated 
with an energy gain for the electrons. If this energy gain is greater than the loss in kinetic
energy which is associated with the two electrons occupying the same spin state then usually the 
electron spin of the material will prefer to align parallel. This is the reason why magnetism in 
material normally originates from the $d$-electrons which are strongly localized at the atoms so
that the hopping to the neighbor atoms is rather small minimizing the loss in kinetic energy.

The most simple model that captures this physics is the Hubbard model. An introduction to the Hubbard 
model can be found for example in Refs. \cite{Auerbach:book:94}, and \cite{Montorosi:book:92}. The 
Hubbard model consists of only two spin-orbitals per atom which can split in energy because of the 
on-site Coulomb repulsion $U$. Atoms are coupled only to their nearest neighbors by a hopping $t$, 
and Coulomb interactions between atoms are neglected. Thus the Hamiltonian of the one-dimensional 
Hubbard model is given by
\begin{equation}
  \label{eq:Hubb-H}
  \hat H = \hat T + \hat V, 
  \, \mbox{ where } \,
  \hat V = U \sum_i \hat n_{i\su} \, \hat n_{i\sd} \, \mbox{ and } \,
  \hat T = -t \sum_{i,\sigma} \hat c_{i\,\sigma}^\dagger \hat c_{i+1\,\sigma} + {\rm h.c.}.
\end{equation}
$\hat T$ is the so-called hopping term, which describes the kinetic energy Hamiltonian in the tight-binding 
approximation, and $\hat V$ is the on-site Coulomb interaction. As pointed out above the Hubbard model is 
employed to describe $d$-band materials, i.e. materials whose electronic properties are determined by the 
$d$-electrons which are responsible for magnetism.

In the following we will study the formation of domain walls in one-dimensional ferromagnetic chains with the 
Hubbard model, and their electronic transport properties. In order to mimic the situation in nanocontacts 
where it is expected that a domain wall builds up in the atomic-size constriction \cite{Bruno:prl:99}, we will
restrict the formation of the domain wall to a finite region of the chain. Outside that region the electronic 
and magnetic structure is assumed to be that of the homogenous chain. The formation of such a constrained 
domain wall will be calculated self-consistently following the Hartree-Fock approximation (HFA) as explained 
in the following section.

%%for open systems given in Sec. \ref{sec:tt:os-hf}.
%%The restriction of the DW to a finite region
%%of the chain is intended to mimic the situation in nanocontacts
%%where it is expected that a sharp DW builds up
%%at the atomic-size constriction \cite{Bruno:prl:99}
%%whose length is given by the length of the constriction.
%%However, this constrained in reality is not natural for a perfect
%%one-dimensional chain studied here.

\subsection{Hartree-Fock approximation -- the Stoner model}
\label{sub:stoner}

The HFA to the Hubbard model is also called the Stoner model 
\cite{Fazekas:book:99}. Here we will treat the one-dimensional Stoner model in the 
non-collinear unrestricted HFA (NC-UHF, see App. \ref{app:NC-UHF})
in order to allow the formation non-collinear magnetization profiles of the domain wall.

In the NC-UHF we find the following Fock matrix for the Hubbard Hamiltonian (\ref{eq:Hubb-H}):
\begin{eqnarray}
  \label{eq:Hubb-HFA}
  F^{\sigma_1 \sigma_2}_{i j} = T_{ij} \delta_{\sigma_1 \sigma_2} 
  + U \, \delta_{i j} \left( 
    \delta_{\sigma_1 \sigma2} (P_{ii}^{\su\su}+P_{ii}^{\sd\sd}) 
    - P_{ii}^{\sigma_1 \sigma_2} 
  \right).
\end{eqnarray}
The first term in the brackets is the Hartree contribution which describes the pure Coulomb repulsion, the second
term is the Fock contribution describing the Coulomb exchange interaction. The Fock terms for $\sigma_1=\sigma_2$
cancel exactly with the Hartree-terms $\propto \rho_{ii}^{\sigma\sigma}$ with $\sigma=\sigma_1=\sigma_2$ which 
present an unphysical self-interaction of the electrons, correcting thereby the self-interaction error. 

The density matrix can be expressed through the creation and  annihilation operators:
\begin{eqnarray}
  \label{eq:Hubb-DMat}
  P_{ij}^{\sigma_1 \sigma_2} 
  &=& \langle \hat c_{i\sigma_1}^\dagger \hat c_{j\sigma_2} \rangle^\ast.
\end{eqnarray}
so the Fock-operator corresponding to the Fock matrix 
(\ref{eq:Hubb-HFA}) can be expressed in 2\raisebox{1ex}{nd} quantization as:
\begin{equation}
  \label{eq:Hubb-F}
  \hat{\mathcal F} = \hat{\mathcal T} + U \sum_{i,\sigma} \langle \hat n_{i\sigma} \rangle \hat n_{i\bar\sigma}
  -  U \sum_{i,\sigma} \langle \hat c_{i\sigma}^\dagger \hat c_{i\bar\sigma} \rangle^\ast \hat c_{i\sigma}^\dagger \hat c_{i\bar\sigma}.
\end{equation}

In the next subsection we will study the situation of a homogeneous chain
in the Stoner model.

\subsection{Homogeneous Chain}
\label{sub:chain}

In an infinite chain all sites are equivalent. 
Therefore the Stoner model Hamiltonian (\ref{eq:Hubb-F})
becomes homogeneous, i.e. the expectation values 
$\langle\hat n_{i\sigma}\rangle$ are the same
at each site $i$,$\langle\hat n_{i\sigma}\rangle = \bar n_\sigma$.
This allows us to choose the quantization axis for
the spin, say $z$, so the spin-mixing terms
in the Fock operator vanish:
\begin{equation}
  \label{eq:HubbChain-F}
  \hat{\mathcal F} = \hat{\mathcal T} + U \sum_{i,\sigma} \bar n_{\bar\sigma} \hat n_{i\sigma}.
\end{equation}

We can diagonalize the Fock matrix of the homogenous chain by introducing 
Bloch states, $\ket{k} = \frac{1}{\sqrt{2\pi}} \sum_j \exp(\imath k j)\ket{j}$:
\begin{equation}
  \label{eq:HubbChain-FDiag}
  \hat{\mathcal F} = \sum_{k, \sigma} F^\sigma(k) \hat c_{k \sigma}^\dagger \hat c_{k \sigma},
\end{equation}
with
\begin{equation}
  \label{eq:HubbChain-F(k)}
  F^\sigma(k) = \sum_{j=-1}^{1} \exp(\imath k j) \bra{0 \sigma} \hat{\mathcal F} \ket{j \sigma}
  = \epsilon(k) + U n_{\bar\sigma}.
\end{equation}
$\epsilon_k$ is the kinetic energy of the Bloch wave $\ket{k}$ (Hamiltonian $\hat T$):
\begin{equation}
  \label{eq:HubbChain-eps(k)}
  \epsilon(k) = 2 t \cos( k ).
\end{equation}
The two resulting spin-bands (energy $F^\sigma(k)$)
have the dispersion relation $\epsilon(k)$ of the non-interacting chain 
but are split by $U(n_\su-n_\sd)$ in energy.
The corresponding DOS $\mathcal{D}$ can be written as a sum of
the spin-up and the spin-down DOS:
\begin{eqnarray}
  \label{eq:HubbChain-DOS}
  \mathcal{D}(\epsilon) &=& \mathcal{D}_\su(\epsilon) + \mathcal{D}_\sd(\epsilon).
\end{eqnarray}
where
\begin{eqnarray}
  \mathcal{D}_\sigma(\epsilon) 
  &=& \left\{ 
    \begin{array}{ll} 
      \frac{1}{\pi}\frac{d}{d\epsilon}
      \arccos( (\epsilon  - U n_{\bar\sigma})/2t ) & ; 
      \mbox{ if } \abs{\epsilon - U n_{\bar\sigma}} \le 2\abs{t} \\ \\
      0 & ; \mbox{elsewhere}
    \end{array}   
    \right.
    \label{eq:HubbChain-SDOS}
\end{eqnarray}

\begin{figure}
  \begin{minipage}{0.5\linewidth}
    \includegraphics[width=\linewidth]{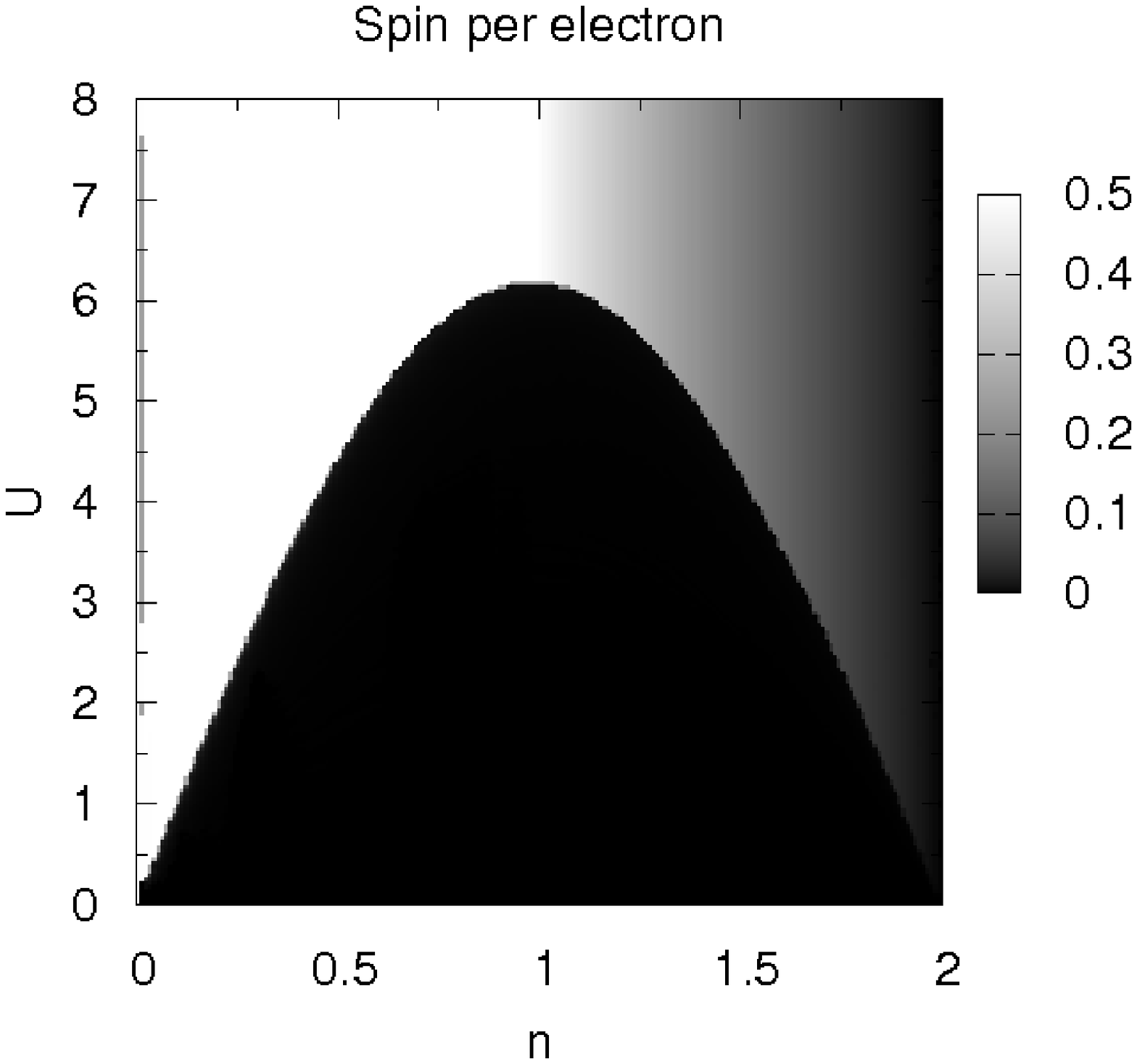}
  \end{minipage}
  \begin{minipage}{0.5\linewidth}
    \includegraphics[width=\linewidth]{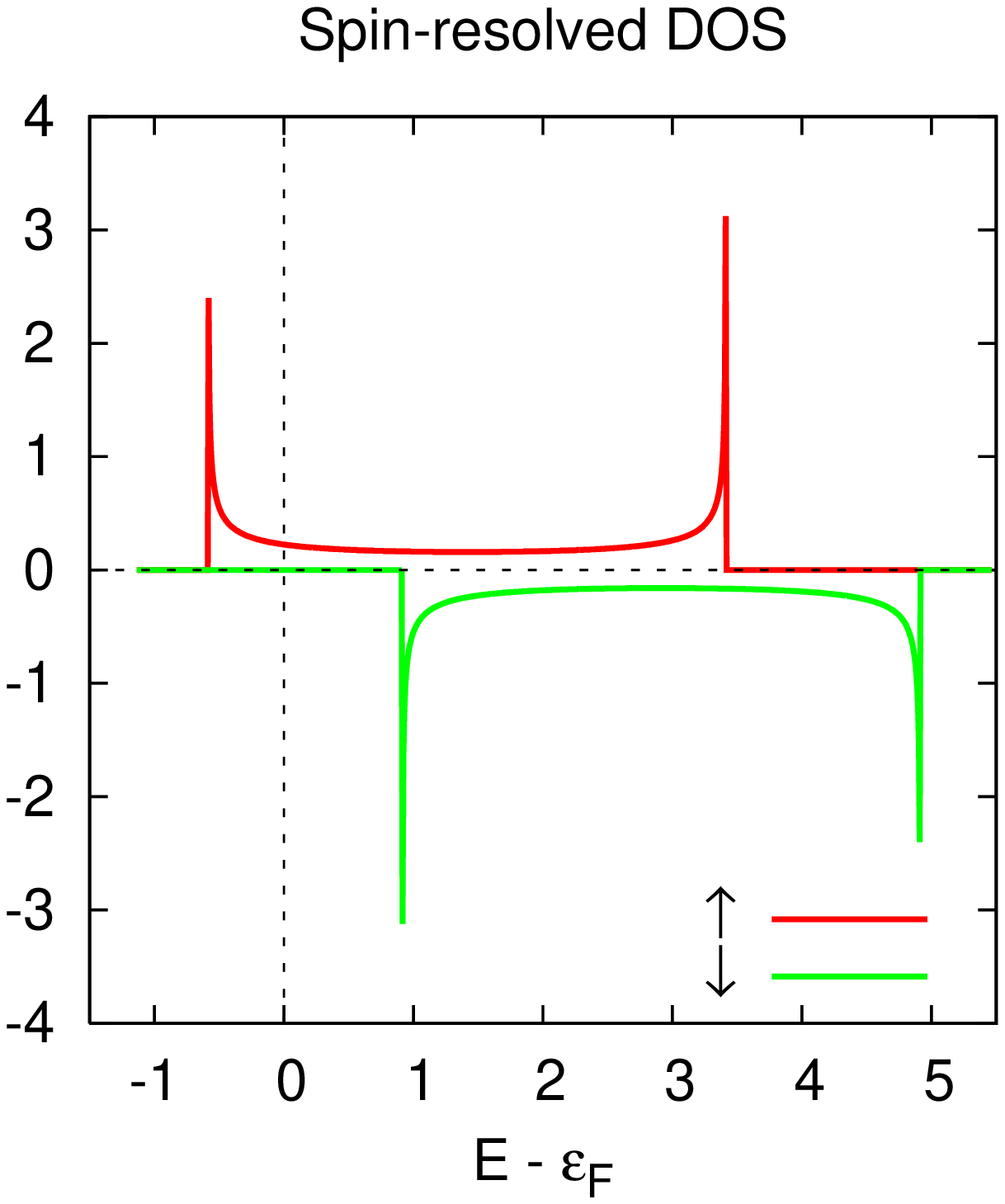}
  \end{minipage}
  \setcapindent{0cm}
  \caption{Self-consistently calculated magnetization phase diagram 
    for the one-dimensional Stoner model (left) 
    and spin-resolved DOS (right) for the ferromagnetic (FM) phase 
    (filling $n=0.25$, Coulomb parameter $U=6$) of the one-dimensional 
    Stoner model. The dotted vertical line indicates the Fermi level.
  }
  \label{fig:hubb_magn}
\end{figure}

The spin-dependent occupation numbers $n_\su$ and $n_\sd$ are given by 
integrating the spin-resolved DOS up to the Fermi energy $\mu$:
\begin{eqnarray}
  n_\sigma(\mu)
  &=& \int_{-\infty}^{\infty} d\epsilon \, 
  \mathcal{D}_\sigma(\epsilon) \,  f( \epsilon - \mu )
  = \frac{1}{\pi} \int_{U n_{\bar\sigma} -2\abs{t}}^{U n_{\bar\sigma}+2\abs{t}} 
  d\epsilon \, \frac{d}{d\epsilon} \arccos 
  \left( (\epsilon  - U n_{\bar\sigma})/2t \right) \, f( \epsilon - \mu ) 
  \nonumber \\ 
  \nonumber \\
  &=& \left\{ 
    \begin{array}{ll} 
      0 & ; \mu \le U n_{\bar\sigma} -2\abs{t}, 
      \\ \\ 
      \frac{1}{\pi} \arccos \left((\mu  - U n_{\bar\sigma})/2t\right) & 
      ; \abs{\mu-U n_{\bar\sigma} }\le 2\abs{t}, 
      \\ \\
      1 & ; \mu >  U n_{\bar\sigma} +2\abs{t}.
    \end{array} 
  \right.
  \nonumber \\
    \label{eq:HubbChain-nspin}
\end{eqnarray}

We fix the number of electrons $n = n_\su + n_\sd$ per unit cell (also called filling factor),
the hopping parameter $t$ and the Coulomb parameter $U$.
The self-consistent solution of the problem then yields the chemical potential
$\mu$, the occupation numbers $n_\su$ and $n_\sd$, and thus the spin-density
$s = n_\su - n_\sd$ and the total energy per unit cell.

%%The self-consistent procedure for the solution is as follows:
%%\begin{enumerate}
%%  \item Guess $n_\su$ and $n_\sd$, so that 
%%    $n = n_\su + n_\sd$ is fulfilled. 
%%  \item $n_\su$ and $n_\sd$  determine the Hamiltonian (\ref{eq:HubbChain-FDiag}), 
%%    and the DOS (\ref{eq:HubbChain-SDOS})
%%  \item Determine $\mu$ by finding the zero of $n_\su(\mu) + n_\sd(\mu) - n$,
%%    where $n_\sigma(\mu)$ is given by ()
%%  \item IF resulting $n_\su$ and $n_\sd$ are converged\\ 
%%    THEN Finished\\
%%    ELSE GOTO 2
%%\end{enumerate}

The left hand side of Fig. \ref{fig:hubb_magn} shows the phase diagram of the self-consistently calculated 
Stoner model for the magnetization (spin per electron) of the chain in dependence of the on-site 
Coulomb repulsion $U$ and the filling $n$. 
Obviously there is a ferromagnetic (FM) and a paramagnetic (PM) phase.
Since we want to study transport in magnetic materials we are only interested in the 
FM phase here. In the FM phase the Stoner model is a so-called half-metal, meaning
that the electrons are completely spin-polarized, as illustrated on the right hand side of
Fig. \ref{fig:hubb_magn}: Only one of the two spin bands is partially filled, while
the other is always empty.
The electrons at the Fermi level and thus the conduction electrons are 100\% spin-polarized.
This has important consequences for the spin transport. First, at the Fermi level only one
transport channel (the majority spin channel) is open, leading to a conductivity of 1 $G_0$.
Second, in the case of a domain, the MR will be 100\% for an abrupt DW since
all traversing electrons will be blocked. The latter will be discussed in the next section
in more detail.

\section{DW formation and scattering}
\label{sub:dw-formation}

%%\emph{Model -}
As said in the beginning of this chapter the DW formation
will be restricted to a finite region of the chain - the device region (D).
We will make use of the self-consistent procedure described in in Ch. \ref{ch:ab-initio}
for computing the electronic structure of an open system in the HFA.
The semi-infinite regions to the left and right of D, i.e. the electrodes L and R,
will have bulk electronic structure which does not change during the self-consistent
calculation of the DW formation.
In order to give the correct magnetic boundary conditions
for the formation of a DW the magnetization of L and R will be aligned
antiparallel (AP).

Following (\ref{eq:GD}), the GF of the device region
is given by 
\begin{equation}
  \M{G}_D(E) = ( E - \M{F}_D - \M\Sigma_L(E) - \M\Sigma_R(E) )^{-1},
\end{equation}
where $F_D$ is the Fock matrix of the device in the NC-UHF approximation
for the Stoner model. The self-energy matrices can be subdivided into the
spin-resolved self-energy matrices $\M\Sigma_{L,R}^\su$ and $\M\Sigma_{L,R}^\sd$
following (\ref{eq:spin-Sigma/Gamma}) which can be calculated from
electrodes' self-energies following (\ref{eq:SigmaL},\ref{eq:SigmaR}).

The left and right lead are described by tight-binding chains with 
a spin-dependent energy shift of $\epsilon_\sigma=U  n_{\bar\sigma}$.
For this case the Dyson equations (\ref{eq:DysonL}) and (\ref{eq:DysonR}) 
can be solved analytically and one obtains for the spin-dependent retarded self-energy of the 
left (l) and right lead (r):
\begin{eqnarray}
  \Sigma^\sigma_{l,r}(E) = 
  \left\{ \begin{array}{ll}
    \frac{E - \epsilon_\sigma}{2} + \frac{\sqrt{ (E - \epsilon_\sigma)^2-4t^2 }}{2} ; \, \mbox{ for } E < \|2t\| \\
    \\
    \frac{E - \epsilon_\sigma}{2} - \frac{i\sqrt{ (E - \epsilon_\sigma)^2-4t^2 }}{2}; \, \mbox{ for } \|E\| \le \|2t\| \\ 
    \\
    \frac{E - \epsilon_\sigma}{2} - \frac{\sqrt{ (E - \epsilon_\sigma)^2-4t^2 }}{2} ; \, \mbox{ for } E > \|2t\|
  \end{array} \right.
\end{eqnarray}
%%

%% -RESULTS-
%%
We follow the procedure described in Sec. \ref{sec:KS-NEGF} adapted to the HFA to calculate 
self-consistently the formation of a DW in the AP configuration. The magnetization of an atom 
is defined as the total atomic spin normalized to the number of electrons of the atom:
\begin{equation}
  \label{eq:M(i)}
  \vec M(i) = \frac{ \langle \vec S(i) \rangle }{n(i)},
\end{equation}
where the expectation value of the atomic spin is obtained from
the density-matrix:
\begin{equation}
  \label{eq:S(i)}
  \langle \vec S(i) \rangle = \frac{1}{2}
  \left( \begin{array}{c}
    2{\rm Re}[\rho_{ii}^{\su\sd}] \\
    2{\rm Im}[\rho_{ii}^{\su\sd}] \\
    \rho_{ii}^{\su\su}-\rho_{ii}^{\sd\sd}
  \end{array}
  \right).
\end{equation}
For a ``smooth'' DW, i.e. a DW where the magnetization angle 
changes smoothly with the position, the initial guess must have non-collinear 
magnetization vectors. In the following we will start with an initial guess 
representing a ``linear'' DW, i.e. a DW where the magnetization angle
changes linearly with the position.
On the other hand if we want to calculate the formation of an abrupt domain 
we have to start with an initial guess with only collinear magnetization vectors,
i.e. the magnetization of each atom of the domain must be along the same axis
(the z-axis) as the magnetization of the electrodes.

\begin{figure}
  \begin{minipage}{0.49\linewidth}
    \includegraphics[width=\linewidth]{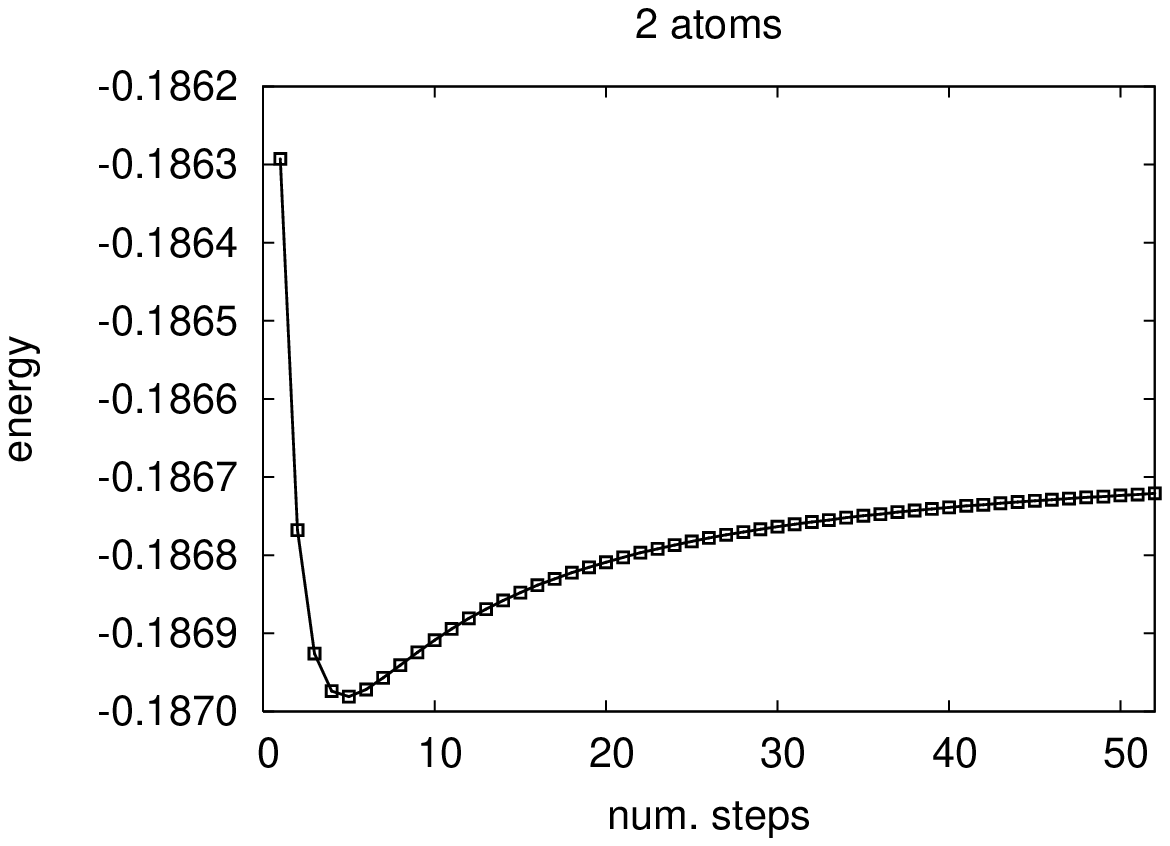}
  \end{minipage}
  \begin{minipage}{0.49\linewidth}
    \includegraphics[width=\linewidth]{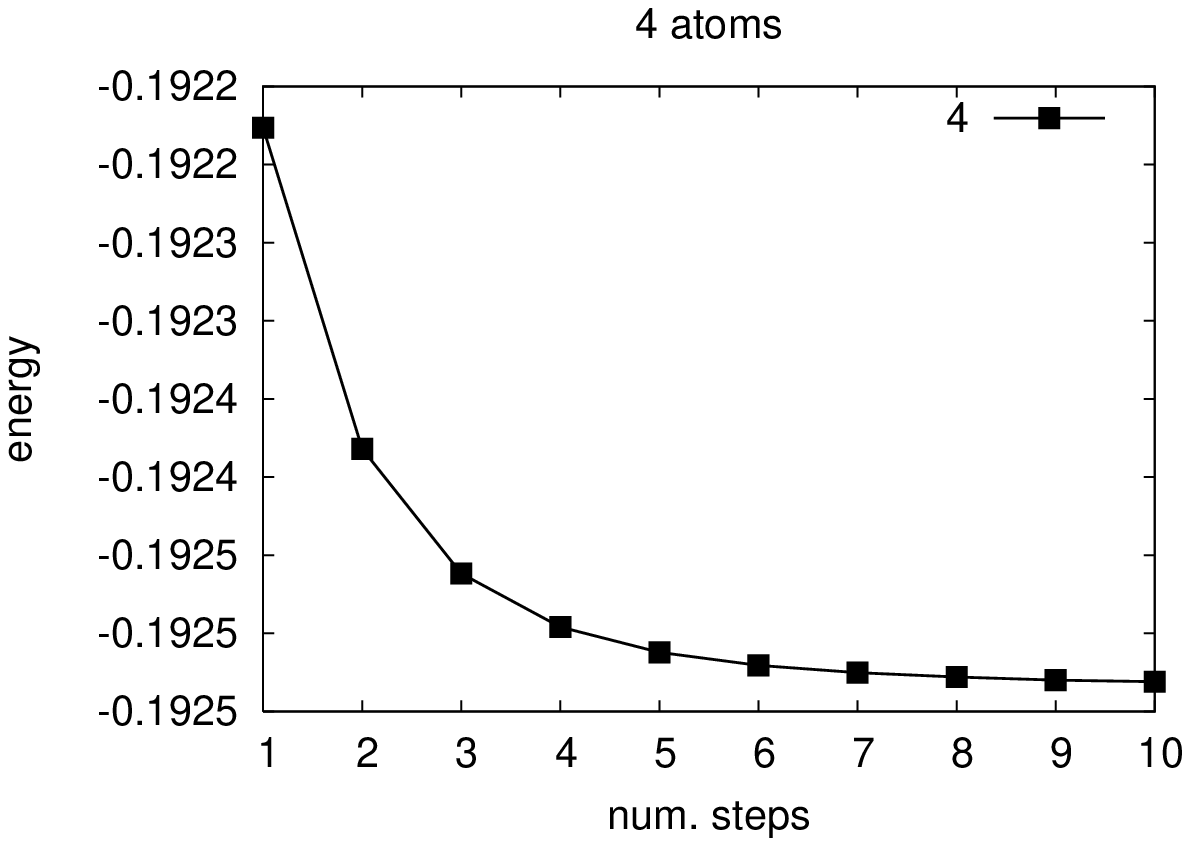}
  \end{minipage}
  \setcapindent{0cm}
  \caption{ Total energy following eq. (\ref{eq:E_tot}) corrected for double-counting
    following eq. (\ref{eq:KS-Energy}) in dependence of number of steps in the 
    self-consistent DW formation for a domain length of (a) 2 atoms and (b) 4 atoms.
    Filling $n=0.1$ and Coulomb parameter $U=3$.}
  \label{fig:conv}
\end{figure}

Fig. \ref{fig:conv} shows the total energy 
of two constrained smooth DWs of different length 
(2 and 4 atoms) after eqs. (\ref{eq:KS-Energy}) and (\ref{eq:E_tot})
for each step in the self-consistent procedure until 
convergence of the density matrix. 
Obviously, in the case of a domain length of only two atoms
the energy is \emph{not} minimized. 
This is not surprising if we consider that for such a short 
domain the electronic structure of the electrodes has surely 
not relaxed to the bulk electronic structure as we have assumed. 
In order to minimize the energy we would have to recalculate 
the electronic structure of the electrodes as well, and not
let it fixed during the self-consistency as we do here.
But here we are interested in the formation of constrained domains,
since in real nanocontacts the DW formation is also
constrained by the geometry of the contact \cite{Bruno:prl:99}.
Since we have neglected the geometrical part of the problem by
considering one-dimensional chains we have to constrain the 
DW artificially. This of course leads to a constrained
variational problem, so that not the energy is minimized but
the free energy defined in a proper way.
We note that already for a domain of 4 atoms (right panel of
Fig. \ref{fig:conv} ) the constrained variational 
search minimizes the total energy of the system. The 
assumption of a fixed bulk electronic structure in the electrodes 
becomes better with increasing DW length.

\begin{figure}
  \begin{minipage}[t][][b]{0.75\linewidth}
    \includegraphics[width=\linewidth]{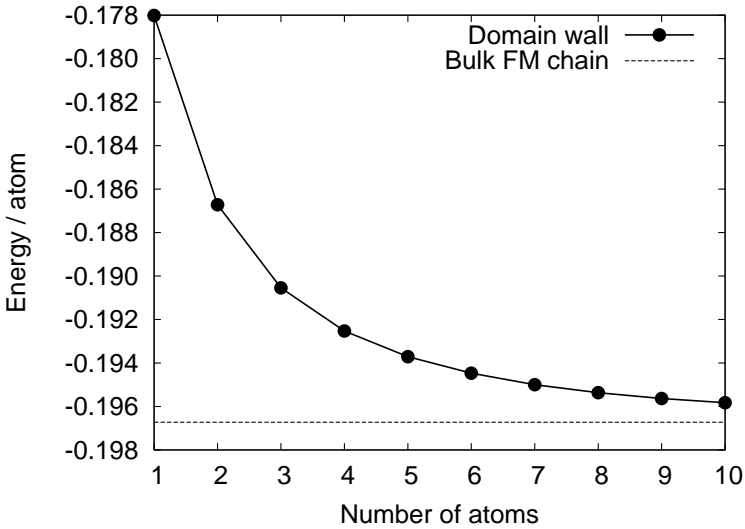}
  \end{minipage}
  \begin{minipage}[t][][b]{0.24\linewidth}
    \setcapindent{0cm}
    \caption{ Total energy following eqs. (\ref{eq:KS-Energy}) and (\ref{eq:E_tot}) 
      as a function of the DW length for the self-consistent 
      DW formation. Filling $n=0.1$ and Coulomb parameter $U=3$. }
    \label{fig:hubb_energy}
  \end{minipage}
\end{figure}

Nevertheless the resulting energies can be compared to each
other and to the energy of the bulk chain. Thus we can obtain
the cost in energy for the formation of a DW of a 
certain length. Fig. \ref{fig:hubb_energy} shows
the DW formation energy per atom as a function of the
DW length in comparison to the energy of the 
FM bulk chain. 
The homogeneous chain has of course the lowest energy since
the ferromagnetic solution favors the parallel alignment
of the electron spins to increase the gain in exchange energy.
As the DW length is increased the energy per atom 
shrinks because locally the ferromagnetic solution 
is recovered as the DW becomes smoother 
with increasing domain length.

\begin{figure}
  \begin{minipage}{0.5\linewidth}
    \begin{flushright}
      \includegraphics[width=\linewidth]{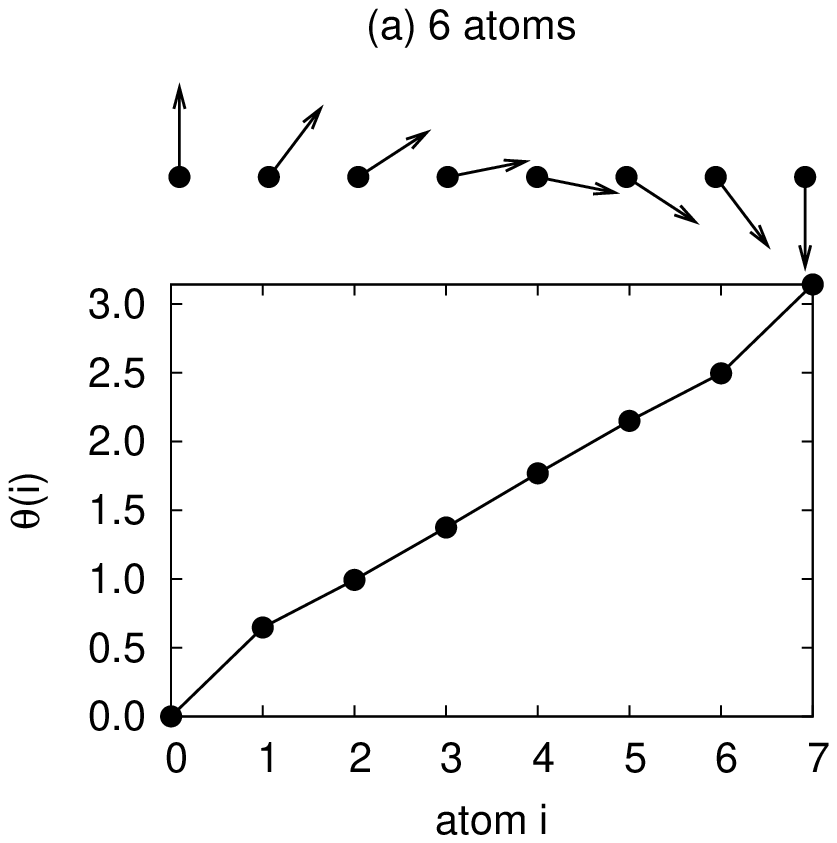}
    \end{flushright}
  \end{minipage}
  \begin{minipage}{0.5\linewidth}
    \begin{flushleft}
      \includegraphics[width=\linewidth]{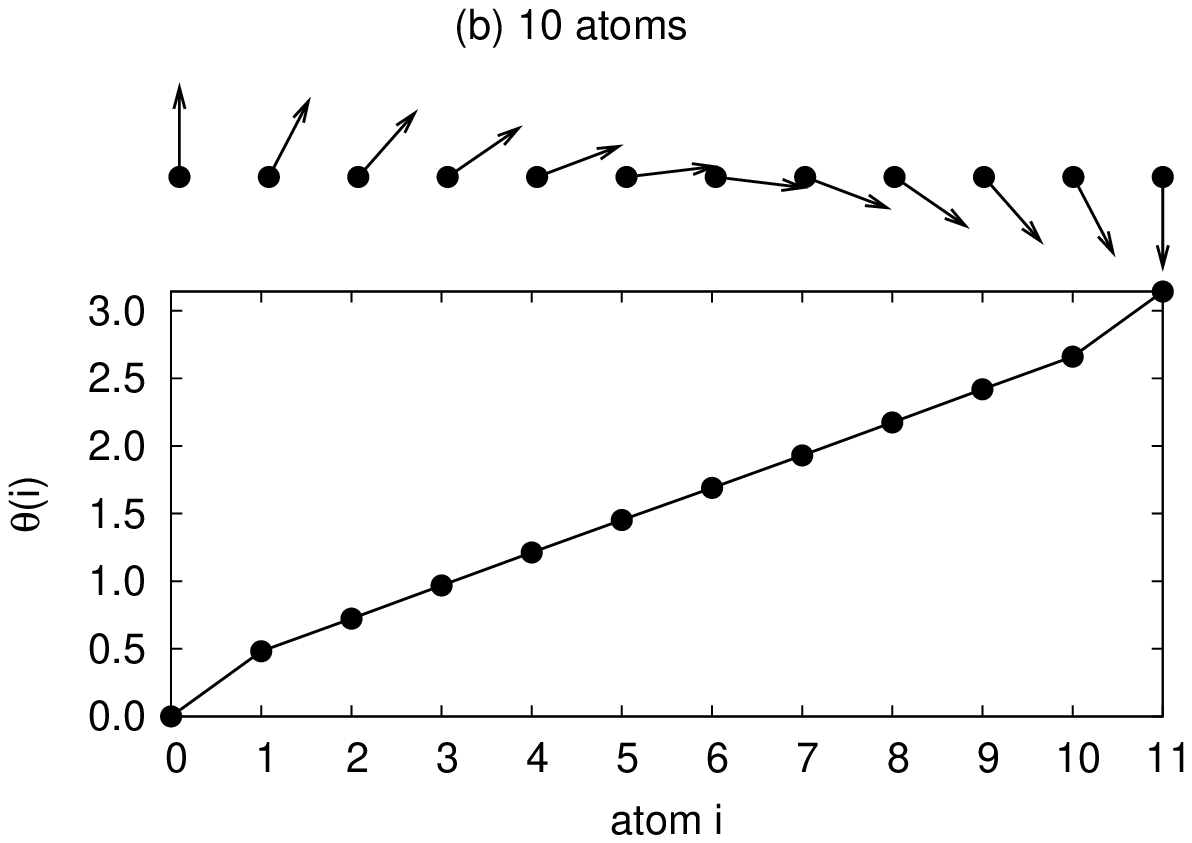}
    \end{flushleft}
  \end{minipage}
  \setcapindent{0cm}
  \caption{ DW profiles for (a) 6 atoms and (b) 10 atoms in the
    device region. The upper panel shows the magnetization vector for each 
    atom in the xz-plane, and the lower panel the rotation angle $\theta(i)$ 
    of the magnetization vector as a function of the atom position $i$.
    The leftmost (0) and the rightmost (7 in (a) and 11 in (b), respectively) 
    atom belong to the left and right electrode, respectively, which have bulk electronic
    structure. Filling $n=0.1$ and  Coulomb parameter $U=3$.}
  \label{fig:hubb_dw_prof}
\end{figure}

The smoothing of the DW with increasing domain length 
is demonstrated in Fig. \ref{fig:hubb_dw_prof}.
The upper panels of  Fig. \ref{fig:hubb_dw_prof} show the
angle $\theta(i)$ of the magnetization vector on the 
$i$-th atom in the device region (=domain) for a domain length of 
(a) 6 atoms and (b) 10 atoms, respectively, while the lower panels 
show the corresponding magnetization vectors.
In the first place, one notes that the  angle $\theta$
for atoms inside the domain (atoms 1 to 6 in (a), and 1 to 10 in (b)) 
depends almost linear on the position, and that this dependence 
approaches a perfect line with increasing domain length.
In the second place, one observes that the \emph{rotation angle} 
$\Delta\theta=\theta(i+1)-\theta(i)$ between the last (first) atom of 
the left (right) electrode and the first (last) atom of the domain 
is bigger than the rotation angles inside the domain.
This difference between the rotation angles $\Delta\theta$ between
electrode and device and inside the device becomes also smaller
with increasing domain length.
Finally, we mention that also the atomic charge $n(i)$ and the length of the magnetization 
vector $S(i)$ (not shown) varies slightly
\footnote{$n(i)$ deviates in both directions by less than 2\% from the bulk value $n=0.1$ for (a) and (b),
  while $S(i)$ is smaller by 4-10\% for (a) and by 2-4\% for (b) than the bulk magnetization $S=n/2$.} 
with the position, and that the deviation from 
the bulk values of the atomic charge and the magnetization becomes also smaller 
with increasing domain length.
In summary, the DW becomes smoother with increasing domain length
approaching a \emph{linear DW} where the magnetization angle
depends linearly on the position and as well the magnitude of the magnetization
as the atomic charge are constant and equal to that of the FM bulk chain
thereby locally recovering the FM state.

%\begin{figure}
%  %%\begin{minipage}{0.5\linewidth}
%  %%  \includegraphics[width=0.98\linewidth]{figures/spin-transport/hubb_dw_prof1_6.eps}
%  %%\end{minipage}
%  %%\begin{minipage}{0.5\linewidth}
%  \includegraphics[width=0.98\linewidth]{figures/spin-transport/hubb_dw_prof2_6.eps}
%  %%\end{minipage}
%  \label{fig:hubb_dw_prof_6}
%  \caption{ 6 atoms, $n=0.1$, $U=3$.}
%\end{figure}

%\begin{figure}
%  %%\begin{minipage}{0.5\linewidth}
%  %%  \includegraphics[width=0.98\linewidth]{figures/spin-transport/hubb_dw_prof1_10.eps}
%  %%\end{minipage}
%  %%\begin{minipage}{0.5\linewidth}
%  \includegraphics[width=0.98\linewidth]{figures/spin-transport/hubb_dw_prof2_10.eps}
%  %%\end{minipage}
% \label{fig:hubb_dw_prof_10}
%  \caption{ 10 atoms, $n=0.1$, $U=3$.}
%\end{figure}

\begin{figure}
  \begin{tabular}{cc}
    (a) Total transmission & (b) Magneto-resistance \\ 
    \includegraphics[width=0.45\linewidth]{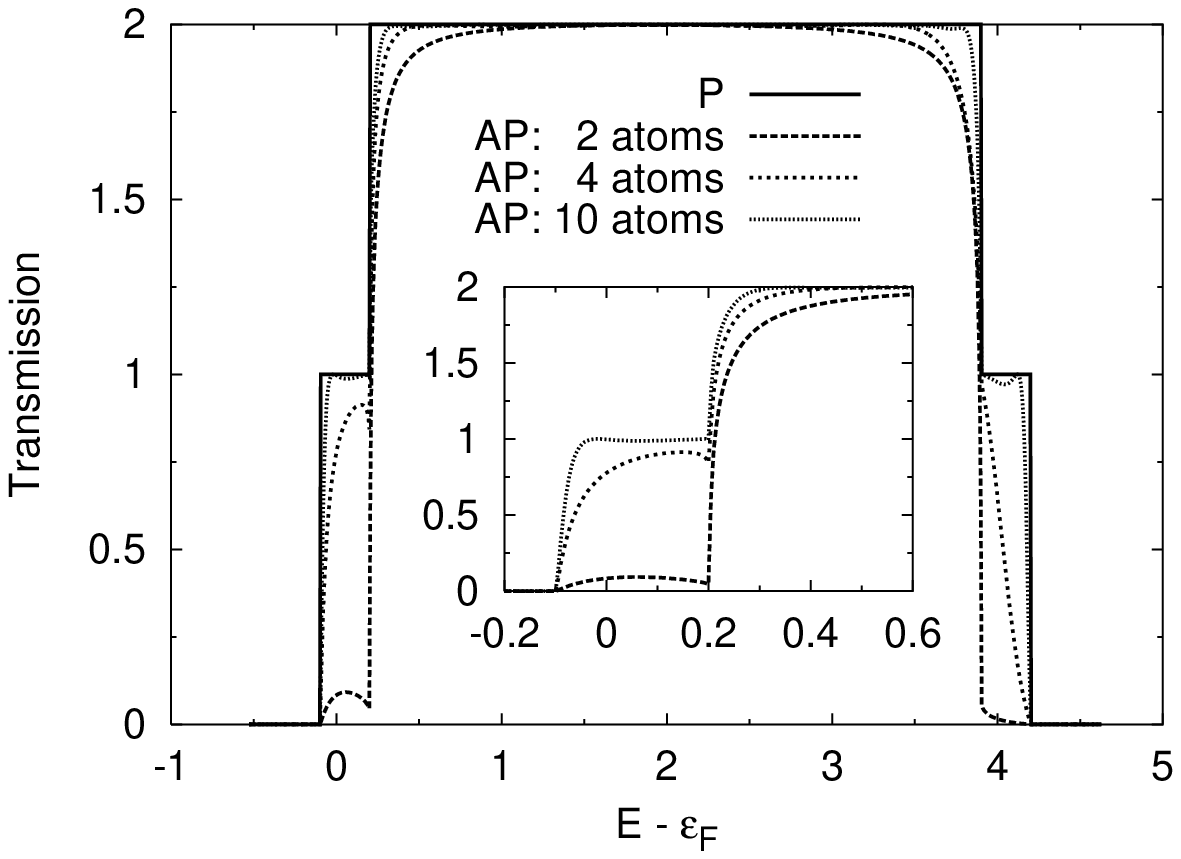} &
    \includegraphics[width=0.45\linewidth]{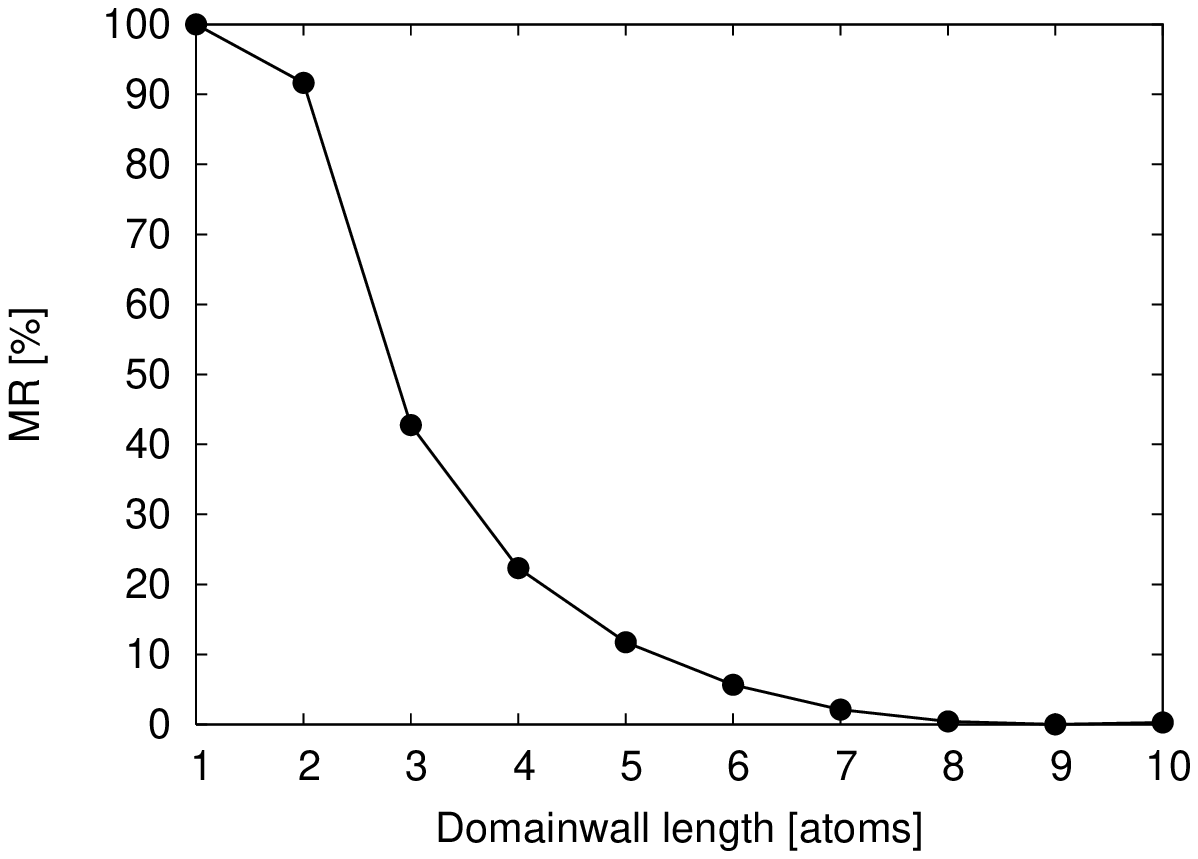}
  \end{tabular}
%  \begin{minipage}{0.5\linewidth}
%    \includegraphics[width=0.98\linewidth]{figures/spin-transport/hubb_trans.eps}
%  \end{minipage}
%  \begin{minipage}{0.5\linewidth}
%    \includegraphics[width=0.98\linewidth]{figures/spin-transport/hubb_mr.ldw.eps}
%  \end{minipage}
  \setcapindent{0cm}
  \caption{(a) Total transmission in dependence of the energy
    (relative to the Fermi energy $\epsilon_F$) of one-dimensional chain in
    the Stoner model for some domain lengths in the AP configuration
    compared to the perfect transmission in the P configuration.
    (b) Magneto-resistance (using def. MR$_1=(T_{\rm P}-T_{\rm AP})/T_{\rm P}$) 
    as a function of the domain length.
    Filling $n=0.1$, Coulomb parameter $U=3$ for both (a) and (b).}
  \label{fig:hubb_trans}
\end{figure}

In turn the magnetization profile of the DW affects the
electron transport through the DW. 
Fig. \ref{fig:hubb_trans}a shows the total transmission 
as a function of the energy relative to the Fermi energy 
in the P case and for some domain lengths (2,4,10 atoms) in the AP
case. One observes that the transmission for the AP case deviates globally 
from the perfect transmission in the P case.
The deviation shrinks with increasing length of the domain.
For a domain length of 10 atoms the perfect P transmission is almost
recovered. The effect is especially strong near the Fermi energy as
the close-up in Fig. \ref{fig:hubb_trans}a shows.
Near the Fermi energy only one spin-band contributes to the
conduction in the P case, since the ferromagnetic chain is a half-metal
as explained in the previous section.
Thus for an abrupt domain the transmission is completely blocked
near the Fermi level: An electron passing the DW
has to change its spin to travel on in the other electrode
since only one spin-state is allowed near the Fermi level, and
the electrodes are oppositely magnetized in the AP case.
But an abrupt domain does not allow the electrons to change their
spin because of the absence of spin-mixing terms in the 
spin-polarized Hamiltonian.
On the other hand a non-collinear magnetization of the domain 
gives rise to spin-mixing terms which can rotate the spin
state of an incoming electron. 
The transmission probability becomes higher the smaller
the rotation angle of the magnetization vector between 
neighboring atoms.
Therefore in the case of a non-collinear domain of 2 atoms the 
transmission is nonzero but very small compared to the P transmission. 
Making the domain larger the rotation angles decrease - the domain 
becomes more smooth - and the transmission probability creases
until recovering the perfect transmission of 1 in the case
of an adiabatic DW which locally resembles the 
ferromagnetic solution.
As a consequence the MR decreases with increasing length of the
domain, as demonstrated in Fig. \ref{fig:hubb_trans}b.
For an abrupt domain the MR is 100\% as the DW 
blocks the transmission completely while for a very smooth
domain the MR goes to zero: The AP conductance becomes the same
as the P conductance.

\begin{figure}
  \begin{tabular}{cc}
    (a) $T_{\su\su}$ & (b) $T_{\su\sd}$ \\
    \includegraphics[width=0.45\linewidth]{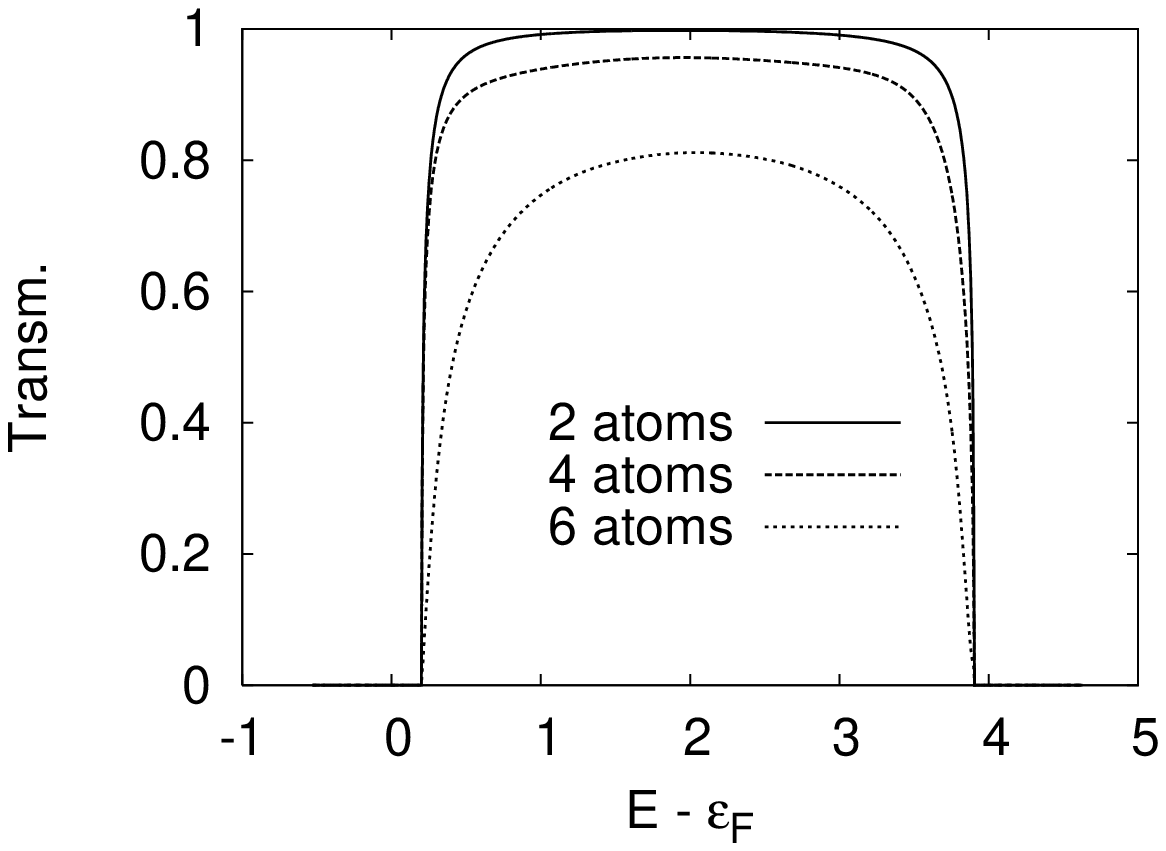} &
    \includegraphics[width=0.45\linewidth]{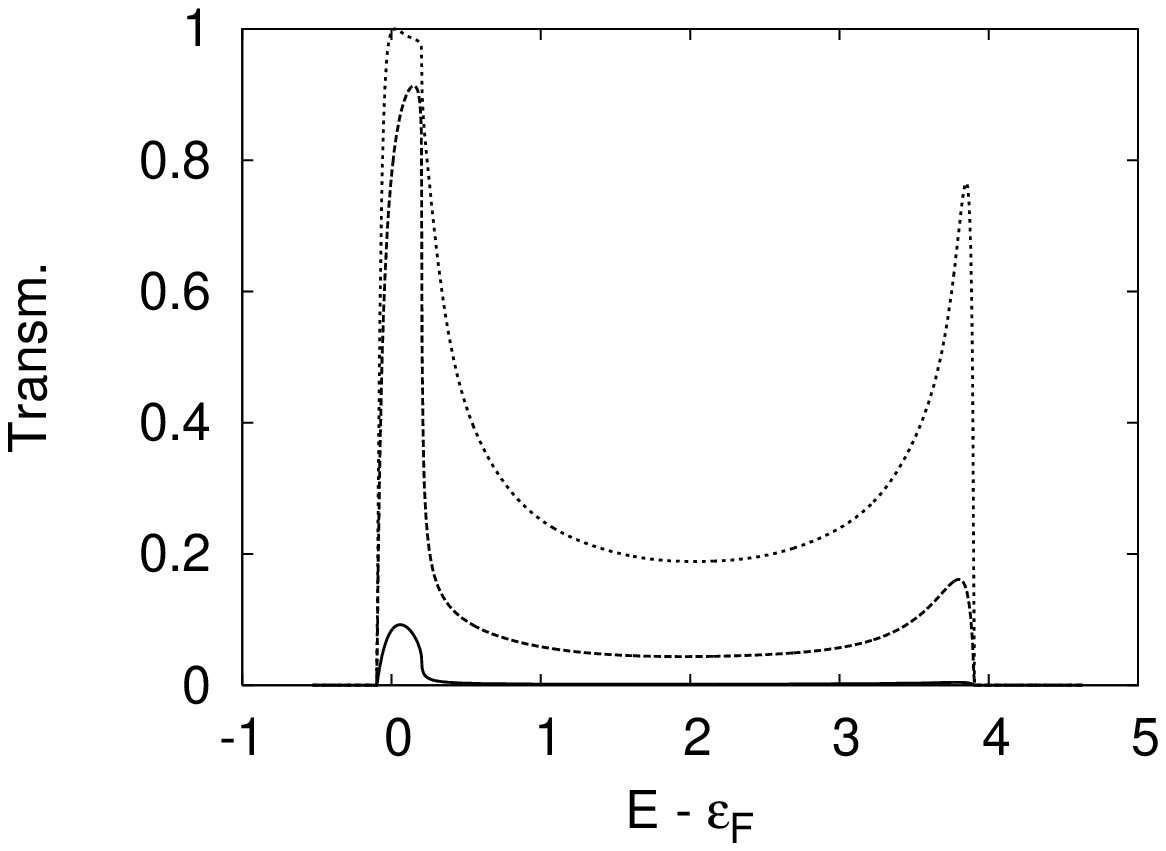} \\
    & \\
    (c) $T_{\sd\su}$ & (d) $T_{\sd\sd}$ \\
    \includegraphics[width=0.45\linewidth]{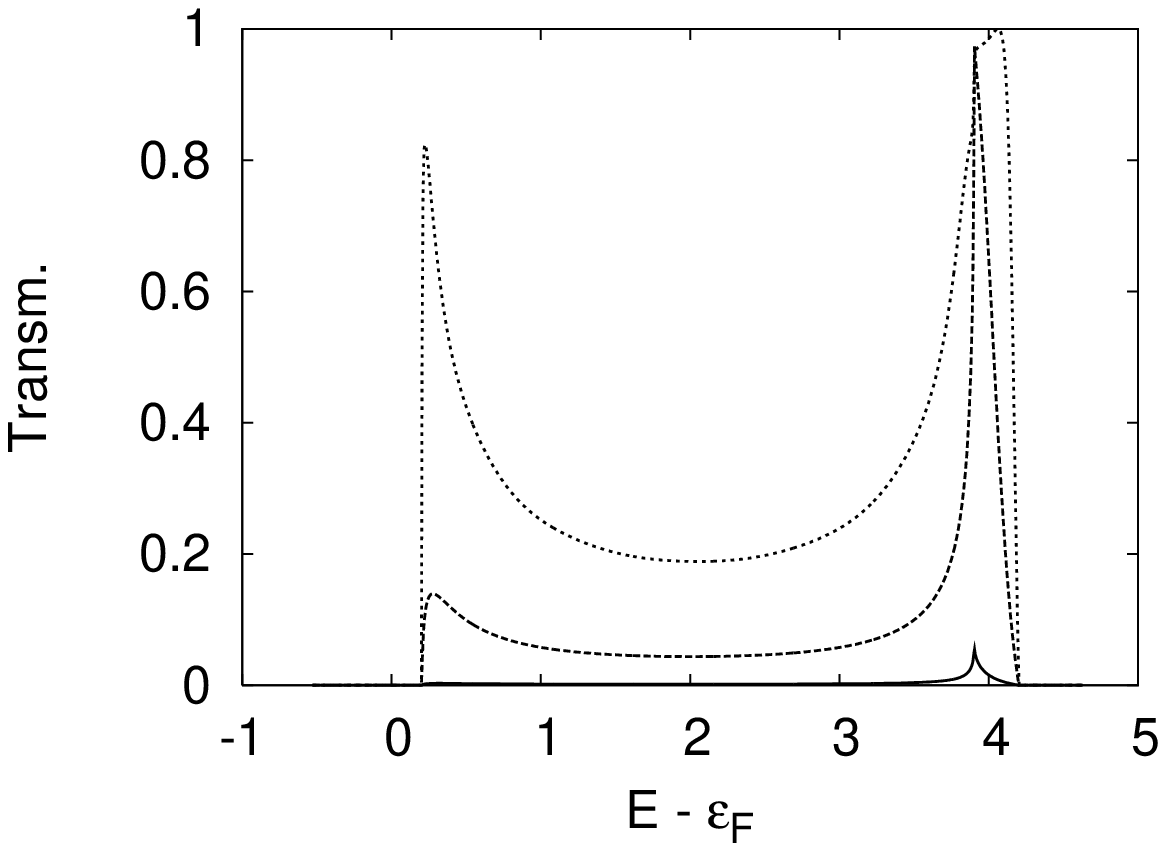} &
    \includegraphics[width=0.45\linewidth]{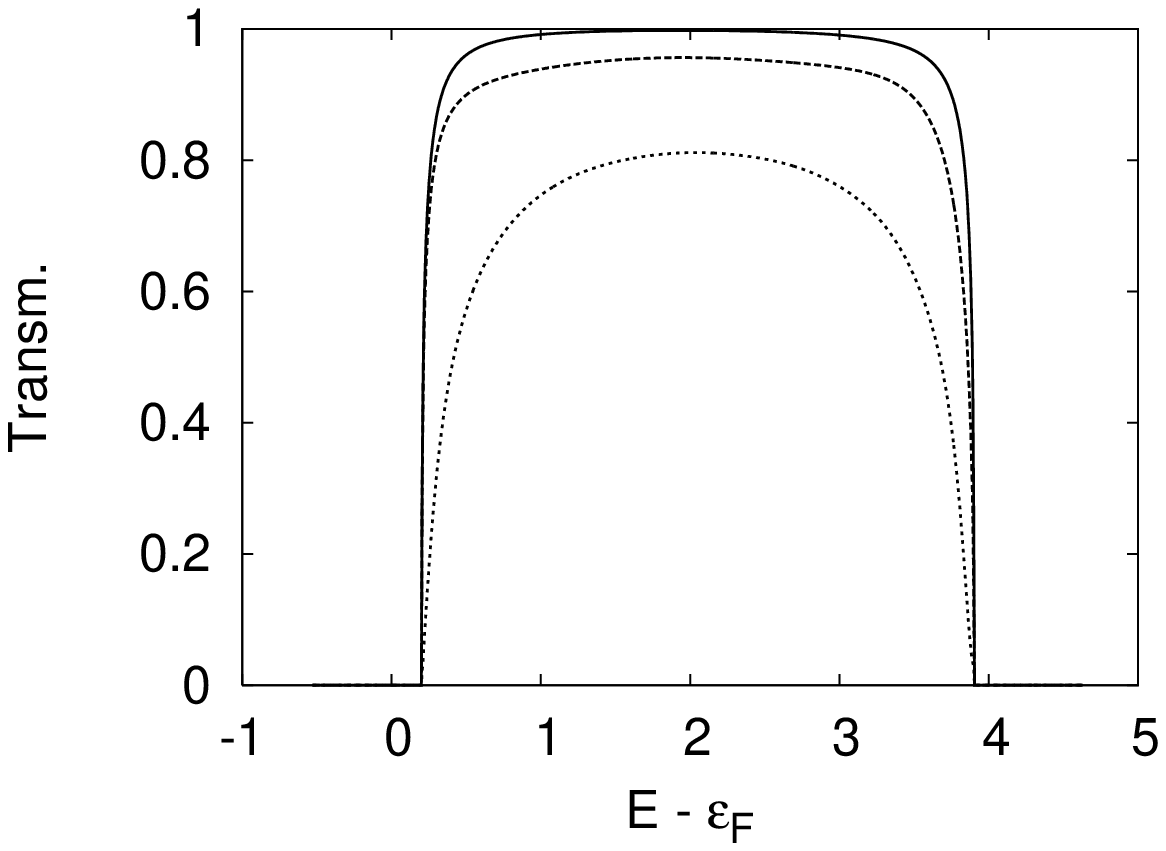} 
  \end{tabular}
  \setcapindent{0cm}
  \caption{Transmission of individual spin-channels in dependence of the energy 
    relative to the  Fermi level for AP configuration and domain lengths of 
    2, 4  and 8 atoms. Filling $n=0.1$, Coulomb parameter $U=3$.}
  \label{fig:hubb_tchan}
\end{figure}

Finally, we will have a look at the spin-resolved conductance channels.
Fig. \ref{fig:hubb_tchan} shows the contributions of the
individual spin-channels (\ref{eq:spin-caroli}) to the total transmission.
As illustrated in Fig. \ref{fig:spin-channels} $T_{\su\su}$ ($T_{\sd\sd}$) 
gives the probability of an electron coming from the left lead with spin-up 
(spin-down) and being transmitted to the right electrode conserving its spin
while $T_{\su\sd}$ ($T_{\sd\su}$) is the probability of an electron entering 
with spin-up (spin-down) and flipping its spin when being transmitted to the right.
The $\su\su$-and $\sd\sd$-channel do not contribute anything to the transmission
near the Fermi energy, in accordance with the above reasoning that in the half-metallic
limit only one spin-band in the electrodes contributes to the conductance, so that
in the AP case an electron can not conserve its spin when being transmitted to the other
electrode.
On the other hand a non-collinear magnetization of the domain opens the $\su\sd$- and
$\sd\su$-channels which do not contribute to the transmission in the P case. 
But only the $\su\sd$-channel gives rise to a non-zero transmission near the 
Fermi level because of the electrons near the Fermi energy being completely
spin-up polarized in the left and spin-down polarized in the right electrodes.
The contribution of the $\su\sd$- and $\sd\su$-channels to the total transmission 
grows with increasing domain length as the spin-mixing becomes more and more important.
For the same reason the contribution of the $\su\su$-and $\sd\sd$-channel vanish
with increasing domain length. 
Finally, in the limit of an adiabatic DW only the $\su\sd$- and $\sd\su$-channels
contribute and have the same transmission as the $\su\su$- and $\sd\sd$-channel in the P case
illustrating once again how the ferromagnetic case is recovered locally for an adiabatic 
DW.

\section{Summary}
\label{sec:summary}

In summary, we have calculated the transmission through domain walls in the classical $sd$-model and the Stoner model
taking into account non-collinear magnetization profile. For the classical $sd$-model we have assumed a simple linear
dependence of the magnetization angle with the position inside the domain like in the work by Tatara {\em et al.}
\cite{Tatara:prl:97}, and obtain similar results. Increasing the length of the domain wall its smoothness 
increases giving rise to a general increase of the transmission until for very smooth domain walls the perfect
transmission of the homogenous ferromagnetic chain is recovered. This can be understood by the fact that a non-
collinear magnetization allows electrons to spin-flip while crossing the domain wall, so that electrons can
be transmitted elastically even at energies where there is no overlap between spin-bands. In the limit of a 
perfectly smooth domain wall the spin of an incoming electron is flipped adiabatically while crossing the
domain thus eliminating the backscattering by the magnetic structure.

On the other hand, we also calculated the formation of a domain wall in the Stoner model. We find that also here
the local magnetization vector changes more smoothly, with increasing domain length, although for a finite domain 
length the DW is not simply ``linear'', i.e. the magnetization angle does not change linearly with the position 
inside the domain as was assumed for the classical $sd$-model in \ref{sec:t-J-model} and by Tatara {\em et al.} \cite{Tatara:prl:97}.
Moreover the magnitude of the magnetization vector is not homogeneous for a finite domain length, but is modulated 
along the domain, and thus is different from the magnetization of the ferromagnetic bulk. The deviation from the 
linear DW with respect to the magnetization angles and the modulation of the magnitude of the magnetization becomes
bigger the smaller the domain, thus introducing extra-scattering in comparison to the perfect linear DW.

\chapter{Ni nanocontacts}
\label{ch:Ni-nanocontacts}

The observation of huge magneto-resistance (MR) in ferromagnetic nanocontacts 
\cite{Garcia:prl:99} comparable or even exceeding the celebrated GMR effect \cite{Baibich:prl:88} 
has given rise to a strong interest in these systems over the last years because of its possible
implications in the context of nanoscale spintronics applications\cite{Oshima:apl:98,Garcia:prl:99,
Ono:apl:99,Chung:prl:02,Viret:prb:02,Chopra:prb:02,Hua:prb:03,Untiedt:prb:04,Sullivan:prb:05,
Gabureac:prb:04,Keane:apl:06,Bolotin:nl:06}. The observed huge MR was explained theoretically by 
the following reasoning: It is expected that the magnetizations of the two sections of the wire 
which are connected via the atomic-size contact align antiparallel (AP), so that the overall 
magnetostatic energy is reduced. Then a fairly sharp domain-wall (DW) should form at the atomic-size 
contact, since the cost in exchange energy for the formation of the DW is minimized due to the low 
coordination of the contact atom \cite{Bruno:prl:99}. It has been further argued that the sharp DW 
should give rise to strong ballistic spin scattering resulting in a large extra contribution to the 
ballistic resistance in the AP configuration \cite{Tatara:prl:99,Imamura:prl:00}. Applying an external 
magnetic field the magnetizations of the two wires will align parallelly (P) erasing the DW at the neck, 
and thus eliminating the extra contribution to resistance. The quantity characterizing the effect is the 
magneto-resistance MR which is defined as the difference between the resistances in the AP configuration 
$R_{\rm AP}$ and the P configuration $R_{\rm P}$ normalized to either $R_{\rm AP}$ (MR$_1$) or $R_{\rm P}$ 
(MR$_2$). The definition of MR as ${\rm MR}_1 = (R_{\rm AP}-R_{\rm P})/R_{\rm AP}$ is bounded, i.e. 
MR$_1 \le 100\%$ while the second definition of MR as ${\rm MR}_2 = (R_{\rm AP}-R_{\rm P})/R_{\rm P}$ 
has no upper limit, i.e. MR$_2 < \infty$. In the literature both definitions are used. Following the 
above reasoning high values of MR ---named ``ballistic'' MR (BMR) for its origin in ballistic electron 
scattering--- in ferromagnetic metal nanocontacts could be expected. 

However, BMR in nanocontacts has been a controversial topic since its first observation: While some 
groups have measured huge values of BMR in Ni nanocontacts, comparable to or even exceeding GMR values
(typically MR$_2\approx 225\%$) \cite{Garcia:prl:99,Chung:prl:02,Chopra:prb:02,Hua:prb:03,Sullivan:prb:05},
other groups have obtained moderate BMR values \cite{Viret:prb:02,Gabureac:prb:04}, do not measure any MR 
effect \cite{Untiedt:prb:04}, or even obtain negative BMR values \cite{Oshima:apl:98}. The question of 
huge BMR in ferromagnetic nanocontacts and its origin is a very important one because of its expected
impact on the magneto-electronics industry.

In this chapter electron transport through atomic-size Ni nanocontacts and the effect of the formation
of a domain wall in the atomic-size constriction on the transport-properties is investigated theoretically
using {\it ab initio} methods. An atomic-size nanocontact is a constriction of just one atom in diameter
connecting two sections of a wire. Within the atomic-size constriction the electrons are confined to 
effectively one dimension. 
In order to understand the behavior of the electrons in a 
material at low dimensions it is therefore reasonable to investigate 
the electronic structure of perfect one-dimensional monatomic wires before 
investigating the more complicated situation of a nanocontact
with a realistic geometry.
The investigation of monatomic nanowires of 3d-transition metals with
{\it ab initio} methods has been performed earlier by Smogunov {\em et al.}
\cite{Smogunov:ss:02,Smogunov:ss:04}.
As a precursor to the work on nanocontacts we have repeated the
{\it ab initio} electronic structure calculations of Ni chains, and
fully reproduce their results.
The insight gained by the calculations on perfect one-dimensional 
monatomic wires will help us to analyze and interprete the results
for nanocontacts with realistic geometries.

Ni is a 3d-transition metal that crystallizes in the face-centered
cubic (FCC) crystal structure. 
The bulk material is ferromagnetic with an atomic magnetic moment 
of 0.6$\mu_{\rm B}$ which is considerably less than that of his 
immediate neighbor to the left in the periodic table, 
Co (1.7$\mu_{\rm B}$).
The magnetism of Ni originates from the electrons in the d-bands
which are spin-split. The s-band is not spin-split and thus
the s-electrons do not contribute to the magnetism.

The electronic structure of bulk Ni is quite well described within the 
local spin density approximation (LSDA)\cite{Vosko:cjp:80} of density-functional theory (DFT)
\cite{Doll:ss:03}. 
Therefore we will at first perform {\it ab initio} calculations of Ni wires and 
nanocontacts at the LSDA level (Sec. \ref{sec:Ni-chain} and Sec. 
\ref{sec:Ni-contact}). Then in Sec. \ref{sec:Ni-self-interaction} we 
will compare the LSDA results with results obtained using the hybrid-functional 
B3LYP which corrects for the self-interaction error inherent in LSDA as discussed 
in Ch. \ref{ch:ab-initio}.

\section{Monatomic Ni chain}
\label{sec:Ni-chain}

In order to illuminate the properties of the electronic structure
of Ni at low dimensions we first study the electronic structure of
infinite monatomic chains of Ni using the quantum chemistry program 
CRYSTAL \cite{Crystal:03}, which allows to calculate the electronic 
structure of periodic systems with standard {\it ab initio} methods like 
the Hartree-Fock approximation (HFA) and DFT.
However, no evidence of chain formation between Ni nanocontacts 
have been reported to date in contrast to Au or Pt. This is also
supported by our {\it ab initio} simulations of breaking Ni contacts
presented in the next section.
Nevertheless the results for the ideal chains will help us
to analyze and interprete the more complex results obtained for 
the more realistic contact geometries of the next section.

\begin{figure}
  \begin{minipage}[t][][b]{0.6\linewidth}
    \includegraphics[width=\linewidth]{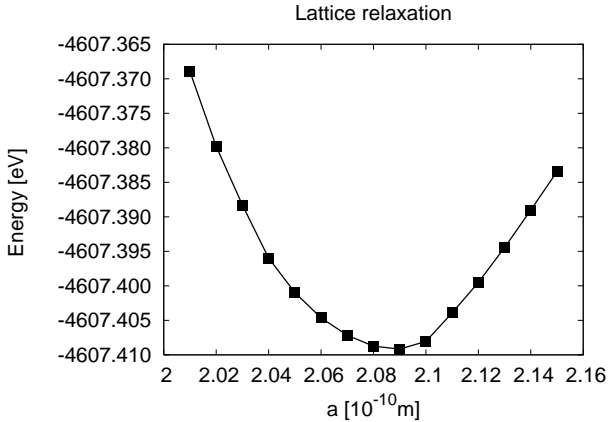}
  \end{minipage}
  \begin{minipage}[t][][b]{0.4\linewidth}  
    \setcapindent{0cm}   
    \caption{Optimization of the lattice spacing $a$ of the
      ferromagnetic monatomic Ni chain calculated with CRYSTAL using LSDA 
      and the CRENBL pseudo-potential + basis set.}
  \end{minipage}
  \label{fig:chain-opt}
\end{figure}

First, we optimize the geometry of the chain, i.e. we search for the
lattice spacing $a$ of the chain which minimizes the energy.
Fig. \ref{fig:chain-opt} shows the energy in dependence
of the lattice spacing $a$ between 2\r{A} and 2.2\r{A} obtained
with the LSDA functional and the CRENBL basis-set with core 
pseudopotential.
The CRENBL \cite{Hurley:jcp:86} effective core pseudopotential (ECP) 
is a small core ECP, describing the 10 innermost electrons only. 
The basis set describes the remaining 18 outer electrons of Ni, 
and is of very high quality.
The optimum lattice constant of 2.09\r{A} is in agreement with that
obtained by Wierzbowska {\em et al.} \cite{Wierzbowska:04}, and is significantly 
higher than the nearest-neighbor distance between bulk atoms of 2.5\r{A}.

\begin{figure}
  \begin{tabular}{cc}
    \includegraphics[width=0.45\linewidth]{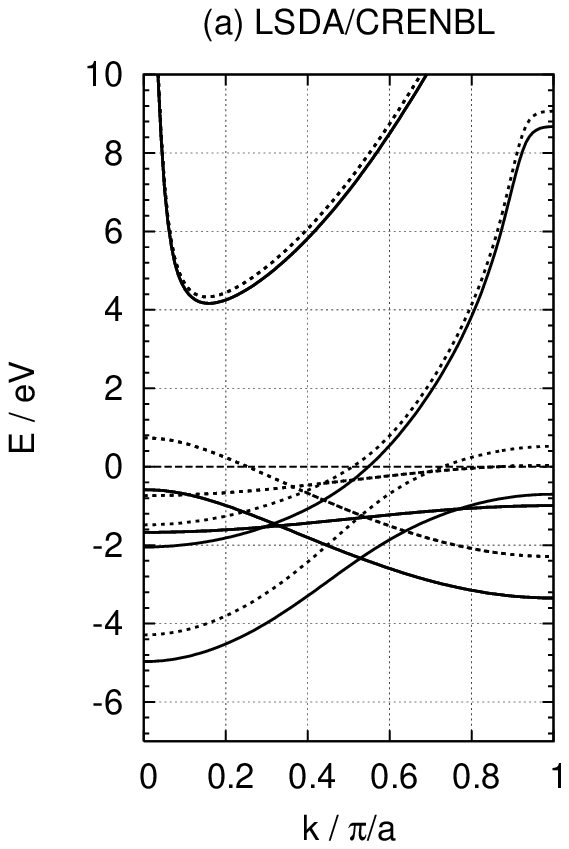} &
    \includegraphics[width=0.45\linewidth]{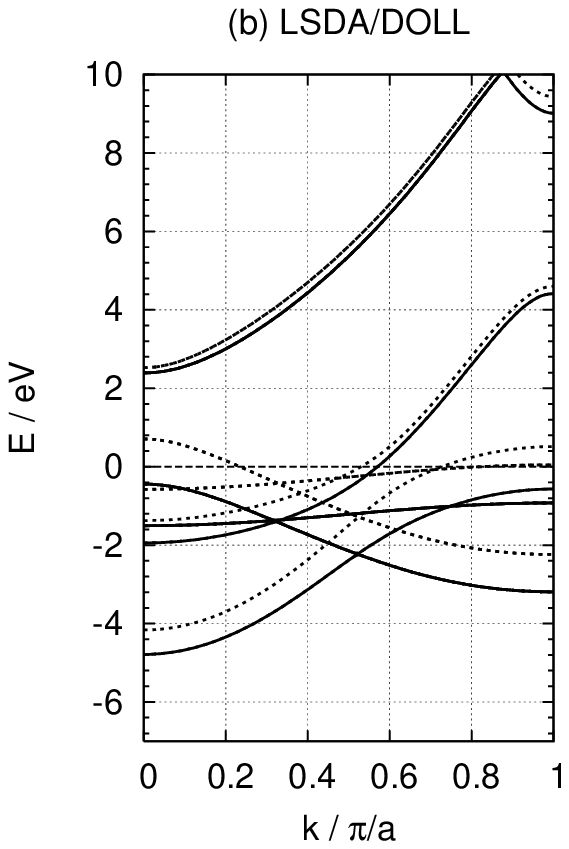}
  \end{tabular}
  \setcapindent{0cm}   
  \caption{Electronic band structure with respect to the Fermi energy 
    (zero line) of monatomic Ni chain at equilibrium 
    lattice spacing $a=2.9$\r{A} calculated with LSDA functional and CRENBL (a) 
    and DOLL (b) basis set, respectively.
    The full lines are majority-spin bands, while the dotted lines are minority-spin.}
  \label{fig:bands}
\end{figure}

Fig. \ref{fig:bands} shows the electronic band-structure of the ferromagnetic (FM) 
Ni chain at equilibrium lattice spacing $a=2.09$\r{A} calculated with the CRENBL ECP basis-set,
and an all-electron basis-set (DOLL) optimized for Ni bulk \cite{Doll:ss:03}.
The results for the two basis-sets are very similar below the Fermi level
and above near the Fermi energy.
Only for energies well above the Fermi level deviate the band structures
appreciably from each other.

The fact that the band structure of the ECP basis-set CRENBL agrees well with the 
all-electron basis-set DOLL below the Fermi energy indicates that the the core 
electrons are very well approximated by the ECP.
The reason for the deviation of the band-structures for the two basis-sets 
at higher energies is that the CRENBL basis-set is far larger than the 
DOLL basis-set and thus provides a better description of the outer
electrons.
Since we are above all interested in transport properties which are
determined by the electrons near the Fermi energy, the deviation of
the ban structures well above the Fermi energy is not important here.

In a perfect chain no elastic scattering occurs so that all available
transport channels at some energy transmit perfectly, i.e. have a transmission
of 1. The number of channels available is given by the number of bands
at some energy, so the zero-bias conductance is given by the number of bands
crossing the Fermi level. 
Thus in the case of the perfect Ni chain there are one majority-spin channel and
six minority-spin channels (two of the bands are doubly-degenerate, see below) 
contributing to the conductance.

The majority-spin and minority-spin bands are split in energy by up to $\approx 1$eV for some
bands, and the magnetic moment per atom of the Ni chain is $\approx$1.1$\mu_{\rm B}$ for 
both basis sets and thus almost twice as high as the magnetic moment 
of Ni bulk which reflects the low coordination of the chain-atoms
compared to the bulk atoms: Due to the lower coordination of the chain atoms,
the bands become narrower than in the bulk so that the cost in kinetic 
energy to pay for the polarization of the bands becomes smaller.
This is partially counteracted by the decrease in nearest-neighbor distance
with respect to the bulk which leads to a broadening of the bands.

\begin{figure}
  \includegraphics[width=0.9\linewidth]{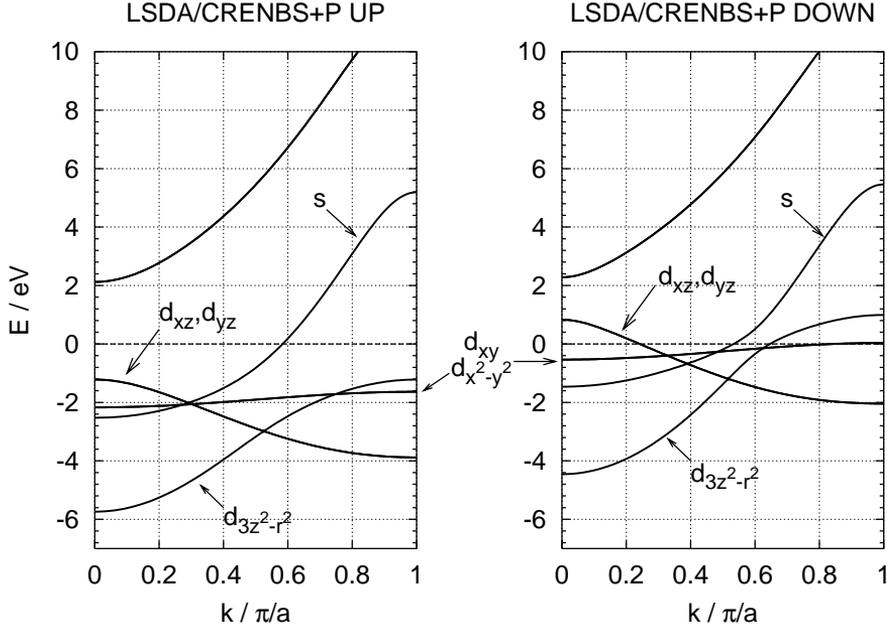}
  \setcapindent{0cm}   
  \caption{Electronic band structure with respect to the Fermi energy 
    (zero line) of monatomic Ni-chain calculated
    with LSDA functional and CRENBS+P basis set. The left panel shows the 
    majority-spin and the right panel the minority-spin bands.}
  \label{fig:band-crenbs}
\end{figure}

%%Analysis: flat bands: $d_{xy}$/$d_{x^2-y^2}$ , broad bands
Next, we analyze the orbital nature of the different bands of
the Ni chain. Fig. \ref{fig:band-crenbs} shows the 
band structure of the Ni-chain separately for the two realizations
of the spin quantum number. In this case the result was obtained
with the CRENBS\cite{Hurley:jcp:86} minimal basis set with large core (the 18 inmost
electrons) ECP. Because of the smaller basis-set compared to CRENBL
or DOLL and the larger core ECP results obtained with this basis
are in general less accurate. However, as Fig. \ref{fig:band-crenbs}
shows, the changes in the band structure compared to the 
results obtained with CRENBL and DOLL are quite moderate, especially
near the Fermi level.
The number of bands crossing the Fermi level remains the same
(1 spin-up band, 6 spin-down bands) and the bands conserve their
overall aspect. 
Only the spin-splitting of some bands has become stronger now (almost 2eV)
their bandwidth changes slightly and the magnetic moment per atom now is $\mu_{\rm B}$.  
On the other hand the minimal basis-set facilitates the analysis of the
band structure a great deal. 

\begin{figure}
  \includegraphics[width=0.9\linewidth]{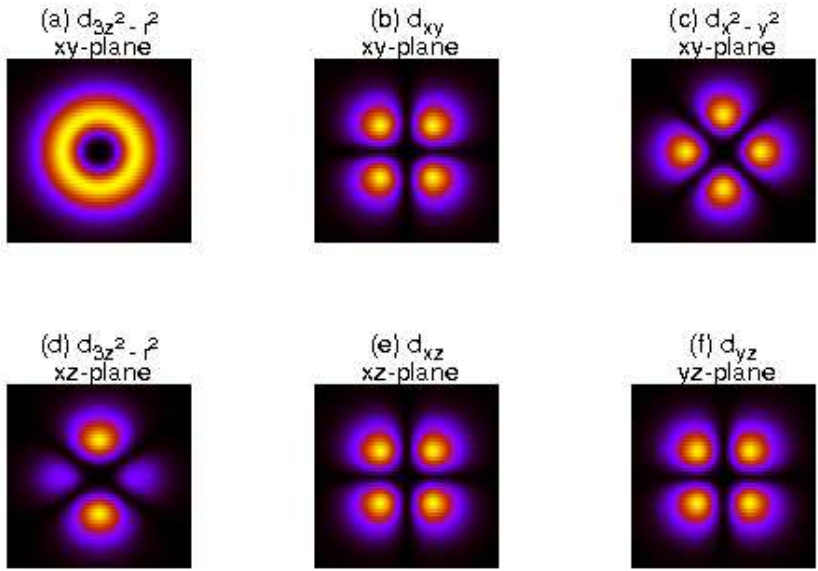}
  \setcapindent{0cm}   
  \caption{Sections of the modulus of the (Cartesian) d-orbitals. 
    The $d_{3z^2-r^2}$-orbital (a,d) has rotational symmetry 
    around the z-axis (vertical symmetry axis in (d)) 
    and is directed along that axis. 
    All other $d$-orbitals (b,c,e,f) are shown in their respective planes
    of extension. In the direction perpendicular to that plane the 
    extension of these orbitals is very small.}
  \label{fig:d-orbitals}
\end{figure}

The orbital nature of the different bands is determined by the
rotational symmetry of the monatomic chain. The eigenstates
of the monatomic chain are therefore simultaneously eigenstates 
of the z-component (for a chain oriented along the z-axis) of the 
angular momentum operator $\hat L_z$.
The $s$-orbital and the $d_{3z^2-r^2}$ orbital both have
angular momentum zero, $l_z=0$, so there are two bands 
with angular momentum zero per spin.
As shown in Fig. \ref{fig:band-crenbs} the spin-bands
with angular momentum zero are the broadest of the $s$ and $d$-bands.
One is mostly $s$-type (thus labeled $s$-band in Fig. \ref{fig:band-crenbs})
 with a small contribution of $d_{3z^2-r^2}$ while the other is mostly $d_{3z^2-r^2}$-type
(labeled $d_{3z^2-r^2}$-band) with a small contribution of $s$.
The $s$-bands are only slightly spin-split while the $d_{3z^2-r^2}$-bands
exhibit quite a big spin-splitting of about 1eV 
(for CRENBL, Fig. \ref{fig:band-crenbs}a).  
The $s$ band is by far the broadest as expected, since $s$-orbitals
give rise to highly delocalized electrons.
The orbitals $d_{xz}$ and $d_{yz}$ can be combined to the rotationally
symmetric orbitals with angular momentum $\pm1$. Thus two degenerate 
bands per spin result from the $d_{xz}$- and $d_{yz}$-orbitals.
These are also spin-split by about 1eV (for CRENBL, Fig. \ref{fig:band-crenbs}a)
and are a bit narrower than the $d_{3z^2-r^2}$-bands, since the extension of 
the $d_{xz}$- and $d_{yz}$-orbitals along the z-direction is 
smaller than that of the $d_{3z^2-r^2}$-orbital (see graphical representation
of $d$-orbitals in Fig. \ref{fig:d-orbitals}) giving
rise to a stronger localization of the electrons in these bands.
Finally, the two degenerate bands per spin with angular momentum $\pm2$ 
are composed of linear combinations of the $d_{xy}$- and $d_{x^2-y^2}$-orbitals 
which are directed perpendicular to the $z$-axis. 
This leads to a very small overlap of the orbitals in the z-direction
and consequently to very narrow bands giving rise to strong electron localization
in these bands. As the other $d$-bands they are also spin-split by about 1eV.

\begin{figure}
  \includegraphics[width=0.98\linewidth]{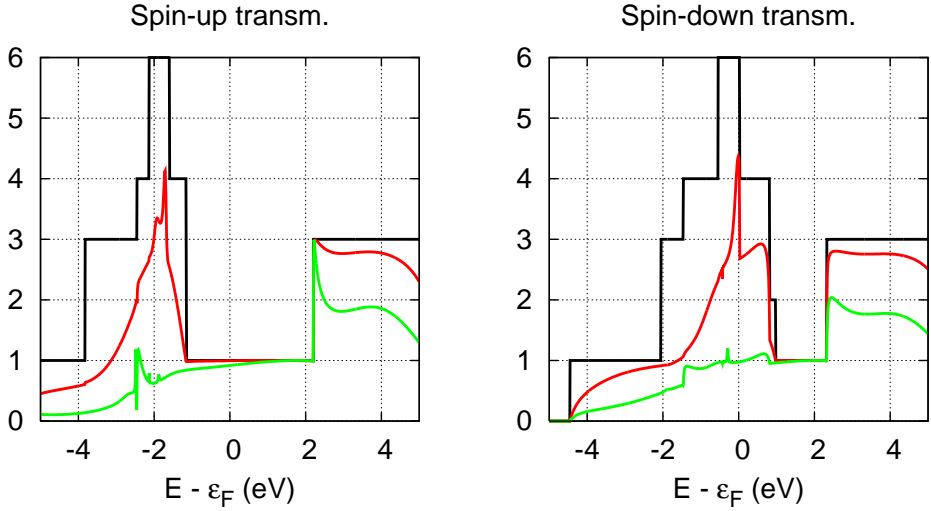}
  \setcapindent{0cm}   
  \caption{
    Effect of electron scattering in the one-dimensional Ni chain. 
    The length of the gap $d$ between the two semi-infinite Ni chains
    is larger than the lattice spacing of $a\approx 2.1$\r{A} of the
    chains and thus gives rise to scattering. The black curves show the 
    spin-resolved transmissions in the case of the perfect chain (no scattering),
    the red curves in the case of $d=1.2a=2.5$\r{A}, and the green curves in the
    case of $d=1.5a=3.1$\r{A}.
  }
  \label{fig:chain-transm}
\end{figure}

Finally, we study the effect of electron scattering on the conductance of the
Ni chain. Therefore we calculate the transmission of two semi-infinite atomic Ni 
chains separated by a gap which is larger than the lattice spacing of $a\approx 2.1$\r{A} 
of the two semi-infinite chains, and compare to the transmission of the perfect chain,
Fig. \ref{fig:chain-transm}. We observe that the transmission of the spin-down channel 
near the Fermi level decreases strongly as we increase the length of the gap $d$. The 
spin-up transmission on the other hand is relatively stable near the Fermi level, so that
the spin-polarization decreases with increasing length of the gap. This can be understood
by the fact that the $s$-electrons are much less susceptible to scattering than the 
$d$-electrons. Therefore the spin-down channel which is composed of $s$- and $d$-electrons
is much stronger affected by the scattering than the spin-up channel which consists solely
of a single $s$-type channel.

In summary, the d-bands are all spin-split by about 1eV while the
$s$-band is only slightly spin-split due to its hybridization with
the $d_{3z^2-r^2}$-band. The spin-splitting of the $d$-orbitals
gives rise to the filling of the spin-up $d$-bands while the 
spin-down $d$-bands are only partially filled. Thus only the 
spin-down $d$-bands and both (spin-degenerate) $s$-bands contribute
to the conduction.
Because of the spin-up $d$-bands being completely filled and the 
spin-down $d$-bands being partially filled the $d$-bands 
behave like a half-metal and can be modeled by the Stoner model
(see Ch. \ref{ch:spin-transport}). 
Consequently a sharp domain wall in the monatomic Ni chain would
lead to a total blocking of the 5 $d$-channels, while the $s$-channels
which behave like a paramagnetic metal are not affected by the domain-wall 
at all. Thus the total transmission of the one-dimensional chain in the AP 
configuration will be 2, while the transmission in the P configuration
is 7. Therefore the MR is given by $(7-2)/7=5/7=$71\% in def. MR$_1$ and
$5/2=$250\% in def. MR$_2$.

These results for the perfect Ni chain have been obtained earlier by Smogunov {\em et al.}
\cite{Smogunov:ss:02,Smogunov:ss:04} by a full {\it ab initio} calculation of the
magnetization reversal in the chain. 
This result could explain some early experiments \cite{Garcia:prl:99}, but not recent 
ones \cite{Sullivan:prb:05}. Furthermore, to date, no evidence of chain formation in Ni 
has been reported. 
Even so, scattering at the electrode-chain contact will always be present, and lead
to a reduction in the transmissions of the $d$-channels which are very susceptible to
scattering.

%%d-orbitals: spin-splitted.
%%s-orbitals: not spin-splitted.
%%Domain-wall: Magnetic blocking of d-orbitals, MR$\approx 250\%$ (Tossati).

%%\section{Monatomic Ni chain with scattering center}

\section{More realistic contact geometries}
\label{sec:Ni-contact}

Now that we have analyzed the perfect one-dimensional case we are prepared to study Ni 
nanocontacts with more realistic contact geometries. The fact that the electronic structure 
of the one-dimensional chain near the Fermi level is quite well described employing the 
CRENBS minimal basis-set with large-core ECP justifies that we restrict our calculations 
here to that basis.\footnote{We actually performed also calculations with the DOLL and CRENBL 
basis-set which confirmed that the results do not change essentially.} The bulk electrodes will 
be described by a Bethe lattice (BL) model as described in Ch. \ref{ch:ab-initio} which provides 
a geometry independent description of the electrodes with a Bulk DOS which is smoother than the 
DOS of the perfect bulk crystal and mimics an average over both disorder realizations and the 
actual electrode crystal orientations

\begin{figure}
  \begin{minipage}[t][][b]{0.5\linewidth}
    \begin{flushleft}
      \includegraphics[width=0.8\linewidth]{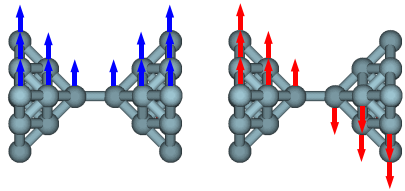}
    \end{flushleft}
  \end{minipage}
  \begin{minipage}[t][][b]{0.5\linewidth}
    \setcapindent{0cm}   
    \caption{Left: Ni nanocontact with parallel (P) magnetization of both electrodes.
      Right: Ni nanocontact with antiparallel (AP) magnetization of both electrodes.
    }
    \label{fig:ni-nanocontact}
  \end{minipage}
\end{figure}

A reference atomic structure of the contact region has been 
initially  taken like that shown in Fig. \ref{fig:ni-nanocontact}. 
Following Viret {\em et al.} \cite{Viret:prb:02}, we consider the  narrowest 
region to consist of two pyramids facing each other,
formed along the (001) direction, and with the
two tip Ni atoms 2.6 \AA \, apart forming a dimer. Bulk atomic distances and
perfect crystalline order are assumed otherwise. {\em Ab initio} simulations of the breaking
process as the one shown in Fig. \ref{fig:geom-stretch} 
support this choice. We stress that the section of the nanocontacts varies 
in the direction of the current flow. 
This is the situation in real nanocontacts and differs from perfect 1-dimensional
systems, studied in Refs. \cite{Smogunov:ss:02,Smogunov:ss:04}, and from bulk systems
studied by Van Hoof {\em  et al.} \cite{vanHoof:prb:99}. In this regard, the geometries 
proposed by Bagrets {\em et al.} \cite{Bagrets:prb:04} are closer to real nanocontacts, 
but are not backed up by experiments or simulations.  

\begin{figure}
  \begin{center}
    \includegraphics[width=0.9\linewidth]{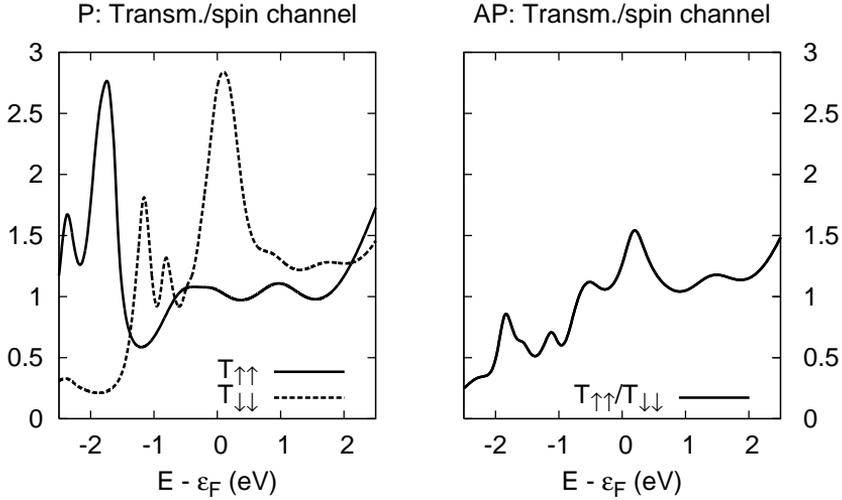}
  \end{center}
  \setcapindent{0cm}   
  \caption{Left: Transmission per spin channel of Ni nanocontact for parallel (P) 
    magnetization of both electrodes as shown on left hand side of Fig. 
    \ref{fig:ni-nanocontact}. Right: Same as before but for the Ni nanocontact with
    antiparallel (AP) magnetization of the two electrodes as shown on right hand side
    of Fig. \ref{fig:ni-nanocontact}.
  }
  \label{fig:lsda-transm}
\end{figure}

Figure (\ref{fig:lsda-transm}) shows the LSDA conductance as a function of energy for both up
and down spin channels in two situations: (a) Parallel (P) and (b) antiparallel (AP) bulk magnetic
arrangements.  In both cases the self-consistent solution has been forced to
respect the high symmetry of the  nanocontact. In the AP case the
self-consistent magnetization reverses abruptly between tip atoms. The resulting
magnetic moment for the contact atoms is $\approx 1.0 \mu_{\rm B}$ in both  
situations. This value is significantly larger than that obtained for bulk or surface atoms
($\approx 0.6 \mu_{\rm B}$ and reflects the low coordination of the tip atoms forming the contact. 

The eigenchannel analysis restricted to the contact atoms as explained in the next section 
reveals that in the  P case the majority channel is, for the most part, 
composed of a single $sp$ orbital channel and conducts perfectly around the  Fermi energy 
(set to zero) while the minority channel is composed of three orbital channels 
(one $s$- and two $d$-like, which conduct roughly the same), 
and exhibits a transmission strongly dependent on the scattering energy. 
In the AP case the system is invariant under the combined transformations that exchange
 L with R and $\uparrow$ with $\downarrow$, resulting in identical  values for the conductance of the two spin
channels, which now are composed of a dominant $s$ channel and a strongly diminished contribution 
of the $d$ channels.
The conductance ratio for this particular case is $x=2.8/3.65=0.77$. This 
yields MR$_1=23\%$ and MR$_2=30\%$, which is clearly below large MR claims\cite{Garcia:prl:99,Sullivan:prb:05}.

\begin{figure}
  \begin{center}
    \includegraphics[width=\linewidth]{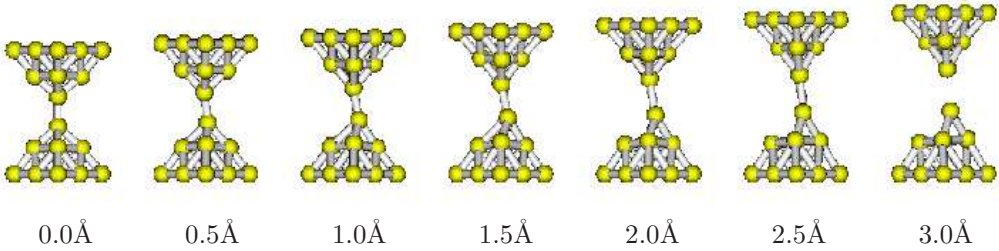} 
    \begin{tabular*}{0.9\linewidth}
      { @{\extracolsep\fill}
        c@{\extracolsep\fill}
        c@{\extracolsep\fill}
        c@{\extracolsep\fill}
        c@{\extracolsep\fill}
        c@{\extracolsep\fill}
        c@{\extracolsep\fill}
        c@{\extracolsep\fill}}
      0.0\r{A} & 0.5\r{A} & 1.0\r{A} & 1.5\r{A} & 2.0\r{A} & 2.5\r{A} & 3.0\r{A}
    \end{tabular*}
  \end{center}
  \setcapindent{0cm}   
  \caption{
    Ab-initio simulation of stretching of atomic-size Ni contact. The outer planes (9+9=18 atoms)
    are displaced in steps of 0.5\r{A} starting from a slightly compressed contact as compared 
    to the ideal contact geometry in Fig. \ref{fig:lsda-transm}. The inner atoms of the nanocontact 
    (10 in total) are allowed to relax to local minimum energy in each step of the stretching while 
    the outer planes are kept fixed during the relaxation.
  }
  \label{fig:geom-stretch}
\end{figure}

Since the minority channel conductance evaluated 
at the LSDA level exhibits a strong dependence on the scattering energy,
we study now whether or not different
geometries  can change the  above results qualitatively. In an attempt to  explore
other realizations of the self-consistent potential compatible with the
magnetic boundary conditions and the experimental information, we  perform 
{\em ab initio} structural relaxations as a function of the displacement between
outer planes in the core cluster. To do so,
 we consider a cluster like that shown in Fig. \ref{fig:lsda-transm}.   The inner atoms in
the cluster (10 in total) are allowed to relax to local minimum energy
configurations as we stretch. This results, logically, in lower energy solutions and in the
loss of symmetry, so that the transmission in the AP case now becomes 
slightly spin-dependent.

\begin{figure}
  \begin{center}
    \includegraphics[width=\linewidth]{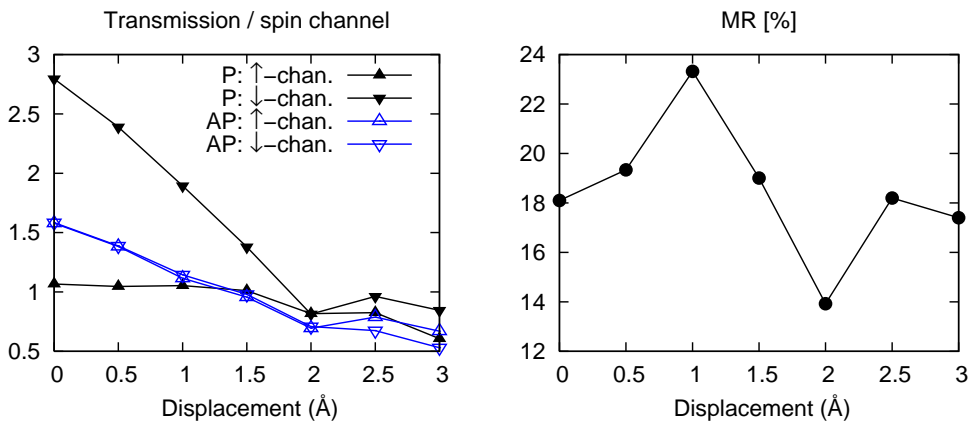}
  \end{center}
  \setcapindent{0cm}   
  \caption{LSDA transmission per spin channel at the Fermi energy for P (black) 
    and AP (blue) configuration in dependence of the stretching of the contact 
    (Fig. \ref{fig:geom-stretch}).}
  \label{fig:results-stretch}
\end{figure}

%\begin{figure}
%  \begin{center}
%    \includegraphics{figures/ni-nanocontacts/stretch.lsda.transm.eps}
%  \end{center}
%  \caption{LSDA total transmission at the Fermi energy for P (black) and AP (blue) configuration
%    in dependence of the stretching of the contact (Fig. \ref{fig:stretch}). The transmission 
%    for the P case is always higher than for the AP case leading throughout to a positive MR, see also Fig. 
%    \ref{fig:mr-stretch}.}
%  \label{fig:transm-stretch}
%\end{figure}

In Fig. \ref{fig:results-stretch} the transmission at the Fermi energy per spin
channel for the P and the AP configurations are shown  as a function of
the stretching up to the break-up point, starting  from a slightly compressed
nanocontact. From this figure we see that the conductance of the minority channel 
for the P configuration changes significantly upon small changes while the 
transmission of the majority channel is quite stable against geometrical
changes. This behavior reflects the fact that the majority channel is mostly
$s$-type, and therefore is quite insensitive to geometrical changes. 
On the other hand the minority channel has a strong contribution from the
$d$-orbitals which are very sensitive to geometrical changes due to their high
directionality (see Fig. \ref{fig:d-orbitals}).
In the AP configuration the transmission of the two spin-channels
shows an intermediate behavior between the majority and the minority channel of the P configuration
upon stretching of the contact.
This can be understood by considering that the two spin-channels are now composed
of $s$- and $d$-type orbitals, where the contribution from the $d$-orbitals
is weaker than in the minority channel of the P case, but not negligible as for
the majority channel in the P configuration.
Consequently, the MR, shown in Fig. \ref{fig:results-stretch}, is small and barely changes 
as the nanocontact is stretched, oscillating between 14\% and 23\%.

% \begin{figure}
%   \begin{minipage}[t][][b]{0.64\linewidth}
%     \includegraphics[width=\linewidth]{figures/ni-nanocontacts/stretch.lsda.mr.eps}
%   \end{minipage}
%   \begin{minipage}[t][][b]{0.35\linewidth}
%     \setcapindent{0cm}   
%     \caption{LSDA magneto-resistance (MR) in dependence of the stretching of the contact
%       where def. MR$_1=(T_{P}-T_{AP })/T_{P}$ is used. }
%     \label{fig:mr-stretch}
%   \end{minipage}
% \end{figure}

%%reaching vanishing 
%%values for the last points in Fig. \ref{fig:chan-stretch}. 
%%The conductance approaches a stable value around $2e^2/h$ for both P and AP configurations.

\section{Orbital eigenchannel analysis}
\label{sec:Ni-eigenchannels}

It is often useful to decompose the total conductance into the contributions 
of the transport eigenchannels, first introduced by B\"uttiker\cite{Buettiker:ibmjrd:88}.
These are defined as the linear combinations of the incoming 
modes in a lead that do not mix upon reflection on the scattering region
and present a unique transmission value\cite{Brandbyge:prb:97}. 
The decomposition of the measurable total transmission
in terms of the transmissions of these
eigenchannels simplifies considerably the interpretation of the
results. Knowledge and analysis of the eigenchannel wavefunctions would, in turn,
allow one to make predictions regarding the behavior of the 
conductance upon distortions of the geometry or other perturbations of 
the scattering region\cite{Jacob:prb:05}.
Unfortunately, in the NEGF approach it is not straightforward to extract the
orbital composition of the transport eigenchannels.
Only the eigenchannel transmissions can be obtained easily in the NEGF approach 
from the non-negligible eigenvalues of the transmission matrix.
However, the associated eigenvectors turn out to be useless as obtained.
The reason is that these eigenvectors contain the contributions to the 
eigenchannel wavefunctions of the atomic orbitals at one of the borders
of the scattering region immediately connected to the leads, but not inside.

In this chapter we present a method for analyzing the orbital contributions to the transport eigenchannels
at an arbitrary cross-section of a nanoscopic conductor by calculating the transmission matrix 
projected onto that cross-section. Our approach generalizes previous work by Cuevas {\em et al.}
\cite{Cuevas:prl:98:80} for tight-binding-type Hamiltonians to non-orthogonal atomic orbitals basis 
sets as those commonly used in quantum chemistry packages. An alternative approach to investigate the 
contributions of certain atomic or molecular orbitals to the conductance consists in directly removing 
the respective orbitals from the basis set \cite{Thygesen:prl:05:94:3}.

\begin{figure}
  \begin{minipage}[t][][b]{0.5\linewidth}
    \begin{center}
      \includegraphics[width=0.75\linewidth]{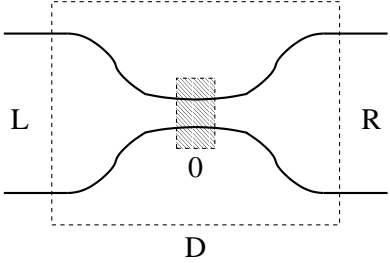}
    \end{center}
  \end{minipage}
  \begin{minipage}[t][][b]{0.5\linewidth}
    \setcapindent{0cm}   
    \caption{Sketch of the scattering problem. 
      L: Left lead. D: device. R: right Lead.
      0: cross-section of interest.}
    \label{fig:constriction}
  \end{minipage}
\end{figure}

For the sake of clarity we will repeat here some of the important formulas of the NEGF presented
in Ch. \ref{ch:transport}. Fig. \ref{fig:constriction}, shows a sketch of a nano-constriction 
connecting two bulk leads. We assume that the leads are coupled only to the constriction but not 
to each other. The Hamiltonian describing this situation is then given by the matrix
\begin{equation}
  {\bf H} = \left( 
    \begin{array}{ccc}
      {\bf H}_{\rm L}  & {\bf H}_{\rm LD} & {\bf 0}      \\
      {\bf H}_{\rm DL} & {\bf H}_{\rm D}  & {\bf H}_{\rm DR} \\
      {\bf 0}          & {\bf H}_{\rm RD} & {\bf H}_{\rm R} 
    \end{array}
  \right).
\end{equation}
Since many density functional theory (DFT) codes work in non-orthogonal basis sets, we also allow 
explicitly for overlap between atomic-orbitals given by the following overlap matrix:
\begin{equation}
  {\bf S} = \left( 
    \begin{array}{ccc}
      {\bf S}_{\rm L} & {\bf S}_{\rm LD} & {\bf 0}      \\
      {\bf S}_{\rm DL}& {\bf S}_{\rm D}  & {\bf S}_{\rm DR} \\
      {\bf 0}         & {\bf S}_{\rm RD} & {\bf S}_{\rm R} 
    \end{array}
  \right).
\end{equation}

The standard approach to calculate the conductance is to calculate the 
self-energies of the leads from the Green's functions (GF) of the isolated leads,
i.e., for the left lead ${\bf\Sigma}_{\rm L}(E) = ({\bf H}_{\rm DL}-E {\bf S}_{\rm DL}) {\bf g}_{\rm L}(E) ({\bf H}_{\rm LD}-E {\bf S}_{\rm LD})$ 
where ${\bf g}_{\rm L}(E) = (E {\bf S}_{\rm L} - {\bf H}_{\rm L})^{-1}$
is the GF of the isolated left lead and analogously for the right lead.
From this we can calculate the GF of the device:
\begin{equation}
  {\bf G}_{\rm D}(E) = (E {\bf S}_{\rm D} - {\bf H}_{\rm D} - {\bf \Sigma}_{\rm L}(E) - {\bf \Sigma}_{\rm R}(E))^{-1},
\end{equation}
which, in turn, allows us to calculate the (hermitian) transmission matrix 
\begin{equation}
  \label{eq:TCarolis}
  {\bf T}(E) = {\bf \Gamma}_{\rm L}(E)^{1/2} {\bf G}_{\rm D}^\dagger(E) {\bf \Gamma}_{\rm R}(E) {\bf G}_{\rm D}(E)  {\bf \Gamma}_{\rm L}(E)^{1/2},
\end{equation}
where ${\bf\Gamma}_{\rm L}=i({\bf\Sigma}_{\rm L}-{\bf\Sigma}_{\rm L}^\dagger)$ and 
${\bf\Gamma}_{\rm R}=i({\bf\Sigma}_{\rm R}-{\bf\Sigma}_{\rm R}^\dagger)$.
Typically the leads are only connected to the left and right borders of 
the device and are sufficiently far away from the scattering region so that they
can be described by a bulk electronic structure.
From the structure of eq. (\ref{eq:TCarolis}) it follows that only 
the sub-matrix of ${\bf T}$ representing the subspace of the device 
immediately connected to one of the leads are non-zero. 
Thus the eigenvectors obtained by diagonalizing ${\bf T}$ only contain 
the atomic orbital contributions to the eigenchannels at one border of 
the device region but not at the center where the resistance is ultimately 
determined.

%%Diagonalization of the ${\bf T}$ yields the eigenchannel transmissions. 
%%Typically the leads are only connected to the left and right borders of the device
%%and are sufficiently far away from the scattering region so that they
%%can be described by a bulk electronic structure. The eigenvectors obtained by 
%%diagonalizing ${\bf T}$ thus only contain the atomic orbital contributions to 
%%the eigenchannels at the left and right borders of the device region but not 
%%at the center where the resistance is ultimately determined.

To investigate the orbital nature of the eigenchannels at an arbitrary part of the device
we can simply calculate the transmission matrix associated to this part.
By choosing this region to be a cross-section, like that indicated in Fig. \ref{fig:constriction},
current conservation guarantees that the so-calculated conductance is \emph{approximately} equal
to the conductance calculated from the transmission matrix of the whole device.
We want to emphasize here that this is really only approximately true for a Hamiltonian beyond 
the tight-binding approximation since hoppings between atoms on both sides beyond the selected 
region are neglected. Of course this approximation becomes better the thicker the chosen cross-section is.
We proceed by further subdividing the device region. The cross-section of interest will be referred to 
as 0 while the regions on either side will be denoted as l and r, respectively:
\begin{eqnarray}
  \label{eq:HD-SD}
  {\bf H}_{\rm D} = \left(
    \begin{array}{ccc}
      {\bf h}_{\rm l}  & {\bf h}_{\rm l0} & {\bf h}_{\rm lr} \\
      {\bf h}_{\rm 0l} & {\bf h}_{\rm 0}  & {\bf h}_{\rm 0r} \\
      {\bf h}_{\rm rl} & {\bf h}_{\rm r0} & {\bf h}_{\rm r} 
    \end{array}
  \right) \hspace{0.1\linewidth}
  {\bf S}_{\rm D} = \left(
    \begin{array}{ccc}
      {\bf s}_{\rm l } & {\bf s}_{\rm l0} & {\bf s}_{\rm lr} \\
      {\bf s}_{\rm 0l} & {\bf s}_{\rm 0}  & {\bf s}_{\rm 0r} \\
      {\bf s}_{\rm rl} & {\bf s}_{\rm r0} & {\bf s}_{\rm r} 
    \end{array}.
  \right)
\end{eqnarray}

As mentioned above we will neglect the hoppings (and overlaps) between the 
left and right layers outside the region of interest so we set 
${\bf h}_{\rm lr} = {\bf h}_{\rm rl} = {\bf s}_{\rm lr} = {\bf s}_{\rm rl} = 0$.
With this approximation the GF matrix of the cross-section $0$
can be written as
\begin{equation}
  \label{eq:G0}
  {\bf G}_{\rm 0}(E) =( E {\bf s}_{\rm 0} - {\bf h}_{\rm 0} - {\bf\Sigma}^\prime_{\rm l}(E) - {\bf\Sigma}^\prime_{\rm r}(E) )^{-1}.
\end{equation}
The self-energy matrices representing the coupling to the left and right lead, 
${\bf\Sigma}^\prime_{\rm l}(E) = ({\bf h}_{\rm 0l}-E {\bf s}_{\rm 0l}) {\bf g}_{\rm l}(E) ({\bf h}_{\rm l0}-E {\bf s}_{\rm l0})$ and 
${\bf\Sigma}^\prime_{\bf r}(E) = ({\bf h}_{\rm 0r}-E {\bf s}_{\rm 0r}) {\bf g}_{\rm r}(E) ({\bf h}_{\rm r0}-E {\bf s}_{\rm r0})$, 
are given by the GF of the left layer $\rm l$ connected only to the left 
lead $\rm L$ and the right layer $\rm r$ connected only to the right lead $\rm R$, respectively:
\begin{eqnarray}
  {\bf g}_{\rm l}(E) &=& ( E {\bf s}_{\rm l} - {\bf h}_{\rm l} - {\bf \Sigma}_{\rm L}(E) )^{-1} \\
  {\bf g}_{\rm r}(E) &=& ( E {\bf s}_{\rm r} - {\bf h}_{\rm r} - {\bf \Sigma}_{\rm R}(E) )^{-1}. 
\end{eqnarray}
The \emph{reduced transmission matrix} (RTM) with respect to the chosen cross-section is now given by
\begin{equation}
  {\bf T}^\prime(E) = {\bf\Gamma}_{\rm l}^\prime(E)^{1/2} {\bf G}_0^\dagger(E) {\bf\Gamma}_{\rm r}^\prime(E) {\bf G}_0(E) {\bf\Gamma}_{\rm l}^\prime(E)^{1/2}
\end{equation}
with ${\bf\Gamma}_{\rm l}^\prime=i({\bf\Sigma}_{\rm l}^\prime-{{\bf\Sigma}_{\rm l}^\prime}^\dagger)$ and ${\bf\Gamma}_{\rm r}^\prime=i({\bf\Sigma}_{\rm r}^\prime-{{\bf\Sigma}_{\rm r}^\prime}^\dagger)$.
Diagonalizing ${\bf T}^\prime(E)$ now yields the contribution of the atomic orbitals within the cross-section $0$
to the eigenchannels. 
%%As we will show in the next section the approximation made in the method is reasonable when the chosen 
%%cross-section located in the bottle-neck of the system, e.g the contact atom of an atomic point-contact.
%%Nevertheless the method should also work well for a cross-section which is not in the bottle-neck
%%but some wider part as long as we choose the length of the cross-section sufficiently big, so that
%%hopping between atoms in the left and right outside layer become negligible.

%%
\begin{figure}
  \begin{minipage}[t][][b]{0.4\linewidth}
    \includegraphics[width=\linewidth]{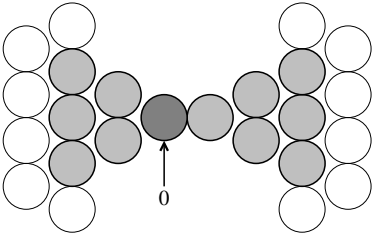}
  \end{minipage}
  \begin{minipage}[t][][b]{0.6\linewidth}
    \setcapindent{0cm}   
    \caption{Sketch of Ni nanocontact consisting
      of two pyramids facing each other along the (001) 
      direction with the two tip atoms forming a dimer bridge.
      The device region (grey circles) consists of 28 Ni atoms
      and the left and right electrodes (empty circles) are modeled by BLs
      with appropriate tight-binding parameters to reproduce Ni Bulk DOS.}
    \label{fig:Ni28}
  \end{minipage}
\end{figure}
%%

%%\subsection{Results and Discussion}
In the following we apply the above described method to analyze the orbital nature of the conducting 
channels of Ni nanocontacts which have recently attracted a lot of interest because of their apparently 
high magneto-resistive properties \cite{Garcia:prl:99,Viret:prb:02}. We consider the nanocontact to 
consist of two ideal pyramids facing each other along the (001) direction and with the two tip atoms 
being 2.6 \r A apart. Bulk atomic distances (2.49 \r A) and perfect crystalline order are assumed for 
each pyramid. Just as in our previous work on Ni nanocontacts\cite{Jacob:prb:05} we perform 
\textit{ab initio} quantum transport calculations for this idealized geometry. To this end we use our 
code ALACANT (ALicante Ab initio Computation Applied to Nano Transport). The electronic structure is 
computed on the LSDA level of DFT with a minimal basis set and the electrodes are described by means of 
a semi-empirical tight-binding BL model.

\begin{figure}
  \begin{center}
    \includegraphics[width=0.9\linewidth]{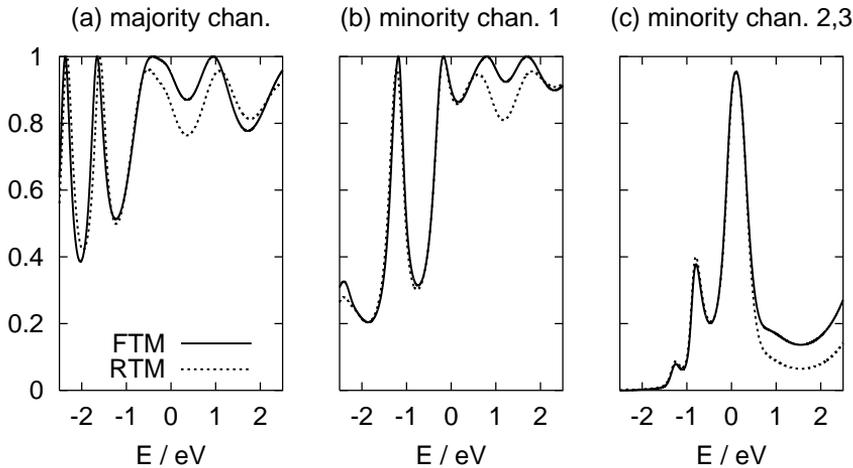}
  \end{center}
  \setcapindent{0cm}   
  \caption{  
    Transmission functions of open transport channels for the Ni nanocontact 
    sketched in Fig. \ref{fig:Ni28} as calculated from the FTM
    ${\bf T}(E)$ (solid line) and from the RTM ${\bf T}^\prime(E)$ (dashed lines).
    (a) shows the only contributing M channel and (b)-(c) the three m
    channels. See text for further discussion.
  }
  \label{fig:Ni28-channels}
\end{figure}
\begin{table}
  \begin{center}
    \begin{tabular}{l|c|c|c|c}
      AO             & majority & minority 1 & minority 2 & minority 3 \\
      \hline 
      $s$            & 97\% & 62\%  &  0   &  0   \\
      $p_x$          &  0   & 0     & 28\% &  0   \\
      $p_y$          &  0   & 0     &  0   & 28\% \\
      $p_z$          &  3\% & 23\%  &  0   &  0   \\
      $d_{3z^2-r^2}$ &  0   & 15\%  &  0   &  0   \\
      $d_{xz}$       &  0   & 0     & 72\% &  0   \\
      $d_{yz}$       &  0   & 0     &  0   & 72\% \\
      $d_{x^2-y^2}$  &  0   & 0     &  0   &  0   \\
      $d_{xy}$       &  0   & 0     &  0   &  0   \\
    \end{tabular}
  \end{center}
  \setcapindent{0cm}   
  \caption{Eigenvectors of the RTM at the Fermi level
    for the contact sketched in Fig. \ref{fig:Ni28}.
    Each column gives the weights of the atomic orbitals 
    (AO) given in the left column on the tip atom in each 
    eigenchannel shown in Fig. \ref{fig:Ni28-channels}.
  }
  \label{tab:1}
\end{table}

As indicated in Fig. \ref{fig:Ni28} we calculate the RTM ${\bf T}^\prime(E)$ 
for one of the tip atoms of the contact (labeled with 0) and diagonalize it
to obtain the eigenchannels and the corresponding transmissions projected
on the tip atom.
In Fig. \ref{fig:Ni28-channels} we compare the individual channel transmissions 
calculated on the one hand from the full transmission matrix (FTM) ${\bf T}(E)$ 
and on the other hand from the RTM  ${\bf T}^\prime(E)$.
Though the electron hopping between regions l and r of the contact has been neglected
in calculating the RTM the so calculated channel transmissions approximate very well 
those calculated using the FTM so that it is very easy to relate the RTM channel 
transmissions with the FTM channel transmission.
This shows that the hopping between the regions l and r on both sides of the tip atom
is almost negligible. Only for the one majority (M) channel we see a small deviation
near the Fermi energy indicating that here 2nd neighbor hopping contributes 
to the transmission of that channel. 
As the eigenvectors of the RTM (see Table \ref{tab:1}) reveal, this channel is mainly s-type.
Since s-electrons are strongly delocalized there is a small but finite contribution 
from second-neighbor hopping explaining the deviation between the FTM and RTM transmission
in that channel.
The first minority (m) channel is also mainly s-type but now it is hybridized with 
$d_{3z^2-r^2}$ and $p_z$ orbitals.
The other two m channels are degenerate and mainly $d_{xz}$- and
$d_{yz}$-type strongly hybridized with $p_x$- and $p_y$-orbitals, respectively. 

%% Discussion
As discussed in our previous work\cite{Jacob:prb:05} the five 
d-type transport channels for the m electrons available in the perfect Ni chain\cite{Smogunov:ss:02}
are easily blocked in a contact with a realistic geometry like that in Fig. \ref{fig:Ni28}
because the d-orbitals are very sensitive to geometry. We have referred to this
as \emph{orbital blocking}.
It is not so surprising that the $d_{x^2-y^2}$- and $d_{xy}$-channel 
which are very flat bands just touching the Fermi level in the perfect chain 
are easily blocked in a realistic contact geometry. These bands
represent strongly localized electrons which are easily scattered in geometries with low symmetry.
Interestingly, even the $d_{3z^2-r^2}$-channel, which for the perfect chain is a very broad band 
crossing the Fermi level at half band width, does not contribute to the conduction as our eigenchannel analysis shows.
This channel is blocked because the $d_{3z^2-r^2}$-orbital lying along
the symmetry axis of the contact is not ``compatible'' with the geometry 
of the two pyramids.
On the other hand the $d_{xz}$- and $d_{yz}$-channels are both open 
in that geometry because their shape is compatible with the pyramid geometry of the
contacts.
This illustrates how the geometry of a contact can effectively block (or open) channels 
composed of very directional orbitals. Of course, for different geometries we can expect 
different channels to be blocked or opened.

\begin{figure}
  \begin{minipage}[t][][b]{0.4\linewidth}
    \includegraphics[width=\linewidth]{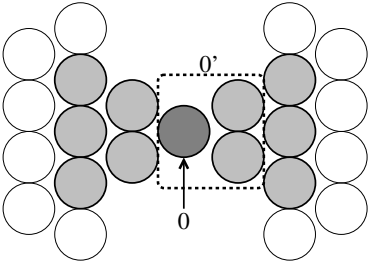}
  \end{minipage}
  \begin{minipage}[t][][b]{0.6\linewidth}
    \setcapindent{0cm}   
    \caption{Sketch of Ni nanocontact (27 atoms). As in Fig. \ref{fig:Ni28}
      the contact consists of two pyramids along the (001) direction but 
      now both pyramids share the same atom at the tip.
      The device region (grey circles) consists of 27 Ni atoms
      and the left and right electrodes (empty circles) are modeled by BLs
      with appropriate tight-binding parameters to reproduce Ni bulk DOS.
    }
  \end{minipage}
  \label{fig:Ni27}
\end{figure}
\begin{figure}
  \begin{tabular}{c}
    \includegraphics[width=0.9\linewidth]{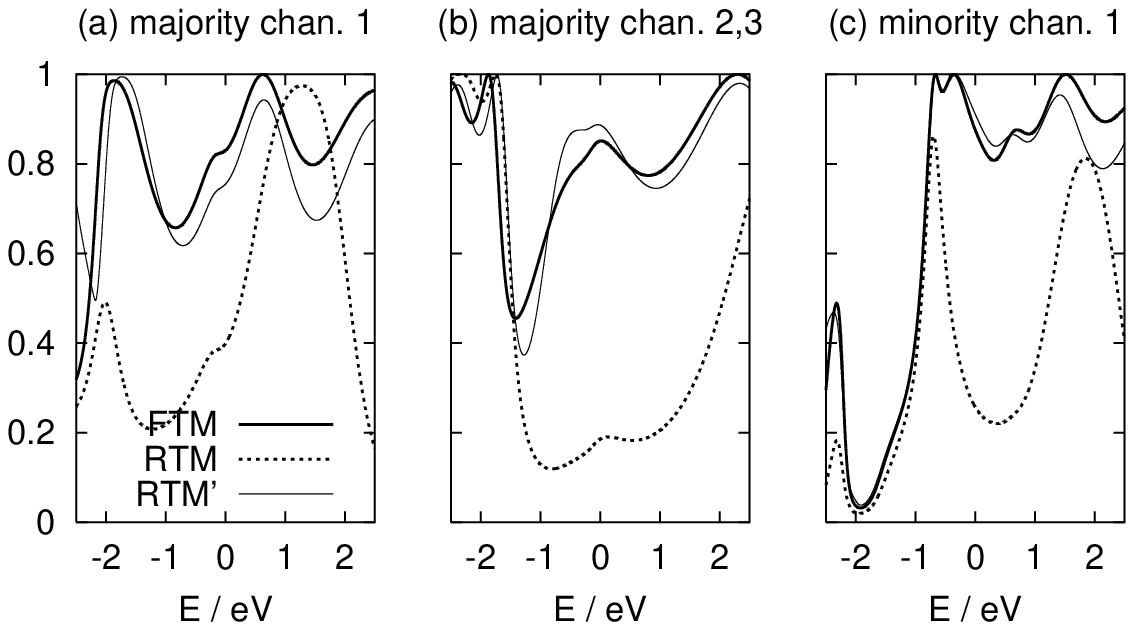} \\
    \includegraphics[width=0.9\linewidth]{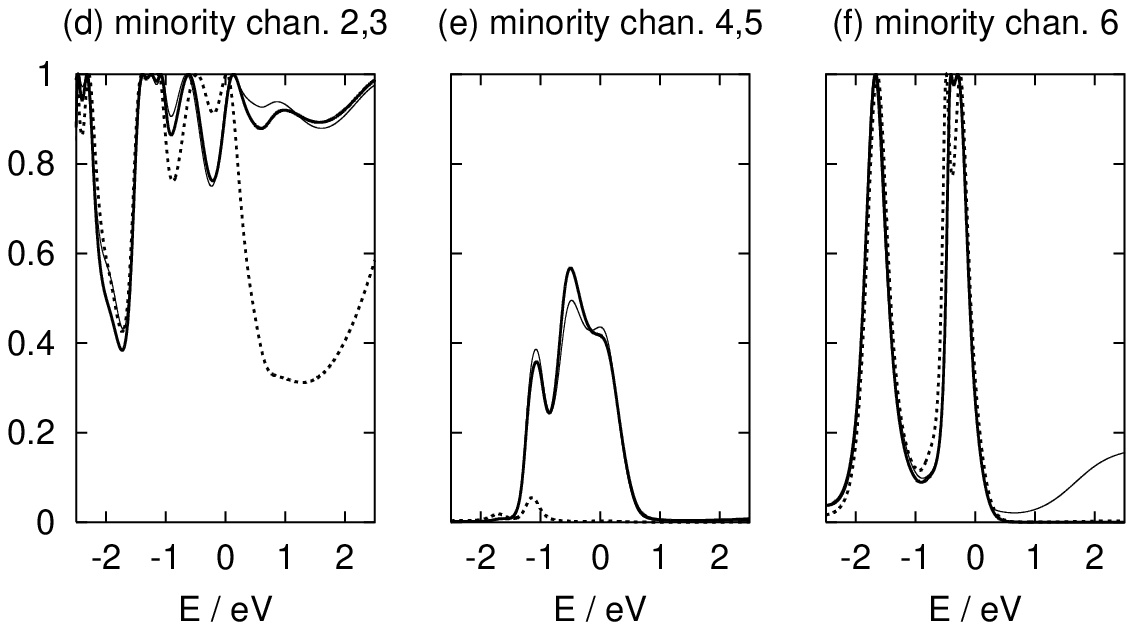} 
  \end{tabular}
  \setcapindent{0cm}   
  \caption{  
    Transmission functions of open transport channels for the Ni nanocontact 
    sketched in Fig. \ref{fig:Ni27} as calculated from the FTM ${\bf T}(E)$ 
    (solid lines) and from the RTM ${\bf T}^\prime(E)$ (dashed lines).
    The RTM transmissions calculated for the cross-section labeled with 0$^\prime$
    in Fig. \ref{fig:Ni27} are given by the thin solid curves (labeled RTM$^\prime$).
  }
  \label{fig:Ni27-channels}
\end{figure}
%%

%%\section{Conclusions}
Obviously, the approximation made in the calculation of the RTM
becomes worse the bigger the hopping between the regions l and r is.
For example, in the contact geometry shown in Fig. \ref{fig:Ni27}
electron hopping from the layers immediately connected to the central atom 
(labeled 0) is certainly bigger than in the geometry of Fig. \ref{fig:Ni28}.
Indeed, Fig. \ref{fig:Ni27-channels} shows that
for almost all channels the RTM transmissions  differ appreciably from FTM 
transmissions, making it difficult in some cases to relate them to each other.
Fortunately, we can judge by exclusion which
RTM transmission relates to which FTM transmission since for the
other channels at least the RTM transmission function mimics the overall
behavior of the FTM transmission function.
However, for more complicated situations it might be impossible
to match the RTM transmission with the FTM transmission for all channels.
The cure to this problem is obvious: One has to choose a bigger 
cross-section, i.e., add an atomic layer to the cross-section 
so that the hopping between l and r becomes small again.
If we choose, e.g., the cross-section labeled with 0$^\prime$ in 
Fig. \ref{fig:Ni27} (including the atomic layer to the right of the central atom) 
the so calculated RTM transmissions now approximate very well the FTM transmissions as
can be seen in Fig. \ref{fig:Ni27-channels}.

%%Again, the channels that yield the highest deviation between FTM and RTM
%%transmission (M channel 1 and m channel 1) are mostly s-type describing
%%very delocalized electrons which thus have a lot of contributions from 
%%second-neighbor hopping.

In summary, we have shown how 
to obtain the orbital contributions to the eigenchannels at an arbitrary
cross-section of a nanoscopic conductor. 
The method has been implemented into our \textit{ab initio} quantum 
transport program ALACANT
and we have illustrated the method by exploring the orbital nature
of the eigenchannels of a Ni nanocontact. 
The method works very well when the chosen cross-section is thick 
enough so that hopping from the layers left and right to the 
cross-section becomes negligible.
Hence in some cases an additional atomic layer has to be included
to the cross-section we are actually interested in.
Taking this into account the method has no limitations and can be readily applied
to \textit{ab initio} transport calculations in all types of 
nanocontacts \cite{Fernandez-Rossier:prb:05} and molecular junctions.

\section{The self-interaction problem}
\label{sec:Ni-self-interaction}

LDA provides a commonly accepted description of the electronic structure of bulk and 
surface ferromagnetism  in transition metals. However, the low coordination of the atoms 
in nanocontacts might give rise to a further localization of the $d$-electrons (compared 
with bulk). But as discussed in Ch. \ref{ch:ab-initio}, LDA fails to describe localized  
electrons properly due to the spurious self-interaction. And although GGA improves 
somewhat on LDA with regard to the self-interaction problem by taking into account 
derivatives of the electron density, this is still not sufficient for materials with 
strongly localized $d$-electrons like e.g. the transition metal oxides. As explained 
at the end of Sec. \ref{sec:DFT} an alternative approach to the electronic structure comes 
from the use of a hybrid functionals which reintroduce some Hartree-Fock exchange (HFX) in order
to correct the spurious self-interaction. B3LYP for example combines HFX
with a certain GGA exchange functional \cite{Becke:jcp:93}, and happens to 
give a reasonable description of the electronic structure and local magnetic moments in NiO 
\cite{Moreira:prb:02} and La$_2$CuO$_4$ \cite{Perry:prb:01}. Popular alternatives to using 
hybrid functionals are e.g. LDA+U \cite{Anisimov:prb:91} and SIC \cite{Perdew:prb:81}. 

\begin{figure}
  \begin{center}
    \includegraphics[width=\linewidth]{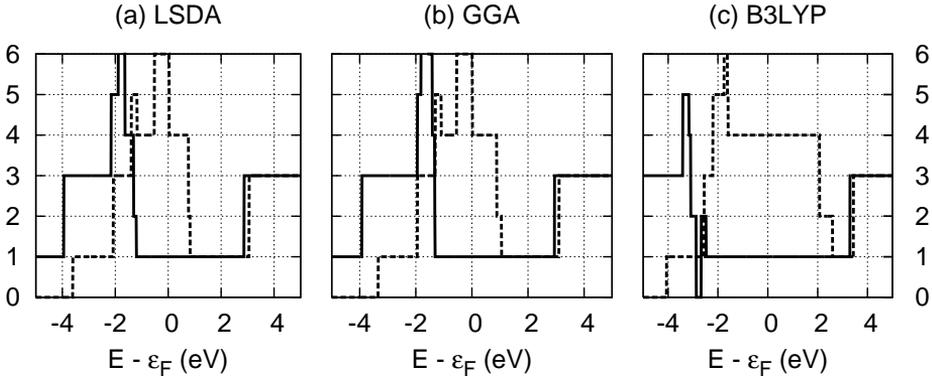}
  \end{center}
  \setcapindent{0cm}   
  \caption{
    Spin resolved transmission function of perfect Ni chain for different 
    functionals: (a) LSDA, (b) GGA, (c) hybrid functional  with 10\% of HFX 
    and (d) hybrid functional with 20\% of HFX. Solid lines are majority-spin
    and dashed lines minority-spin transmissions.
  }
  \label{fig:ni-chain-functionals}
\end{figure}

Here, we explore how the use of different functionals which improve on LDA with regard 
to the spurious self-interaction affects the transport properties of Ni nanocontacts. 
We start by considering again the perfect monatomic Ni chain. Fig. 
\ref{fig:ni-chain-functionals} shows the transmission functions of both spin-channels 
for different density functionals. We note that the transmission of the perfect chain
barely changes when using a GGA functional (b) instead of the LSDA functional (a). Only 
when introducing HFX by using the B3LYP hybrid functional does the 
transmission change appreciably. With B3LYP, the number of conduction channels is reduced to 
1 majority + 4 minority spin-channels at the Fermi level compared to the LSDA results. 
This agrees with recent LSDA+U calculations reported by Wierzbowska {\em et al.} \cite{Wierzbowska:04} 
where the two degenerate flat minority bands $d_{xy},d_{x^2-y^2}$ are shifted downwards in energy 
because of the exchange interaction canceling part of the self-interaction of the strongly 
localized electrons in these flat bands. 
%%However, this is not the reason for the drop in the 
%%minority conductance seen in Fig.\ref{fig:b3lyp-transm} since the 
%%corresponding channel does not contribute to the conductance in the LSDA case either, as will be 
%%discussed in the next section.

\begin{figure}
  \begin{tabular}{cc}
    \includegraphics[width=0.49\linewidth]{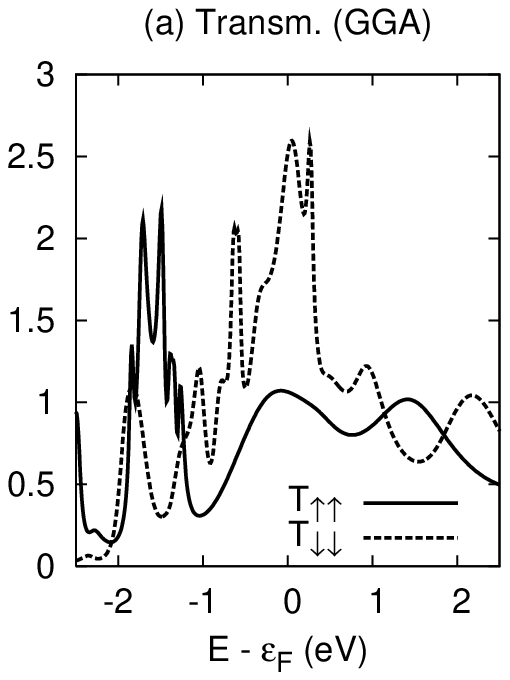} &
    \includegraphics[width=0.49\linewidth]{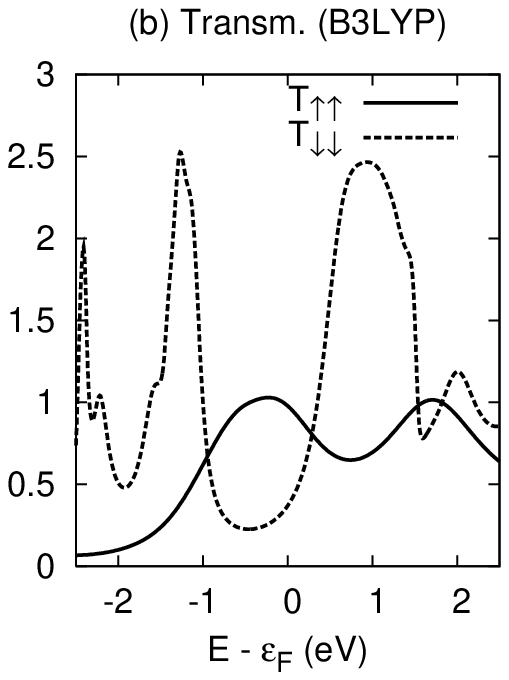}
  \end{tabular}
  \setcapindent{0cm}   
  \caption{(a) Transmission per spin channel calculated with GGA density functional in the 
    P configuration for the model nanocontact shown in Fig. \ref{fig:ni-nanocontact}. (b) 
    same as (a) but calculated with B3LYP hybrid functional.
  }
  \label{fig:ni-nanocontact-functionals}
\end{figure}

As can be from Fig. \ref{fig:ni-nanocontact-functionals}(a) computing the conductance of the Ni 
nanocontact shown in Fig. \ref{fig:ni-nanocontact} with a GGA functional, again the result does 
not change very much with respect to the LSDA result (compare with Fig. \ref{fig:ni-nanocontact}). 
However, we should note that when using GGA more elaborate basis sets than the minimal basis 
set employed in the previous calculations on the LSDA level are required in order to recover
the LSDA result. If the same minimal basis set as in the LSDA calculations is used with the GGA
functional the conductance of the $d$-type minority-channel is considerably reduced. This seems
to have to do with the fact that due to its dependence on the gradient of the electron density 
the GGA functional is more sensitive to the geometry than the LSDA functional and more complete 
basis sets are required to get reliable results. With B3LYP the results for the conductance  
(see Fig. \ref{fig:ni-nanocontact-functionals}(b)) are remarkably different in regard to the 
minority channel conductance which is strongly reduced at the Fermi level. The minority 
$d$-channels are most strongly affected by this reduction of the transmission. Why the minority-spin
channel transmission is so strongly suppressed when using the B3LYP functional in the case of the 
nanocontact but not in the case of the perfect infinite chain might have to do with the 
$d$-electrons of the tip atoms becoming very localized so that they are more strongly affected by the 
HFX, but further investigation of this issue is needed. However, by comparison
with the Ni conductance histogram \cite{Untiedt:prb:04} one can see that the results obtained with 
LSDA and GGA are in very good agreement with the first peak in the histogram at
$\sim2.4-3.0e^2/h$ while the B3LYP value is not. So LSDA and GGA functionals might be more 
appropriate for describing purely metallic nanocontacts than B3LYP. However, this might be different 
when oxygen adsorbates are present in the contact region, as will be discussed in the next Chapter. 
Clearly, LSDA and GGA are not appropriate for describing the conduction through a molecule contacted 
by atomic tips as they do not reproduce well the energy levels of molecules. On the other hand B3LYP
does reproduce molecular energy levels extremely well. So the question of the appropriate functional
to use in DFT based transport calculations is an open one, and it seems that depending on the 
application one has to employ different functionals and justify their use by comparison with 
experiments. This however limits severely the predictive power of DFT based transport calculations.
A systematic improvement of the DFT based transport calculations can only be achieved by employing 
many-body techniques on top of the DFT calculations like the GW perturbation theory or DMFT as explained 
in Sec. \ref{sec:beyond-DFT}.

%%With B3LYP the bulk 
%%and surface magnetic moments are slightly higher than the LSDA ones while the magnetic moment for 
%%the tip atoms is roughly the same ($\approx 1 \mu_{\rm B}$). 

\section{Discussion of results}
\label{sec:Ni:discussion}

As a rule of thumb, the maximum number of conduction channels in atomic-size 
contacts is roughly determined by the number of valence electrons of the contact 
atom(s) \cite{Scheer:nature:98}. However, as shown above, this hypothetical upper 
limit is never reached in Ni nanocontacts since the transmission of the minority-spin 
channel being composed of $s$- \emph{and} $d$-orbitals is strongly reduced by 
scattering due the contact geometry. This result is impossible to predict without a 
full atomistic self-consistent calculation. The majority channel is $s$-type. Thus, 
this channel transmits almost perfectly and evolves smoothly with the stretching of the 
contact giving a stable contribution of $T_{\uparrow\uparrow} \approx$ 1 (see Fig. 
\ref{fig:results-stretch}). The $s$-orbitals in the minority channel are strongly 
hybridized with $d$-orbitals and, therefore, are more sensitive to the contact geometry. 
The contribution to the conductance of the latter, which form narrower bands, disappears 
with the stretching and disorder, as expected (see Fig. \ref{fig:results-stretch}). On 
the other hand, in the AP configuration mostly one $s$-orbital channel per spin 
contributes.  In the AP case the conductance {\em per spin channel} lies thus in 
the vicinity of $e^2/h$, giving $\approx 2 e^2/h$ in total and is fairly stable during 
the last stage of the breaking of the nanocontact. 

Comparing the nature of the eigenchannels contributing to the transmission of the 
nanocontact in the P case with the orbital nature of the bands of the one-dimensional 
chain helps to understand why 3 of the $d$-channels are blocked in the contact which 
transmit perfectly in the one-dimensional case. An eigenchannel analysis restricted to 
the tip atom (see Sec. \ref{sec:Ni-eigenchannels}) for the symmetric cluster (Fig. 
\ref{fig:lsda-transm}) reveals that in the P case the only majority channel contributing 
to the conductance is $s$-type, while of the 3 minority channels contributing appreciably 
to the conductance, one is $s$-type (with a small contribution of the 
$d_{3z^2-r^2}$-orbital), one is $d_{xz}$ and one is $d_{yz}$-type. The three other 
minority channels that contribute in the perfect chain, namely the doubly-degenerate 
$d_{xy}$,$d_{x^2-y^2}$-channel and the $d_{3z^2-r^2}$-channel do not contribute to the 
conductance. It is worth noting that these channels have not ``vanished'' at the Fermi
level, as can be seen from the local DOS (not shown), only their transmission has become 
practically zero. The reason for this lies in the geometry of the $d$-orbitals. As is 
already clear from the discussion of the perfect chain, the  $d_{xy}$- and 
$d_{x^2-y^2}$-orbitals have very small overlap in the direction of the $z$-axis which is 
the main axis of the contact and of the chain. Therefore the electrons in this channel are 
very localized (flat band in the perfect chain) and thus very easily scattered by the 
geometry. On the other hand, it is a bit surprising at first glance, that the 
$d_{3z^2-r^2}$-channel which is the second-broadest band (and thus the electrons in that 
band are almost as delocalized as the $s$-electrons) in the perfect chain does not 
contribute either to the conductance in the nanocontact. However, in the pyramid-geometry 
this orbital which is directed along the z-direction (Fig. \ref{fig:d-orbitals}a,d) does 
not connect well to the orbitals in the pyramid and thus is geometrically blocked for 
transmission. On the other hand the minority bands of $d_{xz}$- and $d_{yz}$-type contribute 
appreciably to the conductance of the nanocontact since these orbitals connect well to the 
corresponding orbitals in the pyramids. Only by distorting the nanocontact upon stretching 
the geometrical matching is destroyed quite rapidly, leading to a rapid decrease in the 
conductance of the minority channel. The reason behind the very small MR values is thus the 
orbital (or geometric) blocking of most of the \textit{a priori} available minority 
channels in the P configuration due to the non-ideal geometry of the nanocontacts. 

The results presented in this chapter have been confirmed by recent experiments on Ni nanocontacts 
under very controlled conditions \cite{Bolotin:nl:06,Keane:apl:06}, and also by recent theoretical 
work employing slightly different methodologies \cite{Smogunov:prb:06,Rocha:prb:06,Pauly:prb:06}. 
Moreover, the total conductance of $\approx 2.5-3.5 e^2/h$ for the single-atom nanocontact
agrees fairly well with the broad peak in the Ni conductance histogram \cite{Untiedt:prb:04}. So 
what else could be the reason behind the huge MR values observed in some experiments when ballistic 
domain wall scattering can be excluded? First, as will be shown in the next chapter, the BMR effect 
can be enhanced considerably (by more than one order of magnitude) due to the presence of oxygen 
surface adsorbates in the contact region \cite{Jacob:prb:06}. Indeed, just the experiments that 
obtain huge MR values have not been performed in ultra-high vacuum conditions in contrast to the 
more recent ones that do not obtain high MR \cite{Untiedt:prb:04,Bolotin:nl:06,Keane:apl:06}. But 
this cannot explain the highest MR results obtained \cite{Chopra:prb:02,Hua:prb:03,Sullivan:prb:05} 
which are several orders of magnitude higher. Another likely explanation for very high MR values are 
magnetostriction and magnetostatic effects which lead to a distortion of the contact geometry in 
dependence of the relative magnetization of the two sections \cite{Gabureac:prb:04,Egelhoff:jap:04}.

%%% Local Variables: 
%%% mode: latex
%%% TeX-master: "~/suficiencia/report/ab-initio"
%%% End: 

%%%%%%%%%%%%%%%%%%%%%%%%%%%%%%%%%%%%%%%%%%%%%

%%%%%%%%%%%%%%%%%%%%%%%%%%%%%%%%%%%%%%%%%%%%%
%% Chapter 6 - NiO chains in Ni            %%
%%             nanocontacts                %%
%%%%%%%%%%%%%%%%%%%%%%%%%%%%%%%%%%%%%%%%%%%%%

\chapter{NiO chains in Ni nanocontacts}
\label{ch:NiO-chains}

%% --- Introduction ---
In going from bulk to lower dimensions material properties often change drastically. 
A recent example is that of interfaces between different insulators which can become 
metallic \cite{Ohtomo:nature:04,Okamoto:nature:04}. Even more recently, it has been 
predicted theoretically that certain oxygen surfaces of some insulating ceramic oxides 
can exhibit magnetism and half-metallicity \cite{Gallego:jpcm:05}. The ultimate limit 
in this respect can be found in atomic chains formed in metallic nanocontacts which 
allow to study the transport properties of one-dimensional systems of atomic size
\cite{Agrait:pr:03}. Due to the lower coordination of the atoms the properties of 
metallic atomic chains formed in nanocontacts can be remarkably different from those 
in the bulk. 
%%For example, Pt nanocontacts can exhibit magnetic order when atomic chains 
%%are formed (see Ch. \ref{ch:Pt-nanowires}).
%%More complex one-dimensional systems like carbon-cobalt atomic chains\cite{Durgun:epl:06} or
%%organometallic benzene-vanadium wires\cite{Maslyuk:05} have even been predicted to be 
%%half-metallic conductors.

However, not all metals form atomic chains in nanocontacts, although recently, it has 
been found that the presence of oxygen favours their formation \cite{Thijssen:prl:06}. 
For example, experiments with Ni nanocontacts \cite{Garcia:prl:99,Sullivan:prb:05,
Viret:prb:02,Untiedt:prb:04,Keane:apl:06,Bolotin:nl:06} have never shown evidence of 
chain formation. Nevertheless, the presence of oxygen in the contact region could 
possibly lead to the formation of NiO chains. In this context it has also been proposed 
that the rather moderate magnetoresistive properties of pure Ni nanocontacts 
\cite{Jacob:prb:05,Bagrets:05,Smogunov:prb:06,Rocha:prb:06} could be enhanced 
considerably by the presence of oxygen adsorbates on the surface of the Ni electrodes
\cite{Papanikolaou:jpcm:03}. On the other hand, bulk NiO is a common example of a 
correlated insulator with antiferromagnetic (AF) order (see e.g., Ref. 
\cite{Sawatzky:prl:84,Moreira:prb:02}), which remains insulating even above the N\'eel 
temperature when the AF order is lost. Thus it is not at all obvious whether or not 
oxidized Ni nanocontacts or NiO chains should be conductors.  

In this chapter we investigate the electronic and magnetic structure and the
transport properties of one-dimensional NiO chains, both idealized infinite
ones and more realistic short ones suspended between Ni nanocontacts.
%% Anticipated results
Anticipating our most important results our \emph{ab initio} quantum transport
calculations show that short NiO chains suspended between Ni nanocontacts can
become half-metallic conductors, i.e.,  carry an almost 100\%
spin-polarized current. This result holds true even for a single O atom in
between Ni electrodes. Consequently, for antiparallel  alignment of the
electrode magnetizations the transport through  the contact is strongly
suppressed resulting in very large MR [difference in resistance 
between antiparallel and parallel alignment of the magnetizations 
of the electrodes normalized either to the higher resistance value (${\rm MR}_1$)
or to the lower one (${\rm MR}_2$)]: ${\rm MR}_1 \approx 90\%$ 
and  ${\rm MR}_2 \approx 700\%$, respectively.

\section{Electronic and magnetic properties of bulk NiO}

Bulk nickel monoxide (NiO) is an example of a strongly correlated material with
insulating character and antiferromagnetic (AF) order, which has been studied
extensively in the past, both experimentally and theoretically (see e.g. \cite{
Terakura:prb:84,Sawatzky:prl:84,Zaanen:prl:85,Leung:prb:91,Anisimov:prb:93,
Towler:prb:94,Aryasetiawan:prl:95,Tjernberg:prb:96,Bredow:prb:00,Moreira:prb:02} 
and Refs. therein). It is now understood that the measured optical gap of 
$\sim$4eV is due to charge-transfer excitations from the O $2p$ band (which 
overlaps with the filled Ni $3d$ band) to the unfilled Ni $3d$ band 
\cite{Sawatzky:prl:84}. Thus it is not a pure Mott-Hubbard insulator where the 
gap is solely determined by excitations from filled to unfilled Ni $3d$ bands. 
Though the AF coupling in bulk NiO is strong, it is found experimentally that the 
magnetic order seems not to affect the electronic structure significantly. Even 
above the N\'eel temperature when the magnetic order is lost, bulk NiO remains an 
insulator and preserves the charge-transfer character and the magnitude of the gap
\cite{Tjernberg:prb:96}.

\begin{figure}
  \includegraphics[width=\linewidth]{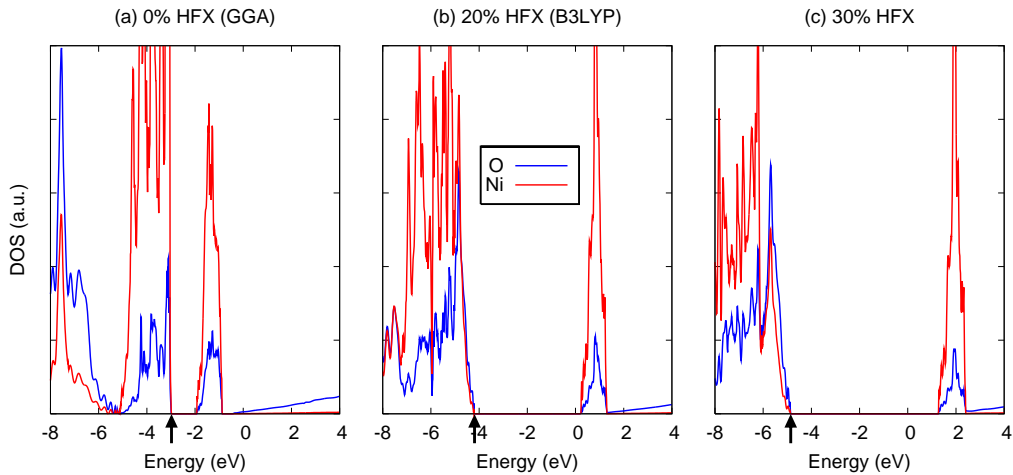}
  \setcapindent{0cm}
  \caption{
    DOS of bulk NiO projected onto O atomic orbitals (red lines)
    and Ni atomic orbitals (green lines) for different density functionals. (a) GGA functional,
    (b) B3LYP hybrid functional with 20\% HFX, and (c) hybrid functional with 30\% HFX. 
    The black arrow marks the top of the valence band in each panel. For the calculations the 
    CRYSTAL {\it ab initio} package has been employed \cite{Crystal:03}. 
    See text and Ref. \cite{Moreira:prb:02} for further discussions.
  }
  \label{fig:NiO-bulk-dos}
\end{figure}

Due to the strong electron correlations of the Ni $3d$-electrons, NiO presents a 
challenge for {\it ab initio} electronic structure calculations. In fact, the standard 
approximations of DFT ---LDA and GGA--- do not give a satisfying description of the 
electronic structure of bulk NiO. Because of the insufficient cancellation of the 
self-interaction in the local exchange functional of the LDA, the occupied narrow 
$3d$-bands are raised in energy. As a result LDA severerly underestimates the gap of 
bulk NiO \cite{Leung:prb:91}. The GGA exchange functional improves somewhat the 
description of NiO but the energy gap is still too small and also the charge-transfer 
character is not captured correctly \cite{Leung:prb:91}, as can be seen from Fig. 
\ref{fig:NiO-bulk-dos}(a).

As explained in Ch. \ref{ch:ab-initio}, the self-interaction error inherent in LDA and GGA 
can be corrected by using hybrid functionals like e.g. the B3LYP functional \cite{Becke:jcp:93} 
which happens to give a reasonable description of the electronic and magnetic structure 
of bulk Ni \cite{Moreira:prb:02}. As we can see from Fig. \ref{fig:NiO-bulk-dos}(b), the 
B3LYP functional which mixes 20\% of HFX to the GGA exchange functional gives just the right 
magnitude of the gap of $\sim$4eV. Also the charge transfer character of the band gap is 
somewhat improved although there is still a strong contribution of Ni $3d$-states at the
upper band edge of the valence band. Increasing the amount of HFX beyond the 20\% of the
B3LYP functional the O $2p$-states become yet more dominant but also the band gap increases
to about 6eV which is above the experimental value (Fig. \ref{fig:NiO-bulk-dos}(c)). With 
regard to the magnetic structure of NiO, B3LYP predicts the correct antiferromagnetic (AF)
order, but slightly underestimates the magnetic moments and coupling constants. On the other 
hand a hybrid functional with 30\% of HFX gives the correct magnetic moment and coupling constants 
and also predicts the correct AF order. A more detailed discussion can be found in Ref. 
\cite{Moreira:prb:02}. Other approaches for correcting the self-interaction like 
the LDA+U method the self-interaction-corrected  LDA (SIC-LDA), and the GW approximation 
lead to similiar results as those obtained from the hybrid functional approach 
\cite{Anisimov:prb:91,Aryasetiawan:prl:95,Moreira:prb:02}.

For the above electronic structure calculations of bulk NiO we have employed elaborate all-electron basis 
sets for Ni and O \cite{Doll:ss:03,Ruiz:jssc:03} similiar to those employed for reported HF and B3LYP 
calculations of bulk NiO \cite{Moreira:prb:02}, but extended with a diffusive $sp$-function in the case of 
Ni and a $d$-polarization function in the case of O which makes them suitable also for the description of
metallic systems in contrast to the original basis set which have been developed for the description of 
insulating NiO. As can be seen from Fig. \ref{fig:NiO-bulk-dos} we reproduce previous GGA and B3LYP results 
for bulk NiO \cite{Moreira:prb:02} with these basis sets.

Given the uncertainty with respect to the functional we will employ for the calculation of 
NiO chains in the following both a GGA functional and the B3LYP functional. We would like 
to emphasize here that it is not clear from the outset how much HFX is necessary in order 
to give a reliable description of atomic NiO chains since the electronic structure of the
one-dimensional NiO chain is quite different from the one of bulk NiO as we will see in the
following.

\section{One-dimensional NiO}

%%The electronic structure of infinite NiO chains has been calculated with the
%%CRYSTAL03 program package  \cite{Crystal:03} while for the electronic structure
%%and quantum transport calculations of the suspended  chains we have used our
%%\emph{ab initio} quantum transport package ALACANT\cite{Palacios:prb:02,ALACANT:06}. The infinite 
%%chain calculations were done with elaborate all-electron basis sets for Ni and O 
%%\cite{Doll:ss:03,Ruiz:jssc:03}, similiar to those employed for reported HF and B3LYP 
%%calculations of bulk NiO\cite{Moreira:prb:02}, but extended with a diffusive 
%%$sp$-function in the case of Ni and a $d$-polarization function in the case of O. 
%%Using these basis sets we reproduce the B3LYP results for bulk NiO \cite{Moreira:prb:02}. 
%%We have also employed minimal basis sets with effective core pseudo-potentials 
%%described in earlier work \cite{Jacob:prb:05} for both Ni and O, and have found 
%%that the results change very little. Thus we have employed these minimal basis 
%%sets for the more demanding transport calculations. 

\begin{figure}
  \begin{minipage}[b][][b]{0.54\linewidth}
    \includegraphics[width=0.9\linewidth]{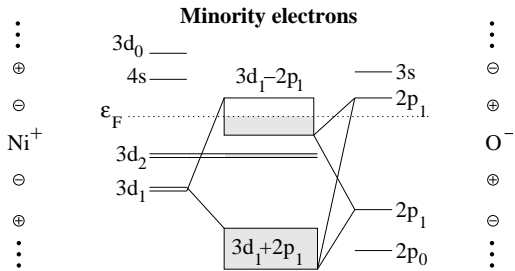}
  \end{minipage}
  \begin{minipage}[b][][b]{0.44\linewidth}
    \setcapindent{0cm}
    \caption{
      Schematic one-electron energies of a one-dimensional NiO chain in z-direction
      for minority-spin. To the left and right the orbital energies of an individual 
      Ni$^+$ cation and O$^-$ anion in the crystal field of a one-dimensional Ni$^+$O$^-$ 
      chain are shown. In the center the formation of valence and conduction bands by 
      hybridization of Ni $3d$ and O $2p$ orbitals is shown.
    }
    \label{fig:scheme}
  \end{minipage}
\end{figure}

% --- Results and discussion ---
The electronic properties of bulk NiO are, to a large extend, determined by the
atomic scale properties, like the crystal field splitting of the Ni and O
energy levels and the amount of electron charge transfered from Ni to O.
The latter, in turn, is determined by the interplay between Madelung binding 
energy, the ionization potential of Ni, and the electron affinity of O. 
Due to the lower coordination and the corresponding decrease in Madelung 
binding energy the electron transfer from Ni to O is less favourable in an
atomic chain than in bulk (where the electron transfer is almost complete 
resulting in an ionic configuration of Ni$^{2+}$O$^{2-}$).
The proper starting point to discuss the formation of energy  bands in the
one-dimensional NiO chain are therefore the univalent ions Ni$^+$ and O$^-$.

In order to understand how the low coordination affects the atomic properties of the 
constituting ions we have first performed B3LYP calculations of both a single Ni$^+$ ion 
and a single O$^-$ ion each in the field of point charges that mimic the crystal field 
of a one-dimensional chain of univalent Ni and O ions.
We find that for both ions the spin-doublet state ($S=1/2$) minimizes the energy
as is the case for the free ions. In Fig. \ref{fig:scheme} we show schematically 
the energy levels of Ni$^+$ and O$^-$ in the presence of the point charges
for minority spin only.
Interestingly, for minority spins, 
the occupied Ni $3d_1$ orbitals ($d_{xz}$ and $d_{yz}$) of the Ni$^+$ 
ion fall energetically in between the occupied and  unoccupied O $2p_1$ orbital 
($p_x$ and $p_y$) of the O$^-$ ion. 
Thus the $3d_1$ and $2p_1$ orbitals can form two filled degenerate bonding bands and 
two degenerate partially filled antibonding bands as indicated in the middle part of
Fig. \ref{fig:scheme}.  The $3d_2$ ($d_{xy}$ and $d_{x^2-y^2}$) doublet is somewhat 
above in energy to the $3d_1$ but cannot hybridize with the oxygen $2p$ orbitals. 
The Ni $3d_0$ ($d_{3z^2-r^2}$) and $4s$ are empty while the O $2p_0$ ($p_z$) and 
$2s$ orbitals are  filled and much lower in energy so that no hybridization takes 
place though symmetry would allow for it.

On the other hand, for majority-spin electrons (not shown) all five Ni $3d$
orbitals are filled while the $4s$ is also  empty, i.e., the Ni$^+$ valence
configuration is $3d^9$ and not $4s^1 3d^8$ as for the free Ni$^+$ ion. 
Moreover, all of the O $2p$ orbitals are filled so that the Ni$^+$ and O$^-$
ions can only form  either completely filled or completely empty bands for the
majority spin.  Thus the ionic picture suggests that a one-dimensional NiO chain 
should become a half-metallic conductor where only the minority-spin levels form
conducting bands. 

%%In this case only the minority-spin $3d_1$ bands would conduct so that, according to the proposed 
%%classification in Ref. \cite{Coey:jphysd:02}, it would be of Type I$_{\rm B}$.

%%
\begin{figure}
  \begin{minipage}[t][][b]{0.68\linewidth}
    \includegraphics[width=0.98\linewidth]{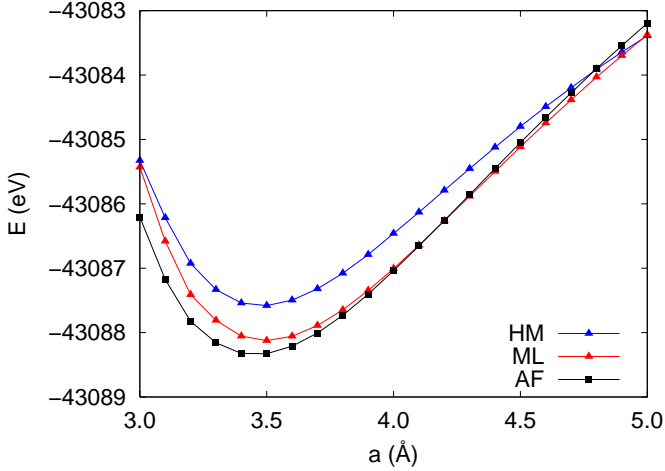}
  \end{minipage}
  \begin{minipage}[t][][b]{0.3\linewidth}
    \setcapindent{0cm}   
    \caption{ 
      Energy per unit cell of infinite NiO chain in dependence 
      of lattice spacing $a$ calculated with B3LYP hybrid functional. 
      Blue triangles indicate the half-metallic state 
      (HM), red triangles the molecule-like insulating state with FM order (ML), 
      and black boxes the insulating state with AF order (AF).
    }
  \end{minipage}
  \label{fig:NiO-chain-stretch}
\end{figure}

Not surprisingly, our calculations for inifinite one-dimensional NiO chains
(Fig. \ref{fig:NiO-chain-stretch}) show that a  univalent ionic configuration as an initial
guess results in a half-metallic state for large separation
of the individual chain atoms (i.e., large lattice spacing of 5\r{A}) as
suggested by the ionic picture. However, as can be seen from
Fig.\ref{fig:NiO-chain-stretch}(a), the half-metallic state is only a metastable state for
most values of the lattice spacing.
This half-metallic state  is ``shadowed'' by a second state with FM  order and
insulating character. By successively decreasing the lattice spacing $a$ of the
chain and using the  (half-metallic) state of the previous step for the initial
guess, the half-metallic state can be generated also for smaller  inter-atomic
distances which points towards its metastability. 
Around the equilibrium lattice spacing ($a \sim 3.4$\r{A}) the ground state of the chain has AF order and is 
of insulating character with a substantial gap of $\sim 4$eV like in bulk. When stretched out of equilibrium 
the FM state and the AF state become comparable in energy until finally, at a lattice spacing of $\sim 4.2$\r{A}, 
the FM state becomes the ground state. 

\begin{figure}
  \begin{tabular}{cc}
    \includegraphics[width=0.49\linewidth]{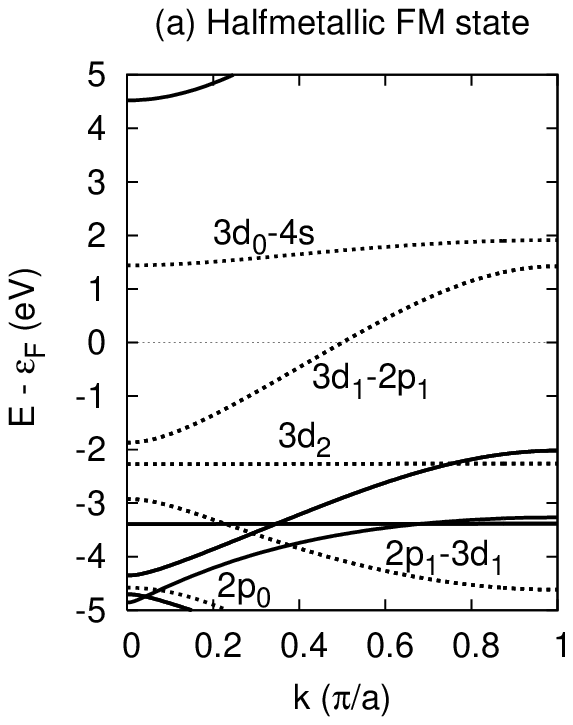} &
    \includegraphics[width=0.49\linewidth]{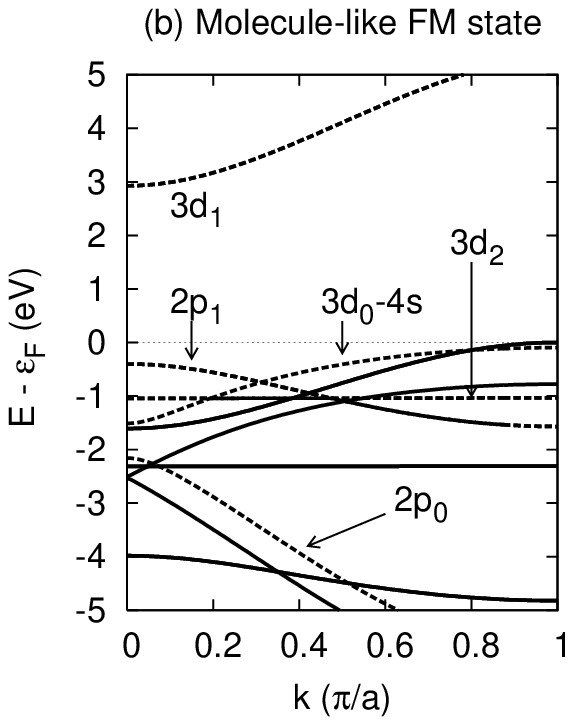} 
  \end{tabular}
  \setcapindent{0cm}   
  \caption{(a) B3LYP band structure of HM state for a lattice 
    spacing of 3.6\r{A}. Solid lines indicate majority-spin bands and 
    dashed lines indicate minority-spin bands. 
    (b) Same as (a) but for ML FM state.
  }
  \label{fig:NiO-chain-bands-b3lyp}
\end{figure}

The band structure diagram in Fig. \ref{fig:NiO-chain-bands-b3lyp}(c) shows that the metastable
state with FM order corresponds indeed to the half-metallic state suggested by
the ionic picture: The half-filled doubly-degenerate conduction band is formed by minority-spin
Ni $3d_1$ orbitals hybridized with O $2p_1$ orbitals while the Ni $3d_2$
orbitals do  not hybridize with O $2p$ orbitals and thus form a flat valence
band. The lowest-lying empty band is formed by the  minority-spin Ni $3d_0$
orbital which is slightly hybridized with the Ni $4s$ orbital. On the other
hand the stable state with FM order and insulating character (see band
structure in Fig. \ref{fig:NiO-chain-bands-b3lyp}(b))  actually corresponds to the ground state
of the NiO molecule which is a ${}^3\Sigma^{-1}$ state\cite{Doll:prb:97}. The 
main difference with the half-metallic state is that now the
\emph{non-degenerate} Ni $3d_0$ and $4s$ orbitals  form a minority-spin valence
band while the minority-spin \emph{doubly degenerate half-filled} antibonding
band  composed of Ni $3d_1$ and O $2p_1$ bands is emptied and a substantial gap
of $\sim 3$eV opens. Thus the infinite chain  behaves like an insulator for
reasonable values of the chain stretching. 

On th other hand using a GGA functional, the one-dimensional NiO 
chain is always conducting for FM order in contrast to the previous 
B3LYP results. In Fig. \ref{fig:NiO-chain-gga}, we show the 
GGA band structure of an ideal one-dimensional NiO chain in the 
ferromagnetic (FM) phase. Compared to the B3LYP band structure for 
the half-metallic state at same lattic spacing shown in Fig. 
\ref{fig:NiO-chain-bands-b3lyp}(a) we see that the occupied bands 
have been raised considerably in energy. In particular, the 
doubly-degenerate flat minority-spin band of type ($d_2$)
composed of Ni $3d_{xy}$ and $3d_{x^2-y^2}$ orbitals (well below
the Fermi level with B3LYP), now actually crosses the Fermi level.
Also the doubly-degenerate and previously half-filled minority-spin band of type
($d_1$) composed of Ni $3d_{xz}$ and $3d_{yz}$ orbitals hybridized with O
$2p_x$ and $2p_y$ orbital (the only conduction band with B3LYP)
calculation are raised somewhat in energy.
Consequently, the previously empty minority-spin band of type ($d_0$)
composed of Ni 3$d_{3z^2-r^2}$ orbitals is lowered in energy with
respect to the other filled or partially filled $3d$ bands, and becomes
a conduction band.

On the other hand also the majority-spin bands are raised in energy. The
doubly-degenerate majority-spin band composed of Ni $3d_{xz}$ and $3d_{yz}$
orbitals hybridized with O $2p_x$ and $2p_y$ orbitals which was well
below the Fermi level in the B3LYP calculation now also crosses the Fermi level
near to its upper band edge.
Thus in GGA the ideal case of the infinite NiO chain in the FM phase
does not represent a half-metallic conductor, although the spin-polarization
of the conduction bands is quite strong (5 minority-spin bands vs.
2 majority-spin bands).

The change in the electronic structure of the one-dimensional NiO chain on the GGA level
with respect to the B3LYP results can be explained by the insufficient cancellation
of the self-interaction by the GGA exchange functional, which causes the occupied
Ni $3d$ orbitals to artificially rise in energy.

\section{O-bridge in Ni nanocontacts}

%% --- Suspended NiO chains ---
Atomic chains formed in break junctions have a finite length and are suspended 
between electrodes. It is well known that the
contact between the atomic chain and the electrode tip will have considerable
effect on  the electronic structure of the chain, especially when $d$-orbitals
are involved like is the case here\cite{Jacob:prb:05}. 

\begin{figure}
  \includegraphics[width=\linewidth]{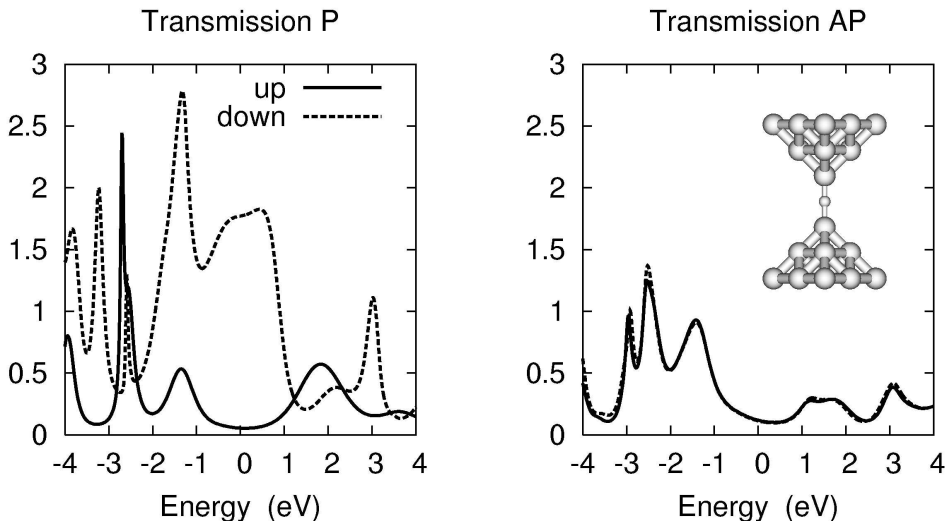}
  \setcapindent{0cm}   
  \caption{
    Transmission per spin channel in the case of P (left)
    and AP (right) alignment of the electrode magnetizations
    for NiO chain consisting of one oxygen atom bridging the
    two Ni tip atoms of the Ni electrodes as shown on the inset 
    in the right panel. The separation of the two Ni tip atoms 
    is 3.6\r{A}.
  }
  \label{fig:NiO-transm-b3lyp}
\end{figure}

We have thus calculated the electronic structure and transport properties
of  both a single oxygen atom and a  O-Ni-O chain bridging the two tips 
of a Ni nanocontact as shown in the insets of the right panels of Fig. 
\ref{fig:NiO-transm-b3lyp} and Fig.\ref{fig:NiO-transm2}.
In the case of the single oxygen atom (Fig. \ref{fig:NiO-transm-b3lyp}) the electron transport 
is almost 100$\%$ spin-polarized around the Fermi level for parallel (P) alignment 
of the magnetizations of the two Ni electrodes. Moreover, an orbital eigenchannel
analysis \cite{Jacob:prb:06} reveals that the  transport is mainly due to two
almost perfectly transmitting minority-spin channels  composed of Ni $3d_1$ and
O $2p_1$ orbitals, i.e. they correspond to the doubly-degenerate conduction
band of the metastable half-metallic state in the  perfect chain. Thus the
half-metallic state which was suppressed in the idealized case  of the infinite
chain emerges in the more realistic situation of a short suspended chain. 
We can understand this phenomenon in terms of the orbital blocking mechanism proposed earlier 
in the context of Ni nanocontacts\cite{Jacob:prb:05,Jacob:prb:06}. The highest minority-spin 
valence band of the insulating state with FM order in the infinitely long chain has a strong 
contribtution from the Ni $3d_0$ orbital which is not ``compatible'' with the geometry of the 
pyramid shaped Ni contacts, so that this band is blocked and thus cannot be occupied. 
Instead, the doubly-degenerate band composed of Ni $3d_1$ orbitals hybridized with
O $2p_1$ orbitals is partially filled resulting in the half-metallic state which in the 
perfect chain is only metastable. Thus the orbital blocking by the contacts actually turns the 
chain into a half-metallic conductor.
Consequently, the conductance is strongly suppressed in the case of antiparallel (AP) alignment 
of the magnetizations of the Ni electrodes as can be seen from the right panel of Fig. 
\ref{fig:NiO-transm-b3lyp} and the MR becomes very large: ${\rm MR}_1 \approx 90\%$ and ${\rm MR}_2 \approx 700\%$.

\begin{figure}
  \begin{tabular}{cc}
    \includegraphics[width=0.49\linewidth]{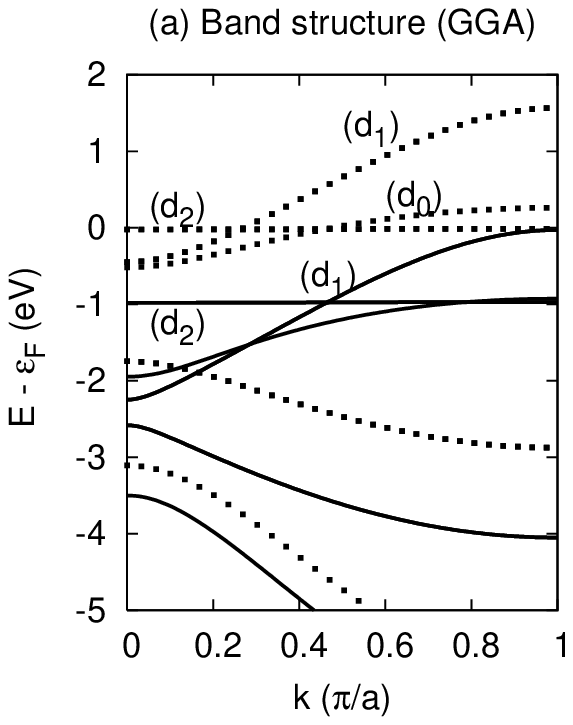} &
    \includegraphics[width=0.49\linewidth]{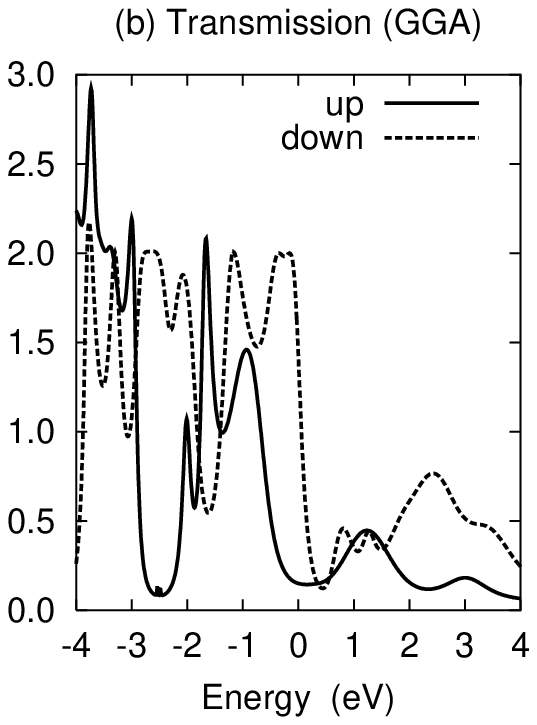} 
  \end{tabular}
    \setcapindent{0cm}   
    \caption{
      (a) GGA band structure of FM groundstate for a lattice spacing 
      of 3.6\r{A}. Solid lines indicate majority-spin bands and 
      dashed lines indicate minority-spin bands. (b) GGA Transmission per 
      for NiO chain consisting of one oxygen atom bridging the
      two Ni tip atoms of the Ni electrodes as shown on the inset 
      in the right panel of Fig. \ref{fig:NiO-transm-b3lyp}. 
      The separation of the two Ni tip atoms is 3.6\r{A}.
    }
  \label{fig:NiO-chain-gga}
\end{figure}
%%

%%Finally, we also calculate the electronic structure and transport properties
%%of the single-oxygen atom bridge on the GGA level.
Fig. \ref{fig:NiO-chain-gga}(b) shows the transmission per spin-channel for
the Ni-O-Ni nanobridge calculated with GGA. Surprisingly, the transmission does
not look very different from the transmission calculated with B3LYP for that case.
Near the Fermi level the transmission is strongly spin-polarized: The transmission
of the majority-spin channel is strongly suppressed while in the minority-spin
channel essentially two perfectly transmitting channels contribute to the conductance.

This seems to be at odds with the band-structure calculated for the infinite
one-dimensional chain, which suggests that there should be 5 minority- and
2 majority-spin channels contributing to the overall conductance.
However, the geometry of the Ni nanocontact blocks the transmission of just
these channels which due to the unphyscial self-interaction of GGA have been
raised to the Fermi level.
Indeed, an orbital eigenchannel analysis \cite{Jacob:prb:06} of the transmission
reveals that only the doubly-degenerate band of type ($d_1$) contributes to the
conductance of the minority-spin channel just as in the case of the B3LYP functional.
The electrons in the flat minority-spin band of type ($d_2$) which actually crosses
the Fermi level in the ideal case of the infinite chain are easily scattered as
they present strongly localized electrons, and thus do not contribute to the
overall conductance.
The other minority-spin channel composed of Ni $3d_{3z^2-r^3}$ orbitals does not
contribute either to the conductance since the symmetry of the orbital is not
compatible with the geometry of the two Ni electrodes - a mechanism to which we have
referred to in previous work as {\it orbital blocking} \cite{Jacob:prb:05}.
The small but finite conductance in the majority-spin channel relates to the
doubly-degenerate majority-spin band of type ($d_1$) of the infinite NiO chain
raised to the Fermi level due to the self-interaction error.
The ($d_1$) band only crosses the Fermi energy near the upper band edge where
the band becomes flat. Thus the the electrons in this channel are quite susceptible
to scattering near the Fermi level resulting in a low transmission.

%% --- Stretching and sensitivity to geometry ---
Varying the distance $d$ between the Ni tip atoms leads to similiar results as those shown in Fig. 
\ref{fig:NiO-transm-b3lyp}. For the P case the current through the chain is almost 100\% spin-polarized with
two open minority channels composed of Ni $3d_1$ and O $2p_1$ orbitals, while
for the AP case it is strongly suppressed, resulting in very high MR values between 80\% and 90\%
for MR$_1$ for $d$  between 3.0\r{A} and 5.0\r{A}. 
Geometry relaxations for different values of the tip-tip distance show that for small distances
the oxygen atom goes into a zigzag position. The bonding angle decreases with increasing distance
 until it becomes zero at 3.6\r{A}. Finally, the chain breaks for $d > 4.8$\r{A}.
Thus the scattering is strong for small distances $d < 3.6$\r{A} when the bonding angle
is appreciable and for large stretching, $d > 4.2$\r{A}, resulting in a considerable reduction in the
conduction of the two minority channels. 

\section{O-Ni-O bridge in Ni nanocontacts}

\begin{figure}
  \includegraphics[width=\linewidth]{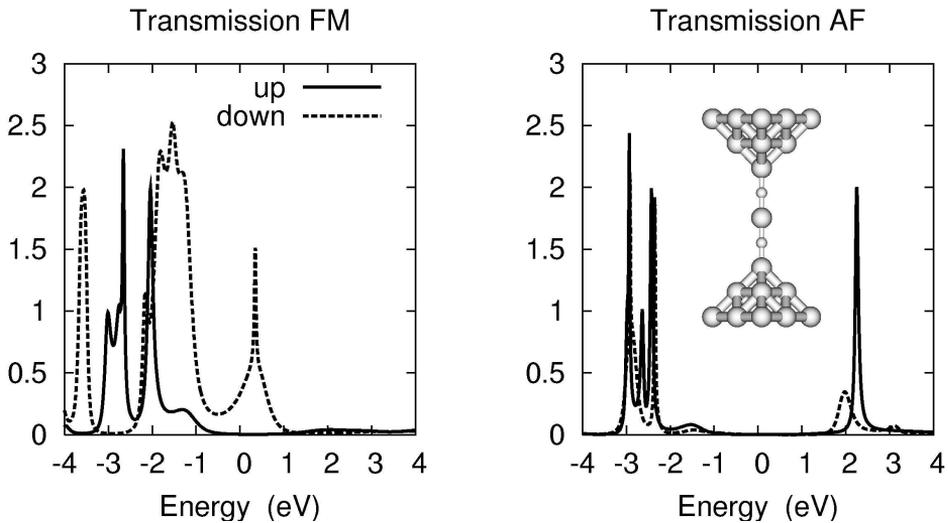}  
  \setcapindent{0cm}   
  \caption{
    Transmission per spin channel for suspended chain consisting of a O-Ni-O bridge and the
    Ni tip atoms shown in inset of right panel in the case of FM order (left) and of AF order 
    (right). For AF order the magnetization of the Ni atom in the center of the chain is 
    reversed with respect to the two Ni tip atoms. Distance between a Ni tip atom and the 
    center atom is 3.6\r{A}.
  }
  \label{fig:NiO-transm2}
\end{figure}

%% --- Longer chains ---
In longer suspended chains the insulating state with FM order starts to emerge inside the chain and
a away from the contacts. As a result the conductance is reduced as can be seen already in the case of the O-Ni-O 
bridge (Fig. \ref{fig:NiO-transm2}). The minority-spin conduction is reduced considerably ($\sim 30\%$) 
compared to the case of the single oxygen bridge. On the other hand the conduction of the majority-spin 
channel becomes practically zero ($<0.2\%$). This can be understood by the fact that the residual 
majority-spin channel conductance of the single oxygen bridge is due to direct hopping of Ni $s$ electrons 
between the electrodes and therefore vanishes when the distance between the electrodes is large. 
The finite minority-spin conductance opens up the possibility to an interesting phenomenon:
When the middle Ni atom reverses its spin, the conductance drops to nearly zero (see right panel of Fig. 
\ref{fig:NiO-transm2}) since the AF chain is insulating. In other words, this system behaves as a 
\emph{single atom spin valve} which presents an extremely large MR even higher than that reported above for
the single oxygen bridge:  ${\rm MR}_1 \approx 99\%$ and ${\rm MR}_2 \approx 10,000\%$.
Apart from controlling the magnetization direction of the central atom by a magnetic field, Fig. \ref{fig:NiO-chain-bands-b3lyp}(a) 
suggests that a mechanical control of the spin valve (by  stretching the chain) would also be possible.

\section{Conlcusions}

%%In conclusion, we have shown that one-dimensional infinite NiO chains are insulating 
%%and have AF order around the equilibrium lattice spacing but can actually become a FM 
%%insulator when stretched out of equilibrium, in contrast to bulk NiO. 
%%In the more realistic case of a short NiO chain suspended 
%%between Ni nanocontacts the chain becomes an almost 100\% spin-polarized conductor. 

In conclusion, we have shown that ideal one-dimensional infinite NiO chains could be either insulating
or halfmetallic depending on the functional used in the calculation. Following the results obtained 
with the B3LYP hybrid functional one-dimensional NiO-chains are always insulating for all reasonable 
values of the lattice spacing but change from AF to FM order when stretched slightly out of equilibrium
in contrast to bulk NiO which has always AF order. A halfmetallic conducting state which is normally
metastable only becomes the ground state when the chain is stretched to unreasonable large values of
the lattice spacing. The GGA functional on the other hand predicts NiO chains to be ferromagnetic 
halfmetallic conductors. Although it seems clear that the GGA results are affected by the insufficient
cancellation of self-interaction error it is a priori not clear whether B3LYP predicts the correct
ground state as it tends to localize electrons and thus tends to predict insulating behaviour. Thus the 
question of the electronic structure of infinite NiO chains remains open at the moment, and more 
sophisticated many-body techniques like GW or DMFT are probably required to answer this question.

However, in the more realistic case of short NiO chains suspended between Ni nanocontacts both 
B3LYP and GGA predict the chains to become strongly spin-polarized conductors which can be related
to the corresponding halfmetallic states of the infinite NiO chain, i.e. the metastable halfmetallic 
state in the case of the B3LYP functional and the halfmetallic ground state in the case of the GGA 
functional. The emergence of almost perfect half-metallicity 
%%(i.e. almost 100\% spin-poalrization of the conduction channel) 
in suspended chains leads to a strong suppression of the current for AP alignment 
of the electrodes resulting in very large MR values of ${\rm MR}_1 \approx 90\%$ and ${\rm MR}_2 \approx 700\%$, 
respectively. This could perhaps explain to some extend the very large MR values in Ni nanocontacts obtained in some 
experiments \cite{Garcia:prl:99,Sullivan:prb:05} where oxygen is likely to be present. Finally, the O-Ni-O 
bridge suspended between Ni electrodes operates as a \emph{single atom spin valve} where the 
currentflow is controlled by the magnetization of a single atom.

%%% Local Variables: 
%%% mode: latex
%%% TeX-master: "~/my_papers/nanocontacts/NiO-nanowires/NiO-nanowire"
%%% End: 

%%%%%%%%%%%%%%%%%%%%%%%%%%%%%%%%%%%%%%%%%%%%%

%%%%%%%%%%%%%%%%%%%%%%%%%%%%%%%%%%%%%%%%%%%%%
%%%% Chapter 7 - Emergence of magnetism on %%
%%%%             the nanoscale             %%
%%%%%%%%%%%%%%%%%%%%%%%%%%%%%%%%%%%%%%%%%%%%%
\chapter{Transport through magnetic Pt nanowires}
\label{ch:Pt-nanowires}

Fabrication of metallic nanocontacts permits to probe the electronic and
mechanical properties of conventional metals with unconventional atomic
coordination\cite{Agrait:pr:03}.  Electron transport in these systems depends 
on the tiny fraction of  atoms in the sample forming the atom-sized neck
which have a reduced coordination and are responsible for the two-terminal
resistance. Transport experiments can thereby probe the atomic and related
electronic structure of these atoms and provide information about a fundamental
question:  
How bulk properties evolve when the system reaches atomic sizes and atoms with
full bulk coordination are no longer majority. A bulk property that is
susceptible to change is magnetism.  Bulk Pt, for instance, is a
paramagnetic metal but a transition to a ferromagnetic state could be expected 
upon reduction of the atomic coordination with the concomitant increase of the 
density of states (DOS) at the Fermi energy beyond the Stoner limit.  Density functional 
calculations \cite{Bahn:prl:01,Delin:prb:03,Nautiyal:prb:04}
for one-dimensional infinite Pt chains support this hypothetical scenario,
resulting in a ferromagnetic transition above a critical lattice spacing which,
depending on the computational approach, can be below the equilibrium lattice
constant\cite{Delin:prb:03}.  The formation of local moments in real Pt
nanocontacts would not be totally unexpected.

Formation of and
electronic transport in finite Pt chains have been extensively
studied experimentally \cite{Sirvent:prb:96,Smith:prl:01,Smith:prl:03,Nielsen:prb:03,Rodrigues:prl:03}. 
Based upon the appearance of a peak at $G=0.5 \times 2e^2/h=0.5G_0$ in the
conductance histogram  Rodrigues {\em et al.} suggested that  Pt and Pd
nanocontacts could be spin polarized\cite{Rodrigues:prl:03}. The origin of this
peak has been later attributed to adsorbates\cite{Untiedt:prb:04} so that
magnetism in Pt and Pd nanocontacts has not been confirmed experimentally yet.
Previous theory work has
addressed the formation of local moments in Pd nanocontacts\cite{Delin:prl:04}
and in Co , Pd and Rh short chains sandwiched between Cu
planes\cite{Stepnyuk:prb:04}. To the best of our knowledge theory work on Pt
nanocontacts\cite{Vega:prb:04,Garcia:prb:04,Thygessen:prb:05,Garcia-Suarez:prl:05}
has overlooked the possibility of local magnetic order so far. In this chapter we
perform density functional calculations of both the electronic structure and
transport. We find that local magnetic order can develop spontaneously in Pt
nanocontacts. Local magnetic moments as high as 1.2 $\mu_B$ in low-coordination
atoms are found. Interestingly,  while transport is definitely spin polarized,  
the calculated  total conductance of magnetic and
non-magnetic Pt nanocontacts is very similar and in agreement with experimental
data, explaining why magnetism  has been unnoticed so far.  

% The rest of the paper is organized as follows. First we discuss the
% methodology of the electronic structure calculation presented here. Second we
% present results for idealized infinite and finite one-dimensional Pt chains.
% Then we calculate both  zero-bias transport and the electronic structure of Pt
% nanocontacts, which turn out to be magnetic beyond attainable values of the stretching. 
% Finally we discuss
% whether magnetism in Pt nanocontacts is robust to  thermal and quantum
% fluctuations, and propose experimental verification of the different
% scenarios. 
% We also comment on the
% effect of spin-orbit interaction, missing in our calculations.

The electronic structure of  various low dimensional structures of Pt which mimic  
actual nanocontacts are calculated in the density functional approximation, using 
either CRYSTAL03\cite{Crystal:03} and our \textit{ab initio} transport 
package ALACANT that interfaces GAUSSIAN03 to implement the NEGF as explained in 
based on DFT electronic structure calculations Ch. \ref{ch:ab-initio}. 
We use scalar relativistic (SR) pseudopotentials for the 60 inner electrons of the
Pt atom and the remaining 18 electrons are treated using generalized gradient 
approximation (GGA) density functionals. The basis set  used for all the calculations 
has been optimized to  describe bulk Pt as well as Pt surfaces\cite{Doll:ss:04}. Other 
basis sets such as LANL2DZ or SDD\cite{Gaussian:03} have occasionally been employed for
comparison. The main results do not depend on the choice of  basis set.

% \cite{Palacios:prb:01,Palacios:prb:02,Palacios:ctcc:05} 
% \cite{Gaussian:03}. ALACANT describes the bulk electrodes using a
% semiempirical tight-binding Hamiltonian on a Bethe lattice. The
% spin-resolved density matrix includes the effect of the electrodes by
% means of self-energies and the spin-polarized transport\cite{Palacios:prb:05,Jacob:prb:05} is
% obtained with the standard Landauer formula.  Both
% GAUSSIAN03/ALACANT and CRYSTAL03 perform electronic structure
% calculations using a basis of  localized atomic orbitals (LAO). CRYSTAL03
% permits to calculate infinite systems with crystalline symmetry whereas
% ALACANT is suitable for systems with no symmetry or periodicity such as nanocontacts in or 
% out-of-equilibrium situations. 

\section{Electronic and magnetic structure of atomic Pt chains}

\begin{figure}
  \vspace{1cm}
  \begin{center}
    \includegraphics[width=0.9\linewidth]{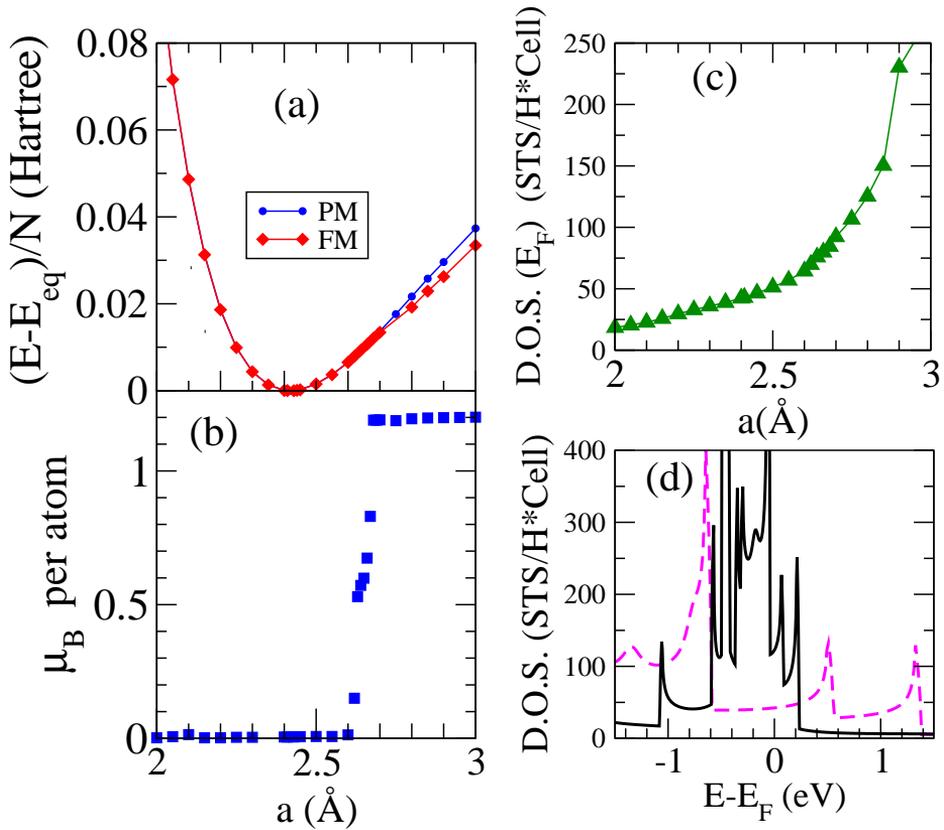}
  \end{center}
  \setcapindent{0cm}   
  \caption{ \label{fig1}(Color online).
    (a) Energy per atom for a perfect monostrand Pt chain. (b) Magnetic moment per atom as a
    function of lattice spacing $a$. (c) DOS of the paramagnetic
    chain at the Fermi energy
    as a function of $a$.  (d) D.O.S. as a function of energy for 
    $a=2.4$\r{A} (dashed) and $a= 2.8$\r{A} (solid) } 
  \label{chain}
\end{figure}

We first consider a perfect one-dimensional mono-strand Pt
chain. Such an idealized system serves as a standard starting point to
understand lower symmetry geometries. It also permits to  test whether our
LAO pseudopotential methodology  reproduces the
results obtained with SR all electron plane-wave calculations
reported  by Delin {\em et al.}\cite{Delin:prb:03}. In Fig. \ref{chain}(a) we
show the energy per atom as a function of the lattice constant $a$ both for
the paramagnetic (PM) and the ferromagnetic (FM) chain. 
They both have a minimum at $a=2.4$\r{A}. The FM chain 
develops a non-negligible magnetic moment when the lattice constant goes
beyond  $a\simeq 2.6$\r{A}. This configuration is clearly lower in energy above
that distance. The energy difference between the FM and the PM configurations
is  16 meV per atom for $a=2.7$\r{A}  and 33 meV per atom for $a=2.8$\r{A}.  The 
magnetic moment per atom reaches a saturation value of $1.2\mu_B$.    The
equilibrium distance, critical spacing, asymptotic magnetic moment and shape of
the phase boundary obtained by us are similar to those obtained by Delin {\em et
al.}\cite{Delin:prb:03} using a SR all-electron plane-wave calculation.   Our results and those of Delin
{\em et al.} underestimate the onset of the
magnetic  transition compared to calculations including spin-orbit coupling\cite{Delin:prb:03,Nautiyal:prb:04} that predict that a magnetic moment forms
already below the equilibrium distance. 

\begin{figure}
  \vspace{1cm}
  \begin{minipage}[t][][b]{0.6\linewidth}
    \includegraphics[width=0.9\linewidth]{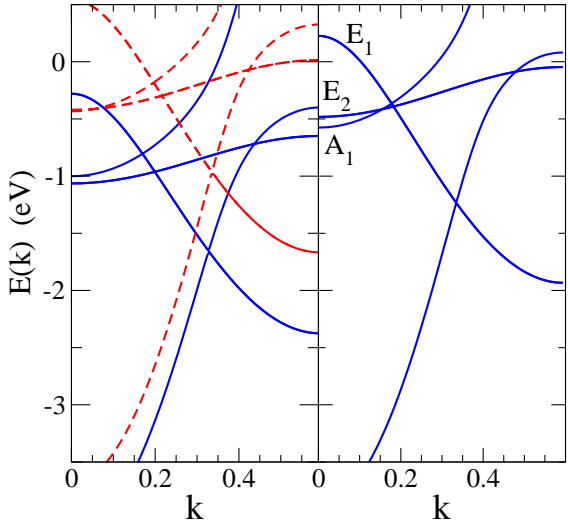}
  \end{minipage}
  \begin{minipage}[t][][b]{0.4\linewidth}
    \setcapindent{0cm}   
    \caption{ (Color online).
      Energy bands for ideal Pt chain with $a=2.8$\r{A}. Left: ferromagnetic phase.
      Right: paramagnetic case.  } 
    \label{fig2}
  \end{minipage}
\end{figure}

The magnetic transition in the phase diagram [Fig. 1(b)]  is compatible with
the Stoner criterion for ferromagnetic instability.  As the chain is stretched,
the atom-atom coupling becomes weaker, the bands narrow down and so does the
DOS ($\mathcal{D}(\epsilon)$).  Since the integrated $\mathcal{D}(\epsilon)$
must be equal to the number of electrons per atom, narrowing of the DOS implies
an increase of the $\mathcal{D}(\epsilon)$ [see Fig. \ref{chain}(d)] and, therefore,
an increase of the spin susceptibility, which is proportional to
$\mathcal{D}(\epsilon_F)$.  In Fig. \ref{chain}(c) we show how the $\mathcal{D}(\epsilon_F)$
of the PM chain increases as a function of the lattice constant. The remarkable
feature of  Pt chains is that the Stoner instability occurs close to the
equilibrium lattice spacing.

The electronic structure of the ideal Pt chain sheds some light on the
electronic structure of the nanocontact.
In Fig. \ref{fig2} we show the energy bands for the ideal Pt chain both in the
FM (left panel) and PM (right panel) configurations, 
for a lattice spacing $a=2.8$\r{A}. We notice that the spectrum at $k=0$
has four resolved energy levels per spin. These correspond to the $6s$ level, and
the 5$d$ levels which, because of the axial potential created by the
neighboring atoms,  split into 2 doublets $E_1$, $E_2$ and one singlet
$A_1$. The $E_1$ and $E_2$ are linear combinations  of  orbitals with 
$L_z=\pm 1$ and $L_z=\pm 2$,
respectively, whereas the $A_1$ singlet is a $L_z=0$ orbital that hybridizes
with the lower energy $6s$ orbital. The largest contribution to the DOS, 
and therefore to the magnetic instability, comes from the $A_1$-like band at the edge
of the Brillouin zone.  However, the prominent
role played by these bands in the magnetic behavior of Pt chains is in stark contrast 
with their role on the transport properties of Pt nanocontacts 
(see below and see also related work on Ni nanocontacts\cite{Jacob:prb:05}). 
Four spin-degenerate bands cross the Fermi energy in the
PM case whereas  7 spin-split bands do it in the FM chain.  In the
FM chains the number of spin minority channels is 6, and the
number of spin majority channels is 1. Although spin-orbit interaction modifies 
significantly the bands \cite{Delin:prb:03},
the number of bands at the Fermi energy is pretty similar in both cases. 
Therefore, one can anticipate that 
the number of open channels in the magnetic and
non-magnetic Pt nanocontacts studied below should be roughly
the same and thereby the conductance should be similar, 
although the spin polarization might well be large in the former case. 
The conductance of the ideal FM chain is $3.5 G_0$, very far
from the value of  0.5$G_0$ that allegedly signals the emergence of magnetism
and also
far away from half of the conductance of the PM chain, so it is very
unlikely that the celebrated half quantum can be attributed exclusively 
to magnetism.

Real  Pt chains are typically less than five atoms long and are connected to
bulk electrodes. Although not surprising, we have verified that magnetism
survives in isolated short chains with $N_A=$ 3, 4 and 5 Pt atoms.  
The equilibrium distance is 2.4\r{A} for all  $N_A=3$, $N_A=4$, and $N_A=5$.
Interestingly, the short chains are always magnetic in the $N_A=3$ and $N_A=4$
cases and show a non-magnetic to magnetic crossover at $a=2.6$\r{A} in the
$N_A=5$ case, already similar to the ideal infinite chain. The total  magnetic moment 
of all the $N_A=3$ chains with $a<3.0 $\r{A} is 4$\mu_{\rm B}$. The outer atoms have a
magnetic moment of $1.36\mu_{\rm B}$ and the central atom with larger coordination has a
smaller magnetic moment of $1.29\mu_{\rm B}$. In the case of $N_A=4$ the total 
magnetic moment is 6$\mu_B$, and their distribution is similar to the $N_A=3$ case.

\section{Transport through suspended Pt chains}

\begin{figure}
  \begin{center}
    \includegraphics[width=0.9\linewidth]{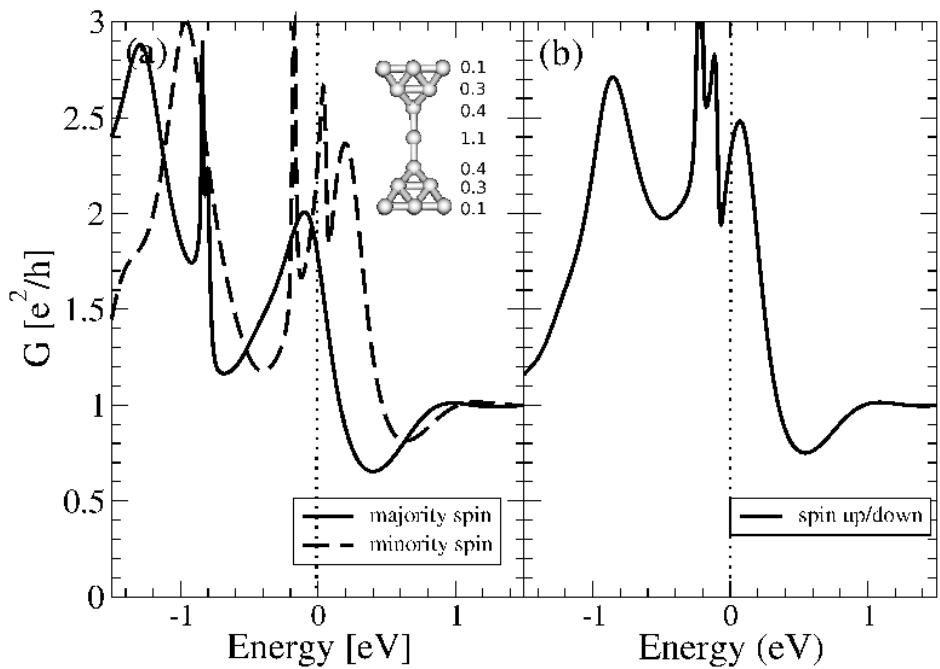}
  \end{center}
  \setcapindent{0cm}   
  \caption{ 
    Conductance per spin channel for a nanocontact with a 
    3-atom Pt chain (see inset) for the magnetic solution (a) 
    and the non-magnetic one (b). 
    The atom-atom distance in the chain is $2.8$\r{A}.} 
  \label{Pt-nanocontact} 
\end{figure}

The calculations above show that magnetism is present both in finite- and
infinite-sized Pt systems with small atomic coordination. It remains to be seen
that this holds true in nanocontacts where none or only few atoms have a small
coordination (like in the case of formation of short chains),  but these are strongly
coupled  to the bulk.  In order to verify  whether or not this is the case, we
have calculated  both electronic structure and transport for a model Pt
nanocontact. It is formed by two opposite pyramids grown in the (001)
crystallographic orientation of bulk Pt and joined by one atom which presents
the lowest possible coordination [see inset in Fig. \ref{Pt-nanocontact}(a)].
Relaxation of the 11 inner atoms of the cluster has been  performed starting
from an equilibrium situation as a function of the distance $d$ of the outer
planes. Zig-zag configurations appear in the chain for small values of $d$ (not
shown)\cite{Garcia-Suarez:prl:05} until the three-atom chain straightens up (see inset in Fig.
\ref{Pt-nanocontact}(a)] followed by a plastic deformation (not shown). This
deformation can be in the form of a rupture or a precursor of the addition of a
new atom to the chain which comes from one of the two 4-atom
bases\cite{Bahn:prl:01}. 

We now compute the transmission before the plastic deformation occurs (left
panel in Fig. \ref{Pt-nanocontact}), where the atom-atom  distance in the short
chain  is 2.82 \r{A}.  The value of the corresponding
atomic-plane-averaged  magnetic moments are also shown in the inset. As
expected, it decreases for atoms in the bulk as the coordination reaches the
bulk value.
Some atomic realizations in the stretching process (like zig-zag ones) 
result in nanocontacts with smaller
Pt-Pt distance and no magnetism, in agreement with the infinite chain phase
diagram in Fig. 1.  
In contrast to the infinite chain, there are only  three channels contributing
to the total conductance for minority electrons and there are more than one
(three) for  majority ones. For the majority electrons these are a perfectly
transmitting $s$-type channel and two partially open ($T=0.4$) $pd$-type
channels (one $p_xd_{xz}$- and one $p_yd_{yz}$- hybridized). The three minority
channels have the same character as the majority channels, except that  here
the $s$-type  channel does not transmit perfectly while the transmission of two
$pd$-type channels is enhanced so that all three minority channels have a
transmission around $0.7$. The other two remaining $pd$-like channels are
responsible for the sharp resonances that appear around the Fermi level. The
total conductance of the nanocontact in the FM case thus turns out
to be around $4 e^2/h = 2 G_0$  which is only slightly larger than the average
experimental value corresponding to the last plateau  (1.75$G_0$), but,
interestingly, barely differs from the value obtained when the possibility of
magnetic  order is ignored [around $2.3 G_0$, see right panel in Fig.
\ref{Pt-nanocontact}].  As in the case of the ideal chain, the FM
conductance is not half  of the PM conductance nor half of $G_0$. To
conclude this discussion we notice that although transport is only weakly spin
polarized, magnetism brings the $pd$-like resonances up to the Fermi level
compared to the non-magnetic case. These resonances may well give features in
the low bias conductance not present if Pt were not magnetic.

\section{Conclusions and Discussion}

The main conclusion of this part is that density functional calculations predict that 
nanochains formed in Pt nanocontacts can be stretched as to become magnetic.
The magnetic moment is localized mainly in the atoms with small coordination and
does not modify appreciably the total conductance, although the transmission is moderately
spin polarized.  How robust are these results? 
It is well known that both   local and gradient-corrected density functionals 
present some degree of  electronic self-interaction, in contrast with the
Hartree-Fock approximation. Self-interaction is larger for localized electrons
and shifts the $d$ bands upwards in energy, as shown  in the case of Co,
Ni and Pd one dimensional chains\cite{Wierzbowska:prb:05}. A number of schemes
to avoid this problem, like LDA+U  and Self-Interaction Correction functionals
have been proposed.  The method of choice between chemists is hybrid
functionals\cite{Becke:jcp:93} in which local and Hartree-Fock exchange (HFX) are
combined and the self-interaction is reduced.   We have calculated  the
magnetic phase diagram of the one dimensional  Pt chain using the hybrid B3LYP
functional\cite{Becke:jcp:93} and found, somewhat expectedly, that magnetism is
enhanced and that B3LYP infinite Pt chains are ferromagnetic down to the
equilibrium distance ($a=2.4$\r{A}). Both non-local exchange and 
spin-orbit coupling \cite{Delin:prb:03} enhance the stability of 
magnetism in Pt nanocontacts.  This and previous results on Ni
nanocontacts\cite{Jacob:prb:05} lead us to believe that self-interaction is an
issue in the electronic structure and transport properties of transition metal
nanocontacts and further work is necessary along these
lines\cite{Ferretti:prl:05}. 

The  mean field picture of the electronic structure describes a static  magnetic moment 
without preferred spatial direction. 
In reality the nanomagnet formed in the break junction is
 exchanged coupled dynamically to the Fermi
sea of the conduction electrons of the electrodes. In the case of a spin
$S=1/2$ this can result in the formation of a Kondo singlet that would yield
an anomaly in the zero bias conductance.  For larger spins, the
conduction electron sea cannot screen the spin completely so that the magnetic
moment survives. The magnetic moment of the 
nanocontact in Fig. 3 is  $S\approx 6$,
% 1.1 + 2* 0.4 + 8*0.3 + 18*0.1 = 1.1+ 0.8 + 2.4 +1.8 = 6.1
comparable to that of single molecule magnets
\cite{Caneschi:jacs:91}, making the formation of a Kondo singlet unlikely. 

In the absence of spin-orbit interactions and external magnetic field a
electronic configuration with total spin $S$ has $2S+1$ degenerate
configurations corresponding to the spin pointing along different directions.  
However, spin-orbit interaction is strong in Pt and produces  spin anisotropy, 
favoring orientation along the  transport direction axis in the case of one
dimensional chains\cite{Delin:prb:03}.  Thermal fluctuations of the magnetic
moment between these two configurations are quenched for  temperatures smaller
than the  anisotropy  barrier.  Departures from the easy axis orientation
will be  damped via electron-hole pair creation across the Fermi energy that
would also result in  small bias features in transport\cite{Untiedt}. The
application of a sufficiently strong magnetic field in the direction
perpendicular to the easy axis moves the  local magnetic moments away from
their easy axis. This is known to change the number of open channels at the
Fermi energy in both the case of Ni ideal chains\cite{Velev:prl:05} and in the
case of ferromagnetic semiconductor tunnel junctions \cite{Brey:apl:04}.  This
effect,  or maybe even larger, can be expected in Pt nanocontacts and  could be
used to detect the nanomagnetism experimentally.

%%% Local Variables: 
%%% mode: latex
%%% TeX-master: t
%%% End: 

%%%%%%%%%%%%%%%%%%%%%%%%%%%%%%%%%%%%%%%%%%%%%

%%%%%%%%%%%%%%%%%%%%%%%%%%%%%%%%%%%%%%%%%%%%%
%% Chapter 8 - The effect of spin-orbit    %%
%%             coupling                    %%
%%%%%%%%%%%%%%%%%%%%%%%%%%%%%%%%%%%%%%%%%%%%%
%%\include{spin-orbit}
%%%%%%%%%%%%%%%%%%%%%%%%%%%%%%%%%%%%%%%%%%%%%

%%%%%%%%%%%%%%%%%%%%%%%%%%%%%%%%%%%%%%%%%%%%%
%% Chapter 9  - Conclusions and outlook    %%
%%%%%%%%%%%%%%%%%%%%%%%%%%%%%%%%%%%%%%%%%%%%%
\chapter{Summary and Outlook}
\label{ch:conclusions}

\section{Overview of developed computational tools}

In this thesis, I have studied spin transport through nanocontacts and nanowires,
both with calculations of simplified models and with full {\it ab initio}
calculations based on density functional theory (DFT). To this end various computational
tools have been developed either from scratch or by extension of already existing
code:

i) {\bf SpinTrans}: This program for computing spin transport in simple models with 
Coulomb interaction (e.g. Hubbard model) in the {\it non-collinear} unrestricted 
Hartree-Fock approximation (NC-UHF), see also App.\ref{app:NC-UHF} has been developed 
from scratch in the first stage of the thesis. This program was applied for the 
self-consistent calculation of toy models of magnetic nanostructures to study the 
effect of non-collinear magnetization profiles on the transport presented in Ch. 
\ref{ch:spin-transport}.

ii) {\bf ALACANT}: The ALACANT (ALicante Ab-initio Computation Applied to Nano Transport) 
project was started in the year 2000 by professor J. J. Palacios in the Applied Physics 
department of the University of Alicante. It implements the NEGF formalism in connection
with {\it ab initio} electronic structure calculations based on DFT, as explained in 
Ch.\ref{ch:ab-initio}. As a part of this thesis, the ALACANT package has been further 
developed in various aspects:
\begin{itemize}
\item The spin-unrestricted transport formalism was implemented into the package
\item The entire program code which had been written in FORTRAN77 has been ported
  to FORTRAN90 in order to make use of modern programming techniques like the use
  of modules, dynamic memory allocation etc.
\item A module for calculating self-energies for one-dimensional electrodes described 
  by Hamiltonian and overlaps matrices which can be taken from ab-initio calculations
  has been implemented. 
\item An interface to the CRYSTAL {\it ab initio} program for crystalline systems 
  has been developed.
\end{itemize}

\section{Summary of results}

With the developed computational tools various calculations of the electronic and magnetic 
structure and the transport properties of nanocontacts and nanowires have been performed.
First, we have studied spin transport with simple models, namely a one-dimensional Hubbard 
chain in the ferromagnetic phase, in order to gain an understanding of the fundamental 
mechanisms of nanoscale spin transport. Neglecting the geometric aspects as well as the 
much more complex Hamiltonian and Hilbert space of real nanocontacts allowed to concentrate 
on the pure spin aspects of electron scattering by the magnetic structure. We find that
domain walls form self-consistently under the appropriate magnetic boundary conditions.
Furthermore, we find that the longer the atomic chain the smoother the domain wall becomes,
and the scattering by the domain wall is reduced since the spin-mixing transforms the
spin of an incoming electron adiabatically into the opposite spin. Vice versa, for short 
necked nanocontacts (i.e. no chain formation) like is typical for Ni, the domain walls will 
be rather sharp, and thus the spin scattering by a domain wall formed in the neck of the 
nanocontact becomes maximal.

In order to investigate the question of the possibly huge BMR values in Ni nanocontacts 
we have performed {\it ab initio} calculations of the electronic structure and transport 
properties of Ni nanocontacts on the level of DFT. By comparing solutions 
with and without formation of a domain wall in the atomic neck of the nanocontact, we find
that BMR is certainly not large in pure Ni nanocontacts, but rather moderate. This is due to the 
fact that a grand part of the a priori available spin-polarized channels which could in principle 
give rise to a large BMR are blocked by the geometry of a real nanocontact. On the other hand the 
spin-unpolarized $s$-type channel which does not give rise to any BMR is not affected by the 
geometry. Thus even in the DW configuration the conductance is appreciable ($\approx 2e^2/h$
compared to the $<4e^2$ for the ferromagnetic solution). Another important conclusion that 
we can draw from our calculations is that disorder of the atoms in the nanocontact strongly 
reduces the spin-polarization of the current since it has a stronger effect on the transmission 
of the spin-polarized $d$-type channels than on the unpolarized $s$-type channels. Moderate
BMR values for ferromagnetic nanocontacts has been confirmed recently by experiments with 
very clean samples under controlled conditions (ultra-high vacuum conditions,exclusion of any
magnetostriction effects) \cite{Keane:apl:06,Bolotin:nl:06}, and also by a number of 
theoretical papers\cite{Smogunov:prb:06}.

Next, we have studied the electronic structure and transport properties of one-di\-mensional NiO 
chains, both ideal infinite ones, and short ones suspended between Ni nanocontacts. Indeed, it 
has been observed experimentally, that Ag which like Ni does not form atomic chains in a nanocontact, 
actually does form chains when oxygen is present \cite{Thijssen:prl:06}, and it has been argued
that the incorporation of oxygen atoms into the a chain actually stabilizes the chain 
\cite{Novaes:prl:06}. In fact we find that NiO chains bridging the tip atoms of a Ni nanocontact 
are stable, and so NiO chains could in principle form when nanocontacts are fabricated in an oxygen 
atmosphere. Most importantly, we have found that short NiO chains can actually become almost  perfect 
half-metallic conductors when suspended between Ni nanocontacts, i.e. the current becomes almost 
100\% spin-polarized. In fact, already a single oxygen atom bridging the tips of a Ni nanocontact 
can have such an effect. Consequently, this leads to a huge BMR of the order of the GMR effect, and 
could perhaps explain to some extent the large MR values obtained in some experiments 
\cite{Garcia:prl:99,Sullivan:prb:05}. Another very important conclusion we can draw is that 
already a single atom can completely change the conductance behavior of an atomic-scale device.

The emergence of magnetism at the nanoscale in materials that are otherwise paramagnetic in bulk
is a fascinating topic of Nano science, and has been observed for example in Au nanoclusters 
\cite{Crespo:prl:04}, and predicted for atomic chains of Pt and Pd with {\it ab initio}
electronic structure calculations \cite{Delin:ss:04,Delin:jpcm:04}. In order to see whether 
conduction measurements of Pt nanocontacts could possibly reveal the magnetism of an atomic
chain formed between the tip atoms of the nanocontact, we have performed {\it ab initio}
electronic structure and transport calculations of atomic Pt chains suspended between the tips
of a Pt nanocontact. We find that although suspended Pt chains are also magnetic, the conductance
is very similar to the case when there would be no magnetism in the chain. Thus it seems not possible 
to decide experimentally whether Pt chains become magnetic or not by simple conductance measurements
of Pt nanocontacts.

Furthermore, we can draw the following general conclusions:

i) The DFT based transport approach on the LDA level at small bias voltages seems to work reasonably 
well for metallic nanocontacts and nanowires, even when $d$-electrons are involved in the conduction. 
This is in agreement with theoretical studies which show that DFT calculations on the LDA or GGA level 
give a reasonable description of the electronic structure of bulk metals and metallic surfaces 
\cite{Moroni:prb:97,Doll:ss:03}, and corrections by methods like the GW approximation or DMFT which 
take into account electron correlations are often relatively small in the vicinity of the Fermi level 
\cite{Grechnev:06,Belashchenko:prb:06}. However, when going away from the purely metallic systems 
things become more complicated. NiO for example is a strongly correlated material, and both LDA and 
GGA fail in describing its electronic structure correctly. Also the problems of LDA/GGA of describing 
the electronic structure of molecules are notorious. Thus it must be doubted whether LDA or GGA 
can give a good description of nanoscale NiO e.g. oxidized Ni nanocontacts or one-dimensional NiO chains,
or molecular conductors. On the other hand the semi-empirical DFT methods (e.g. hybrid functionals or 
LDA+U) often describe quite well the electronic structure of some bulk oxides with strongly correlated 
electron systems, and especially hybrid functionals are extremely successful in describing molecules.
However, these methods require an adjustment of the empirical parameters by fitting with experimental
results, which limits their predictive power. Also hybrid functionals often do not yield a satisfying
description of metallic systems. Thus the electronic structure of a molecule in a molecular electronics 
device probably needs to be described by a hybrid functional while the metal electrodes for contacting 
the molecule are not well described by hybrid functionals but needs to be described on the LDA or GGA
level.
 
ii) Magnetic nanocontacts and nanowires are quite sensitive to the atomic structure, i.e. distortions and 
defects often have a strong effect on the transport properties. This can be understood by the fact that 
the $d$-orbitals which are in fact responsible for the magnetism are very sensitive with respect to the 
geometry and thus $d$-electrons are easily scattered. On the other hand, the conduction channels composed
of $d$-orbitals are the only channels that are actually spin-polarized and thus are the ones responsible
for the magneto-resistive behavior. So exactly the channels that are interesting in the context of spintronics
applications are not stable against variation of their atomic structures. Thus it is arguable whether
magnetic atomic-size nanocontacts could one day serve as ingredients for nanoscale electronics. But rather 
they allow us to gain a fundamental understanding of conduction and the interplay with other physical 
processes at the atomic scale, and to develop an appropriate methodology for the description of nanoscale 
conductors. This might well proof essential for the development of new electronic devices in the near future.

%%3) Thick magnetic nanowires of several atoms in diameter could perhaps
%%Nevertheless the research conducted in atomic-scale conductors is essential

% \begin{itemize}
% \item DFT based transport seems to work quite well for metallic systems. More
%   tests needed. But not so good for molecule or correlated materials.
% \item Magnetic nanocontacts very sensitive to distortions, defects, adsorbate 
%   (a single oxygen atom makes the difference): Opportunity or Challenge?
% \item More stable systems needed nanoscale electronics: Thick nanowires?
% \item Sensitivity to adsorbates: problem for reliable electronics. 
%   But maybe good for sensors?
% \end{itemize}

\section{Outlook}

For the results presented in this thesis the spin-orbit coupling of the electrons has been
neglected. This can be justified by the fact that the spin-orbit coupling is relatively 
small for not too heavy elements like Ni. For heavier elements like Pt on the other hand the
spin-orbit coupling becomes comparable to the other energies and thus probably affects the 
electronic and magnetic structure of the material: As pointed out at the end of Ch. 
\ref{ch:Pt-nanowires} the spin-orbit interaction introduces spin-anisotropy favoring the 
magnetization of the Pt chain to be oriented along the transport axis. Furthermore the 
spin-orbit coupling favors the onset of magnetism in atomic Pt chains as has been shown
by {\em ab initio} calculations including the spin-orbit coupling 
\cite{Delin:prb:03,Nautiyal:prb:04}. 

By applying a sufficiently strong magnetic field, the orientation of the magnetization of the 
chain atoms can be changed from the one favored by the spin-orbit coupling called the 
{\em easy axis}. This however affects the electronic structure of the atomic chain due to 
the coupling of the orbital degree of freedom to the spin degree of freedom and possibly 
alters the transport properties of the system resulting in {\em anisotropic}
magneto-resistance (AMR) defined as the maximal change in resistance when rotating the magnetic field
\cite{Velev:prl:05}. Despite the relatively small spin-orbit coupling of Ni and Fe in the order of 50meV, 
recent experiments have measured an appreciable AMR of 10-15\% in the case of Ni nanocontacts 
\cite{Keane:apl:06,Bolotin:prl:06} and huge AMR of up to 75\% in the case of Fe nanocontacts
\cite{Viret:epb:06}.

We have therefore implemented spin-orbit coupling into our {\em ab initio} transport program
ALACANT and obtained first results for one-dimensional Ni chains with a scattering center as 
simplified models of real Ni nanocontacts. The calculated AMR values are similar to the 
experimentally measured AMR values. However, further investigations with more realistic geometries
are needed in order to compare with the experiments as the spin-orbit coupling solely affects the
$d$-electrons which are very sensitive to geometrical effects as has been discussed extensively
in this thesis. Due to the strong spin-orbit coupling of Pt one might also expect that the magnetism
of atomic Pt chains should give rise to a strong AMR signal which would thus give experimental evidence 
for the emergence of magnetism in atomic Pt chains. Thus it could be worthwhile to calculate the AMR
of Pt nanocontacts from {\em ab initio} quantum transport calculations including the spin-orbit
coupling.

An important problem for future nanoscale spintronics applications is the injection of 
spin-polarized currents from magnetic into non-magnetic nanoscopic conductors like paramagnetic 
nanowires, carbon nanotubes and organic molecules. The possibility of spin-injection into
a non-magnetic material will very likely depend strongly on the atomic-scale properties of the 
contact or between the magnetic and non-magnetic material. It is thus of fundamental importance 
to study interfaces between magnetic and non-magnetic nanoscale conductors. Related to the 
question of spin injection is the problem of spin relaxation which designates the phenomenon 
that the spin-polarization of a current through a non-magnetic material decreases due to the 
precession of the electron spin caused by the spin-orbit coupling. Studying spin injection and
spin relaxation in nanowires, nanotubes and organic molecules with {\em ab initio} quantum 
transport calculations thus defines an interesting future line of work.

%%of some 10-15\% in the contact or ballistic regime 
%%(BAMR) and of up to 25\% in the tunneling regime (TAMR) and in the case of Fe nanocontacts huge BAMR of up to 75\% 
%%and huge TAMR of up to 100\% 
%%which is much higher than the AMR of bulk
%%ferromagnets which is usually below 1\%.

% Outlook:
% \begin{itemize}
% \item Study spin injection into carbon systems: CNTs, graphene [experiments].
% \item Effect of surface adsorbates on conductance of nanoscopic conductors.
% \item Interfaces between different materials (MTJs and GMR spinvalves).
% \item Extend DFT based transport to take into account electron correlations.
%   GW, CI or perturbative quantum chemistry methods, DMFT.
% \item Noncollinear spin, Spin-Orbit for ab-initio computation.
% \end{itemize}
%%% Local Variables: 
%%% mode: latex
%%% TeX-master: t
%%% End: 

%%%%%%%%%%%%%%%%%%%%%%%%%%%%%%%%%%%%%%%%%%%%%

%%%%%%%%%%%%%%%%%%%%%%%%%%%%%%%%%%%%%%%%%%%%%
%% Appendix                                %%
%%%%%%%%%%%%%%%%%%%%%%%%%%%%%%%%%%%%%%%%%%%%%

\begin{appendix}

\chapter{Representation of operators in non-orthogonal basis sets}
%% and orthogonalization}
\label{app:NOBS}

The natural definition for the matrix $\M A$ of a one-body operator $\Op A$
in a non-orthogonal basis set (NOBS) $\left\{\ket{\alpha}\right\}$ is simply by its matrix elements:
\begin{equation}
  \label{eq:NOB-Mat}
  \M A = ( A_{\alpha\beta} ) = \left( \bra{\alpha} \Op A \ket{\beta} \right).
\end{equation}
However, the representation of an operator in a NOBS is not that simple:
\begin{equation}
  \label{eq:NOB-Op}
  \Op A = \sum_{\alpha,\beta} \ket{\alpha} (\M{S}^{-1} \M{A} \M{S}^{-1} )_{\alpha\beta} \bra{\beta}
\end{equation}
where $\M{S}=(S_{\alpha\beta})=\left(\bra{\alpha}\ket{\beta}\right)$ is the overlap matrix for the 
basis set. It is easy to see that this definition leads results in the matrix elements $A_{\alpha\beta}$ 
defined above. Then the identity operator in the NOBS representation is given by
\begin{equation}
  \label{eq:NOB-id}
  \Op{\mathbb 1} = \sum_{\alpha,\beta} \ket{\alpha} (\M{S}^{-1})_{\alpha\beta} \bra{\beta},
\end{equation}
which is also easy to proof.

Now we define a second matrix 
\begin{equation}
  \label{eq:MatTilde}
  \tilde{\M{A}} := \M{S}^{-1} \M{A} \M{S}^{-1},
\end{equation}
which is the matrix that appears above in the representation of the operator in a NOBS. 

One should take care when using the representation of an operator in a NOBS. For example, the matrix element
of an operator between two non-orthogonal orbitals $\ket{\alpha}, \ket{\beta}$ can be zero, 
$\bra{\alpha}\Op{A}\ket{\beta}=0$, but the corresponding matrix element of the matrix $\tilde{\M{A}}$ does 
not necessarily vanish due to the multiplication with the inverse of the overlap matrix on both sides. 
Thus there is actually a non-zero contribution of the two orbitals to the operator although the corresponding 
matrix element of the operator is zero 

Orthogonalizing the basis set by the L\"owdin orthogonalization scheme \cite{Szabo:book:89}, the 
matrices $\M{A}$ and $\tilde{\M{A}}$ are transformed to the matrix $\M{A}^\perp=\left(\bra{i}\Op{A}\ket{j}\right)$ 
defined in the new orthogonal basis set $\left\{\ket{i}\right\}$ according to:
\begin{equation}
  \M{A}^\perp = \M{S}^{-1/2} \M{A} \M{S}^{-1/2} = \M{S}^{+1/2} \tilde{\M{A}} \M{S}^{+1/2}.
\end{equation}
Though there are also other orthogonalization schemes, the L\"owdin scheme is particularly useful in the
context of quantum chemistry methods based on atomic orbitals as the center of the orthogonalized orbital 
remains centered on the same atom as the original non-orthogonal orbital.

\chapter{Partitioning method}
\label{app:partitioning}

As explained in Ch. \ref{ch:transport} we model the transport 
problem by dividing the system in three
parts. Two semi-infinite leads (L) and (R) with bulk electronic structure
are connected to a finite region called device (D).
In a local basis set the Hamiltonian and the overlap matrix of the system are given
by (\ref{eq:HLDR}) and (\ref{eq:SLDR}). Dividing the F matrix into sub-matrices 
in a similar manner we obtain the following matrix equation:
\begin{eqnarray}
  \label{eq:GTilde-Matrix-eq}
  \lefteqn{
    \left(
      \begin{array}{ccc}
        z\,{\bf S}_{L} -{\bf H}_{L}  & z\,{\bf S}_{LD}-{\bf H}_{LD} & {\bf 0}_{RL}                 \\
        z\,{\bf S}_{DL}-{\bf H}_{DL} & z\,{\bf S}_{D} -{\bf H}_{D}  & z\,{\bf S}_{DR}-{\bf H}_{DR} \\
        {\bf 0}_{RL}                 & z\,{\bf S}_{RD}-{\bf H}_{RD} & z\,{\bf H}_{R} -{\bf H}_{R}     
      \end{array}
    \right) \times }
  \nonumber \\
  \nonumber \\
  & & \hspace{2.2cm} \times \left(
    \begin{array}{ccc}
      {\bf \widetilde{G}}_{L}(z)  & {\bf \widetilde{G}}_{LD}(z) & {\bf \widetilde{G}}_{LR}(z) \\
      {\bf \widetilde{G}}_{DL}(z) & {\bf \widetilde{G}}_{D}(z)  & {\bf \widetilde{G}}_{DR}(z) \\
      {\bf \widetilde{G}}_{RL}(z) & {\bf \widetilde{G}}_{RD}(z) & {\bf \widetilde{G}}_{R}(z)       
    \end{array}
  \right) 
  = \left(
  \begin{array}{ccc}
    \M 1_L    & \M 0_{LD} & \M 0_{LR} \\
    \M 0_{DL} & \M 1_D    & \M 0_{DR} \\
    \M 0_{RL} & \M 0_{RD} & \M 1_R
  \end{array}
\right).   
\nonumber\\
\end{eqnarray}
This yields 9 equations for the 9 sub-matrices of the GF $\widetilde{\M G}$. We can resolve this matrix
equation columnwise. Multiplying all rows of $E\M S - \M H$ with the first column of $\widetilde{\M G}$ 
yields three equations for $\widetilde{\M G}_{L}$, $\widetilde{\M G}_{DL}$ and $\widetilde{\M G}_{RL}$ 
which yield:
\begin{eqnarray}
  \label{eq:gf:GTilde_L}
  \widetilde{\M G}_{L}(z) &=& ( z \M S_{\rm L} - \M H_{L} - \widetilde{\M\Sigma}_{D+R}(z) )^{-1} \\
  \label{eq:gf:GTilde_DL}
  \widetilde{\M G}_{DL}(z) &=& \widetilde{\M g}_{D+R}(z) \, (\M H_{DL}-z \M S_{DL}) \, \widetilde{\M G}_{L}(z) \\
  \label{eq:gf:GTilde_RL}
  \widetilde{\M G}_{RL}(z) &=& \widetilde{\M g}_{R}(z) \, ({\M H}_{RD}-z {\M S}_{RD}) \, \widetilde{\M G}_{DL}(z)
\end{eqnarray}
Similarly we obtain from multiplication with the second column:
\begin{eqnarray}
  \label{eq:gf:GTilde_D}
  \widetilde{\M G}_{D}(z) &=& ( z {\M S}_{D} - {\M H}_{D} - \widetilde{\M\Sigma}_{L}(z) - \widetilde{\M\Sigma}_{R}(z) )^{-1} \\
  \label{eq:gf:GTilde_LD}
  \widetilde{\M G}_{LD}(z) &=& \widetilde{\M g}_{L}(z) \, (\M H_{LD}- z \M S_{LD}) \, \widetilde{\M G}_{D}(z) \\
  \label{eq:gf:GTilde_RD}
  \widetilde{\M G}_{RD}(z) &=& \widetilde{\M g}_{R}(z) \, (\M H_{RD}- z \M S_{RD}) \, \widetilde{\M G}_{D}(z)
\end{eqnarray}
And finally from multiplication with the third column, we obtain:
\begin{eqnarray}
  \label{eq:gf:GTilde_R}
  \widetilde{\M G}_{R}(z) &=& ( z \M S_R - \M H_{R} - \widetilde{\M\Sigma}_{D+L}(z) )^{-1} \\
  \label{eq:gf:GTilde_DR}
  \widetilde{\M G}_{DR}(z) &=& \widetilde{\M g}_{D+L}(z) \, (\M H_{DR}-z \M S_{DR}) \, \widetilde{\M G}_{R}(z) \\
  \label{eq:gf:GTilde_LR}
  \widetilde{\M G}_{LR}(z) &=& \widetilde{\M g}_{L}(z) \, (\M H_{LD}-z \M S_{LD}) \, \widetilde{\M G}_{DR}(z)
\end{eqnarray}

We have introduced the Green's functions of the isolated left and right lead $\widetilde{\M g}_{L}$ and 
$\widetilde{\M g}_{R}$ and the corresponding self-energies $\widetilde\Sigma_{L}$ and $\widetilde\Sigma_{R}$:
\begin{eqnarray}
  \label{eq:gf:gTilde_L}
  \widetilde{\M g}_{L}(z) &:=& (z \M S_{L} - \M H_{L})^{-1} \\ 
  \label{eq:gf:SigmaTilde_L}
  \widetilde{\M\Sigma}_{L}(z) &:=& (\M H_{DL}-z \M S_{DL}) \, \widetilde{\M g}_{L}(z) \, (\M H_{LD}-z \M S_{LD}) \\
  \label{eq:gf:gTilde_R}
  \widetilde{\M g}_{R}(z) &:=& (z \M S_{R} - \M H_{R})^{-1} \\
  \label{eq:gf:SigmaTilde_R}
  \widetilde{\M\Sigma}_{R}(z) &:=& (\M H_{DR}-z \M S_{DR}) \, \widetilde{\M g}_{R}(z) \, (\M H_{RD}-z \M S_{RD})
\end{eqnarray}

Furthermore, we have defined the Green's function of the device plus the left lead only, $\widetilde{\M g}_{D+L}$,
of the device plus the right lead only, $\widetilde{\M g}_{D+R}$, and the corresponding self-energies 
$\widetilde{\M\Sigma}_{D+L}$ and $\widetilde{\M\Sigma}_{D+R}$ each one representing the coupling of one of the leads to 
the device and the other lead:
\begin{eqnarray}
  \label{g_D+L}
  \widetilde{\M g}_{D+L}(z) &:=& (z \M S_{D} - \M H_{D} - \widetilde\Sigma_{L}(z) )^{-1} \\
  \label{g_D+R}
  \widetilde{\M g}_{D+R}(z) &:=& (z \M S_{D} - \M H_{D} - \widetilde\Sigma_{R}(z) )^{-1} \\
  \label{eq:Sigma_D+L}
  \widetilde{\M\Sigma}_{D+R}(z) &:=& (\M H_{RD}-z \M S_{RD}) \widetilde{\M g}_{D+L}(z) (\M H_{DR}-z \M S_{DR}) \\
  \label{eq:Sigma_D+R}
  \widetilde{\M\Sigma}_{D+L}(z) &:=& (\M H_{LD}-z \M S_{LD}) \widetilde{\M g}_{D+R}(z) (\M H_{DL}-z \M S_{DL})
\end{eqnarray}

\chapter{Self-energy of a one-dimensional lead}
\label{app:self-energy-1D}

Here we will derive the Dyson equation (\ref{eq:DysonR}) for the calculation of
the self-energy of the semi-infinite right lead. The derivation of the Dyson equation
for the left lead (\ref{eq:DysonL}) goes in a completely analogous way.

The Hamiltonian matrix $\M H_R$ of the (isolated) semi-infinite right electrode is defined in eq.
(\ref{eq:H_R}) as:
\begin{equation}
  \M{H}_R 
  = \begin{pmatrix}
    \M H_0         & \M H_1         &        &        & \M 0   \\
    \M H_1^\dagger & \M H_0         & \M H_1 &        &        \\
    \,             & \M H_1^\dagger & \M H_0 & \M H_1 &        \\
    \M 0           &                & \ddots & \ddots & \ddots
    \end{pmatrix}.
\end{equation}
and the overlap matrix is given in eq. (\ref{eq:S_R}) as:
\begin{equation}
  \label{S_R}
  \M{S}_R 
  = \begin{pmatrix}
    \M S_0         & \M S_1         &        &        & \M 0   \\
    \M S_1^\dagger & \M S_0         & \M S_1 &        &        \\
    \,             & \M S_1^\dagger & \M S_0 & \M S_1 &        \\
    \M 0           &                & \ddots & \ddots & \ddots
    \end{pmatrix}
\end{equation}
To obtain the self-energy of the lead we have to calculate the GF
of the lead from its defining equation:
\begin{equation}
  ( z \M S_R - \M H_R ) \widetilde{\M g}_R(z) = \M 1.
\end{equation}
In the same way as the Hamiltonian and the overlap matrix we subdivide the GF matrix $\widetilde{\M g}_R$ 
into sub-matrices corresponding to the unit cells of the lead. Now the above equation for the right lead's
GF reads:
\begin{eqnarray}
  \begin{pmatrix}
    z \M S_0 - \M H_0                 & z \M S_1 - \M H_1                 &                      &                      \\
    z \M S_1^\dagger - \M H_1^\dagger & z \M S_0 - \M H_0                 & z \M S_1 - \M H_1    &                      \\
%%    \,                                & z \M S_1^\dagger - \M H_1^\dagger & z \M S_0 - \M H_0    & z \M S_1 - \M H_1    &      \\
%%    \M 0                              & \hspace{0.5cm}\ddots              & \hspace{0.5cm}\ddots & \hspace{0.5cm}\ddots &
    \hspace{0.5cm}\ddots              & \hspace{0.5cm}\ddots              & \hspace{0.5cm}\ddots &                      
  \end{pmatrix}  
  \left( 
    \begin{array}{ccc}
      \widetilde{\M g}_{1,1} & \widetilde{\M g}_{1,2} & \ldots \\
      \widetilde{\M g}_{2,1} & \widetilde{\M g}_{2,2} & \ldots \\
%%      \widetilde{\M g}_{3,1} & \widetilde{\M g}_{3,2} & \ldots \\
      \vdots                 & \vdots                 &
    \end{array}
  \right) 
  &=&  
  \left( 
    \begin{array}{ccc}
      \M 1   & \M 0   & \cdots \\
      \M 0   & \M 1   & \ddots \\
      \vdots & \ddots & \ddots
    \end{array}
  \right). 
  \nonumber \\
\end{eqnarray}
As explained in Sec.\ref{sec:NEGF} it suffices to calculate the ``surface'' GF, i.e. $\widetilde{\M g}_{1,1}$.
From multiplication of the 1st, the 2nd and so on until the $n$-th line of $(z \M S_R - \M H_R)$ with the 1st column of 
$\widetilde{\M g}_R(z)$ we get the following chain of equations:
\begin{eqnarray}
  \label{eq:g11}
  (z \M S_0 - \M H_0) \, \widetilde{\M g}_{1,1}(z) + (z \M S_1 - \M H_1) \, \widetilde{\M g}_{2,1}(z) &=&  \M 1 
  \\
%%  \label{eq:g21}
  (z \M S_1^\dagger - \M H_1^\dagger) \, \widetilde{\M g}_{1,1}(z) + (z \M S_0 - \M H_0) \, \widetilde{\M g}_{2,1}(z) 
  + (z \M S_1 - \M H_1) \, \widetilde{\M g}_{3,1}(z) &=& \M 0
  \\
  &\vdots& 
  \nonumber \\
%%  \label{eq:gn1}
  (z \M S_1^\dagger - \M H_1^\dagger) \, \widetilde{\M g}_{n-1,1}(z) + (z \M S_0 - \M H_0) \, \widetilde{\M g}_{n,1}(z)
  + (z \M S_1 - \M H_1) \, \widetilde{\M g}_{n+1,1}(z) &=& \M 0
\end{eqnarray}
For $n>1$ the equations for determining $\widetilde{\M g}_{n,1}(z)$ all have the same structure:
\begin{eqnarray}
   \label{eq:gn1}
  (z \M S_0 - \M H_0) \, \widetilde{\M g}_{n,1}(z) &=& 
  (\M H_1^\dagger - z \M S_1^\dagger) \, \widetilde{\M g}_{n-1,1}(z) + (\M H_1 - z \M S_1) \, \widetilde{\M g}_{n+1,1}(z).
\end{eqnarray}
%%
%%Since $\widetilde{\M g}_{n+1,1}(z)$ is 
%%%%Thus a matrix element
%%$\widetilde{\M g}_{n,1}$ of the lead's GF is completely determined by the matrix element $\widetilde{\M g}_{n-1,1}$, so that we
We define a transfer matrix for $n>1$ by:
\begin{eqnarray}
  \label{eq:T-matrix}
  \M{T}_{n-1,n}(z) \widetilde{\M g}_{n-1,1}(z) = \widetilde{\M g}_{n,1}(z).
\end{eqnarray}
The transfer matrix thus transfers information from site $n-1$ to site $n$ of 
the lead, i.e. from the left to the right. Multiplying Eq. (\ref{eq:gn1}) by 
$(\widetilde{\M{g}}_{n-1,1})^{-1}$ we obtain:
\begin{equation}
  (z \M S_0 - \M H_0) \, \M{T}_{n-1,n}(z) =
  (\M H_1^\dagger - z \M S_1^\dagger) + (\M H_1 - z \M S_1) \, \M{T}_{n,n+1}(z) \, \M{T}_{n-1,n}(z)
\end{equation}
Reordering we obtain the following iterative equation for the transfer matrices:
\begin{equation}
  \label{eq:Dyson-TMat}
  \M{T}_{n-1,n}(z) = (z \M S_0 - \M H_0 - (\M H_1 - z \M S_1) \, \M{T}_{n,n+1}(z) )^{-1}
  \, (\M H_1^\dagger - z \M S_1^\dagger) 
\end{equation}
Since the electrode is semi-infinite it looks the same from each unit cell 
when looking to the right. Thus a given $\widetilde{\M{g}}_{n-1,1}$, 
results always in the same $\widetilde{\M{g}}_{n,1}$ {\it independent} of $n$.
Thus the transfer matrix must be independent of $n$: $\M{T}_{n-1,n}(z)\equiv\M{T}(z)$,
and Eq. (\ref{eq:Dyson-TMat}) allows to determine the $\M{T}(z)$ self-consistently. 

%% %%
%% \begin{eqnarray}
%%  \M T(z) &=& \widetilde{\M g}_{n,1}(z) (\widetilde{\M g}_{n-1,1}(z))^{-1}
%%  \nonumber\\
%%  &=& (z \M S_0 - \M H_0)^{-1} \left[ (\M H_1^\dagger - z \M S_1^\dagger)  
%%    + (\M H_1 - z \M S_1) \, \widetilde{\M g}_{n+1,1}(z)(\widetilde{\M g}_{n-1,1}(z))^{-1} \right]
%%  \nonumber\\
%%  &=& (z \M S_0 - \M H_0)^{-1} \left[ (\M H_1^\dagger - z \M S_1^\dagger)  
%%    + (\M H_1 - z \M S_1) \,  \M T^2(z) \right],
%% \end{eqnarray}
%%
%% where in the last step we have made use of the fact that $\widetilde{\M g}_{n+1,1}(z) = \M T(z) \widetilde{\M g}_{n,1}(z) = \M T^2(z)
%% \widetilde{\M g}_{n-1,1}(z)$. So we have found a quadratic matrix equation for the transfer matrix which can be solved iteratively.
%% We reorder the terms to get a linear dependence in the iterative relationship:
%%
%% \begin{eqnarray}
%%  && (z \M S_0 - \M H_0 - (\M H_1 - z \M S_1) \, \M T(z) ) \, \M T(z) =  (\M H_1^\dagger - z \M S_1^\dagger)
%%  \nonumber\\
%%  \nonumber\\
%%  && \Rightarrow \M T(z) = (z \M S_0 - \M H_0 - (\M H_1 - z \M S_1) \, \M T(z) )^{-1} \, (\M H_1^\dagger - z \M S_1^\dagger)
%% \end{eqnarray}
%%

We define the self-energy as $\M\Sigma(z):=(\M H_1 - z \M S_1) \, \M T(z)$, and obtain the Dyson equation for the 
self-energy:
\begin{eqnarray}
  \M\Sigma(z) = (\M H_1 - z \M S_1) \, (z \M S_0 - \M H_0 - \M\Sigma(z) )^{-1} \, (\M H_1^\dagger - z \M S_1^\dagger).
\end{eqnarray}
We will now see that this self-energy is indeed identical to the one defined for the right lead 
in eq. (\ref{eq:Sigma}), i.e. $\M\Sigma(z)\equiv\widetilde{\M\Sigma}_r(E)$. By plugging in the 
definition of the transfer matrix, eq. (\ref{eq:T-matrix}), into eq. (\ref{eq:g11}) for determining
the surface GF, $\widetilde{\M g}_{1,1}$ we find:
\begin{eqnarray}
 (z \M S_0 - \M H_0) \, \widetilde{\M g}_{1,1}(z) + (z \M S_1 - \M H_1) \, \M T(z) \, \widetilde{\M g}_{1,1}(z)  &=&  \M 1  
 \nonumber\\
 \Rightarrow (z \M S_0 - \M H_0 + \M\Sigma(z) ) \, \widetilde{\M g}_{1,1}(z)  &=&  \M 1,
\end{eqnarray}
where in the last step we have made use of the definition of the self-energy. 
Thus we obtain for the surface GF of the right lead:
\begin{eqnarray}
  \widetilde{\M g}_{1,1}(z) &=& (z \M S_0 - \M H_0 + \M\Sigma(z) )^{-1}.
\end{eqnarray}
And vice-versa the self-energy can be expressed in terms of the surface GF:
\begin{eqnarray}
  \M\Sigma(z) = (\M H_1 - z \M S_1) \, \widetilde{\M g}_{1,1}(z) \, (\M H_1^\dagger - z \M S_1^\dagger).
\end{eqnarray}
This proofs that the self-energy $\M\Sigma(z)$ defined above in terms of the transfer matrix is identical 
to the self-energy $\widetilde{\M\Sigma}_r(z)$ defined earlier in Sec.\ref{sec:NEGF} so that
the self-energy $\widetilde{\M\Sigma}_r(z)$ can be calculated iteratively by the Dyson equation (\ref{eq:DysonR}).

The proof for the left lead runs completely analogously. The surface GF of the left lead is now:
\begin{eqnarray}
  \widetilde{\M g}_{-1,-1}(z) &=& (z \M S_0 - \M H_0 + \widetilde{\M\Sigma}_l(z) )^{-1}.
\end{eqnarray}

\chapter{Bethe lattices}
\label{app:bethe-lattices}

In this appendix we discuss how self-energies for  Bethe lattices (BL) 
used to describe the leads are calculated. A BL is generated by 
connecting a site with $N$ nearest-neighbors in directions that
could be those of a particular crystalline lattice. The new $N$ sites are 
each one connected to $N-1$ different sites and so on and so forth.
The generated lattice has the actual local topology (number of neighbors
and crystal directions) but has no rings, and thus does not describe
the  long range order characteristic of real crystals. Let 
$n$ be a generic site connected to one preceding neighbor $n-1$
and $N-1$ neighbors of the following shell ($n+i$ with $i=1,..,N-1$). 
Dyson's equation
for an arbitrary non-diagonal Green's function is
\begin{equation}
\label{eq:Dyson-BL}
  (E{\bf I}-{\bf H}_0){\bf G}_{n,k}={\bf V}_{n,n-1}{\bf G}_{n-1,k}+
  \sum_{i=1,...,N-1} {\bf V}_{n,i}{\bf G}_{i,k}
\end{equation}
where $k$ is an arbitrary site, $E$ the energy, 
and ${\bf V}_{i,j}$ is a matrix
that incorporates interactions between orbitals at sites $i$ and $j$
(bold capital characters are used to denote matrices). 
${\bf H}_0$ is a diagonal matrix containing the orbital levels
and ${\bf I}$ is the identity matrix. 
Then, we define a transfer matrix as
\begin{equation}
\label{eq:T-Mat-BL}
{\bf T}_{i-1,i}{\bf G}_{i-1,j}={\bf G}_{i,j}
\end{equation}
Multiplying Eq. (\ref{eq:Dyson-BL}) by the inverse of ${\bf G}_{n-1,n}$ we obtain,
\begin{equation}
\label{eq:Dyson-TMat-BL}
  (E{\bf I}-{\bf H}_0){\bf T}_{n-1,n}={\bf V}_{n,n-1}+\left(\sum_{i=1,...,N-1}
    {\bf V}_{n,i}{\bf T}_{n,i}\right){\bf T}_{n-1,n}
\end{equation}
Due to the absence of rings the above equation is
valid for any set of lattice sites, and, thus, solving
the BL is reduced to a calculation of a few transfer matrices.
Note that a transfer matrix such as that of Eq. (\ref{eq:T-Mat-BL}) could
also be defined in a crystalline lattice but, in that case it would 
be useless. 

Eq. (\ref{eq:Dyson-TMat-BL}) can be solved iteratively,
\begin{equation}
  \label{eq:Dyson-TMat-BL-2}
  {\bf T}_{n-1,n}=\left[E{\bf I}-{\bf H}_0-\sum_{i=1,...,N-1}
    {\bf V}_{n,i}{\bf T}_{n,i}\right ]^{-1}{\bf V}_{n,n-1}
\end{equation}
If the orbital basis set and the lattice have full
symmetry (including inversion symmetry) the different transfer
matrices can  be obtained from just a single one through
appropriate rotations. However this is not always the case (see below).

Before proceeding any further we define self-energies that can be (and
commonly are) used in place of transfer matrices,
\begin{equation}
  {\bf \Sigma}_{i,j}={\bf V}_{i,j}{\bf T}_{i,j}
\end{equation}
Eq. (\ref{eq:Dyson-TMat-BL-2}) is then rewritten as,
\begin{equation}
  {\bf \Sigma}_{n-1,n}={\bf V}_{n-1,n}\left[E{\bf I}-{\bf H}_0
    \sum_{i=1,...,N-1}
    {\bf \Sigma}_{n,i}\right ]^{-1}{\bf V}_{n-1,n}^{\dagger}
\end{equation}
\noindent where we have made use of the general property 
${\bf V}_{n,n-1}= {\bf V}_{n-1,n}^{\dagger}$.

As discussed hereafter, in a general case of no  symmetry this would be 
a set of $N$ coupled equations ($2N$ if there is no inversion symmetry).
Symmetry can be broken due to either
the spatial atomic arrangement,  the orbitals on
the atoms that occupy each lattice site, or both. When no symmetry exists,
the following procedure has to be followed to obtain 
the self-energy in an arbitrary direction. The method is valid
for any basis set or lattice. Let ${\bf \tau_i}$ be the $N$
nearest-neighbor directions of the lattice we are interested in
and ${\hat V}_{\bf \tau_i}$ the interatomic interaction matrix in 
these directions. To make connection with the notation used above note that
the vector that joins site $n-1$ to site $n$, namely,
${\bf r_n-r_{n-1}}$ would necessarily be one of the lattice directions
of the set ${\bf \tau_i}$.
The self-energies associated to each direction have to be obtained
from the following set of $2N$ coupled self-consistent equations,
\begin{equation}
  {\mathbf \Sigma}_{\mathbf \tau_i}={\mathbf V}_{\mathbf \tau_i}
  \left [ E{\mathbf I}-{\mathbf H}_0-
    ({\mathbf \Sigma}_{\bar T}-
    {\mathbf \Sigma}_{\bar {\mathbf \tau_i}})\right ]^{-1}
  {\mathbf V}_{\mathbf \tau_i}^{\dagger}
\end{equation}
\begin{equation}
  {\mathbf \Sigma}_{\bar {\mathbf \tau_i}}={\mathbf V}_{\bar {\mathbf \tau_i}}
  \left [ E{\mathbf I}-{\mathbf H}_0-({\mathbf \Sigma}_{T}-
    {\mathbf \Sigma}_{\mathbf \tau_i})\right ]^{-1}
  {\mathbf V}_{\bar {\mathbf \tau_i}}^{\dagger},
\end{equation}
where $i=1,...,N$ and ${\bar {\mathbf \tau_i}}=-{\mathbf \tau_i}$. 
${\mathbf V}_{\mathbf \tau_i}$ is the interatomic
interaction in the ${\mathbf \tau_i}$ direction, and 
${\mathbf \Sigma}_{T}$
and ${\mathbf \Sigma}_{\bar T}$ are the sums of  the self-energy 
matrices entering through all the Cayley tree branches attached to an 
atom and their inverses, respectively, {\it i.e.},   
\begin{equation}
  {\mathbf \Sigma}_{T}=\sum_{i=1}^{N}{\mathbf \Sigma}_{\mathbf \tau_i}
\end{equation}
\begin{equation}
  {\mathbf \Sigma}_{\bar T}=\sum_{i=1}^{N}{\mathbf \Sigma}_
  {\bar{\mathbf \tau_i}}.
\end{equation}
This set of $2N$ matricidal equations has to be solved iteratively.
It is straightforward to check that, in cases of full symmetry,  
it reduces to the single equation.
The local density of states can be obtained from the diagonal
Green matrix,
\begin{equation}
  {\mathbf G}_{n,n}=\left [ E{\mathbf I}-{\mathbf H}_0
    -\sum_{i=1,..,N}{\mathbf \Sigma}_{\tau_i}\right ]^{-1}
\end{equation}
%%

%%Finally, as regards the tight-binding parameters, which should 
%%include only nearest-neighbors 
%%interactions, there is no unique way of choosing them. 
%%The ones we generically use
%%were obtained through fittings to the electronic
%%bulk band structures\cite{Palacios:prb:02}, 
%%but the reader is free to choose any other criterion that fits better the
%%problem at hand. 

\chapter{Non-Collinear Unrestricted Hartree-Fock Approximation}
\label{app:NC-UHF}

The standard Unrestricted Hartree-Fock theory (UHF) only allows
for spin-polarized ground states, i.e. the UHF Hamiltonian commutes
with one component of the total spin operator, e.g. $\hat S_z$, so that
the ground state $\ket{\Psi_0}$ is an eigenstate to $\hat S_z$.
For atoms and molecules this is a good property of the UHF Hamiltonian.
However, in transport problems, this symmetry might be broken 
if the magnetizations of the two macroscopic leads are aligned non-collinear
or anti-parallel.

Therefore, we will derive a non-collinear formulation of unrestricted
Hartree-Fock theory (NC-UHF).
As in UHF we want to minimize the total energy of a single Slater-determinant
of an $N$-electron system,
\begin{equation}
  \label{eq:B:Psi0}
  \ket{\Psi_0} = \ket{\chi_1,\ldots,\chi_N}
\end{equation}
described by the many-body Hamiltonian
\begin{equation}
  \label{eq:B:H}
  \hat{H} := \hat{H}_0 +  \hat{V}_{\rm c} 
  := \sum_i \hat{h}_0(i) + \sum_{i<j} \hat{v}_{\rm c}(i,j).
\end{equation}
Only now the molecular orbitals are a linear combination
of atomic orbitals with different spins, i.e. we explicitly allow
for spin-mixing:
\begin{equation}
  \ket{\chi_a} = \sum_{i,\sigma} c_{ia}^\sigma \ket{\phi_i^\sigma},
\end{equation}
where $\ket{\phi_i^\sigma}=\ket{\phi_i}\otimes\ket{\sigma}$.

We assume that the molecular orbitals and the atomic orbitals are
orthonormal:
\begin{equation}
  \label{eq:B:ONS}
  \bracket{\chi_a}{\chi_b} = \delta_{ab} \mbox{ and } \bracket{\phi_i}{\phi_j} = \delta_{ij}
\end{equation}

In order to find the molecular orbitals that minimize the total energy with the constrained
of orthogonal molecular orbitals we have to minimize the Lagrangian
\begin{equation}
  \label{eq:B:Lagrangian}
  \mathcal L[\{\chi_a\}] = E_0[\{\chi_a\}]-\sum_{a,b}\epsilon_{ab}\left( \bracket{\chi_a}{\chi_b} - \delta_{ab} \right),
\end{equation}
where $E_0[\{\chi_a\}]=\bra{\Psi_0} \hat{\mathcal{H}} \ket{\Psi_0}$.

From the minimization condition $\delta \mathcal{L} = 0$ and the expression for the total energy
of a single Slater determinant
\begin{eqnarray}
  \label{eq:B:E0}
  E_0 &=& \sum_a \bra{\chi_a} \hat{h}_0(1) \ket{\chi_a} 
  +   \frac{1}{2} \sum_{a,b} \bra{\chi_a,\chi_b}\hat{v}(1,2)\ket{\chi_a,\chi_b} \nonumber \\
  &-& \frac{1}{2} \sum_{a,b} \bra{\chi_a,\chi_b}\hat{v}(1,2)\ket{\chi_b,\chi_a}
\end{eqnarray}
we find the following effective single particle Hamiltonian - the Fock operator
\begin{eqnarray}
  \label{eq:B:FOp}
  \hat{\mathcal{F}} &=& \hat{\mathcal{H}}_0 
  + \sum_{i,j,\sigma} \sum_{k,l} \rho_{lk}^{\sigma\sigma} \bra{\phi_i,\phi_k}\hat{v}(1,2)\ket{\phi_j,\phi_l} 
  \hat{c}_{i\sigma}^\dagger \hat{c}_{j\sigma}
  \nonumber \\
  &-& \sum_{i,j,\sigma_1,\sigma_2} \sum_{k,l} \rho_{lk}^{\sigma_1\sigma_2} \bra{\phi_i,\phi_k}\hat{v}(1,2)\ket{\phi_l,\phi_j} 
  \hat{c}_{i\sigma_1}^\dagger \hat{c}_{j\sigma_2}
\end{eqnarray}
which depends on the density operator
\begin{eqnarray}
  \label{eq:B:rho}
  \hat\rho &=& \sum_{a} \ket{\chi_a}\bra{\chi_a} 
  = \sum_{i,j,\sigma_1,\sigma_2} \sum_{a=1}^N c_{ia}^{\sigma_1} 
  (c{ja}^{\sigma_2})^\ast \ket{\phi_i^{\sigma_1}}\bra{\phi_j^{\sigma_2}}
  \nonumber \\
  &=& \sum_{i,j,\sigma_1,\sigma_2} \rho_{ij}^{\sigma_1\sigma_2} 
  \ket{\phi_i^{\sigma_1}}\bra{\phi_j^{\sigma_2}}.
\end{eqnarray}

The second term in eq. (\ref{eq:B:FOp}) is the Hartree-term $\hat{v}_H$ which describes the classical 
(direct) Coulomb repulsion of the electrons. 
It depends only on the (local) charge density $\rho^{\rm T}(x)=\rho^{\su\su}(x)+\rho^{\sd\sd}(x)$:
\begin{equation}
  \hat{v}_{\rm H} = \int {\rm d}x \int {\rm d}y \rho^{\rm T}(y) v(x,y) 
  \sum_\sigma \hat{\psi}_\sigma^\dagger(x) \hat{\psi}_\sigma(x),
\end{equation}
while the third term really depends on the density matrix. This is the Fock-term $\hat{v}_F$ describing
the exchange energy:
\begin{equation}
  \hat{v}_{\rm F} =\sum_{\sigma_1,\sigma_2} 
  \int {\rm d}x \int {\rm d}y \rho^{\sigma_1\sigma_2}(x,y) v(x,y) \hat{\psi}_{\sigma_1}^\dagger(x) \hat{\psi}_{\sigma_2}(y).
\end{equation}
Clearly, only the Fock-term can give rise to non-collinear spin-arrangements.
Moreover, only if the initial guess contains non-diagonal terms with respect 
to the spin, i.e. only if $\rho^{\sigma_1\sigma_2}(x,y)\ne 0$ for $\sigma_1\ne\sigma_2$, the self-consistent Hartree-Fock solution can have a non-collinear spin configuration.
 
\end{appendix}

%%% Local Variables: 
%%% mode: latex
%%% TeX-master: t
%%% End: 

%%%%%%%%%%%%%%%%%%%%%%%%%%%%%%%%%%%%%%%%%%%%%

%%%%%%%%%%%%%%%%%%%%%%%%%%%%%%%%%%%%%%%%%%%%%%%%%%%%%%%%%
%% Bibliography                                        %%
%%%%%%%%%%%%%%%%%%%%%%%%%%%%%%%%%%%%%%%%%%%%%%%%%%%%%%%%%
\bibliographystyle{prsty}
\bibliography{matcon}
\addcontentsline{toc}{chapter}{\numberline{}Bibliography}
%%%%%%%%%%%%%%%%%%%%%%%%%%%%%%%%%%%%%%%%%%%%%%%%%%%%%%%%%

\markboth{}{}

%%%%%%%%%%%%%%%%%%%%%%%%%%%%%%%%%%%%%%%%%%%%%%%%%%%%%%%%%%%%%%%%%
%% List of abbrevations                                        %%
%%%%%%%%%%%%%%%%%%%%%%%%%%%%%%%%%%%%%%%%%%%%%%%%%%%%%%%%%%%%%%%%%
\chapter*{List of abbreviations}
\addcontentsline{toc}{chapter}{\numberline{}List of abbreviations}
%%%%%%%%%%%%%%%%%%%%%%%%%%%%%%%%%%%%%%%%%%%%%%%%%%%%%%%%%%%%%%%%%
\begin{tabular}{ll}
AMR  & Anisotropic magneto-resistance \\
BL   & Bethe lattice \\
BMR  & Ballistic magneto-resistance \\
DFT  & Density functional theory \\
DOS  & Density of states \\
EXX  & Exact exchange \\
GF   & Green's function \\
GGA  & Generalized gradient approximation \\
GMR  & Giant magneto-resistance \\
HF   & Hartree-Fock \\
HFA  & Hartree-Fock approximation \\
HFX  & Hartree-Fock exchange \\
IC   & Integrated circuit \\
KS   & Kohn-Sham \\
LDA  & Local density approximation \\
LDOS & Local density of states \\
LSDA & Local spin density approximation \\
NOBS & Non-orthogonal basis set \\
PDOS & Projected density of states \\
MCBJ & Mechanically controllable break junction \\
MR   & Magneto-resistance \\
NEGF & Non-equilibrium Green's function \\
STM  & Scanning tunneling microscope \\
TMR  & Tunneling magneto-resistance
\end{tabular}
%%%%%%%%%%%%%%%%%%%%%%%%%%%%%%%%%%%%%%%%%%%%%%%%%%%%%%%%%%%%%%%%%

%%%%%%%%%%%%%%%%%%%%%%%%%%%%%%%%%%%%%%%%%%%%%%%%%%%%%%%%%%%%%%%%%
%% List of publications                                        %%
%%%%%%%%%%%%%%%%%%%%%%%%%%%%%%%%%%%%%%%%%%%%%%%%%%%%%%%%%%%%%%%%%
\chapter*{List of publications}
\addcontentsline{toc}{chapter}{\numberline{}List of publications}
%%%%%%%%%%%%%%%%%%%%%%%%%%%%%%%%%%%%%%%%%%%%%%%%%%%%%%%%%%%%%%%%%
\begin{list}{}{\setlength{\leftmargin}{0cm}}
\item F. Mu\~noz-Rojas, D. Jacob, J. Fern\'andez-Rossier, and J. J. Palacios\\
  {\em Coherent transport in graphene nanoconstrictions}\\
  Phys. Rev. B \textbf{74}, 195417 (2006)
\item  D. Jacob, J. Fern\'andez-Rossier, and J. J. Palacios\\
  {\em Electronic structure and transport properties of atomic NiO spinvalves}\\
  J. Magn. Magn. Mater. \textbf{310}, e675-e677 (2007)
\item  D. Jacob, J. Fern\'andez-Rossier, and J. J. Palacios\\
  {\em Emergence of half-metalicity in suspended NiO chains} \\
  Phys. Rev. B \textbf{74}, 081402(R) (2006).
\item D. Jacob and J. J. Palacios \\
  {\em Orbital eigenchannel analysis for ab initio quantum transport calculations} \\
  Phys. Rev. B \textbf{73}, 075429 (2006)
\item J. Fern\'andez-Rossier, D. Jacob, C. Untiedt, and J. J. Palacios \\
  {\em Transport through magnetically ordered Pt nanocontacts} \\
  Phys. Rev. B \textbf{72}, 224418 (2005)
\item  D. Jacob, J. Fern\'andez-Rossier, and J. J. Palacios \\
  {\em Magnetic and orbital blocking in Ni nanocontacts}\\
  Phys. Rev. B \textbf{71}, 220403(R) (2005)
\item  B. Wunsch, D. Jacob, and D. Pfannkuche\\
  {\em Isospin blockade in transport through vertical double quantum dots} \\
  Physica, E \textbf{26}, 464 (2005)
\item  D. Jacob, B. Wunsch, and D. Pfannkuche \\
  {\em Charge localization and isospin blockade in vertical double quantum dots } \\
  Phys. Rev. B \textbf{70}, 081314(R) (2004)
\end{list}

\section*{In preparation:}
\begin{list}{}{\setlength{\leftmargin}{0cm}}
\item D. Jacob, J. Fernandez-Rossier and J. J. Palacios\\
  {\em Anisotropic magneto-resistance in nanocontacts}
\item D. Soriano, D. Jacob and J. J. Palacios\\
  {\em A localized basis set description of free electrons}
\item D. Jacob and J. J. Palacios\\
  {\em Comparison of electrode models for {\it ab initio} quantum transport calculations}
\item D. Jacob, J. Fern\'andez-Rossier, and J. J. Palacios\\
  {\em Domainwall formation and scattering in one-dimensional chains}
\end{list}
%%%%%%%%%%%%%%%%%%%%%%%%%%%%%%%%%%%%%%%%%%%%%%%%%%%%%%%%%%%%%%%%%%%

\end{document}